\documentclass[preprints,article,accept,pdftex,moreauthors]{Definitions/mdpi} 

\firstpage{1} 
\pubvolume{1}
\issuenum{1}
\articlenumber{0}
\pubyear{2025}
\copyrightyear{2025}
\datereceived{} 
\dateaccepted{} 
\datepublished{} 
\hreflink{https://doi.org/} % If needed use \linebreak
\usepackage{subfig}
\usepackage{comment}
\usepackage{booktabs}
\usepackage{overpic}
\usepackage{soul}
\newcommand\ddfrac[2]{\ensuremath{\frac{\displaystyle #1}{\displaystyle #2}}}

\newcommand*{\SAPO}{}%\let\sapoversion=true
\ifdefined\SAPO
\newcommand{\aco}[1]{}
\newcommand{\md}[1]{}
\newcommand{\sa}[1]{}
\newcommand{\mg}[1]{}
\else
\newcommand{\aco}[1]{\textcolor{olive}{[\textbf{ACO:} #1]}}
\newcommand{\md}[1]{\textcolor{red}{[\textbf{MD:} #1]}}
\definecolor{azure}{rgb}{0.0, 0.5, 1.0}
\newcommand{\sa}[1]{\textcolor{azure}{[\textbf{SS:} #1]}}
\newcommand{\mg}[1]{\textcolor{orange}{[\textbf{MG:} #1]}}
\fi

\newcommand{\sapocol}[1]{\textcolor{black}{#1}} % highlight changes made for the referee 
\newcommand{\mgcol}[1]{\textcolor{black}{#1}} 
\newcommand{\mdpi}[1]{\textcolor{black}{#1}}

\usepackage[style=super,nopostdot]{glossaries} % acronyms %it appears in the table of contents [toc]
\makeglossaries
\makeglossaries

\newglossaryentry{pbrl}
{   name=pBRL,
    description={pathfinder Barcelona Raman LIDAR}
}

\newglossaryentry{clue}{
    name=CLUE,
    description={Cherenkov Light Ultraviolet Experiment}
}

\newglossaryentry{orm}{
    name=ORM,
    description={Roque de los Muchachos Observatory}
}

\newglossaryentry{ctao}{
    name=CTAO,
    description={Cherenkov Telescope Array Observatory} 
    }

\newglossaryentry{ctao-n}{
    name=CTAO-N,
    description={Northern site of the Cherenkov Telescope Array Observatory}
}

\newglossaryentry{ctao-s}{
    name=CTAO-S,
    description={Southern site of the Cherenkov Telescope Array Observatory} 
    }

\newglossaryentry{eso}{
    name=ESO,
    description={European Southern Observatory}
}

\newglossaryentry{eas}{
    name=EAS,
    description={Extended Atmospheric Showers}
}

\newglossaryentry{vhe}{
    name=VHE,
    description={Very High Energy}
}

\newglossaryentry{asl}{
    name=a.s.l.,
    description={Above sea level}
}

\newglossaryentry{aod}{
    name=AOD,
    description={Aerosol Optical Depth}
}

\newglossaryentry{rl}{
    name=RL,
    description={Raman LIDAR}
}

\newglossaryentry{magic}{
    name=MAGIC,
    description={Major Atmospheric Gamma-ray Imaging Cherenkov Telescope} 
    }

\newglossaryentry{pbl}{
    name=PBL,
    description={Planetary Boundary Layer}
}

\newglossaryentry{ifae}{
    name=IFAE,
    description={Institut de Física d’Altes Energies}
    }

\newglossaryentry{uab}{
    name=UAB,
    description={Universitat Autònoma de Barcelona}
    }

\newglossaryentry{infn}{
    name=INFN,
    description={Istituto Nazionale di Fisica Nucleare} 
    }

\newglossaryentry{llg}{
    name=LLG,
    description={Liquid Light Guide} 
    }

\newglossaryentry{lupm}{
    name=LUPM,
    description={Laboratoire Univers et Particules de Montpellier} 
    }

\newglossaryentry{iact}{
    name=IACT,
    description={Imaging Atmospheric Cherenkov Telescope} 
    }

\newglossaryentry{fov}{
    name=FoV,
    description={Field of View} 
    }

\newglossaryentry{psf}{
    name=PSF,
    description={Point Spread Function} 
    }

\newglossaryentry{na}{
    name=NA,
    description={Numerical Aperture} 
    }
    
\newglossaryentry{lc}{
    name=LC,
    description={Lens Couple} 
    }

\newglossaryentry{dm}{
    name=DM,
    description={Dichroic Mirror} 
    }

\newglossaryentry{if}{
    name=IF,
    description={Interference Filter}
}

\newglossaryentry{vrr}{
    name=VRR,
    description={Vibrational-Rotational Raman line}
}

\newglossaryentry{pmt}{
    name=PMT,
    description={Photomultiplier Tube}
}

\newglossaryentry{sipm}{
    name=SiPM,
    description={Silicon Photomultipliers}
}

\newglossaryentry{lst}{
    name=LST,
    description={Large Size Telescope}
}
\newglossaryentry{dac}{
    name=DAC,
    description={Digital-to-Analogue Converter}
}

\newglossaryentry{hv}{
    name=HV,
    description={High Voltage} 
    }

\newglossaryentry{lotr}{
    name=LOTR,
    description={Licel Optical Transient Recorder} 
}

\newglossaryentry{licli}{
    name=LICLI,
    description={LIDAR Client} 
    }

\newglossaryentry{nd}{
    name=ND,
    description={Neutral Density} 
    }

\newglossaryentry{se}{
    name=SE,
    description={Shaft Encoders} 
    }
    
\newglossaryentry{cli}{
    name=CLI,
    description={Command Line Interface} 
    }

\newglossaryentry{vaod}{
    name=VAOD,
    description={Vertical Aerosol Optical Depth} 
    }

\newglossaryentry{pde}{
    name=PDE,
    description={Photon Detection Efficiency}}

\Title{A 1.8~m class pathfinder Raman LIDAR for the Northern Site of the Cherenkov Telescope Array Observatory -- Technical Design}

\TitleCitation{Barcelona Raman LIDAR}

\Author{
O.~Ballester$^{1}$\orcidO,
O.~Blanch$^{1}$\orcidB,
J.~Boix$^{1}$,
P.~G.~Calisse$^{2}$,
A.~Campoy-Ordaz$^{3,\dagger}$\orcidC,
S.~M.~\c{C}olak$^{1}$\orcidI,
V.~Da~Deppo$^{4}$\orcidD,
M.~Doro$^{5,6,\dagger}$\orcidM,
L.~Font$^{3}$\orcidF,
E.~Font-Pladevall$^{3}$,
R.~Garcia$^{1}$,
M.~Gaug$^{3,\dagger}$\orcidG,
R.~Grau$^{1}$\orcidR,
D.~Kolar$^{8}$\orcidK,
A.~L\'opez-Oramas$^{7}$\orcidL,
C.~Maggio$^{3}$\orcidJ,
M.~Martinez$^{1}$\orcidT,
\`O. Mart\'inez$^{1}$,
V.~Riu-Molinero$^{3}$,
D.~Roman$^{1}$\orcidV,
S.~Stani\v{c}$^{8}$\orcidS,
J.~Tartera-Barber\`a$^{1}$,
S.~Ubach$^{3}$\orcidU,
M.~Zavrtanik$^{8}$\orcidZ and
M.~Živec$^{8}$\orcidX
}

\AuthorNames{O.~Blanch, J.~Boix, V.~Da Deppo, A.~Campoy-Ordaz, M.~Doro, E.~Font~Pladevall, \\ 
    L.~Font~Guiteras, M.~Gaug, A.~L\'opez-Oramas, C.~Maggio and M.~Mart\'inez}

\AuthorCitation{O. Ballester \textit{et al.}}
\address{%
$^{1}$ \quad Institut de Fisica d'Altes Energies (IFAE), Barcelona, Spain\\
$^{2}$ \quad Cherenkov Telescope Array Observatory gGmbH, (CTAO gGmbH), Saupfercheckweg 1, 69117 Heidelberg,  Germany\\
$^{3}$ \quad Departament de F\'{i}sica, Universitat Aut\`{o}noma de Barcelona and CERES-IEEC, 08193 Bellaterra, Spain\\
$^{4}$ \quad CNR-Institute for Photonics and Nanotechnologies UOS Padova LUXOR, Via Trasea 7, 35131 Padova, Italy\\
$^{5}$ \quad Department of Physics and Astronomy, University of Padova, I-35131 Padova, Italy\\
$^{6}$ \quad Istituto Nazionale di Fisica Nucleare (INFN), sez. Padova, I-35131 Padova, Italy\\
$^{7}$ \quad Instituto de Astrofísica de Canarias and Dpto. de Astrofísica, Universidad de La Laguna, E-38200, La Laguna, Tenerife, Spain\\
$^{8}$ \quad Center for Astrophysics and Cosmology, University of Nova Gorica, Vipavska 13, 5000 Nova Gorica, Slovenia
}

\firstnote{Corresponding author}
\corres{Correspondence: \href{mailto:campoy@ieec.cat}{campoy@ieec.cat}, \href{mailto:michele.doro@unipd.it}{michele.doro@unipd.it}, \href{mailto:markus.gaug@uab.cat}{markus.gaug@uab.cat}}

\abstract{This paper presents the technical design of the pathfinder Barcelona Raman LIDAR (\gls{pbrl}) for the Northern site of the Cherenkov Telescope Array Observatory (\gls{ctao-n}) located at the Roque de los Muchachos Observatory (\gls{orm}). The pBRL is developed for continuous atmospheric characterization, %to monitor atmospheric aerosol extinction profiles, 
essential for correcting high-energy gamma-ray observations captured by Imaging Atmospheric Cherenkov Telescopes (IACTs). The LIDAR consists of a steerable telescope with a 1.8~m parabolic mirror and a pulsed Nd:YAG laser with frequency-doubling and tripling. It emits at wavelengths of 355~nm and 532~nm to measure aerosol scattering and extinction through two elastic and Raman channels. Built upon a former Cherenkov Light Ultraviolet Experiment (\gls{clue}) telescope, the pBRL’s design includes a Newtonian mirror configuration, a coaxial laser beam, a near-range system, a liquid light guide and a custom-made polychromator. During a one-year test at the ORM, the stability of the LIDAR and semi-remote-controlled operations were tested. This pathfinder \sapocol{leads} the way to designing a final version of a \gls{ctao} Raman LIDAR which will provide real-time atmospheric monitoring and as such ensure the necessary accuracy of scientific data collected by the CTAO-N telescope array.}

   %(List three to ten pertinent keywords specific to the article; yet reasonably common within the subject discipline.)} 

\begin{document}

\tableofcontents
%%%%%%%%%%%%%%%%%%%%%%%%%%%%%%%%%%%%%%%%%%
%\setcounter{section}{-1} %% Remove this when starting to work on the template.

\section{Introduction}
\label{sec:intro}
The Cherenkov Telescope Array \sapocol{Observatory (\gls{ctao})}~\citep{ctaconcept,ScienceCTA:2019} is %a project in the Preparatory Phase for a future 
the next generation observatory of ground-based Imaging Atmospheric Cherenkov Telescopes (\glspl{iact}). %\sapocol{\st{and will be constructed and operated by the Cherenkov Telescope Array Observatory (CTAO)}}. 
The CTAO will observe high-energy cosmic photons for high-energy astrophysics research; the widely used term 'gamma rays' will be used from now on throughout this article.
%, a recently approved European Research Infrastructure Consortium (ERIC). %operating of world-wide scale. 
%CTAO will observe cosmic gamma rays for high-energy astrophysics research.
%\sa{I propose to get rid of "gamma rays" everywhere and use "high energy cosmic photons". Despite historical reasons, the term is misleading as "rays" imply many photons in a "ray" and here we deal with single photons. I can rewrite it if you agree.} \aco{I would like to keep the term "gamma-rays" as commonly used in gamma-ray astronomy. I propose to include a sentence introducing the term for "clarifying" the possible confusion. See suggestion.}
%\sa{OK.}
The observatory is composed of more than 70 telescopes at two locations: in the northern hemisphere, CTAO-N is found at the Observatorio del Roque de Los Muchachos (ORM, La Palma, Canary Islands, Spain, 28$^\circ$N 17$^\circ$W), and in the southern hemisphere, \gls{ctao-s} will be constructed at a site belonging to the European Southern Observatory (\gls{eso}, Cerro Paranal, Chile, 24$^\circ$S  70$^\circ$W). The telescope arrays are spread over an area of approximately one square kilometre and are located at altitudes of around 2.200~m above sea level.

%\begin{figure}[h!t]
%    \centering
%    \includegraphics[width=0.9\linewidth]{Figures/sec1/showers.png}
%    \caption{Particle showers in the atmosphere \md{I am wondering whether adding such a figure for non-experts, maybe including in the picture a telescope array and the lidar can be of help. If so I can try and edit one different from the one proposed.} \mg{I would at least remove the right one, which is irrelevant for the RL. Actually, I rather prefer this one from \protect\cite{Naurois:2000a}, which shows the altitudes and variations of 300 GeV gamma-ray showers (green: photons, blue: e$^+$/e$^-$, vertical axis: height a.s.l.): }\md{I agree, but your image is missing axes. Also, there is no scheme of telescope and lidar that could be an option. \mg{I agree, maybe this is the opportunity to create a nice new figure with one of these showers, highlighting also the range of shower development, its dependency on shower energy, the observable Cherenkov radiation and a sketch of a LIDAR and some aerosol extinction measurement. We could produce two figures: one for zenith and another for 60 deg., we have all numbers, which we gathered from simulations for the second MAGIC LIDAR paper. That figure can then be used everywhere, particularly in conferences. I'll talk with Anna to make it together.}}}
%    \vspace{0.5cm}
%        \includegraphics[width=0.7\linewidth]{Figures/Showers_deNaurois.png}
%    \label{fig:showers}
%\end{figure}

\begin{figure}[h!t]
%\begin{adjustwidth}{-\extralength}{0cm}
    \centering
    \includegraphics[width=0.475\linewidth,trim={0.7cm 0 0.5cm 0},clip]{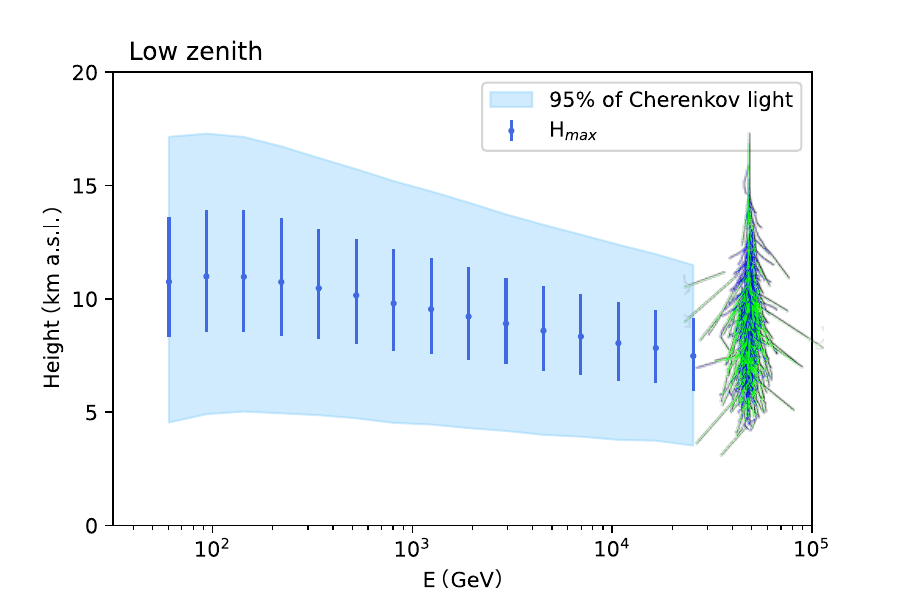}
    \includegraphics[width=0.515\linewidth,trim={0.7cm 0 0.6cm 0},clip]{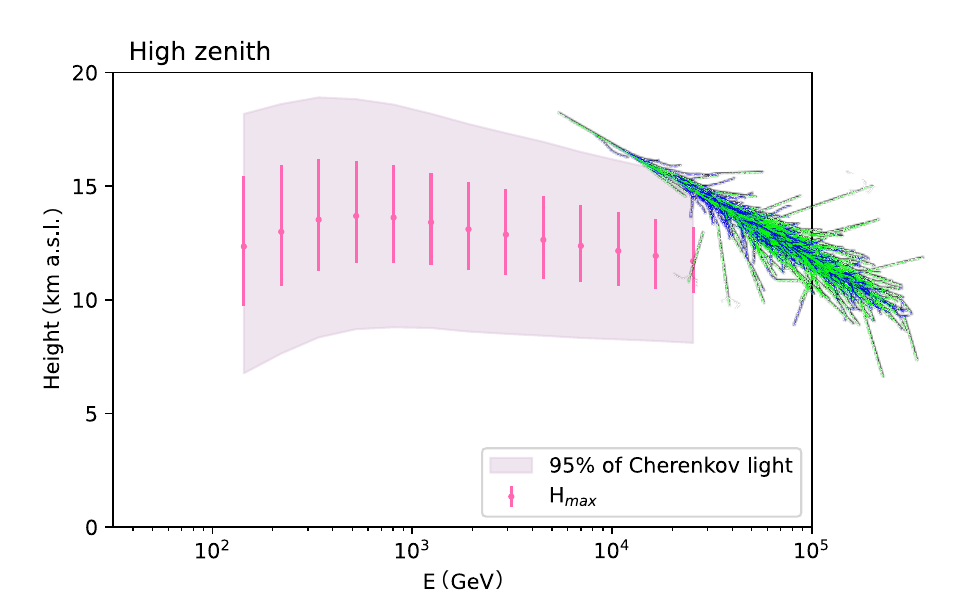}
%\end{adjustwidth}
    \caption{Ranges of shower development as a function of gamma-ray energy; left for vertical incidence, right for shower inclined by \sapocol{60$^\circ$ with respect to the zenith (30$^\circ$ elevation). }
    %60$^\circ$ from zenith. 
    The points show the mean location of the shower maximum, lines the ranges within which 95\% of the showers exhibit the maximum. The bands show the range within which 95\% of the observable Cherenkov light is emitted. The data used to produce this figure have been adopted from~\citep{Schmuckermaier:2023huo,fruck:phd}. On the right side, a simulated particle shower of 300~GeV energy is shown (image adopted from~\citep{Naurois:2000a}). Green lines denote photons and blue lines electrons and positrons. }
    \label{fig:showers}
\end{figure}

IACTs indirectly detect cosmic gamma rays with energies ranging from tens of GeV to several tens of TeV\footnote{1~GeV=$10^{9}$ electronvolt (eV), 1~TeV=$10^{12}$~eV}. At these energies, %the atmosphere is opaque and this radiation does not reach the ground because it 
gamma rays interact with atmospheric nuclei disintegrating into extended atmospheric showers (\gls{eas}) of elementary particles (see Fig.~\ref{fig:showers}). With a very-high-energy (\gls{vhe}) photon interaction length of about 47.1~g/cm$^2$ in air~\citep{NavasPDG}, for vertical incidence, at a pressure level of 32~mbar (corresponding to $\sim$23~km~\gls{asl}), half of the impacting gamma rays have converted to an electron-positron pair that produces the particle shower. % \md{Add reference Markus}.
%&The EAS has a longitudinal and lateral development. 
The length of an EAS depends on the primary energy of the gamma-ray \sapocol{on average}, so that higher-energy showers penetrate deeper into the atmosphere than lower-energy ones (see Fig.~\ref{fig:showers}). Furthermore, the penetration depth depends on the incidence direction of the original gamma ray, and hence the particle shower. Charged particles within the EAS are ultra-relativistic during most of the shower development and emit Cherenkov radiation~\citep{HILLAS:JPGPP1990a}. That radiation is observed on the ground as a brief burst (few nanoseconds) of mainly UV (300--400~nm) light, which illuminates an area of $10^5$--$10^6$~m$^2$.  %\sapocol{\st{However,} 
Most of the observed Cherenkov light originates from altitudes of 5--17~km a.s.l. (for vertical incidence) and from 8--18~km a.s.l. (for \sapocol{low elevation angles of} 
%high zenith angles of 
observation)~\cite{HILLAS:JPGPP1990a,Schmuckermaier:2023huo}.
%propagating from the showers and illuminating an area at the ground of several thousand of square meters. If a telescope is located in this light pool, the properties of the primary gamma rays (energy, direction and time) can be estimated.

The development of an EAS is affected by the refractive index of air (modulated by the density profile of the atmosphere)~\citep{Bernloehr:2000,Munar:2019}, 
%because the Cherenkov emission is strongly dependent on the medium refraction index, 
whereas propagation of Cherenkov light to ground is strongly influenced by atmospheric extinction: molecular and aerosol scattering, as well as scattering by clouds~\citep{Ebr:2019ICRC,Fruck:2022igg,Schmuckermaier:2023huo} and by aerosols in the lower stratosphere (15--20~km~a.s.l.), as a consequence of, for example, strong stratovolcanic eruptions~\citep{GarciaGil:2010} or large-scale vertical advection~\cite{Kremser:2016}.
%The emitted Cherenkov spectrum can be attenuated, or modulated according to the atmospheric particle population, and temperature and humidity due to scattering. 
%Scattering of light may be strongly isotropic in case of molecules or peaked in the forward direction for aerosols(Rayleigh versus Mie scattering). Photons scattered in the forward direction may not be lost and remain in the field-of-view of the telescope and blur the image and/or create halos around the main image. Multiple scattering of light will even enhance this effect. 

%\sapocol{\st{In fact,} 
The dominant contribution to the systematic uncertainty in the energy and flux reconstruction of IACTs results from an inaccurate determination of the atmospheric transmittance of Cherenkov light~\citep{Magic:performance,Hess:performance}. 
%For instance, the MAGIC telescopes, one of the current operating major IACT installations, estimates that atmospheric conditions contribute nearly 10\% in the uncertainty of their energy scale, even in optimal observing conditions~\cite{Magic:performance}. Similarly, H.E.S.S., a second major IACT, confirms that the uncertainty in the reconstructed flux, associated with the assumed atmospheric profile is of the order of 10\%~\cite{Hess:performance}. 
%Moreover, both installations find approximately $12-15$\% additional flux reconstruction uncertainty, estimated from run-by-run variations additional to what 
%is expected from pure statistical variations only. These excess of variation can be attributed mainly to atmospheric variations. 
For this reason, CTAO has chosen~\citep{Gaug:2017Atmo} to continuously monitor and assess the aerosol extinction profile along the line of sight of the observing telescopes~\citep{Ballester:2019}, together with the monitoring of the aerosol optical depth (\gls{aod}) across the observed field of view~\citep{Ebr:2021}. 
For the first part, specifically designed Raman LIDARs (\glspl{rl}) have been proposed~\citep{doro:2013,Ballester:2019}.
The RLs are supposed to point approximately in the direction of the CTAO science target and therefore need to be designed to be fully orientable within a cone \sapocol{from zenith to 20$^\circ$ elevation. }%70$^\circ$ from the zenith. 
Astronomical target tracking is %\sapocol{\st{, nevertheless,}}
not necessary if the duration of data collection is within seconds to a few minutes. RL operation should not interfere %\sapocol{\st{, however,}} 
with data taking or even the operation of other close-by astronomical installations~\cite{otarola:2019,Gaug:2018zfj}. 

%Cherenkov Telescope Array Observatory (CTAO) is a European Research Infrastructure Consortium,  world-class consortium to construct several tens of Cherenkov telescopes to be 

%The CTA telescopes at each site can be operated simultaneously as a singular array or as smaller sub-arrays. 
%Differently than current installations, CTA will be operated as an observatory, with external scientists using CTA data served with high-quality data.  
%To achieve this goal, the atmosphere must be monitored continuously and accurately so that telescope data can be corrected offline before dissemination. 
%This requires the extensive use of remote-sensing instrumentation and CTA has a wide program for this purpose~\cite{Doro:2012}. Remote-sensing instruments, like elastic or Raman LIDARs (RLs), have already been proven powerful tools in environmental studies. 

%RLs emit light through a powerful laser pointed upward, the light interacts with the atmospheric constituents and is backscattered. If the time lag is measured, the backscattered light carries information on the atmospheric condition at each altitude.
%Therefore, a LIDAR installed and operated at the CTA site can
%be used to determine the transmission at each height and hence correct for 
%the systematic biases in reconstructed energy and flux. 

In recent years, several astroparticle experiments that use the atmosphere as part of their detector have committed themselves to use LIDARs~\citep{Bregeon:2016,Auger:2012,Rizi:2019,Tomida:2011cb}. In particular, the \gls{magic} Collaboration has developed a custom-fit elastic LIDAR as well as a dedicated algorithm for IACT data correction using aerosol extinction information~\citep{Fruck:2022igg,Schmuckermaier:2023huo}. Although that LIDAR has been absolutely calibrated, achieving correlated calibration-period-wise accuracies of the \mdpi{vertical aerosol optical depth (\gls{vaod})} of the ground layer better than $\pm$ 0.01 (uncorrelated ones of $\sim\pm$0.015)~\citep{Gaug:2022}, \citet{Schmuckermaier:2023huo} have also shown the limitations of an elastic system based on only one wavelength. 
%\sa{There are a few instances of citet, Schmuckelmeier on line 88 Fruck on line 118, Alexanreas on line 349. I propose to cite them with numbers as everything else.} \aco{I think it is ok to keep them.}
%\sa{It will not kill us, but I am not a fan of mixing citation styles.}\md{I did it to emphasize specific papers relevant for this work. I would also prefer to keep it.}
Elastic LIDAR systems without absolute calibration %\sapocol{\st{, in turn,}} 
reach accuracies of only 20\% for extinction profiles, even with the help of auxiliary sun photometer data that provide estimates of the LIDAR ratio~\citep{Bosenberg:2017,Berjon:2019}. 

%They allow to monitor~\citep{Fruck:2022igg} the
%aerosol properties of the atmosphere and hence to improve the quality and
%knowledge of the recorded data~\citep{Schmuckermaier:2023huo}. 
%They constitute a crucial element to provide the
%requested improvement in reducing systematic uncertainties and
%increasing the duty cycle for the CTA observatory with respect to the
%current generation of Cherenkov telescopes. 

Given the nature of the observed Cherenkov light, LIDARs at IACT installations shall characterize the entire troposphere and the lower stratosphere, i.e. reach at least 20~km height a.s.l. or $\sim$45~km range for \sapocol{elevation angles of 25$^\circ$}
%zenith angle limit of 70$^\circ$ 
and an observatory altitude of $\sim$2~km a.s.l. 
%However, 
Modern astronomical sites are characterized by extremely clear skies~\citep{Hellemeier:2019}, no clouds or only a few clouds~\citep{Otarola:2017} and low dust content~\citep{Laken:2016} or few episodes of dust intrusions~\citep{Lombardi:2010}. Moreover, very bad atmospheric conditions lead to abandoning scientific operations and hence do not require atmosphere characterization. IACT science data cannot be reasonably analysed with AODs larger than about 0.7, even if the optical properties of the atmosphere are well characterized~\citep{Schmuckermaier:2023huo}. 
The heights of the nocturnal planetary boundary layer (\gls{pbl}) at these sites normally reach below 800~m above ground~\citep{Fruck:2022igg,Sicard:2010} and their fine structure need not be resolved~\citep{Garrido:2013} for gamma-ray energies below about 1~TeV, because Cherenkov light is emitted entirely above it~\citep{HILLAS:JPGPP1990a}. Above these energies, gamma-ray showers penetrate down to the ground, and part of the Cherenkov light is emitted within the nocturnal boundary layer; nevertheless, such Cherenkov light is not observable for the CTAO Telescopes if the shower impacts at distances larger than 100~m from the telescopes~\citep{Sobczynska:2014}. \sapocol{Therefore, aerosol profiling for CTAO is acceptable with a %\st{poor} 
range resolution below a few hundred meters, as long as the absolute AOD of the ground layer is determined with accuracies better than 0.03.}
%Therefore, aerosol profiling for CTAO can live with a poor range resolution below a few hundred meters, as long as the absolute AOD of the ground layer is determined with accuracies better than 0.03. 
This can already be achieved with elastic lines only if decent efforts are made to continuously maintain absolute LIDAR calibration~\citep{Fruck:2022igg,Gaug:2022}. 
Finally, optically thick and low (cumulus) clouds are of no interest for precise characterization, since observations will be aborted anyhow under such conditions. Only optically thin clouds, and possibly optically thick clouds at high altitudes~\citep{Sobczynska:2014}, require detailed monitoring. Given the large longitudinal extension of the Cherenkov light-emitting particle showers, clouds above a geometrical thickness of 4~km need to be characterized according to their measured profile~\citep{Prester:2024}. Below that value, a standard average profile might be used. Note that \citet{Fruck:2022igg} have shown that an elastic LIDAR can already determine the optical depth of such clouds with sufficient accuracy; hence Raman capabilities are strictly required only up to the end of the PBL. 

\medskip
RLs designed to reach the stratosphere and even the lower mesosphere using pulsed Nd:YAG lasers and receiver telescopes of $\sim$1~m diameter class have been used for a few decades already~\citep{Keckhut:1990,Rees:2000,Wandinger:2005,Avdikos:2015,Gerding:2016,Rizi:2019}, although recently significant advances have been made using order-of-magnitude more powerful excimer lasers~\citep{Klanner:2021}. These systems rely on static LIDARs, so, the design of pointable LIDARs means an additional challenge. 
%Other, but similar solutions for CTA have been proposed in~\citep{Pallota:2014,Ballester:2019}.

\medskip
In this report, we discuss a RL pathfinder for CTAO-N. The instrument is dubbed pBRL (pathfinder Barcelona Raman LIDAR). The pBRL is designed, maintained and operated by the Institut de Física d'Altes Energies (\gls{ifae}, Barcelona, Spain) and the Universitat Auton\`oma de Barcelona (\gls{uab}, Barcelona, Spain) in collaboration with the University of Nova Gorica (Slovenia), the University of Padova, and the Istituto Nazionale di Fisica Nucleare (\gls{infn}, sez. Padova, Italy). 
It has been designed for a 4-channel (2 elastic, 2 Raman) orientable RL, composed of a 1.8~m parabolic  mirror, with $f/1$, with a Newtonian alt-az mount and a 532~nm frequency doubled and 355~nm frequency tripled Nd:YAG pulsed laser. 

Atmospheric LIDARs~\cite{WinklerCALIPSO:2003,Mueller:2010} often use Nd:YAG lasers with its  1064~nm wavelength and the 2$^\mathrm{nd}$ and 3$^\mathrm{rd}$ harmonic at 532~nm and 355~nm, respectively. In particular, the fundamental line provides stronger backscattering signals from \sapocol{coarse} aerosols compared to the molecular background.  
%This line gives rapidly what is called
%cloud-base (CB) and cloud-height (CH), which are the heights where a
%cloud starts and stops. In addition, it also gives a rough estimation
%of the Aerosol Optical Depth (AOD) by simply calculating the ratio
%between the return power after the cloud and before the
%cloud.  
For our purpose, however, 1064~nm lies rather far from the typical Cherenkov light emission band and may cause interference with the \sapocol{optical telescopes at the ORM} observing in the $I$ and $J$ bands. Finally, strong illumination of the Liquid Light Guide (\gls{llg}) by infrared light may cause faster degradation of it. 
%and rough assumption have to be made about the wavelength dependency of aerosol scattering in the cloud. This dependency is expressed by the \AA ngst\"om index which depends very much on relative humidity and composition of the aerosol layer. Moreover, in the case of our lidar, the optical fiber~\ref{sec:llc} does not stand the UV light, and 
Therefore, our choice fell on the 355~nm and the 532~nm lines, both found well within the wavelength range of the observed Cherenkov light spectrum~\cite{cta}. %, as discussed also in Sect.~\ref{sec:intro}. 
%From the lidar equation theory, we know that the return power connected to an elastic channel is related to the atmospheric extinction through the lidar ratio, which gives an uncertainty of
%about 25\% on the AOD. By adding a second elastic channel, one is able to recostruct the so-called color-ratio, i.e. the relation of the extinction parameters at different wavelengths, so both wavelengths are of interest.
%An important step toward a precise estimation of the optical extinction is
%achieved when using one or more Raman channels.
In addition to that, a Raman channel allows one to discriminate relatively well between aerosols of different LIDAR ratios and achieve accuracies well below 10\%  for aerosol extinction coefficients~\cite{Pappalardo:2004}, 
%and particulate interactions in the atmosphere and the 
%because the laser
%lights that is back-scattered to the lidar has interacted only with
%molecules and not with aerosols. 
%Moreover, the 
%aerosol extinction coefficient, 
as required for CTAO science data analysis (see Sect.~\ref{sec:requirements}). 
A natural choice are the relatively strong vibrational-rotational Raman (\gls{vrr}) Stokes lines of 355~nm scattering on N$_2$~\cite{Zenteno-Hernandez:2021} centred at 387~nm. 
%With one Raman channel, one is able
%to reconstruct the optical transmission with a precision level of 5\%,
%where most of this uncertainty is related to the unknown \AA ngstr\"om
%coefficient which depends on the particulate population. However, the
%dependency on the \AA ngstr\"om coefficient can be modeled well for clear sites and is typically quite slow. 
Adding a second Raman channel at 607~nm (the VRR Stokes line of 532~nm) allows us to retrieve the \AA ngstr\"om extinction exponent with the required precision.
%that best fits the Raman data set at two different wavelengths, therefore further reducing the uncertainties in the determination of the extinction coefficients. 
%at a spectrum of wavelengths to below 5\%. \mg{re-write with MAGIC experience. Markus will check.}
Note that the Raman backscatter cross-section for 532~nm is %\sapocol{\st{, however,}} 
only about 20\% that at 355~nm~\cite{She:2001,Gau:phd,Zenteno-Hernandez:2021}. The inclusion of further lines, like CO$_2$, water vapour, additional elastic and Raman channels, is not strictly needed for the purpose of CTAO and was discarded.

The pBRL structure has been built on a \sapocol{recycled} %dismissed 
Cherenkov Light Ultraviolet Experiment (\gls{clue}) telescope enclosed in a 20~ft standard maritime container~\citep{Alexandreas:1995}, already equipped with a large parabolic reflector of 1.8~m diameter. CLUE was a gamma-ray detector array installed at the ORM, %with a Multi Wire Photon Counting chambers, 
sensitive to the UV light from Cherenkov showers in the range 190--230~nm~\citep{Bartoli:2001gt}.
%The entire system was installed inside a mechanized container that could be opened remotely. 
The array was dismantled in 2002, but some of the individual containers still exist with the telescope inside. The pBRL group acquired two CLUE containers for the purpose of building LIDARs. A third container was purchased by the Laboratoire Univers et Particules de Montpellier (\gls{lupm}), France, which also renovated it as an RL for CTAO-S~\citep{Vasileiadis:2020}. At the moment of writing this document, only one of the two Barcelona containers has been refurbished and has become a pathfinder Raman LIDAR for CTAO.

\bigskip
This paper is structured as follows. In Sect.~\ref{sec:requirements} we introduce the requirements the instruments have to satisfy to be used for the purpose of CTAO. In Sect.~\ref{sec:tech_design} we detail the current technical solutions adopted for the refurbishment of the CLUE container to build the pBRL. In Sect.~\ref{sec:calibration} we calibrate the LIDAR performance. In Sect.~\ref{sec:operation} we discuss the way pBRL is operated within the CTAO framework. We summarize and conclude in Sect.~\ref{sec:conclusions}.

%\newpage
\section{Technical Requirements \label{sec:requirements}}%\md{Anna to do it.} \aco{In progress}
\aco{To be checked before submission.}
%\md{Numbers and structure are under revision, to be checked before submission.} 

This section collects the technical requirements that a RL shall meet in accordance with the CTAO guidelines \sa{A citation here perhaps?}\mg{I believe, these are all internal documents only, unfortunately.}. CTAO has defined a set of level~B product requirements\footnote{Level~A products requirements are those that apply to the product as a whole, i.e. the CTAO observatory, whereas level~B products requirements apply to its different systems.}, separated into scientific and performance requirements, operation and survival conditions for the two array sites, and finally reliability and availability requirements. 
%A case study on the influence of an increasing aerosol content of the atmosphere, both global as layered, on the energy reconstruction of gamma-ray showers can be found in~\cite{Fruck:2022igg,Schmuckermaier:2023huo} \md{Maybe the previous sentence is not need in this section?} \aco{Agree, it does not fit here.}

%The scientific and performance requirements are discussed in Sect.~\ref{sect:ScientificAndPerformance}. The operating and  survival conditions requirements are introduced in Sect.~\ref{sect:OperCond} and Sect.~\ref{subsect:survival}, respectively, while the lifecycle and maintenance are presented in Sect.~\ref{subsect:reliability}. In Sect.~\ref{sect:SiteConsiderations}, the site considerations are displayed.

\subsection{Scientific and Performance requirements \label{sect:ScientificAndPerformance}}

\textbf{Energy scale and flux scale. \label{subsect:EnergyScale}}
%From the document TPC-SPECS/110331a (Level-A requirements), 
CTAO requires an accuracy, both on the energy of reconstructed gamma-ray photons and the reconstructed gamma-ray flux of $<10\%$ at 90\% confidence level, at energies between 50~GeV and 300~TeV. 
%Systematic errors or biases in the energy of reconstructed gamma-ray photons must be $<10$\% at energies between 50~GeV and 300~TeV (at 90\% confidence level).
Several processes contribute to these systematic uncertainties, among which limitations in the understanding of the telescopes and their degradation with time, and the precise state of the atmosphere. The latter is allowed to contribute to the uncertainty of the energy scale by $<8$\% and consists of several individual contributions related to the accurate modelling of the development of air showers, the molecular density and refractive index profile \md{Georgios Voutsinas proceedings to be cited after atmohead https://agenda.infn.it/event/38476/contributions/237456/.} and the light-absorbing molecules. Uncertainties further include limitations in the knowledge of the extinction of Cherenkov light through scattering processes with molecules, clouds, and dust layers. Clouds and dust layers are allowed to contribute the largest uncertainty, about $<$3\% each, for the Cherenkov light yield on the ground. A similar breakdown leads to a requirement of $<$5\% for the estimated contribution of aerosols to the systematic uncertainty of the reconstruction of gamma-ray fluxes. 

%The measurement of flux and energies of gamma-rays is related to the amount of Cherenkov light emitted by the atmospheric showers. The relationship between the number of emitted photons and the measured signal is determined by a calibration process. To achieve the required 10\% and taking into account all different contributions, the uncertainty on the energy scale coming from the measurement of the atmospheric extinction should be at the 5\% level only. 

%To achieve this accuracy, a Raman LIDAR is needed that operates in the wavelength range that is relevant for the Cherenkov light. Moreover, it needs at least two Raman lines and at least one (better two) elastic lines in order to disentangle the molecular and aerosol scattering components. 
\medskip
\textbf{Wavelengths. \label{subsect:Wavelengths}}
The CTAO telescopes are sensitive to Cherenkov light in the wavelength range from 300~nm to $\sim$700~nm. The Cherenkov light spectrum falls with the square of the inverse wavelength, and hence UV and blue photons are more frequent than those approaching the red end of the spectrum. Moreover, the photosensors employed for the telescopes are also more sensitive at shorter wavelengths. After weighting the Cherenkov light spectrum with the detection sensitivities of the MAGIC Telescopes, \citet{Schmuckermaier:2023huo} have found average wavelengths of detected Cherenkov light ranging from $\sim$390~nm to $\sim$410~nm, according to \sapocol{elevations decreasing from 90$^\circ$ to 30$^\circ$.}
%increasing zenith angles from vertical to 60$^\circ$. 
A RL shall hence have at least one laser wavelength located close to the central wavelength between the short wavelength limit and the average, i.e. (350$\pm$10)~nm, and another located centrally between the average and the long wavelength limit, i.e. (550$\pm$20)~nm. The frequency-doubled and tripled wavelengths of a Nd:YAG laser fit these requirements very nicely.

\medskip
\textbf{Elevation range. \label{subsect:Elevation}}
The CTAO telescope arrays are required to be able to observe %\textcolor{teal}{from zenith to 70$^\circ$ in the zenith range} 
at elevations ranging from 20$^\circ$ \sapocol{to 90$^\circ$}
%zenith 
and the full azimuth range for standard observations.
Therefore, an RL shall be able to characterize any line of sight within this cone.

\medskip
\textbf{Range. \label{subsect:Range}}
Case studies~\citep{Doro:2012,Garrido:2013,Sobczynska:2014} have revealed that cloud layers above $\sim$15~km a.s.l. have a negligible impact on the energy and flux reconstruction of vertically incident gamma rays, except for very low-energy gamma rays near the telescope detection threshold (see also Fig.~\ref{fig:showers}). \sapocol{At low gamma-ray incidence elevation angles,}
%At high gamma-ray incidence zenith angles, 
that altitude moves about 2--3~km higher. This is because most of the Cherenkov light is emitted below that height, see Fig.~\ref{fig:showers}.
In the absolutely worst case, the RL points to \sapocol{20$^\circ$ elevation}
%70$^\circ$ zenith angle 
and needs to characterize a cloud found at around 18~km a.s.l.\footnote{Note that such clouds actually do exist above the Canary Islands, as shown in~\protect\citet{Fruck:2022igg}.}. 
%This leads to required maximum ranges of up to $\sim$45~km \md{this number may change, also because the pBRL will not get that far} distance for a RL.  
%
%IACTs primarily focus their observations within a range of about 10~km, corresponding to the altitude where EAS originate at zenith incidence. However, observational practices often encompass wider zenith angles, reaching up to 45 degrees, and occasionally as high as 60 degrees. Considering this broader range of observation angles, it becomes necessary for the LIDAR system to cover a minimum distance of approximately 14~km. 
%
%Moreover, to determine atmospheric extinction characteristics accurately, the use of a Raman LIDAR system with a wide measurement range is essential. The potential for observations at zenith angles of up to 60 degrees suggests that extending the LIDAR's measurement range to approximately 20~km would be advantageous, ensuring comprehensive coverage of relevant atmospheric conditions.
%
At the same time, it has been shown~\citep{Fruck:2022igg,Barreto:2022} that ground-layer aerosols during clear nights are typically concentrated at altitudes below 2--3~km  above ground. Therefore, 
%due to the significance of the first few hundred meters of the atmosphere, 
the design of the RL system shall also ensure sufficient sensitivity for the first few kilometres.
%, particularly ensuring a low enough range of full overlap is achieved, such that the aerosol extinction of first kilometer of atmosphere is still sufficiently well characterized. \md{The previous sentence is vague, it's more quantified in the introduction chapter}

\medskip
\textbf{Range resolution.}
Given the longitudinal extension of air showers, spanning $O$(10)~km, the location or even fine structure of the measured extinction profiles does not need to be measured to better than $\sim$150~m. Even in the worst case of an optically thick, fine layer of aerosols cutting through the air shower at its shower maximum, such a limitation would worsen the achievable accuracy of the aerosol optical depth affecting the detected Cherenkov light by $\lesssim$2\%. However, this entails that the signal sampling of a RL must be a fraction of that value. 

\medskip 
\textbf{Aerosol transmission ranges:}
Since gamma-ray observations will be aborted in any way by the observatory if the atmospheric conditions are so bad that an important fraction of the Cherenkov light is getting lost due to clouds or aerosol (i.e., in the case of optically thick cumulus layers), the RL needs to operate in the case of integrated aerosol and cloud optical depths of less than 0.7.

\medskip 
\textbf{Aerosol transmission accuracy:}
Aerosol transmission shall be reconstructed with an absolute accuracy of better than 0.03, leading to a requirement of the aerosol optical depth profile reconstructed with similar accuracy. Note that the range-resolution requirement applies at the same time. 

\medskip
\textbf{Pulse accumulation time.}
By design and requirements, the laser wavelengths of the RL are found within the sensitive wavelength range of the CTAO telescopes, and hence the RL shall not cross their field of view (\gls{fov}) while observing. This can be avoided by propagating the laser a few degrees outside the observed FoV. Nevertheless, it is often convenient to characterize the atmosphere exactly along the line-of-sight of the observed target, shortly before starting or after ending an observation, or during the repositioning time of the telescopes, which normally amounts to at least a minute. This leads to a requirement for the signal acquisition time of less than about a minute.

\medskip
\textbf{Duty cycle of operation.}
The RL will operate during observable nights at a frequency of 5--10 min, depending on how often the targets need to be changed or even characterized simultaneously. Given roughly 2,000~hours of total observable night time during the year, this leads to about 20,000 profiles taken every year during an expected lifetime of 15~years. 

%\medskip
%\textbf{Induced dead time.}
%The monitoring of the atmosphere is intended to improve the uncertainties of the assisted IACT, as well as to increase the duty cycle during non-perfect weather conditions. 
%Doing this with a LIDAR will imply propagating a powerful laser in the area of the observatory. 
%This should not reduce the available time for data taking during good dark nights by more than 1\%.  \textcolor{red}{MD where does this limit come from?} \mg{remove} \aco{Do we have a limit on that or should I remove the whole 'section'?}

\subsection{Operation and Survival Conditions \label{sect:OperCond}}
\medskip
\textbf{Ambient light of the night sky.}
IACTs generally operate during dark nights or nights with moderate moon and in regions with low or absent anthropogenic light pollution.  CTAO requires full capabilities of the RL for background light levels up to 
10$^{-9}$~W~m$^{-2}$~nm$^{-1}$~sr$^{-1}$ in the wavelength range from 300~nm to 600~nm.

%The standard light of the night sky is about one magnitude 22 star per square arc second, although IACTs may operate with light levels up to 10 times brighter during moon time. \textcolor{red}{MD what does this have to do with operation? computation of background signal at RL?} \mg{Sí.}

\medskip
\textbf{Observation conditions.} In the context of performance requirements, the RL shall guarantee its functionality and accuracy for relative humidities in the range from 2\%~to~90\%, an atmospheric pressure in the range of 770~$\pm$~50~mbar and an ambient temperature range that extends from $-5^\circ$C to $25^\circ$C. 
%; encompassing the typical humidity, pressure and temperature variations encountered during observational operations at ORM. 
Furthermore, air temperature gradients of up to $\pm$7.5$^\circ$/h may occur, and exposure to winds with average wind speeds up to 36~km/h over a 10-minute period during operation.  

\begin{comment}
    the environmental conditions that lead to each of those operation modes are described. The LIDARs installed in the CTAO observatories should fulfil the conditions for the survival mode. For the other operation modes, it is desirable that they are able to operate in the normal operation conditions and to go to safe position during critical situations.
\end{comment}

% \subsection{Survival Conditions}
%\label{subsect:survival}
In addition to the given operating conditions, the RL must withstand all weather extremes that can occur at the northern CTAO site. 
These so-called "survival conditions" apply to the RL\sapocol{, when it is in a "safe state", that is, closed and not operating, minimizing the use of power whilst still providing basic status monitoring}. They are summarized in the subsequent set of requirements. 
%Trigorous environmental requirements are essential to ensure the reliability and longevity of the LIDAR infrastructure at the CTA-N site, supporting its scientific objectives and operational stability.

\medskip
\textbf{Temperature tolerance.} In \sapocol{safe state}, the RL shall withstand ambient temperatures ranging from -15$^\circ$C to +35$^\circ$C and air temperature gradients of up to $\pm$0.5$^\circ$C/min for 20~minutes  without suffering damage. In the event of a power outage, the temperature ranges from -10$^\circ$C to +30$^\circ$C apply. 

\medskip
\textbf{Humidity resistance.} The RL shall not suffer any damage under conditions of relative humidity ranging from 2\% to 100\%, whether in \sapocol{safe state} or during a power outage.

\medskip
\textbf{Precipitation tolerance.} The RL shall withstand precipitation in the form of rain, snow, or hail with average wind speeds of up to 90~km/h. In addition, damage shall not occur due to precipitation with a maximum of 200~mm of rainfall in 24~hours or 70~mm of rainfall in a single hour, or snow loads up to 200~kg/m$^2$ or ice accumulation of up to 20~mm on all surfaces. Lastly, the RL shall withstand the impact of hailstones up to a diameter of 5~mm.

\medskip
\textbf{Wind speeds.}
The RL shall withstand, in a safe state, average wind speeds of up to 120~km/h and wind gusts of up to 200~km/h. 

\begin{comment}
The infrastructure should withstand 10-minute average wind speeds of up to 100 km/h without damage and up to 120 km/h when in the Safe State.
Wind gusts (1-second duration) of up to 200 km/h should not lead to damage.
\end{comment}

\medskip
\textbf{Atmospheric contaminant resistance.} The RL shall not be damaged due to atmospheric concentrations of NO, NO$_{2}$ and SO$_{2}$ of up to 3~ppb or extreme conditions of calima, i.e., an environment with coarse-mode particles of up to $O(10^5)$ per m$^3$ of air.
%for 90\% of the time at 2~m above ground. 
%This limit adheres to ISO Class 9 as defined in ISO 14644-1 for particles of this size.

\medskip
\textbf{Solar radiation resilience.} The RL should withstand solar radiation of up to 1200~W/m$^2$ 
%(averaged over 1 hour) 
at a maximum ambient temperature of 35$^\circ$C in a safe state. All components exposed to direct solar radiation shall be UV resistant.

\medskip
\textbf{Seismic resilience.} No damage shall occur due to peak %horizontal 
ground acceleration 
%and peak vertical ground acceleration 
up to 0.05~$g$.
%at the CTA-N site.

\subsection{Reliability and Availability}
\label{subsect:reliability}

\medskip
\textbf{Reliability.} Although a non-operative LIDAR will not prevent the observatory from taking data, it may seriously degrade the quality of the data taken since the quality of the atmosphere cannot be assessed. Therefore, the RL shall be reliable and %able to
operate with full functionality during 97.5\% of the observation time. 
%; having a minimum operational lifetime of 15 years. 

%\medskip
%\textbf{Lifecycle}
%The LIDARs installed at the CTAO sites will have to operate remotely and with a minimum maintenance for the CTAO expected life time of 20 years. \textcolor{red}{very vague}

\medskip
\textbf{Maintenance.} Preventive maintenance of all auxiliary equipment installed at the site (to which the RL will eventually belong) shall not exceed two person-hours per week
and corrective maintenance less than four person-hours per week, in order to limit the operational cost of the observatory. %ensure the requested reliability and availability. 
%Those requirements are established to the overall set of elements forming CTA-N site. \aco{??}

\medskip
\textbf{Safety.} The RL shall comply with all the requirements listed in the European Machine Directive. 
%During remote operation, the LIDAR must automatically shift into Safe State from Off State once power is provided. On the other side, the RL must not transition from Safe State into Standby State unless instructed by the ACADA. 
In case of sudden power outage, it shall not result in damage beyond the serviceability limit. 

\begin{comment}
The short term periodic maintenance of the LIDAR should not exceed 2 hours per week of 2 people to ensure the requested reliability and availability. Similarly, any long term periodic maintenance should not exceed one day every 6 month of two people.

A part from periodic actions to ensure the full functionality of the LIDAR, there could be interventions due to broken hardware. Those interventions should be minimized and not expected to happen more than once every 5 years. They should not take longer than a couple working days of two people before the systems becomes operational again. \mg{Lifecycle and Maintenance: re-formulate. Needed??}  
\end{comment}

%\mg{I'm done up to here}

%\subsection{Site Considerations}
%\label{sect:SiteConsiderations}
%Finally, the RL has to withstand several situations 
%that can occur at the northern site of the CTAO and fulfil the %corresponding onsite security and safety requirements.

%\newpage
\section{pBRL Technical Design}
\label{sec:tech_design}

LIDARs for science-orientated atmospheric remote sensing studies are often custom-made for the specific use of data~\citep{Keckhut:1990,spinhirne1995,Eisele:1997,Rees:2000,Mueller:2014,Gerding:2016,Klanner:2021}. Only a few companies around the world provide standard or custom products~\citep{Avdikos:2015}. For the purpose of CTAO, strong requirements exist concerning mirror size and laser power, in order to be sensitive to large distances (see Sect.~\ref{sec:requirements}), together with the need for a pointable system. This comes in addition to relatively strong budget constraints. We resolved to buy and adapt a \sapocol{disassembled} %dismissed 
telescope, formerly belonging to the Cherenkov Light Ultraviolet Experiment (\gls{clue})~\cite{Peruzzo:1990ia,Alexandreas:1995}, used to serve as a ground-based Cherenkov telescope array of nine telescopes. 
%The dismissed containers and telescope structures satisfied our purpose, and we committed to refurbish it to the needs of a Raman LIDAR.
%, as discussed in the Introduction. 
The refurbishment involved mostly the focal plane instrumentation and the readout optics and electronics, plus the inclusion of the laser and its structure, while most of the mechanics and the telescope chassis were kept with original pieces.

\begin{figure}[htp]
%\begin{adjustwidth}{-\extralength}{0cm}
\centering
\includegraphics[width=0.65\linewidth]{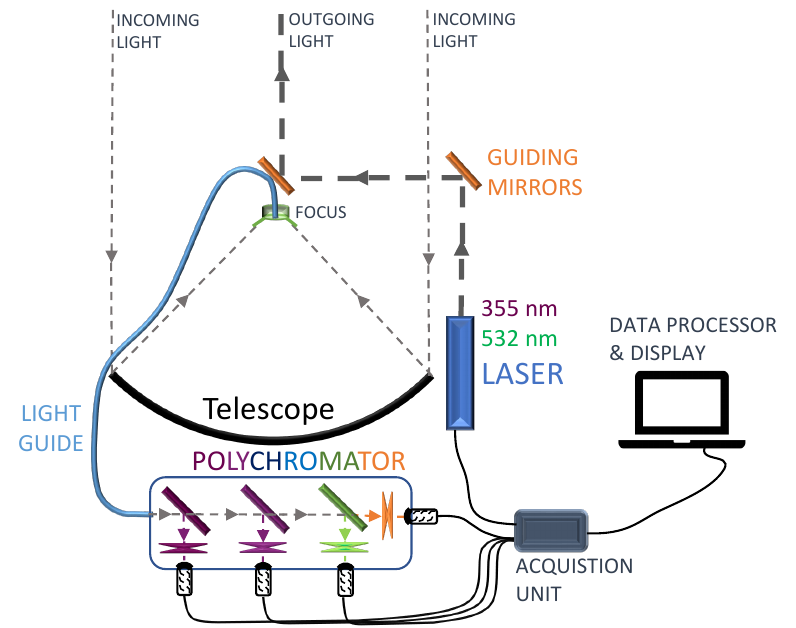}
\includegraphics[width=0.7\linewidth]{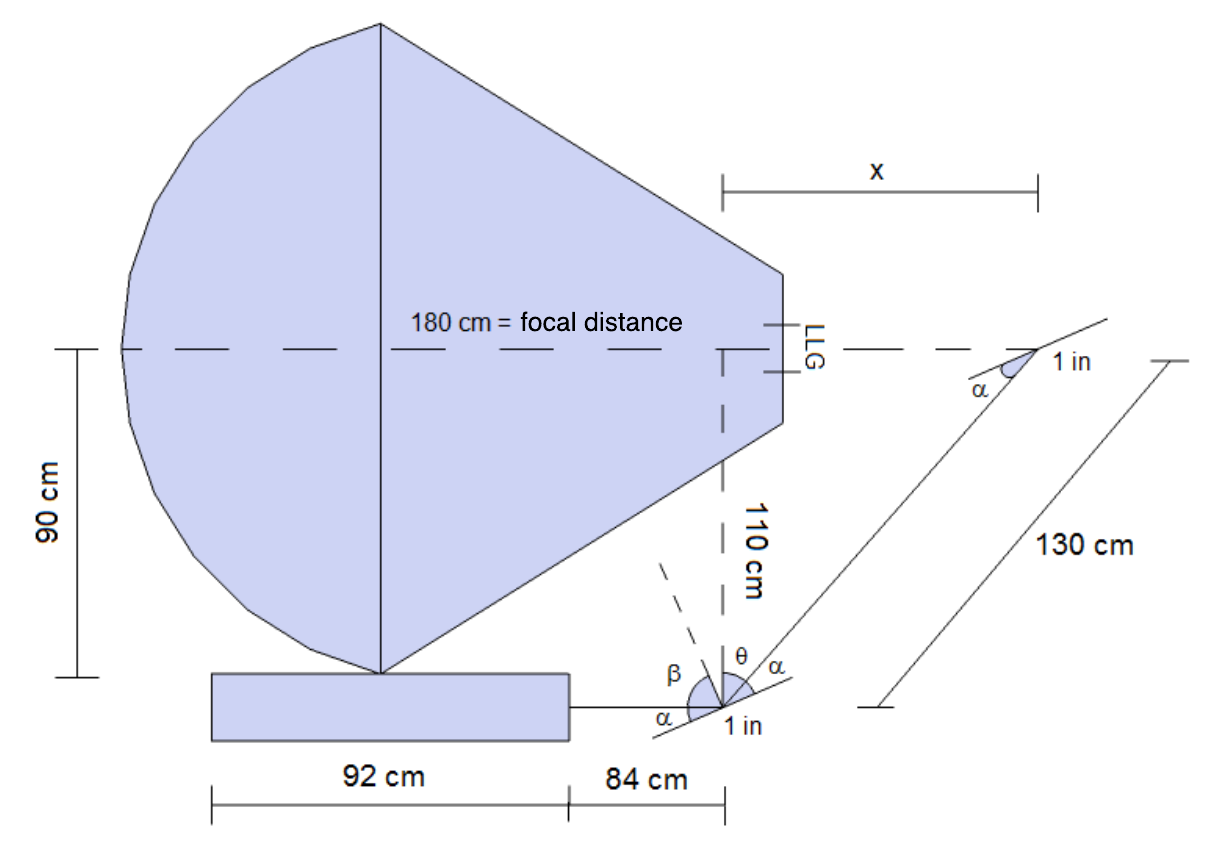}
%\includegraphics[width=0.6\linewidth]{Figures/sec3/lidar_scheme.png}
%\end{adjustwidth}
\caption{Top: A schematic drawing of the pBRL and its main components: the receiver, comprising of the telescope, the polychromator and the data acquisition unit, and the transmitter, comprising of the laser and guiding mirrors. Below: A top view of the system is shown, drawn to scale and turned to horizontal pointing. The angles have been measured to $\alpha = (28.9\pm 0.3)^\circ$ and $\beta = (61.1\pm 0.3)^\circ$.
%\aco{The image at the bottom seems a bit blurry, do we have it in pdf?} \mg{I've exchanged it by a pdf}
%and the main distances are shown. 
%\sa{schematic drawing of the pBRL and its main components: the receiver comprising of the telescope, polychromator and the data acquisition unit and the transmitter comprising of the laser and guiding mirrors.}
\label{fig:lidar_scheme2}}
\end{figure}
%\begin{figure}[h!t]
%\centering
%\includegraphics[width=0.485\textwidth]{Figures/sec3/lidar_scheme.png}
%\caption{Sketch of the Raman LIDAR with the main distances drawn. All distances have an uncertainty of 5~mm, except for the focal distance, which is preciser. \md{Missing complete description. Maybe the polychromator, LLC guide and LICEL shutter, could be added. Also, is this better placed here or in the introduction? Or one can use the image below or modify it }\label{fig:lidar_scheme}}
%\end{figure}

The general scheme of the pBRL is depicted in Fig.~\ref{fig:lidar_scheme2}. Schematically, it is composed of a primary mirror, a focal plane equipped with a LLG that transports the signal to an optical bench (called a polychromator). On the side of the telescope structure, a laser and two guiding mirrors are placed to make the laser beam coaxial with the telescope optical axis. 
%All details below. 

\medskip
This section describes the technical design, discussing the choices made and the solutions adopted. As will be shown later, most of these choices were forced by the initial decision to use the CLUE telescope with its given mechanical and optical properties. 
%\sapocol{\st{Sect.~\protect\ref{sec:design-clue}}\st{presents the characteristics of the CLUE container.} 
%\st{Furthermore, Sect.~\protect\ref{sec:chassis_petals}}\st{shows the telescope chassis, the slewing system, and protective petals, and Sect.~\protect\ref{sec:mirror}}\st{ the primary mirror.} 
%\st{In Sect.~\protect\ref{sec:design_optics}}\st{, we discuss our decision to use a Newtonian mount in comparison to a Cassegrain solution and in Sect.~\protect\ref{sec:biaxiality}}\st{to choose a coaxial laser beam.}
%\st{Sect.~\protect\ref{sec:llg}}\st{ presents the characteristics of the LLG and Sect.~\protect\ref{sec:polychromator}}\st{those of the polychromator.} 
%\st{Finally, we discuss readout electronics in Sect.~\protect\ref{sec:licel}}\st{, the laser in Sect.~\protect\ref{sec:laser}}\st{, dichroic guiding mirrors in Sect.~\protect\ref{sec:dichroicmirrors}}\st{and the focal point shutter in Sect.~\protect\ref{sec:shutter}}. 
%\st{An auxiliary near-range system is explained in Sect.~\protect\ref{sec:shortrange}.}}

\subsection{The Original CLUE Container}
\label{sec:design-clue}
%The Cherenkov Light Ultraviolet Experiment (CLUE) experiments \cite{Alexandreas:1995,Peruzzo:1990ia} was a ground-based Cherenkov telescope array of XX telescopes housed in commercial container. The pBRL is an adaptation of one of the CLUE containers and its telescope, optimized for Cherenkov telescopes' atmosphere monitoring.

\begin{figure}[h]
 \centering
 %\subfloat[Left: Container closed during transport. Right: Container opening.]{
% \label{fig:container}
 \includegraphics[width=0.495\textwidth,trim={12cm 7.5cm 6cm 0cm},clip]{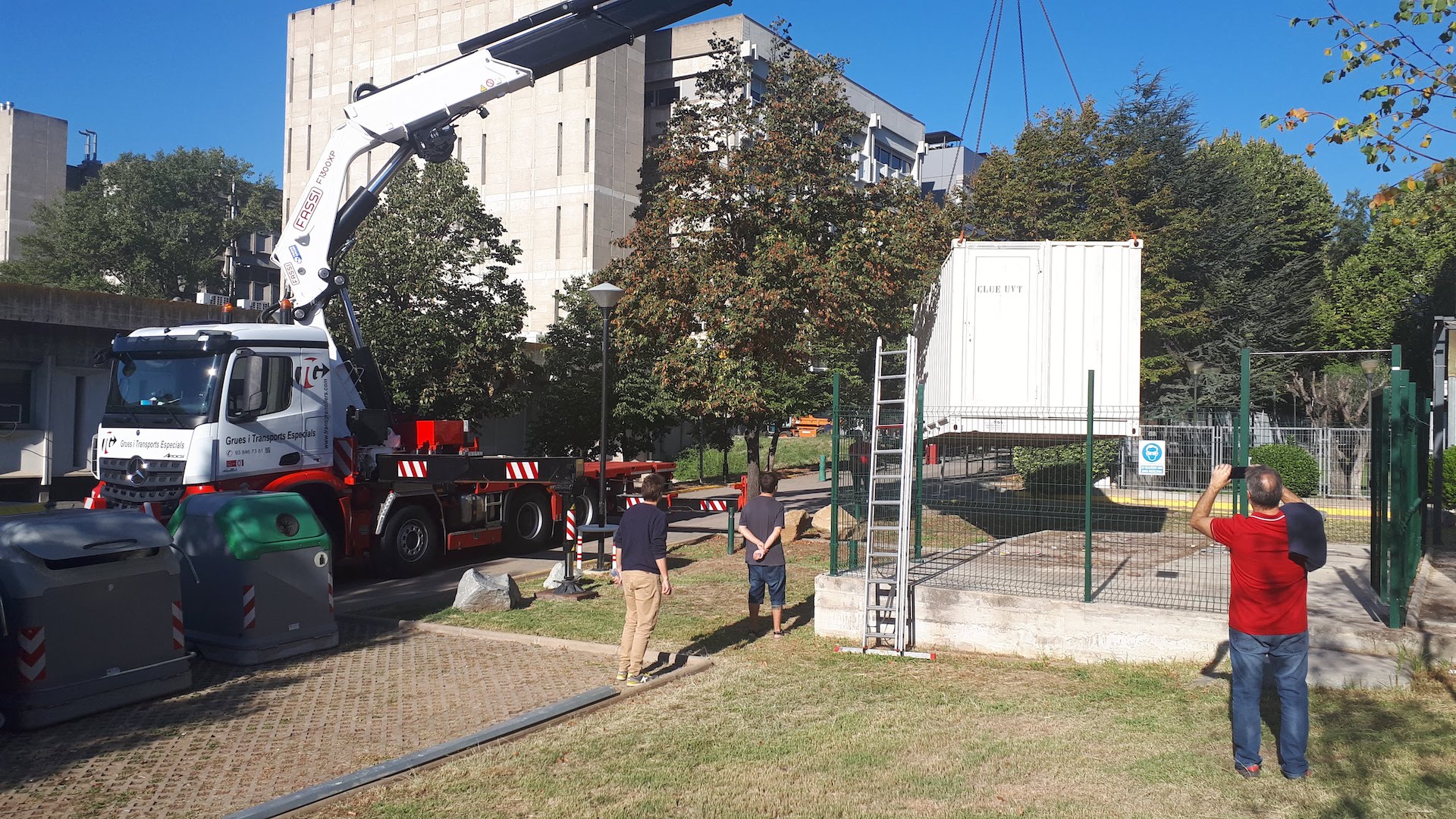}  
 \includegraphics[width=0.455\textwidth]{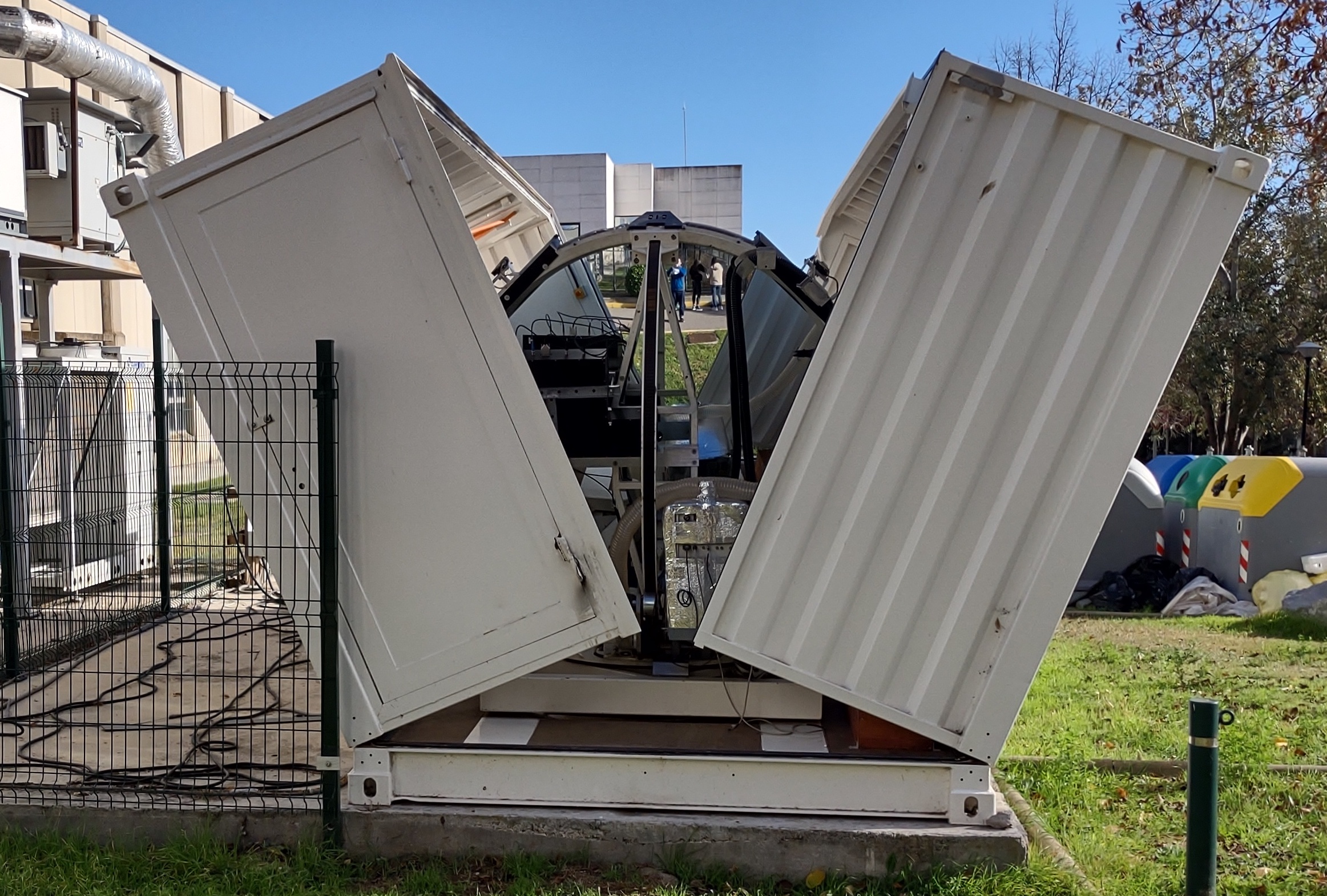}  %} 
 \\
%\subfloat[Inside view of the container.]{
 %  \label{fig:inside_container}
    \includegraphics[width=0.5\textwidth] {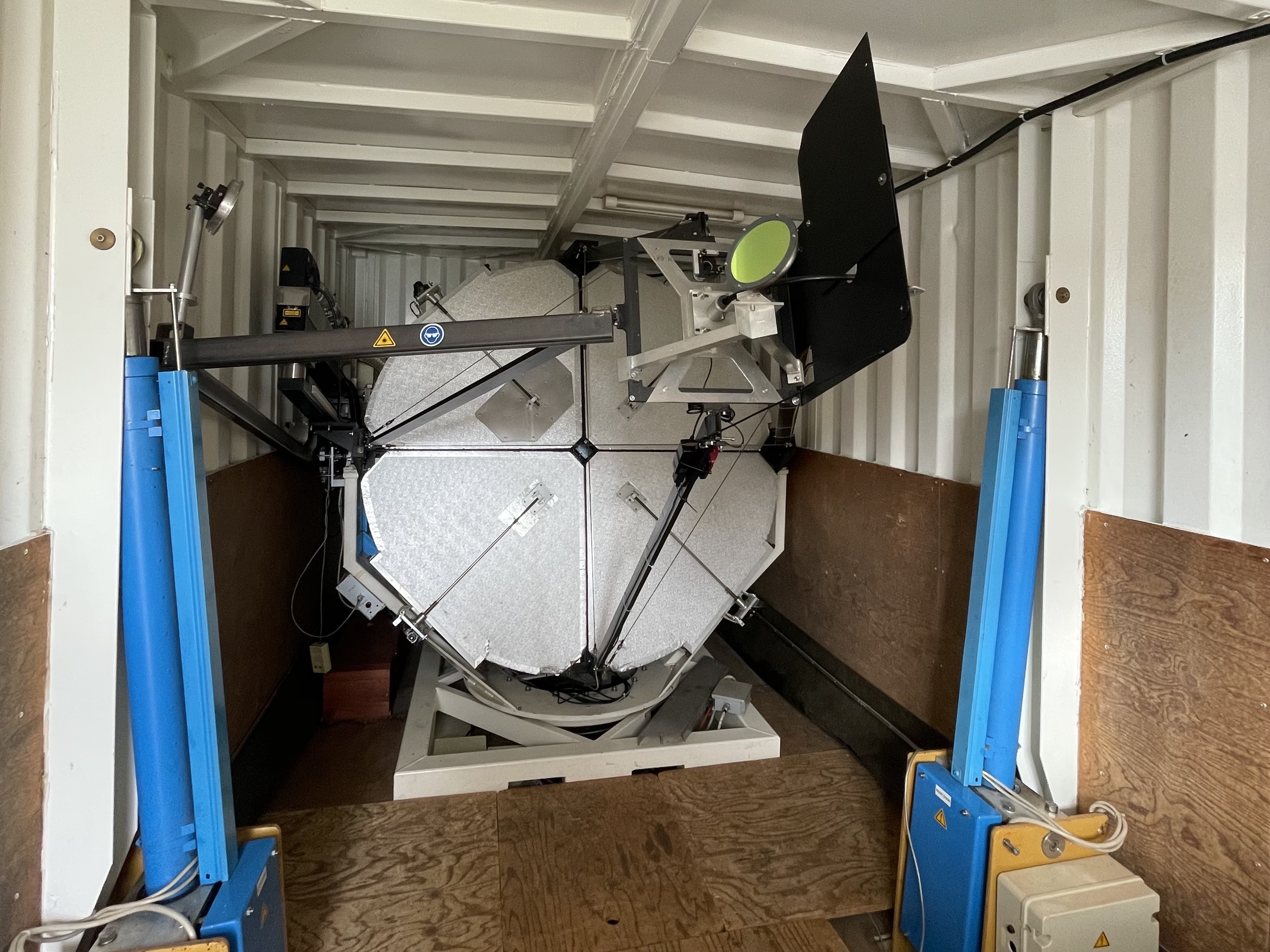} %}
 % \subfloat[Actuator and control box.]{
   %\label{fig:actuator}
    \includegraphics[width=0.47\textwidth,trim={8.2cm 3cm 3cm 0cm},clip]{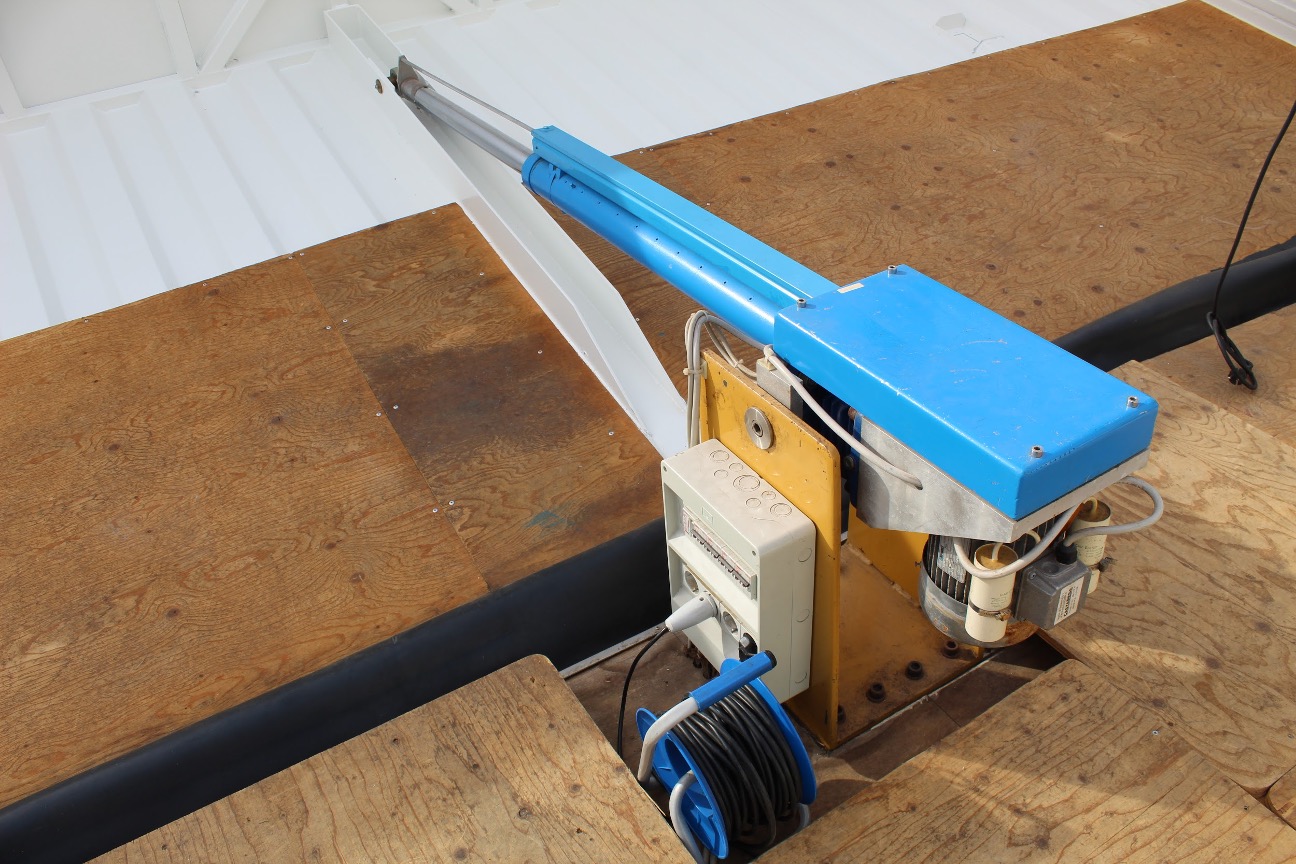} %\hfill %}
 \caption{CLUE container hosting the pBRL. Top left: Container closed during transport. Top right:~Container opening. Below left: Inside view of the container. Below right: Actuator with motor and control box.} 
 \label{fig:container_all}
\end{figure}

The CLUE container (see Fig.~\ref{fig:container_all}) is a 20~ft standard maritime container, of dimensions $5.90~\times2.35~\times~2.39$~m. It provides shelter to the entire system and weighs $\sim$3~tons (2.3~tons from the container and 700~kg from the telescope). The container protects all instrumentation from rain and dust, as well as from light. It did not suffer any damage during transport and was recently de-rusted and repainted. The container walls can be opened sideways in two halves in a single automated movement through two hydraulic motors of the model \textit{Servomech}~\textit{106301} and actuators that can be operated remotely or using a handheld control. An individual wall can also be opened and closed. The complete opening and closing takes about 60~s.  The hydraulic motors have been taken over from the original experiment and have not been replaced so far. They are powered by 230~VAC and consume each $\sim$1.5~kW peak power. Hardware limit switches prevent wall positions from damaging telescope components. The container has a mechanical locking mechanism for transport. To be operational, the container itself must be connected to both a power line and Ethernet. 
%It is powered using two industrial BS4343~IP67 connectors. 
The container was equipped with a false wooden floor with cables passing below and a cabinet rack for control. The container also has a door on the short side that allows an operator to enter and work from within the container. That door is equipped with a locking system connected to a fail-safe switch during operation.

\subsection{The Telescope Chassis and Petals}
\label{sec:chassis_petals}
\begin{figure}[h]
 \centering
 %\subfloat[Telescope chassis (front).]{
 \includegraphics[width=0.32\textwidth]{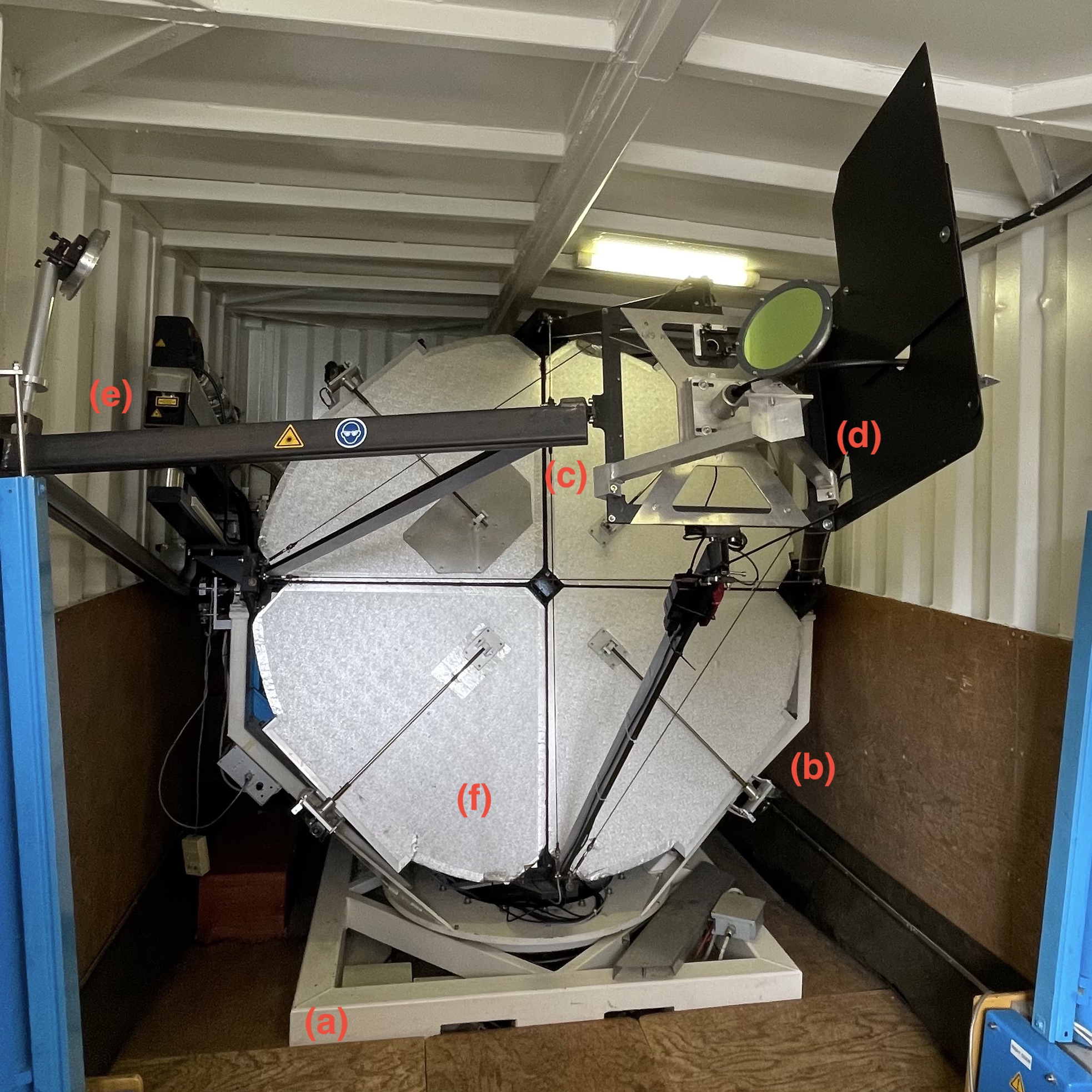} \hfill %}
% \subfloat[Telescope chassis (rear).]{
 \includegraphics[width=0.32\textwidth]{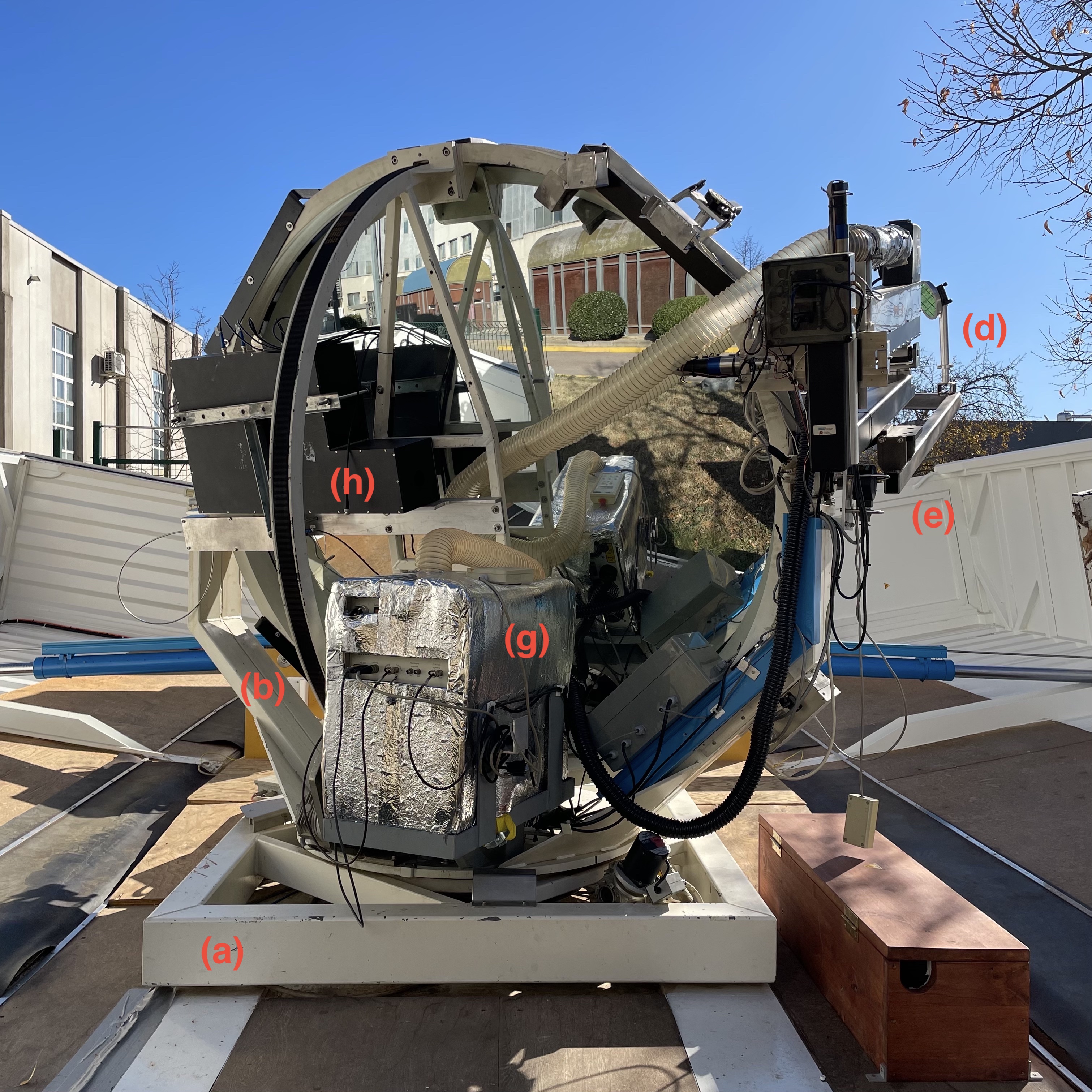} \hfill %}
 %\subfloat[Telescope laser arm.]{
 \includegraphics[width=0.32\textwidth]{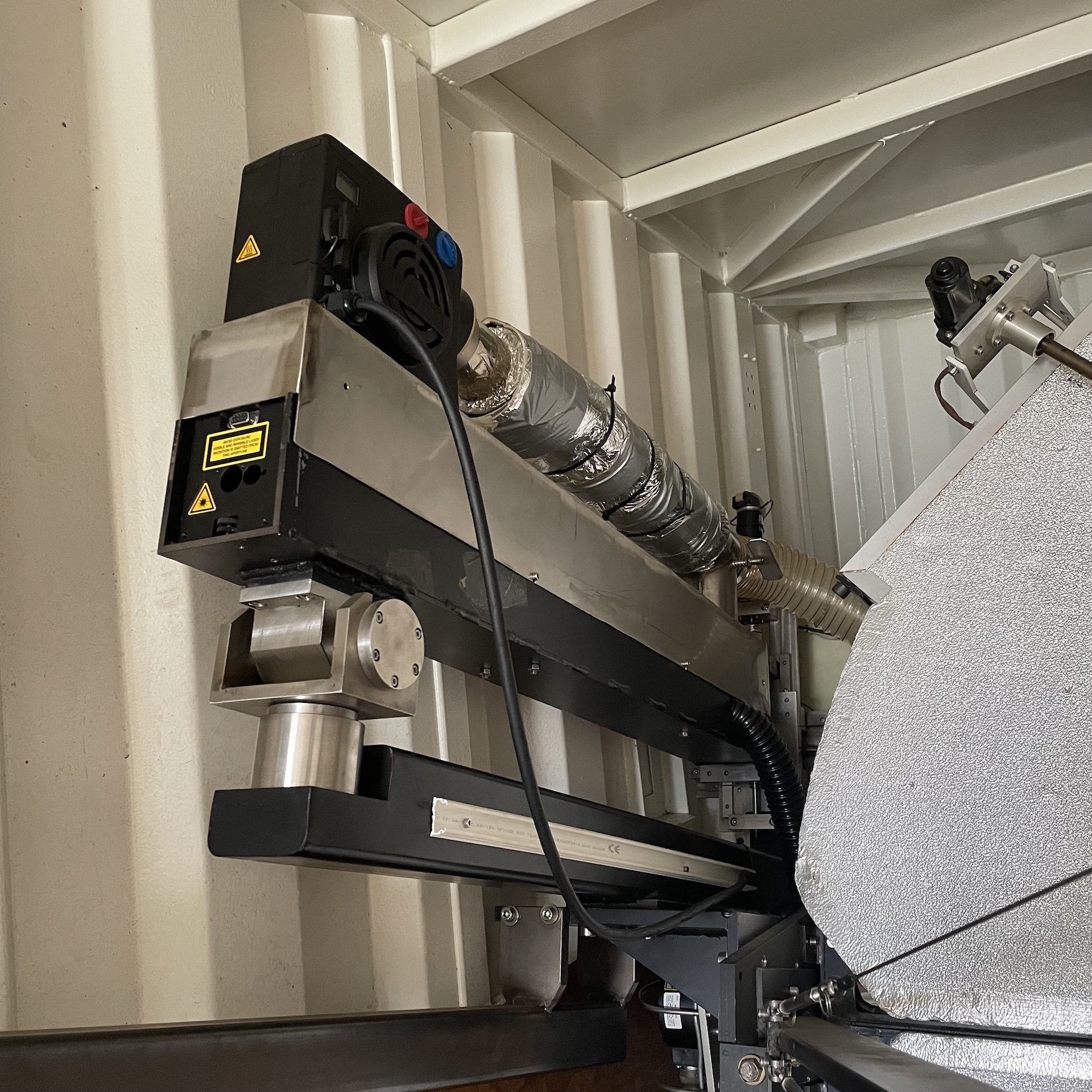}
 %}
  \caption{Left and middle pictures show the telescope chassis seen from the front and rear, respectively. On the right, the telescope laser arm. The main components of the telescope chassis and petals: the metal platform that supports the azimuth movement (a), the U-form structure that supports the \sapocol{elevation }%zenith 
  movement (b), the focal plane support (c), the support for the small mirrors to align the laser beam (d), the structure for the laser arm (e), the petals covering the primary mirror (f), the cooling system of the laser (g) and the polychromator (h).}
\label{fig:telescope_structure}
 \end{figure}

The CLUE telescope chassis is an alt-az mount designed to support a 1.8~m mirror and CLUE's rather bulky focal plane instrumentation, see Fig.~\ref{fig:telescope_structure}.  The movement of the telescope in both degrees of freedom is actuated by stepper motors of the model PH299-F4.0, manufactured by the company VEXTA\footnote{Now called Oriental motor; \url{https://www.orientalmotor.eu/}.} (Japan) and mounted on the chassis. % (Fig.~\ref{fig:telescope_structure}). 
The motors actuate on different reductors, manufactured by the company Bonfiglioli\footnote{\url{www.bonfiglioli.com}} (Italy). Motor power is directed to the telescope through a timing belt. A toothed pulley is fixed to the axis of the reduction. The motion control was adapted from the original CLUE design, keeping the power drivers for the stepper motors.

The chassis also supports a motorized protective foldable petals system and the optical system in the focal plane; see Fig.~\ref{fig:telescope_structure}. %Figs.~\ref{fig:telescope_structure}(a) and~\ref{fig:telescope_structure}(b). 
The empty metal square in the focal plane once held the CLUE multiwire camera, which has been removed. 
The petals are used to protect the primary mirror from dust. They are made of polystyrene and are actuated by four 12~V motors; the movement is controlled by eight series--connected limit--switches. When the motor is operated, a long endless screw opens or closes the petal until the limit switch issues a signal. 
The laser arm visible on the right %left 
side of Fig.~\ref{fig:telescope_structure} %(c) 
was added afterwards.

\subsection{The Primary Mirror}
\label{sec:mirror}
%\md{MG evaluate if there is too much information.} 

\begin{figure}[h!]
\centering
\includegraphics[width=0.485\linewidth]{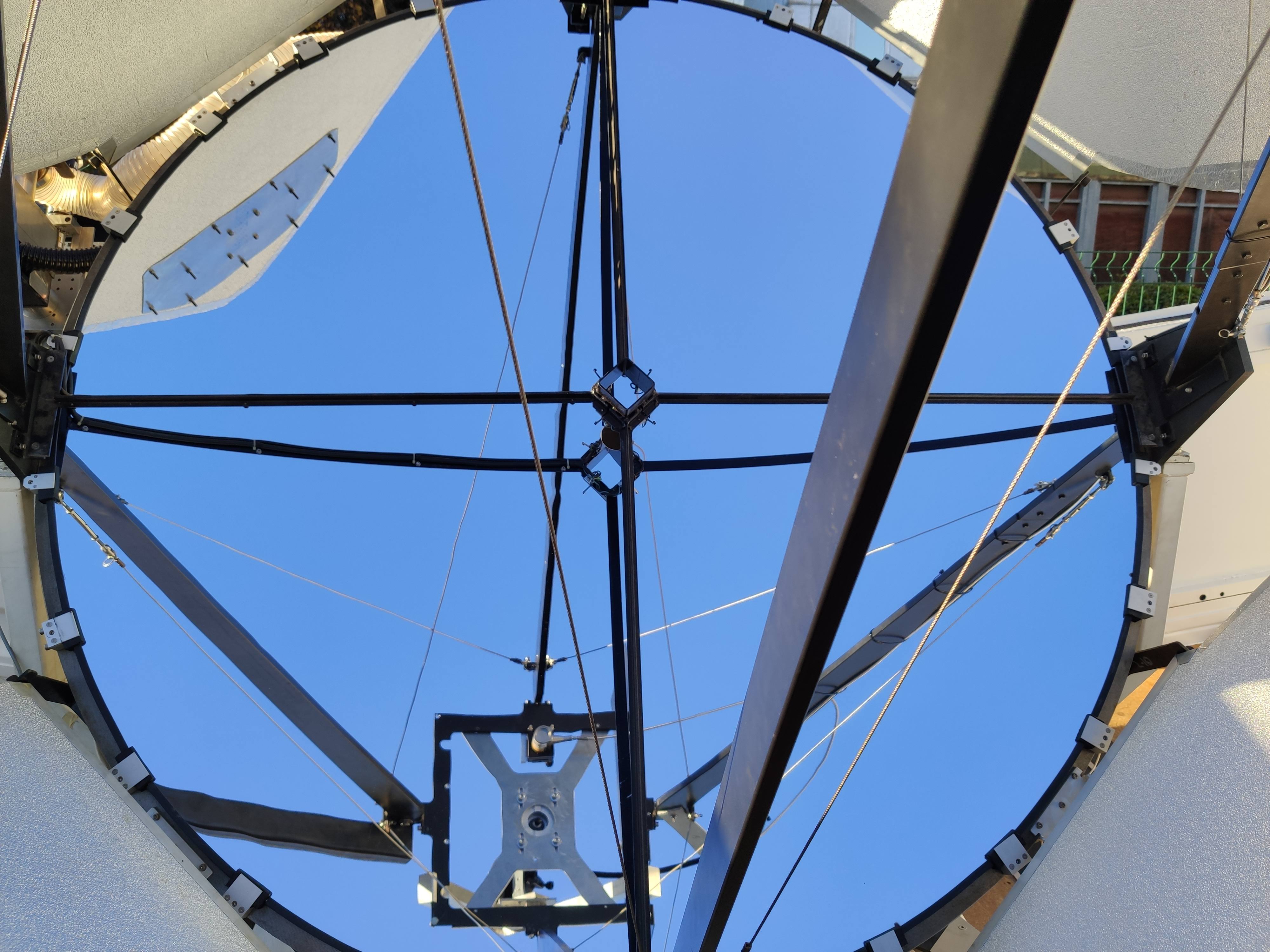}
\caption{\label{fig:primarymirrorpic}
%\sa{I suggest to re-write the caption as follows:} 
A 1.8~m diameter parabolic mirror with $f$-number of one serves as the lidar's primary mirror.}
\end{figure}

The CLUE container was equipped with a 1.8~m diameter parabolic mirror of the same focal length \sapocol{($f$/1)}
%($f-$number=1)
, described by \citet{Alexandreas:1995}. 
It was produced following a hot slumping technique, first invented at CERN~\citep{Baillon:1989a,Baillon:1989b}, and later adapted to the requirements of CLUE, especially the large mirror size. 
The mirror is made of smooth float glass produced by the company Societ\`a Italiana Vetri\footnote{Now called Pilkington Italia; \url{https://www.pilkington.com/it-it/it}.} (SIVET, Porto Marghera, Italy). The glass has a low carbon content and a high chromium (0.5\% C, 13\% Cr) content and a thermal expansion coefficient of $8.5\cdot10^{-6}/^\circ$C. The glass is placed over a mould, both heated in a large electric oven that reaches 600$^\circ C$ at the company Sunglass\footnote{\url{https://www.sunglass-industry.com/}} (Villafranca Padovana, Italy) so that the glass retains the shape of the mould. For this reason, the mould must be very precisely shaped. The mould was cast in a special stainless-steel alloy STAVAC ESR AISI~420 with %low 
% MG: all steel has ~13x10^-6...
thermal expansion coefficient ($13\cdot10^{-6}/^\circ$C), comparable to that of the glass. The mould was machined to a concave parabolic shape with a digitally controlled lathe with nominal accuracy better than 20~$\mu$m. 
The deviation of the mould from the nominal parabolic surface and the glass plate defects introduced differences in the slope of the parabolic mirror of less than 1.6~mrad. These effects enlarged the image in the focal plane by 5.8 mm at maximum, when realized~\citep{Alexandreas:1995}. The mirror thickness amounts to 6~mm for a total weight of about 30~kg.
%and is currently located at the 
%INFN Legnaro laboratories (Padova, Italy).   
Figure~\ref{fig:primarymirrorpic} shows a picture of the primary mirror.

\bigskip
\textbf{Point spread function.}
%At the times of CLUE, the peak-to-valley due to thickness and parallelism imperfections of the glass and the mould were measured to 1.3~mrad~\citep{Peruzzo:2011}. 
\citet{Alexandreas:1995} had reported that 80\% of the light of the eight original mirrors was contained within a square of 5--8~mm, depending on the mirror (more than 9 were produced).
Considering that the mirror had been inoperative for several years, we carried out new measurements to characterize the point spread function (\gls{psf}). The mirror spot size was then characterized 
%through its PSF 
by different methods, producing all compatible results. These were reported in more detail in~\citep{sanchez:tesina,lopezphd}. 

\begin{figure}[h!]
\centering
%\subfloat[Spot size image]{\label{fig:mirror}
\raisebox{0cm}{
\includegraphics[trim={0.3cm 0cm 0cm 0.3cm},clip,width=0.48\linewidth]{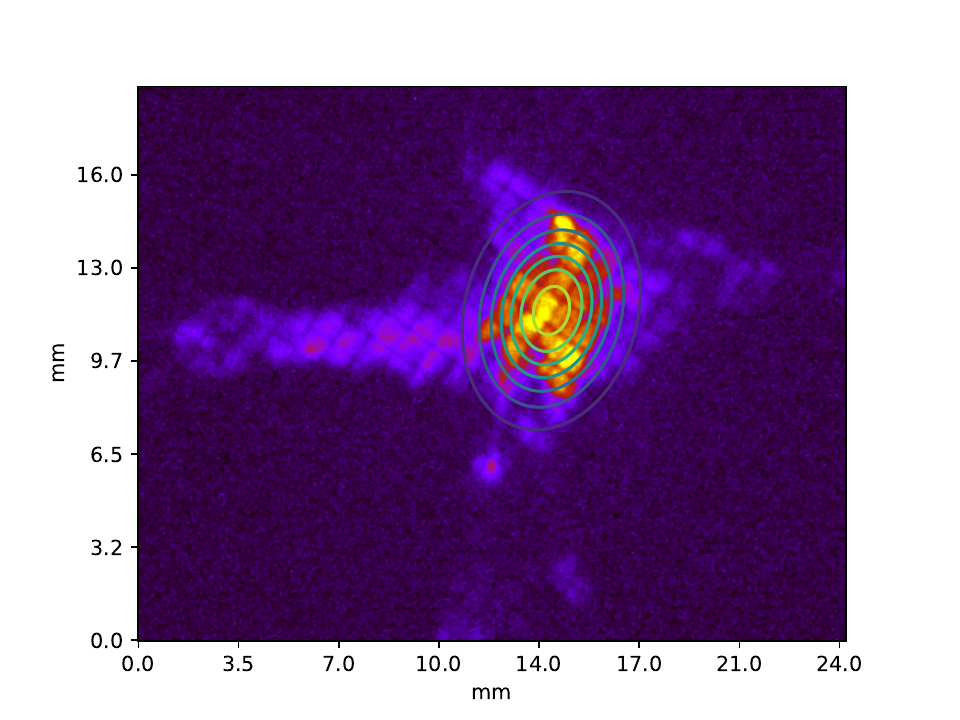}} 
%\subfloat[Spot size profile]
\includegraphics[width=0.45\linewidth]{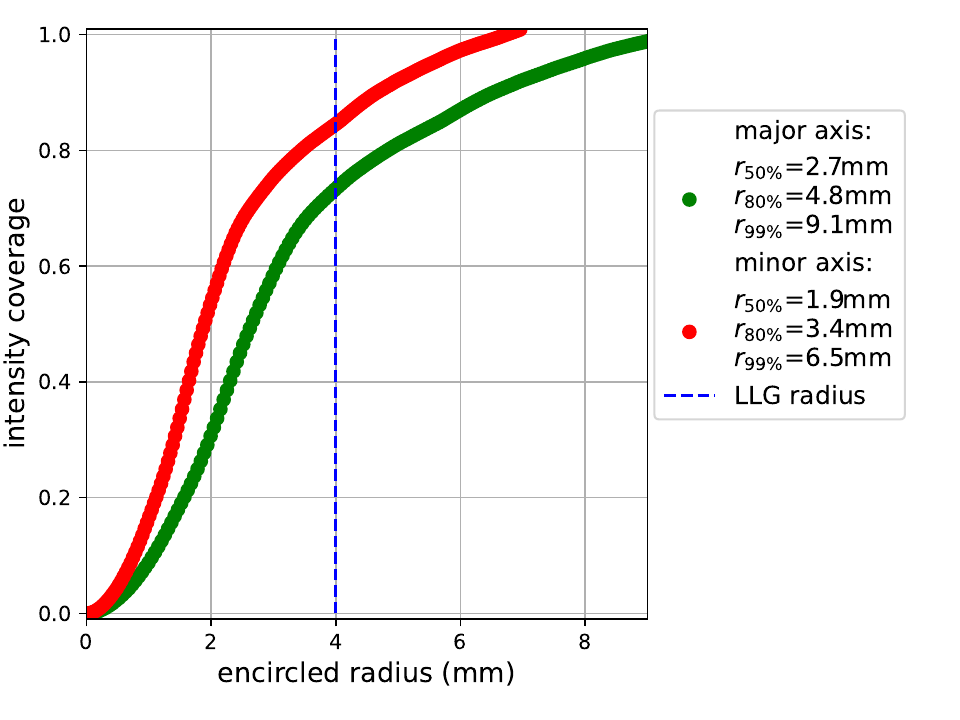}\\ %}
\includegraphics[trim={0.3cm 0cm 0cm 0.3cm},clip,width=0.48\linewidth]{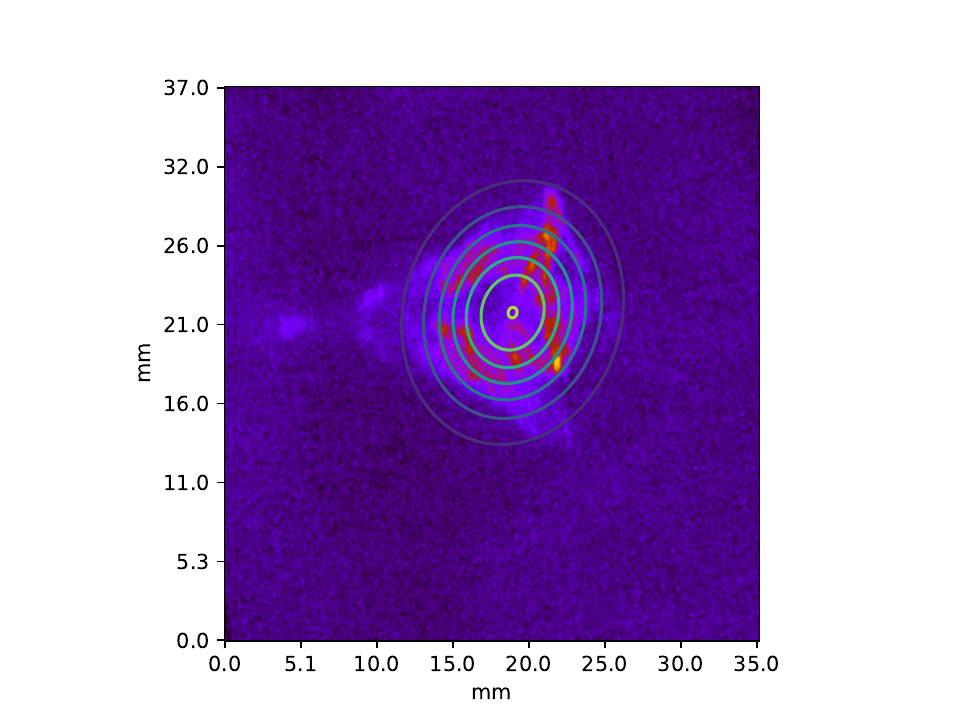}%} 
%\subfloat[Spot size profile]
\includegraphics[width=0.45\linewidth]{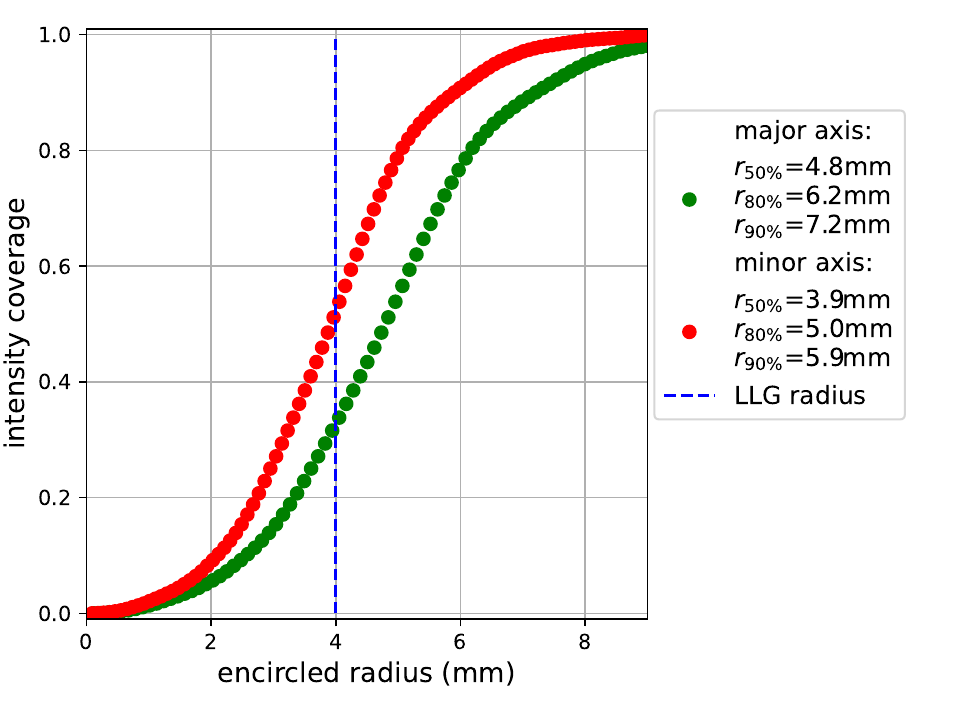} %}
%\subfloat[Spot size profile Alicia]
%\includegraphics[width=0.53\linewidth]{Figures/sec3/mirror_spotsize.png} %}
\caption{%Top: pBRL mirror spot size image. %and profile. 
\mgcol{Left: Spot size profiles with the elliptic contours of the two-dimensional Gaussian fitted to the image. Top: Image of a 65~m distant laser spot, from 2014. Below: Image of Venus, after the LIDAR had returned from its test campaign on La Palma. Note the different scales on the top and below plots. 
Right: Plots containing the relative amount of light enclosed within increasing-sized ellipses, characterized by a major axis (green) and minor axis (red), of the same rotation angle and aspect ratio, for each measurement respectively.  %Alicia. 
See the text for more details on the measurement.}
%\mg{Have to discuss these figures with Alicia} \sa{If we keep these figures they should be larger, they are illegible.}\aco{Yes.}
%\aco{Graph to be reproduced with larger font after the new measurement.}
\label{fig:mirror_psf}} 
\end{figure}

For the most accurate of our tests, we created an artificial light source by pointing a green laser at a wall 65~m from the telescope to observe the backscattered light. The size of the laser spot on the wall was $12\times24$~mm which corresponds to 0.2$\times$0.4~mrad at 65~m, similar to the 0.5~mrad of the actual pulsed laser employed by the pBRL. %(see Sect.~\ref{sec:laser}). 
%This setup simulates that of a parallel beam at infinity, and the eccentric image of the laser spot does not affect our results.
A diffusive millimetre paper was attached to the focal plane and a CANON ESO 1000D camera was placed 25~cm behind the paper.
\mgcol{The image was then fitted to a constant background plus a two-dimensional asymmetric Gaussian (see Eq.~\ref{eq:ellipse} in a similar discussion in Sect.~\ref{sec:laser}), characterized by two widths $\sigma_a$ and $\sigma_b$, and a rotation angle. We then calculated the relative amount of light enclosed in ellipses of the same rotation angle and aspect ratio, with increasing distance from the centre (Fig.~\ref{fig:mirror_psf}, top). We found that $\sim$80\% of the reflected light was contained in a pinhole of 4~mm radius (see Sect.~\ref{sec:llg}), roughly compatible with the measurements made by the CLUE collaboration~\citep{Alexandreas:1995}. Unfortunately, after the pBRL returned from its one-year test period at the ORM (see Sect.~\ref{sec:operation}), the point spread function had become significantly degraded (Fig.~\ref{fig:mirror_psf}, below), probably due to tensions applied to the mirror after inappropriate handling of the container by the various involved transport companies. }

%\begin{figure}[h!]
%\centering
%\includegraphics[width=13cm]{Figures/Polaris_IMG_9017.eps}
%\caption{ Measurement of the primary mirror PSF by using an artificial star. All light is enclosed in the liquid light guide perfectly. In this case, D{$_{90}$} $\sim$ 6.2 mm}
%\end{figure}

%\textbf{Total and focused reflectivity\\} 
%\mg{revisar + update}
%The focal reflectivity tells how much light the mirror reflects as a whole and projects it onto the focal plane. 
%This is a an important parameter for the performance of every telescope, because it plays a main role on the absolute scale of collected light.

\bigskip
\textbf{Mirror Reflectivity and re-aluminization}
The original reflective coating of the mirror was a 50--100~nm layer of aluminum vacuum evaporated at the Osservatorio Astrofisico of the University of Padova (Asiago, Italy).  After four years of operation, the mirror had lost reflectivity from originally 95\% to 50--60\%~\citep{Pesci:1995} due to a missing protective coating and was realuminized at the William Herschel Telescope\footnote{William Herschel Telescope (WHT) homepage: \url{https://www.ing.iac.es/Astronomy/telescopes/wht/}} (Observatorio Roque de Los Muchachos, La Palma, Spain). The mirrors had again considerably degraded in reflectivity when acquired. The surfaces appeared milky and dusty. Focused reflectivity measurements were made showing values of only $\sim$(60--70)\% at 350~nm~\cite{lopez:master}. The surface reflectivity was measured with a spectrophotometer. Many points in the mirror area were sampled locally, producing relative deviations $<$5\%. Focused reflectivity was measured instead by
estimating the sun light on a clear day concentrated on an $0.34 \mathrm{~m}\times 0.34 \mathrm{~m} \times 0.015 \mathrm{~m}$ aluminium plate placed in the focal plane and by evaluating the increase and decrease in plate temperature after opening and closing the mirror petals~\citep{lopez:master}. 
%md{Maybe add more after Alicia's Tesina}

Although the focused reflectivity might have been acceptable for the pBRL, we decided to realuminize the mirror and add a quartz-protective coating. The mirror was sent to the company ZAOT\footnote{\url{https://www.zaot.com/en/}} (Milan, Italy) in November 2020 for refurbishment, together with a second mirror belonging to IFAE and a third from the Laboratoire Univers et Particules de Montpellier~(LUPM, France). %Special wooden boxes and supporting systems were developed for the delicate transport. 
With a coating of 150~nm on a substrate of aluminium, after realuminization, we expected\footnote{Computed using the on-line tool \url{https://www.filmetrics.com/reflectance-calculator}} reflectivity values of 85\%, 87\%, 92\%, 90\% respectively for 355~nm, 387~nm, 532~nm and 607~nm for an angle of incidence of 10~degrees which represents the average tilt of photons impinging the mirror. The coating thickness was optimized for the best reflectivity. We explored thickness between 100 and 250~nm and found maximum differences (peak to peak) in reflectance of 10\%~\cite{victortfg}. 
%Overall, the procedure of coating mirrors of such a large size is lengthy, delicate and costly, because of the risk of breaking them. 

\subsection{Telescope Optics Design}
\label{sec:design_optics}
The CLUE telescope has a parabolic mirror surface of 1.8~m diameter
and $f/1$ with a PSF of about 5~mm diameter for 80\% containment (see also Sect.~\ref{sec:mirror}). This reflects into a telescope angular
acceptance (or telescope ``effective'' field-of-view FoV) of PSF$/f~=~6/1800 = 3.3$~mrad, which contains very well the light coming from the laser with a beam divergence angle of $0.5$~mrad.  

\begin{figure}[h!t]
 \centering
 \includegraphics[width=0.45\textwidth]{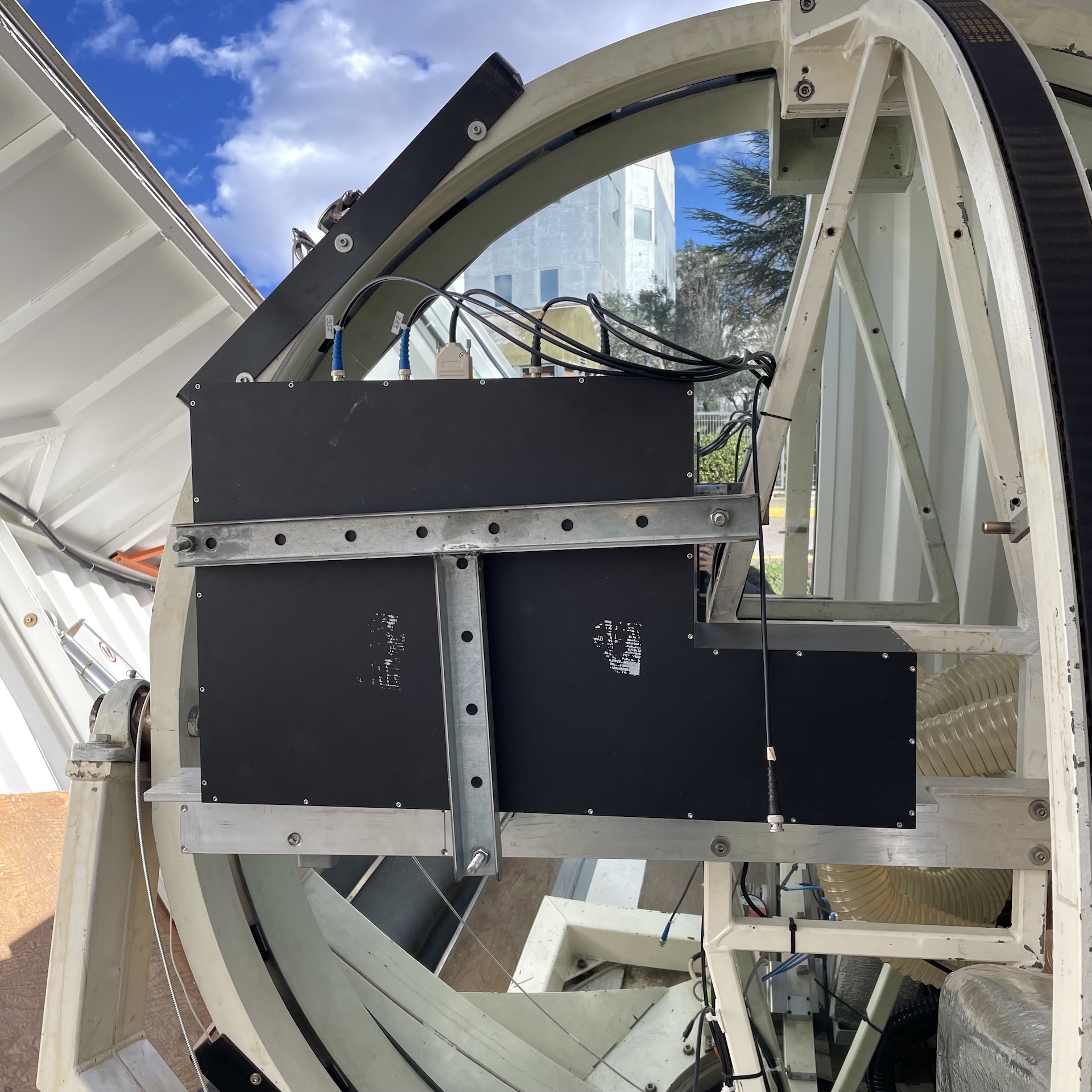}
 \caption{
 %\sa{I suggest a rewrite of the caption as it is now.} 
 The polychromator readout unit (black) is attached to the rear of the mechanical structure, behind the primary mirror.
   \label{fig:lidar_rear}}
\end{figure}

For the pBRL design, one possibility could have been placing
the polychromator unit directly at the focal plane, but this was discarded because the size of the polychromator would have caused significant shadowing on the primary mirror. Instead, the most appropriate place to allocate the polychromator appeared to be the rear of the lidar mirror, where the mechanical structure is already well adapted to hold devices, as shown in Fig.~\ref{fig:lidar_rear}. 

\begin{figure}[h!t]
 \centering
 \includegraphics[width=0.8\linewidth]{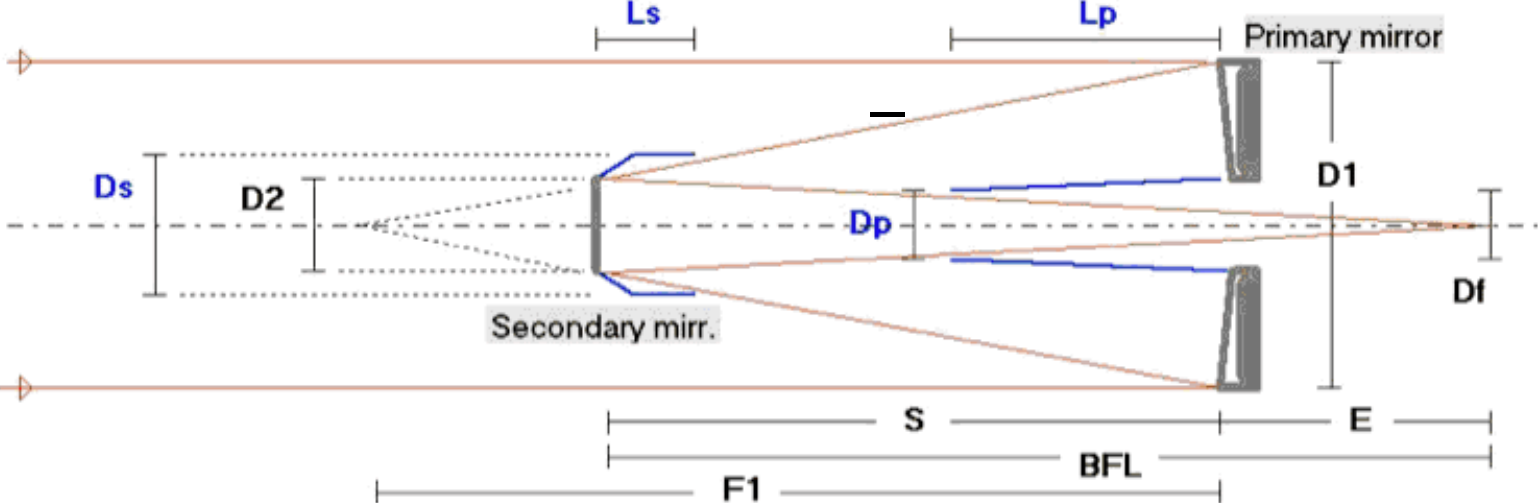}\\
%\subfloat[Cassegrain design with magnification M=8]{\includegraphics[width=0.7\linewidth]{Figures/sec3/cassegrain-m8.png}}
 \caption{Design scheme of a Cassegrain-like configuration for the
   CLUE telescope obtained with the public \texttt{ATMOS} code. All the input and derived parameters are shown in the Table~\protect\ref{tab:cassegrain}.  \label{fig:Cassegrain}}
\end{figure}

The CLUE mirror exhibits, in its centre, a drilled hole of 50~mm diameter. The hole may allow to design a double-mirror
Cassegrain-like configuration, which re-directs the light towards the rear of the primary mirror. This option has been explored in more detail. In that setup, the relatively small size of the hole may support a polychromator with a relatively small field-stop, which in turn allows to use moderately-sized optical elements and the use of normal-sized photosensors,  typical for lidars. We used a public version of \texttt{ATMOS}\footnote{\url{https://www.atmos-software.it/Atmos8_9.html}} that implements simple analytic calculations for the pre-design and later performed a full ray-tracing analysis using \texttt{ZEMAX OpticStudio}\footnote{\url{https://www.ansys.com/products/optics-vr/ansys-zemax-opticstudio}}. 
Fig.~\ref{fig:Cassegrain} shows two investigated designs of Cassegrain-like solutions. The input and output parameters of \texttt{ATMOS} are reported in Table~\ref{tab:cassegrain}.

\medskip
\begin{table}[h!t]
    \centering
    \begin{tabular}{cc|cc}
    \toprule
     & & $f/D=30$ & $f/D=8$  \\
         \midrule
Effective Focal Length & [m] & 54& 14.4 \\ 
Primary Focal Length F1 &  [m]&1.8 & 1.8\\ 
Primary Diameter D1 & [m] & 1.8& 1.8\\ 
Primary Focal Plane E & [cm] & 100& 100\\ 
Field Diameter D$_f$ & [mm]& 50& 50\\ 
\midrule
Primary Curvature Radius & [mm] & -3600 & -3600\\ 
Secondary Curvature Radius & [mm] &-127 & -483\\ 
Primary-Secondary distance & [mm] &-1739 & -1589\\ 
Secondary Mirror Diameter & [mm] & 63 & 217\\ 
Magnification & & 30 & 8\\ 
Back Focal Length & [mm] & 1839 & 1689\\ 
Linear Obstruction & \% & 3.5 & 12.0\\ 
Aperture Ratio & & 30 & 8\\ 
Light Shield Length Ls & [mm] & 48.5 & 53.8\\ 
Light Shield Length Lp&  [mm]& 859& 824\\ 
Light Shield Diameter Ds& [mm]& 112& 271\\ 
Light Shield Diameter Dp& [mm]& 57 & 144\\ 
\bottomrule
    \end{tabular}
    \caption{Input and output parameters for the optical Cassegrain design obtained with the public \texttt{ATMOS} code.\label{tab:cassegrain} }
\end{table}

Despite Cassegrain-like configurations being useful to make compact
telescopes and reduce spherical aberration, they act like a zoom into the FoV of the primary mirror, i.e. they magnify the system: the spot size is seen enlarged with respect to the PSF of the primary mirror alone. Two magnification designs (8 and 30) are calculated in Table~\ref{tab:cassegrain} as the ratio between the focal point of the system to the secondary distance $S+E$ and the distance from the secondary
surface to the focal point of the primary $F1-S$, or the ratio between the
effective focal length $F_\mathrm{eff}$ and the focal length of the primary mirror $F1$:
\begin{equation}\label{eq:magnification}
M=\frac{S+E}{F1-S}=\frac{F_{eff}}{F1}
\end{equation}
In the design with magnification $M=30$, we attempted to minimize obscuration by having a small secondary mirror, thus providing a final PSF of about 18~cm, which not only would have required additional optics or a too-large polychromator unit but also would have implied considerable loss of light given that the central aperture on the primary mirror is only 5~cm. Finally, the diameter of the secondary mirror would have been 6~cm in this configuration, causing additional light shadowing of 3.5\%. 

Reversing the argument: To obtain an image of the primary PSF smaller than the rear hole, the magnification of the secondary mirror must not exceed $M \sim 8$ and therefore the effective focal length should not exceed about 15~m. In this configuration, the diameter of the secondary mirror would be as large as 22~cm and hence produce a linear obstruction of about 12~\% of the light. In addition, the FoV of the telescope would be as large as $D/f_{\rm eff} = 1.8/15 = 12$~mrad, probably covering a nonoptimal, too large fraction of the sky. Given the problems in the two limit configurations considered and the additional mechanical problems related to the aligning of both mirrors in a stable way, in such a light structure as the one of the CLUE telescope, we decided to discard the Cassegrain-like configuration. 

The alternative solution chosen consists of placing an optical fibre to collect light in the focal plane and transmit it to the
polychromator unit mounted at the rear of the telescope (Fig.~\ref{fig:lidar_scheme2}). Given the size of the PSF and the dimensions of the telescope, a thick optical fibre of at least 6~mm diameter and over 3~m length was needed. 
Given that $f/D=1$, a fibre with a numerical aperture (\gls{na}) greater
than $0.45$, with good transmission properties, was chosen in the
wavelength range between 350~nm and 600~nm. The actual choice for this fibre and the optical elements required for its use with the polychromator system are deferred to Sect.~\ref{sec:polychromator}. 

\subsection{Coaxial Laser Beam}
\label{sec:biaxiality}
LIDARs can be built in a coaxial (i.e. the laser beam is coincident with the telescope's optical axis) or a biaxial design (the laser
beam moves parallel to the optical axis of the light-capturing telescope). Coaxial LIDARs require dedicated steering optics to guide the laser beam towards the optical axis of the telescope. However, biaxial lasers require less hardware to align the laser beam and are easier to design at the expense of larger light loss for close-range sensing. 

The range at which the atmosphere can be sensed is determined, at the near end, by the \emph{range of full overlap} between the laser beam and the telescope's field of view. The starting range of full overlap determines the minimum sounding range in the atmospheric boundary layer, a key design parameter for a lidar. It is usually lower for a coaxial lidar and higher for a biaxial lidar, typically several hundred meters to 1~km~\citep{Biavati:2011}. 
Below the full overlap range, the overlap function becomes smaller than one and can in principle be corrected~\citep{wandinger:2002}, however, the corrected data become increasingly noisier. 

For a perfectly aligned biaxial lidar, the following formula can be used to estimate the distance of the full overlap range $R$~\citep{Biavati:2011}:

\begin{equation}
%f \cdot \frac{d_0}{R} + f \cdot (\frac{r_0}{R} + \theta) = D/2 
%\frac{d_0}{R} + \frac{r_0}{R} + \theta = \frac{D}{2f}
%\frac{d_0+r_0}{R} = \frac{D}{2f} - \theta
R = \frac{2x+d}{D/f - 2\theta}  \qquad,
\label{eq:R}
\end{equation}

where $x$ is the distance of the laser to the centre of the telescope mirror, $d$ is the primary mirror diameter, $D$ the radius of the focal plane detector,  $f$ the focal length of the telescope ($D/f$ is also called the \emph{field of view} of the detector). $\theta$ is the opening angle of the laser beam in radians. 
Note that in the case of $x=0$, $R$ does not get reduced to zero, due to the confusion circle of the source image, which becomes large at small distances if the telescope is focused at large distances or infinity. 
With our setup, with the assumption of a perfectly aligned biaxial laser located just at the edge of the telescope at $x=1$~m), with $d=1.8$~m, $D=4$~cm, $f=1.8$~m, $\theta\approx0.5$~mrad, Eq.~\ref{eq:R} yields $R\approx$~200~m. 
Nevertheless, also taking into account the aberrations of the telescope (not reflected in Eq.~\ref{eq:R}), makes a minimum achievable range of 400--500~m more realistic. 

In the case of a coaxial lidar, $R$ reduces to 80~m and 150~m for without and with aberrations, respectively. 
The choice of a coaxial system, as chosen for the pBRL, thus opens the range from 150~m to 500~m above ground for atmospheric characterization. Note that at astronomical sites during clear nights,  the most important aerosol contributions are found within the first 500~m above ground. These aerosols are usually not turbulently mixed~\citep{Fruck:2022igg}, and hence a constant aerosol density does not correctly model the aerosol extinction profile. %The light from air showers of all energies is equally absorbed by this layer, resulting in common shifts of reconstructed energy. For this reason, the boundary layer should be monitored precisely and we opt for a co-axial lidar system.

\subsection{Liquid Light Guide} %\md{Markus+Oscar to revise after final numbers from absolute measurements with the lasers}
\aco{To revise after final results from absolute measurements with the lasers.}
\label{sec:llg}
%\md{There was a discussion with Vincenzo about fluorescence light. There was measurement by Merve which showed this effect especially for 355 e 387 lines, but of the order of $1e-7$, so almost negligible. Check Alicia's thesis. If this discrepancy of order 10\% is confirmed, it's an important fact that should be considered in data-reconstruction. 
%If really the transmission depends on T, we should consider doing an absolute calibration as function of T during data reconstruction as done in Fruck for MAGIC.
%}

\begin{figure}[h!t]
\centering
%\subfloat[LLG pic]
    \includegraphics[width=0.32\textwidth]{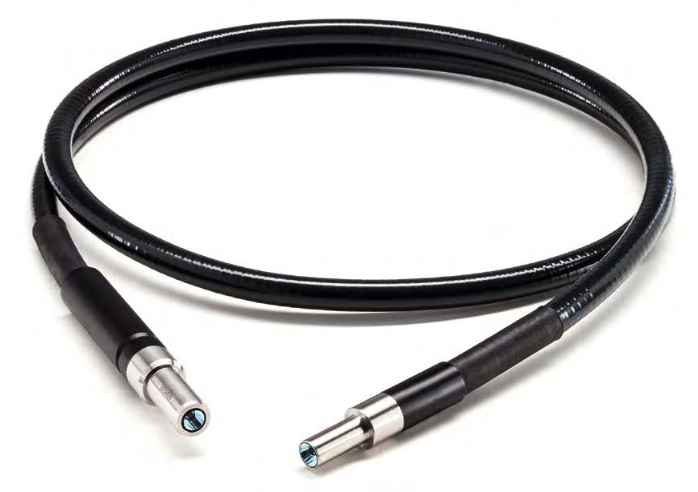} \hfill %}
%\subfloat[LLG principle]
    \includegraphics[width=0.32\textwidth]{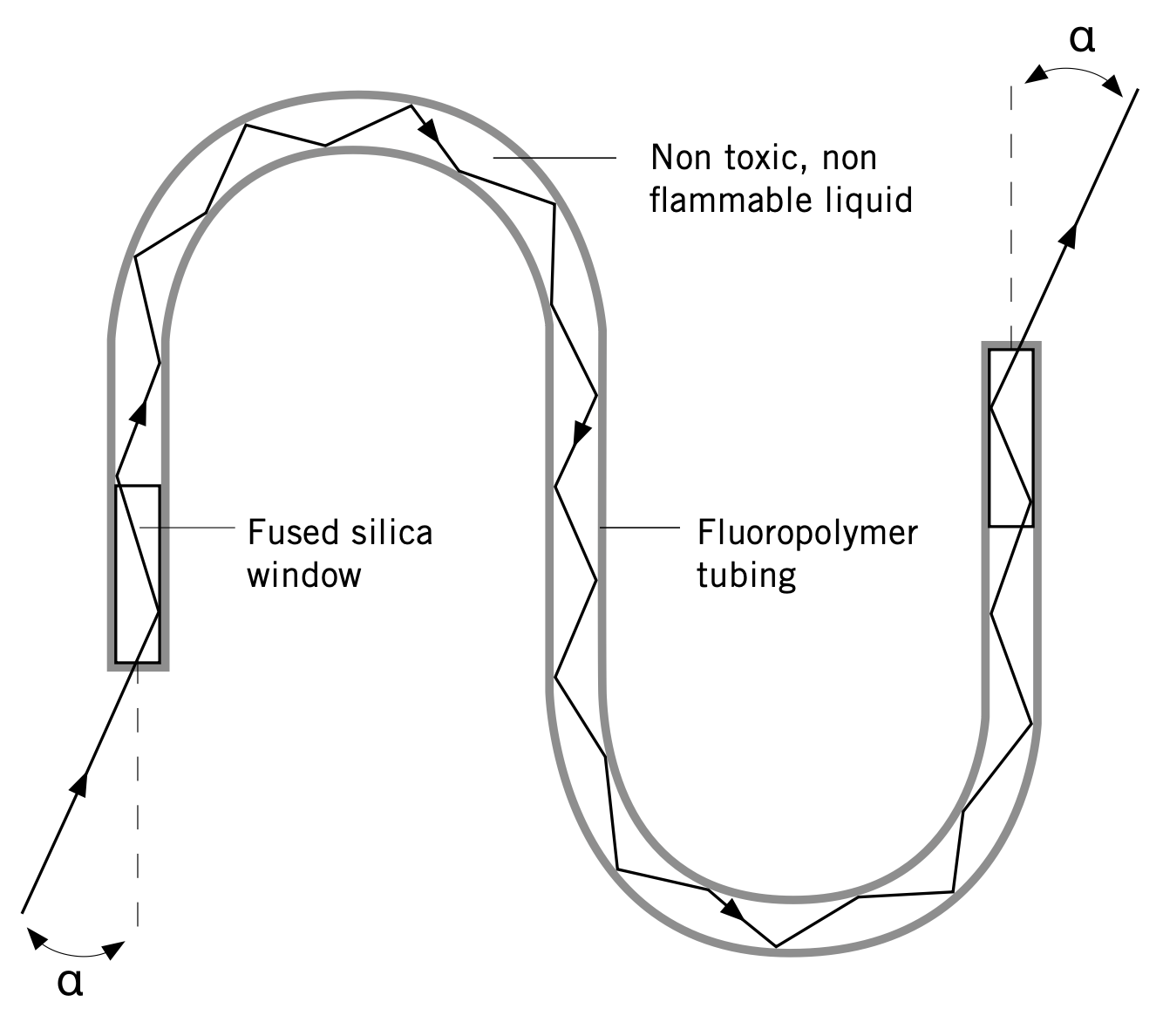} \hfill %} 
%\subfloat[LLG size]
    \includegraphics[width=0.32\textwidth]{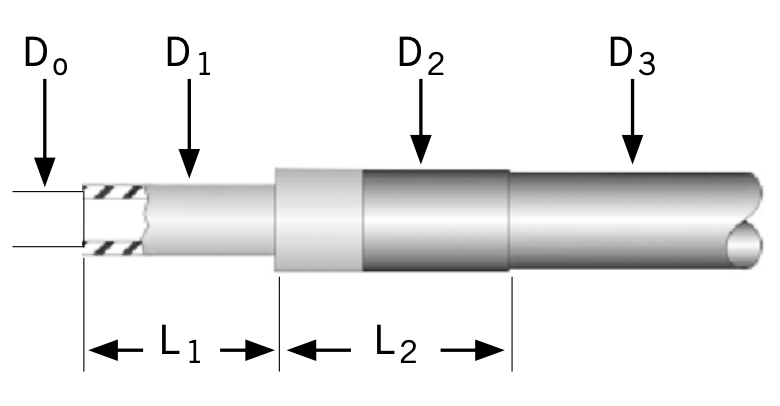} \hfill %}
%\subfloat[Dimensions of the end fittings of the LLG: Active core (d0): 8~mm, end fittings: (d1): 10~mm, (l1): 20~mm, (d2): 15~mm, (l2): 40~mm, protective sleeve (d3): 12.5~mm.]{
%\includegraphics[width=0.72\textwidth]{Figures/sec3/Ftgs2.pdf}}
    \caption{The Lumatec series 300 liquid light guides (LLG). The principle is illustrated in the central plot. The dimensions of the end fittings of the LLG: Active core (d0): 8~mm, end fittings: (d1):~10~mm, (l1): 20~mm, (d2): 15~mm, (l2): 40~mm, protective sleeve (d3): 12.5~mm. All images from \url{https://www.lumatec.de/}. \label{fig:llg}}    
\end{figure}

A liquid light guide (LLG, see Fig.~\ref{fig:llg}) of type Lumatec Series 300\footnote{\url{https://www.lumatec.de/en/products/liquid-light-guide-series-300/}}, with fused silica window and fluoropolymer tubing, of 8~mm diameter and 3.2~m length, is used to transport the light from the focal plane to the polychromator unit.
%(see Fig.~\ref{fig:lidar_scheme2}. 
This model is optimized for the spectral range from 320~nm to 650~nm. 
%, within which radiation is transmitted without optical degradation. 
The LLG has a numerical aperture (NA) of 0.59 (acceptance cone with a diameter of 72$^\circ$, see Fig.~\ref{fig:llg}). 
%\mg{This is very confusing: NA is normally given as a unitless number and for the half angle. This number instead is given in degrees and for the full angle. I checked in Lumatec's brochure that the value is correct and the definition is 2$\alpha$. Nevertheless, we should use a common and unambiguous definition.}
The liquid inside the LLG remains stable for many years if the LLG is not exposed to radiation with wavelengths below 320~nm or above 650~nm. 
Shorter wavelengths may destroy the transmission properties of the liquid, whereas strong illumination by longer wavelengths may overheat the liquid and cause bubbles.
Too sharp bending should be avoided; otherwise, the optical tube may kink and transmission will degrade permanently.
%by a minimum of 20\%. 
The minimum bending radius is about 100~mm, coincident with the mechanical limit. 
The optical transmission may also degrade through improper handling, i.e. kinking, overheating, or by exposing it to a vacuum, arid climate, etc. 
%Hence, proper shipment, storage and operation conditions are
%required. 
%Under normal conditions, the LLG operates for many years without degradation, as we verified with 10 years of operation. % of its transmission properties.

%Transmission is a property of the material itself, while transmittance takes into account the material’s thickness and the wavelength of the light being transmitted. 

\begin{figure}[h!t]
\centering
%\subfloat[LLG Linear transmittance.]{\label{fig:llc_transmittance}
%\includegraphics[width=0.45\linewidth]{Figures/sec3/llc_transmission.png}
\includegraphics[width=\linewidth]{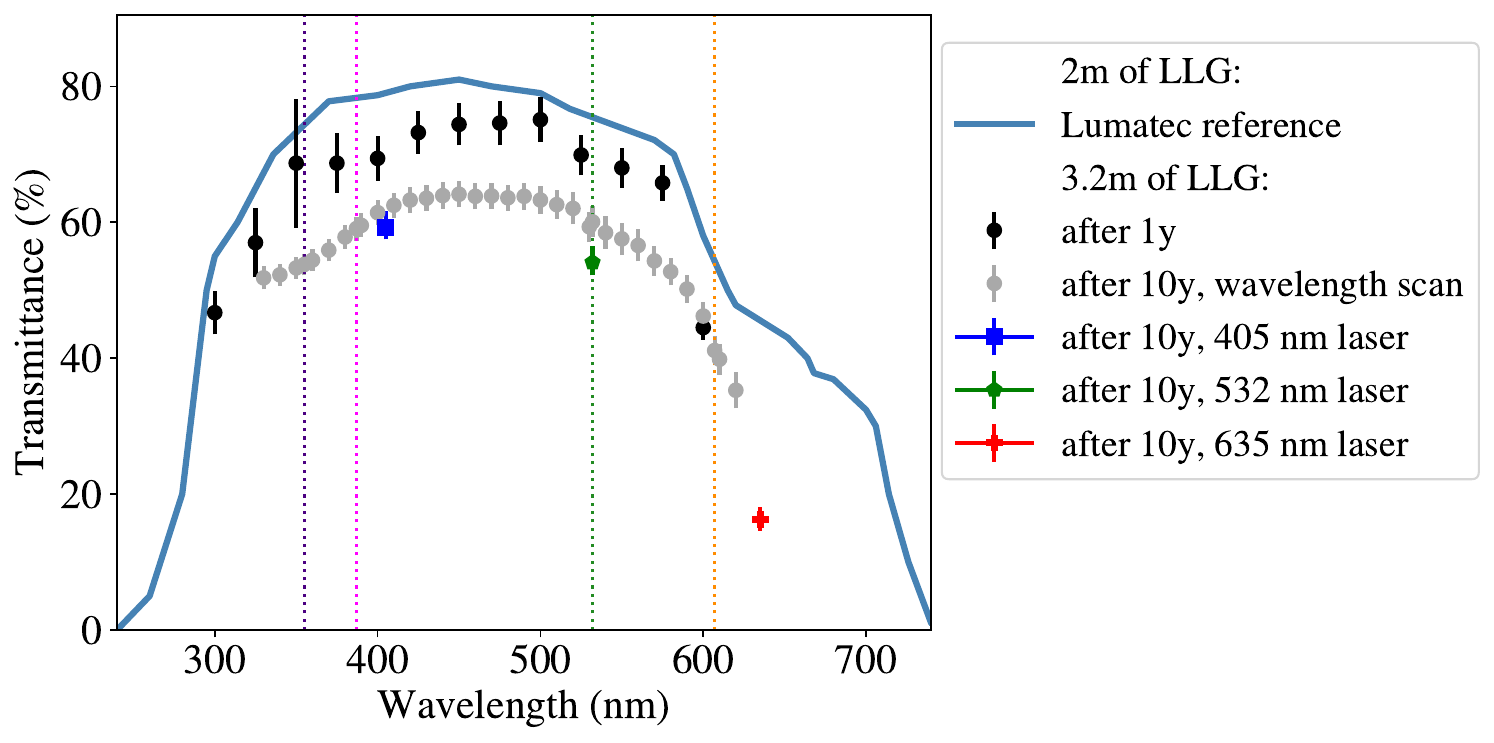}
     \caption{%LLG Linear Transmission
    % Could be transmitivity, but Lumatec uses Transmission
    Linear transmittance of the LLG. The solid blue line is taken from the datasheet of Lumatec for a reference LLG length of 2~m. Black points refer to our measurements for a LLG length of 3.2~m \sapocol{after one year from its purchase and in grey after 10~years of use.} \sapocol{The blue, green and red markers indicate a separate measurement carried out with continuous wavelength lasers of 405~nm, 532~nm and 635~nm}. %\mg{Plot needs to be updated with error bars on these two points after analysis from Julia and Oscar.}
    The dotted coloured lines indicate the four wavelengths used for the pBRL.
       \label{fig:llglinear}}
\end{figure}

\bigskip
\textbf{LLG linear transmission}
Peak transmission values of up to 80\% can be obtained for LLG reference lengths of 2~m (see Fig.~\ref{fig:llg} (left)). We carried out transmittance measurements with the 3.2~m LLG using a deuterium source, a Minolta CM100 monochromator and a Newport 818-UV light sensor in a dark room under controlled temperature and humidity~\cite{sanchez:tesina}. The transmittance was measured from 300~nm to 600~nm
%in steps of 25~nm, 
in ON/OFF mode. The measurement was repeated 10~years after the purchase of the LLG. Fig.~\ref{fig:llglinear} shows an overall agreement with the reference Lumatec data (measured at 2~m fibre length) and our measurements.  Transmittances of $\sim$70\% are found in the relevant wavelength range from 350~nm to 550~nm, and $\sim$45\% for 607~nm. After ten years of operation, the transmittance has reduced to 50\%-60\% for the UV and green wavelengths, but it has remained stable at 607~nm. 
%which reduces to about 50\% outside this range.
%\aco{Need to add the new measurements and compare them. Show separate plots for measurements of transmittance (old + new) and angular+temperature dependency transmissivity. Markus wants to repeat the absolute measurements for the transmittance. }

%\subfloat[LLG Angular-dependent transmissivity.]{\label{fig:llg_transmittance_angle}
%\includegraphics[width=0.485\linewidth]{Figures/sec3/llg_transmission_angle.png}
\begin{figure}[h!t]
\centering    \includegraphics[width=0.6\linewidth]{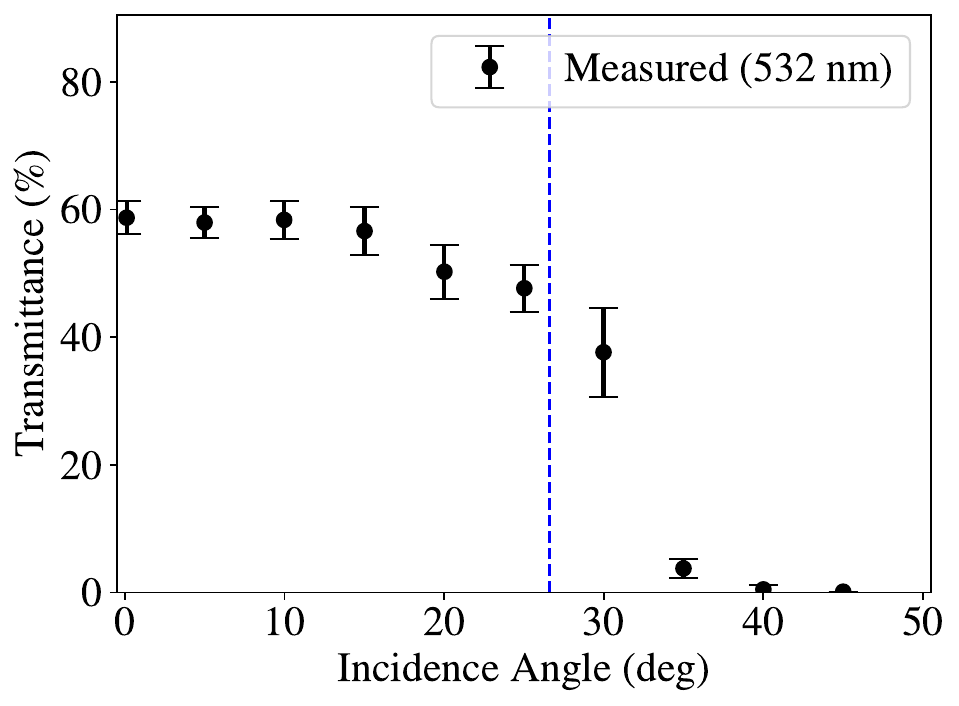} \hfill 
%}
  \caption{
     Measured angular dependency of the transmissivity of green light through the LLG. 
    The dashed blue line shows the maximum incidence angle of light from the primary mirror for reference.
    %\aco{Plots to be updated after final results from the absolute measurements. }
    \label{fig:llgangular}}
\end{figure}

\bigskip
\textbf{LLG angular transmission} 
After testing the transmissivity of the LLG with respect to the incident angle. For these measurements, a class III green laser pointer has been used as a light source. The laser is attached to a rotating support and placed at the entrance of the LLG. The output is measured from 0$^\circ$ to 60$^\circ$, see Fig.~\ref{fig:llgangular}. It has been found that transmission from 15$^\circ$ to 30$^\circ$ decreases slightly and decreases by one order of magnitude from 30$^\circ$ to 35$^\circ$~\cite{sanchez:tesina}. This is sufficient for the design of the pBRL, since the maximum incidence angle of light reflected by the mirror toward the LLG is exactly 26.6$^\circ$. 
%\aco{Maybe it would be nice to add a line/square at 26,6 deg in fig 10 (right).}

\bigskip
\textbf{LLG temperature-dependent transmission} 

\begin{figure}[h!t]
\centering
\includegraphics[width=0.6\linewidth]{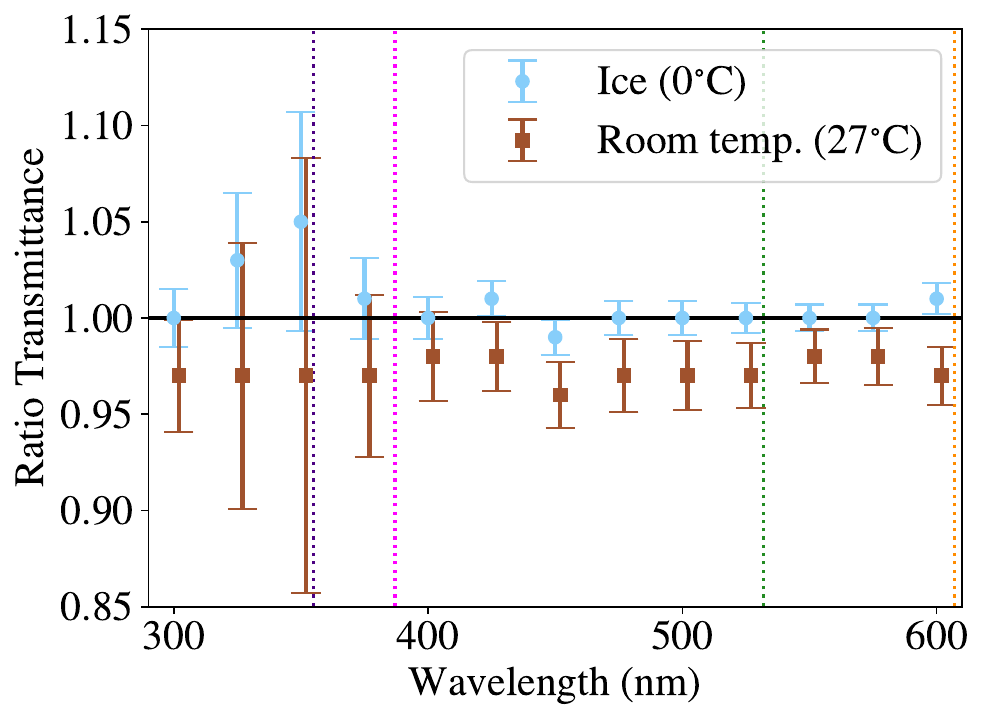}
\caption{\label{fig:llc:temp}Ratio of linear transmittance concerning original one at room temperature, as a function of wavelength. In blue, the transmittance ratio at 0$^\circ$C is shown, whereas in brown the ones obtained after reestablishing room temperature. The error bars show statistical uncertainties only. \sapocol{The dotted coloured lines indicate the four wavelengths used for the pBRL.}}
\end{figure}

For this test, the linear transmittance of the LLG was measured several times at room temperature, in order to obtain a reference for each wavelength. The LLG was then cooled to zero degrees with the help of ice and again measured several times. Finally, the LLG's transmittance was measured again at room temperature. Figure~\ref{fig:llc:temp} shows the transmittance ratios obtained for both temperatures. Although the cooling process did not affect the transmittance, reestablishing room temperature seemed to have reduced the transmittance by about 3\%. 
%\mg{Have to check with Alicia, whether I understood the data correctly. Should also calculate a systematic uncertainty of this measurement. I suppose the LLG was moved before and after measurements and the ice.}

%\md{This part is very relevant, but its description in Alicia's thesis is very unconclusive. Data should be reanalyzed, but hints are of very strong temperature dependence}

\bigskip
\textbf{LLG angular deviation and stability}
The relation between the exit angle and the input angle for the LLG was measured using the same system as for the angular dependence of the transmissivity. A positive linear dependence was measured between these two angles, with a relation $\alpha_{out}=(-1.8\pm1.8)+(1.09\pm0.07)\cdot\alpha_{in}$, consistent with expectations.

%showing that the LLG increases the divergence of the incident beam of a factor $0.2$. This relation is important for the design of the optical bench. The stability of the LLG output was measured also by tilting and bending the LLG with no significant effect measured.
\bigskip
\textbf{LLG fluorescence emission}
LLGs can show fluorescence light emission shifted by \break (3474$\pm$375)~cm$^{-1}$ \cite{Wiencke:2017} when illuminated by strong UV light. In our case, this would result in undesired secondary emission lines of (405$\pm$6)~nm 
% see slides from V. rizi, "experimental_setup...pdf", page 7
% Double_t l1 = 354.8E-9*1E2
% 1./(1./l1 - 3474)
% 1./(1./l1 - 3474 - 375)
% 1./(1./l1 - 3474 + 375)
and (653$\pm$16)~nm. 
% Double_t l1 = 532.3E-9*1E2
% 1./(1./l1 - 3474)
Although no Raman line is found within these ranges, undesired fluorescence light leakage into these channels must be carefully avoided. 
% 355 nm = 28169.01 cm–1
% 24695 cm–1 = 404.94 nm
% 25070 cm–1 = 398.88 nm
% 24320 cm–1 = 411.18 nm
% 532 nm = 18797 cm-1
% 15323 cm–1 = 652.61 nm 
% 15698 cm–1 = 637.02 nm
% 14948 cm–1 = 668.99 nm

\begin{figure}[h!t]
 \centering
 \includegraphics[width=0.4\textwidth]{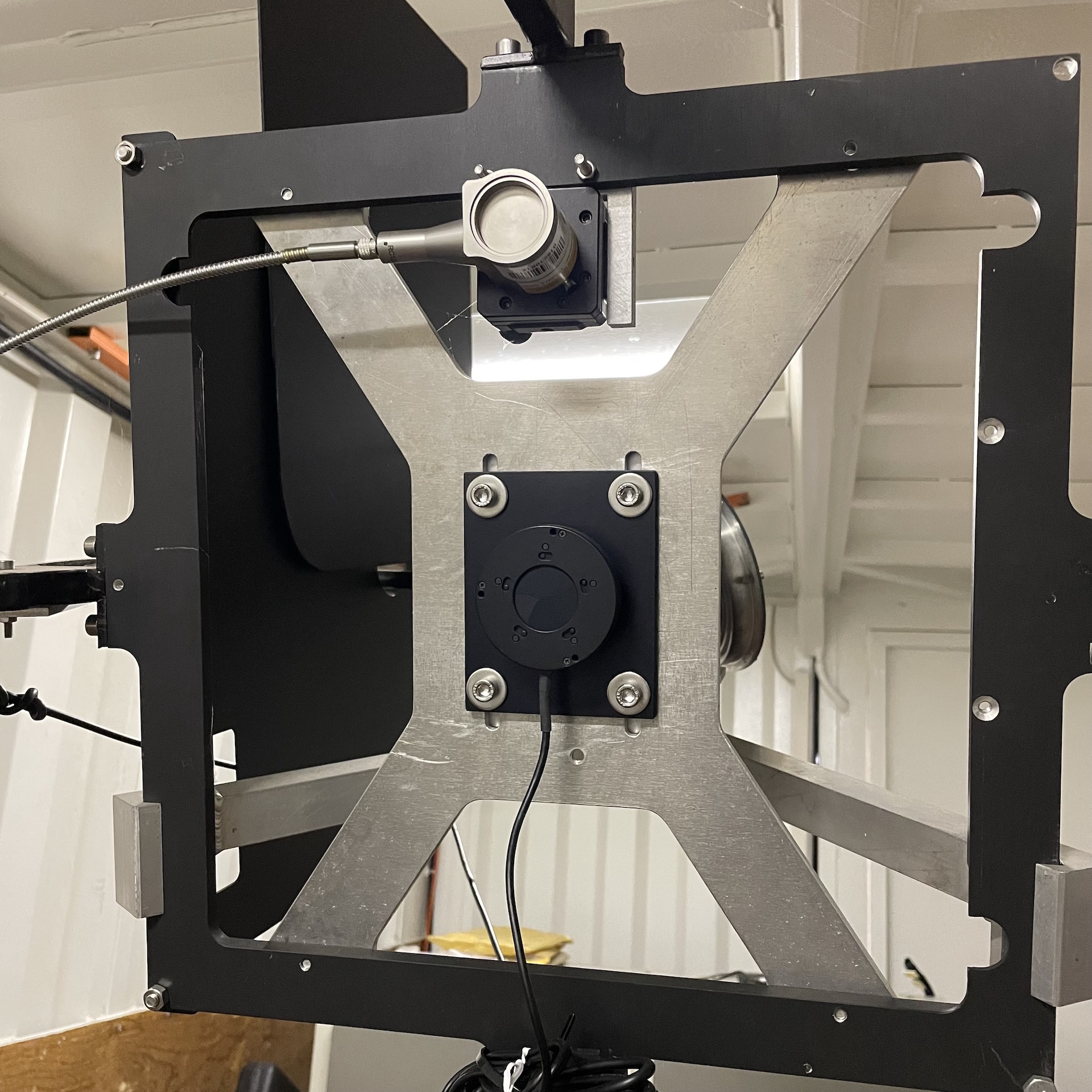}
 \caption{
 %\sa{I suggest a more descriptive caption here.} 
 Picture of the LLG shutter system installed in the center of the focal plane of the lidar. The metallic structure holds the shutter in the center. On the top part of the picture, the rear side of the near-range optics can be seen.
 \label{fig:llg_shutter}}
\end{figure}

\bigskip
\textbf{LLG shutter} \label{sec:shutter}
\sapocol{The pBRL polychromator is operated such that the PMT high voltages (\glspl{hv}) are ramped down whenever operations are finished, which is the case when ambient light levels become too high for the safety of the PMTs.} In order to \sapocol{additionally} protect the entrance of the LLG at the focal plane from spurious light and dust during non-operational times (e.g., during the day), a shutter system was included (see Fig.~\ref{fig:llg_shutter}). We designed a remote-controlled shutter that can be opened at the start of operations and closed at the end. We used the commercial Thorlabs stainless steel diaphragm optical beam SHB1, 1$^{\prime\prime}$ diameter, equipped with a remote controller with TTL inputs and interlocks, and coupled to a
1$^{\prime\prime}$ long SM1S10 lens tube spacer. The system is mounted on a plate placed in the focal plane in front of the LLG entrance. The shutter can be operated up to a rate of 15~Hz, and is guaranteed up to 15~Mcycles. \sapocol{While the operating temperature specified by the manufacturer of the shutter is between 15$^\circ$C and 40$^\circ$C, the shutter was tested to perform flawlessly during a commissioning period at La Palma even at much lower temperatures.}
%and operation between 15$^\circ$C and 40$^\circ$C. 
The LLG shutter has an angular acceptance of 30~degrees, which nicely coincides with the maximum angle under which the primary mirror is viewed, thus further reducing the stray light.

\begin{comment}
Samo said:
While the operating temperature specified by the manufacturer of the shutter is 15-40 degrees, the shutter was tested to perform flawlessly during the 1.5-year-long commissioning period of the BRL at La Palma even at much lower temperatures, so we are confident it is adequate for the task. Despite extensive testing of this particular model we search for an alternative device with more suitable design specifications and test it.

The shutter opens and closes the LLG inlet in the focal point of the primary lidar mirror. Its purpose is to protect the LLG and the PMTs from ambient light, in particular, if the PMT high voltage is ramped up and the lidar is not in operation (container closed, daytime testing, etc.). While the potential not closing of the shutter does not compromise safety, it could cause damage to the PMTs. Not opening of the shutter on the other hand would prevent lidar from receiving back-scattered signal.
\end{comment}
%\md{Missing figure} 

\subsection{Optical Bench Polychromator}
\label{sec:polychromator}

The optical bench (hereafter the polychromator) must collect and collimate the light transported by the LLG and successively separate the different wavelengths using dichroic mirrors and narrow-band filters. All glasses must be transparent to the four wavelengths used, particularly in the UV regime. Fused silica glass or N-BK7 are both good choices, while flint glasses normally used to design achromatic doublets cannot be adopted due to their poor UV transmission.
The drivers for the design of the polychromator unit were: a)~to cope with the large aperture of the LLG, b)~to maintain optical elements within dimensions that are easily commercially available while limiting their number, and c)~to maximize the collective effectiveness and
global efficiency. Design choices were also driven by cost
issues.

\begin{figure}[h!t]
 \centering
 \includegraphics[width=0.8\linewidth]{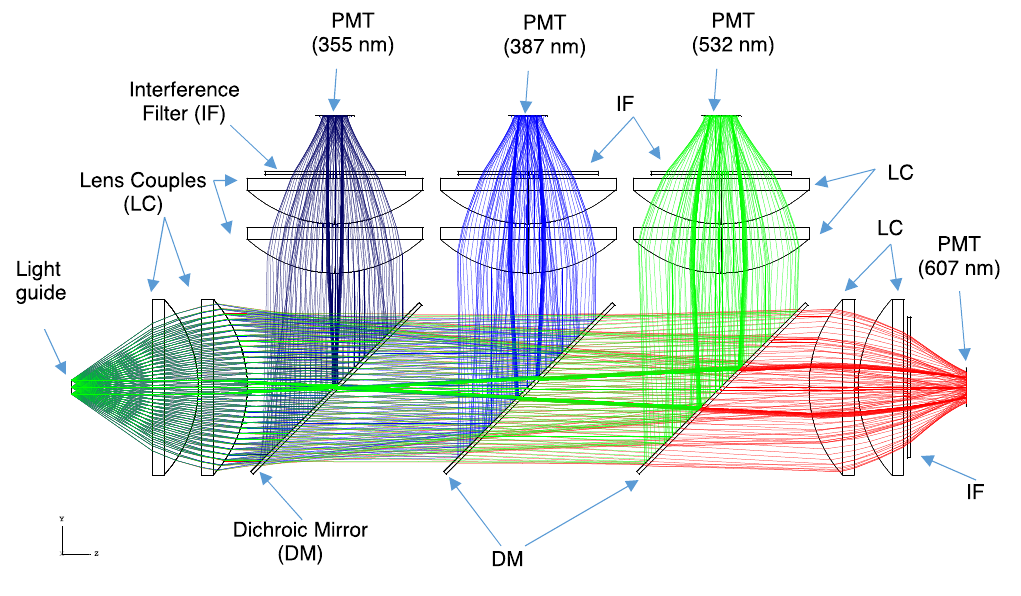}\\
 \scalebox{-1}[1]{\includegraphics[width=0.7\linewidth]{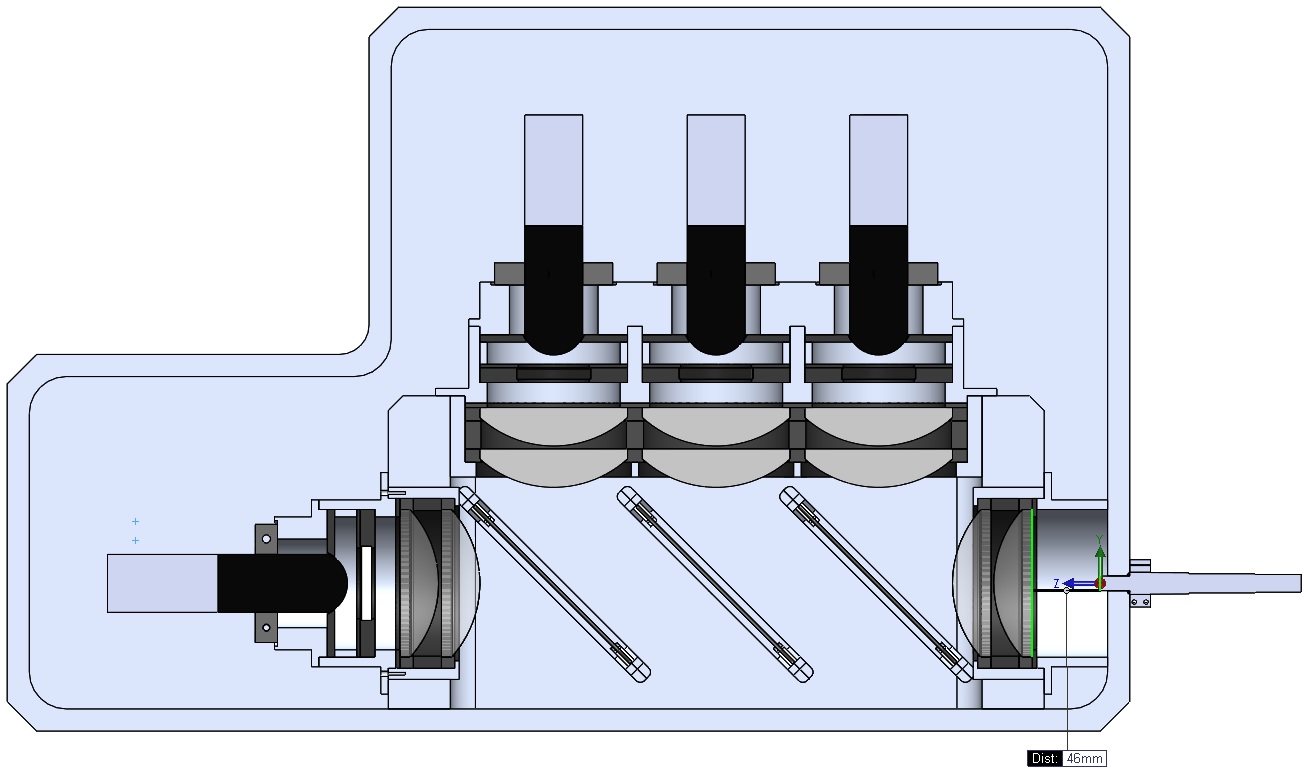}}
 \caption{The polychromator \texttt{ZEMAX} design and sketch of the optical bench. %interior. 
 The dichroic mirrors, the interference filters, and the converging lens doublet system are shown. The design was described in \citet{Deppo:2012}. %\aco{I would remove the bottom pic, as this "scheme" of the polychromator is already shown in Fig. 17. Moreover, Fig 13 bottom is not mentioned over the text.} \md{I mildly disagree, the bottom plot picture well how the design is implemented, but I leave this to MG}
 \label{fig:polychromator}}
\end{figure}

\bigskip
\textbf{Polychromator design}
We therefore defined the baseline design of the pBRL for four read-out channels: two elastic, at 532 and 355~nm, and two Raman at 387 and 607~nm. %Successively, we deferred the 607~nm channel until 
%because of the unavailability of 
%a suitable PMT, sensitive to that wavelength, could be found. 

\begin{table}
\centering
\begin{tabular}{p{5cm}|p{6cm}}
\toprule
\multicolumn{2}{l}{Input specifications} \\
\midrule
Input FoV              & 70$^\circ$ \\
Input source diameter  & 8~mm \\
%Encircled Energy (355 nm) & $>70$\% inside the detector \\
%Encircled Energy (other wavelengths) & $>80$\% inside the detector \\
Wavelength coverage    & 355/387/532/607 nm \\
%Filters                & 4 filters: one per each sub-channel \\
Single photosensor size & 1.5$^{\prime\prime}$ hemispherical PMT \\
\midrule
\multicolumn{2}{l}{Design solutions} \\
\midrule
Optical concept &  Sequential beam filtering with dichroic mirrors, plano-convex lens couples and interference filters.\\ 
Single lens diameter &  100~mm \\
Single lens focal length & 150~mm \\ %F/1.5 \\ %\md{Is this obtained by previous two rows? Should be 60/100?} \\
Single lens curvature radius & 74.5~mm \\
Single lens thickness & 26~mm \\
Material & N-BK7 \\
Single lens transmission at 355~nm & 0.95 \\
Lens couple focal length & 60 mm \\
Distance between two lenses of couple & 2~mm \\
Distance between LLG and first lens couple & 46~mm \\
%Detector & PMTs \\
%Dichroic mirror band pass & 10~nm \\
Filter band pass & 10~nm \\  
\bottomrule
\end{tabular}
\caption{\label{tab:polychromatorreqs} Requirements and design specifications for the optical design of the polychromator unit.}
\end{table}
%\mg{To be cross-checked in Rafa's designs. }

\begin{figure}[h!t]
\centering
    \includegraphics[width=0.325\linewidth]{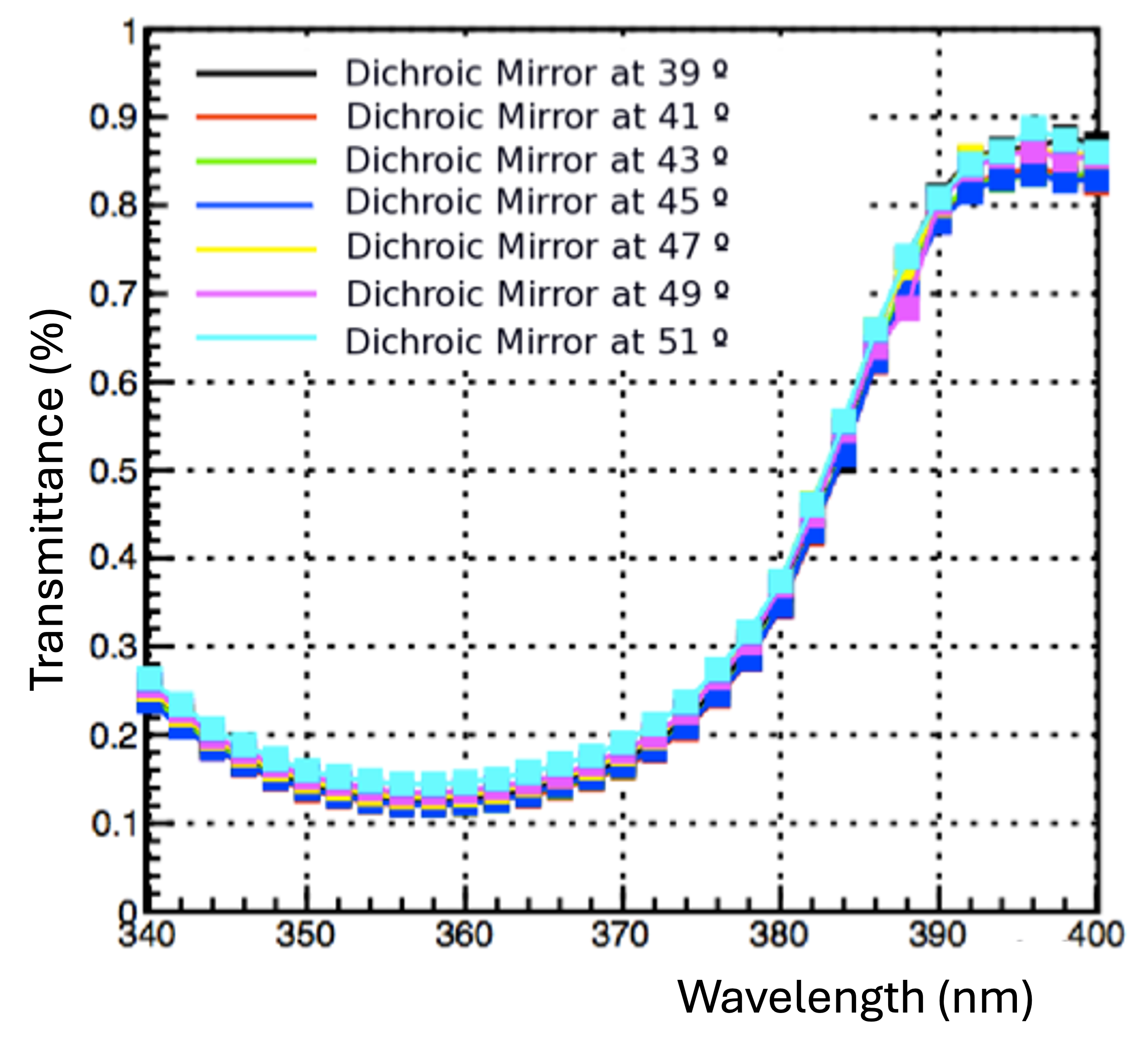} \hfill
    \includegraphics[width=0.325\linewidth]{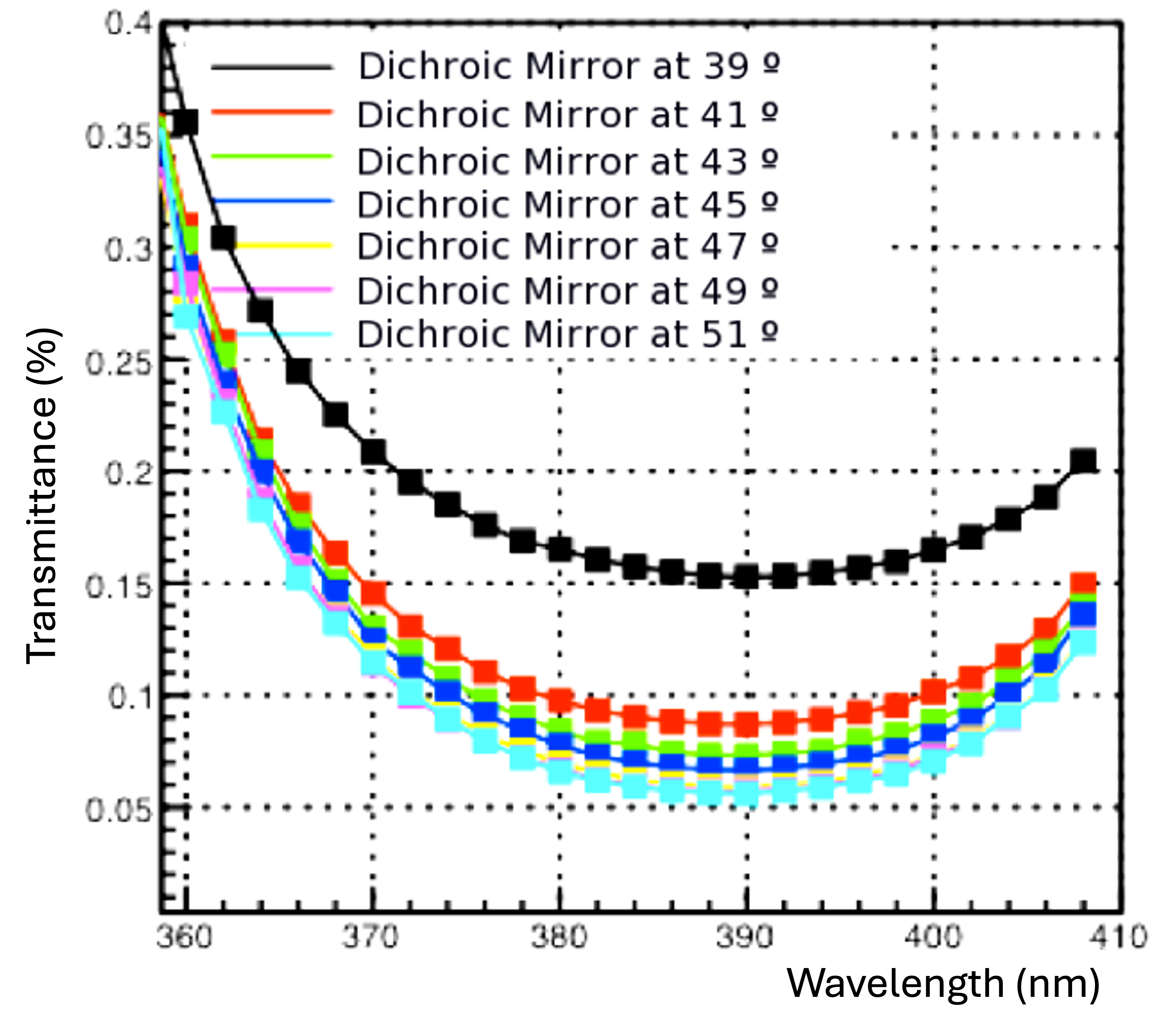} \hfill 
    \includegraphics[width=0.325\linewidth]{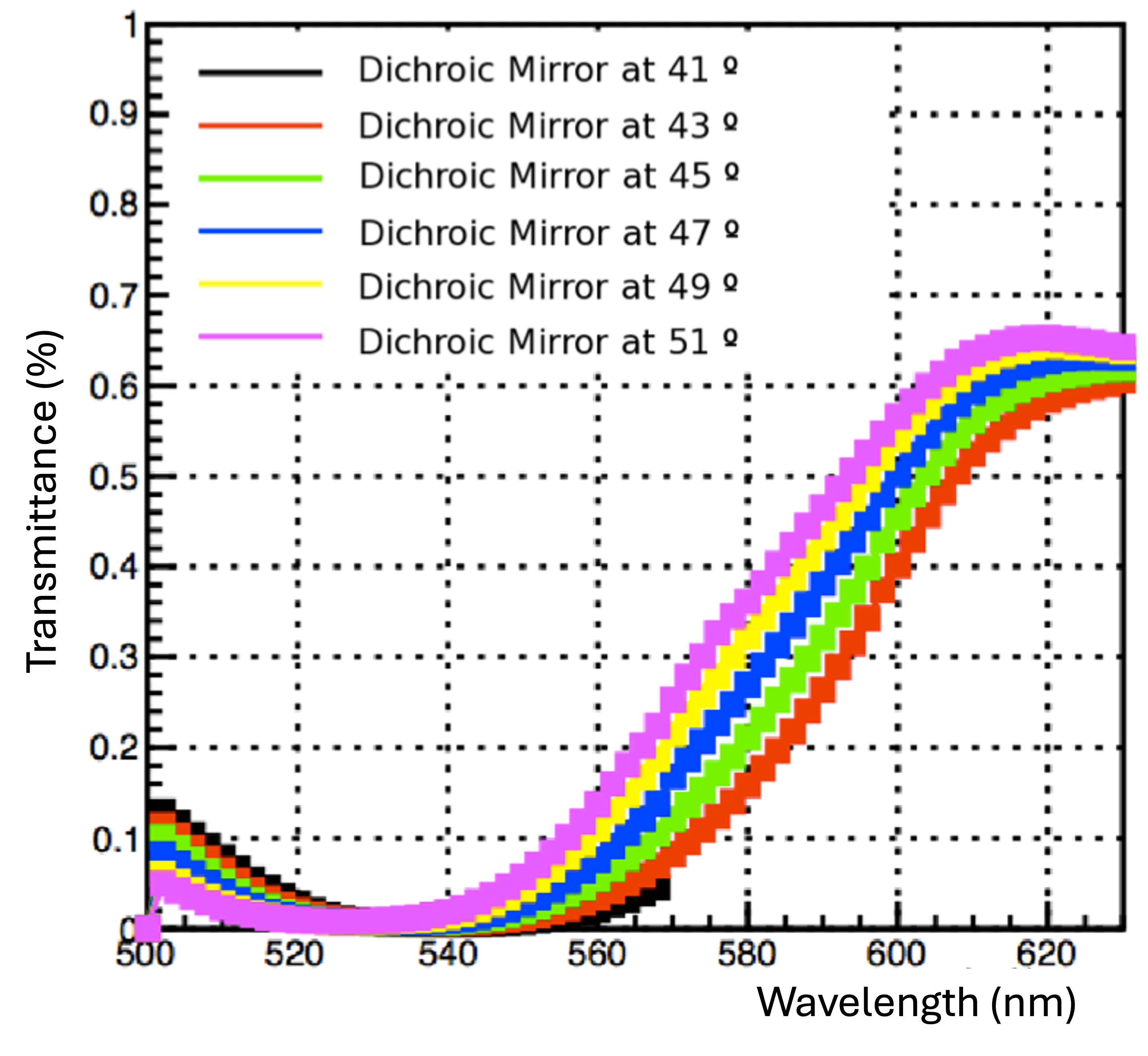} \hfill 
    \caption{Dependency of the transmission properties of the dichroic mirrors on the incidence angle. Left: for 355~nm, centre: for 387~nm and right: for 532~nm.} 
    \label{fig:DM}
\end{figure}

Optical simulations were performed with the \texttt{Zemax} ray-tracing code. 
Table~\ref{tab:polychromatorreqs} shows the requirements for these simulations and the specifications of the optical elements used. 
The layout of the optical design of the polychromator is
shown in Fig.~\ref{fig:polychromator} (top). The light received by the telescope and transported by the LLG exits with a 70$^\circ$ aperture
angle and is then collected by a system of lenses. To collimate such
a diverging beam, a two-lens system is necessary. A three-lens
system might also be used, at the expense of a decrease in the total efficiency of the system. 
For the sake of simplicity of procurement and to save costs, all lenses in the system are identical plano-convex lenses,  made of N-BK7 glass, 
%\mg{are we sure it's made of BK7 ? The transmission of 25mm of BK7 at 355 nm is 0 (see \url{https://www.newport.com/n/optical-materials}). On this page, it seems to be OK: \url{https://www.glassdynamicsllc.com/bk7.html}. We need the bill and manufacturer in any case, I don't have it.}
with a focal length of 150~mm and a diameter of 100~mm\footnote{\url{https://www.lobre.it/en/home-english/}}. 
%\md{Do we mention here the commercial model of the lenses?} 
Fused
silica glass would also have been an option with higher transmission but was discarded because it was more expensive. 
After collimation with the lens couple (\gls{lc}) system, the light is
separated into its different wavelength components via dichroic mirrors (\glspl{dm}). A DM reflects light below a certain wavelength and transmits it above; therefore, the three DMs visible in Fig.~\ref{fig:polychromator} are not identical. With three DMs, the
four wavelengths can be completely separated. After this, a second LC
%, identical to the polychromator entrance LC, 
focalizes the beam of each channel towards its
photon detector, with an interposed interference filter (\gls{if}) for further suppression of light outside the desired respective wavelength window. 
Care was taken in the definition of the wavelength acceptance band for the
DMs and the IFs since the light impinges onto them at various
angles because of the aperture of the light beam. All IFs have been designed with a transmission $>$85\% centered on the corresponding channel wavelength and a transmission FWHM at 10$\pm$2~nm around the central wavelength. 
Transmission outside the passband was specified to lie below 5\%. IFs were acquired from the company Optics Balzer\footnote{\url{https://www.materionbalzersoptics.com/en/}} (Liechtenstein).
Measurements post delivery showed even better performance of $<$2\% transmission outside the passband and $>$99\% at the central wavelength plus/minus 2~nm. 
% 
%in order to cut unwanted spurious light coming from reflection and diffusion inside the instrument, particularly taking into account that the Raman light istwo to three orders of magnitude dimmer than the elastic one. 
The DMs were custom-made by the company BTE\footnote{https://www.bte-born.com/} (Elsoff, Germany). They were required to have a transmittance better than 85\% for the wavelengths of interest and absorption higher than 95\% for the respective range of interest, as well as being optimized for $<$45$^\circ$ incident unpolarized light. The required size was $150\times100$~mm. BTE proposed a Borofloat
solution with a thickness of $2.0\pm0.2$~mm, edges cut, chamfered and a single-side coating matching our requirements.
Figure~\ref{fig:DM} shows the transmission properties of the DMs as a function of the incident angles.

\begin{figure}[h!t]
\centering
\includegraphics[width=0.485\linewidth,trim={0.2cm 0cm 7.9cm 0.3cm},clip]{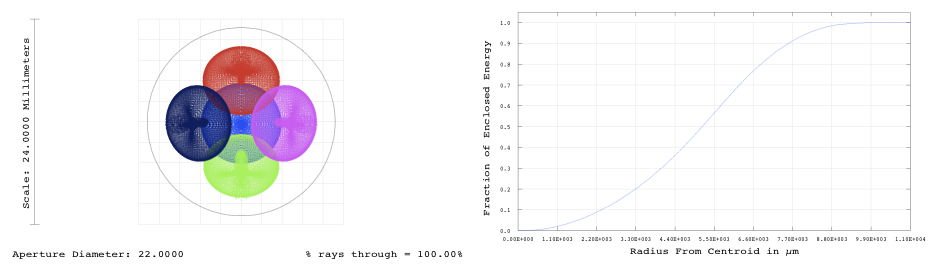} 
\includegraphics[width=0.485\linewidth,trim={0.1cm 0cm 0.1cm 0.1cm},clip]{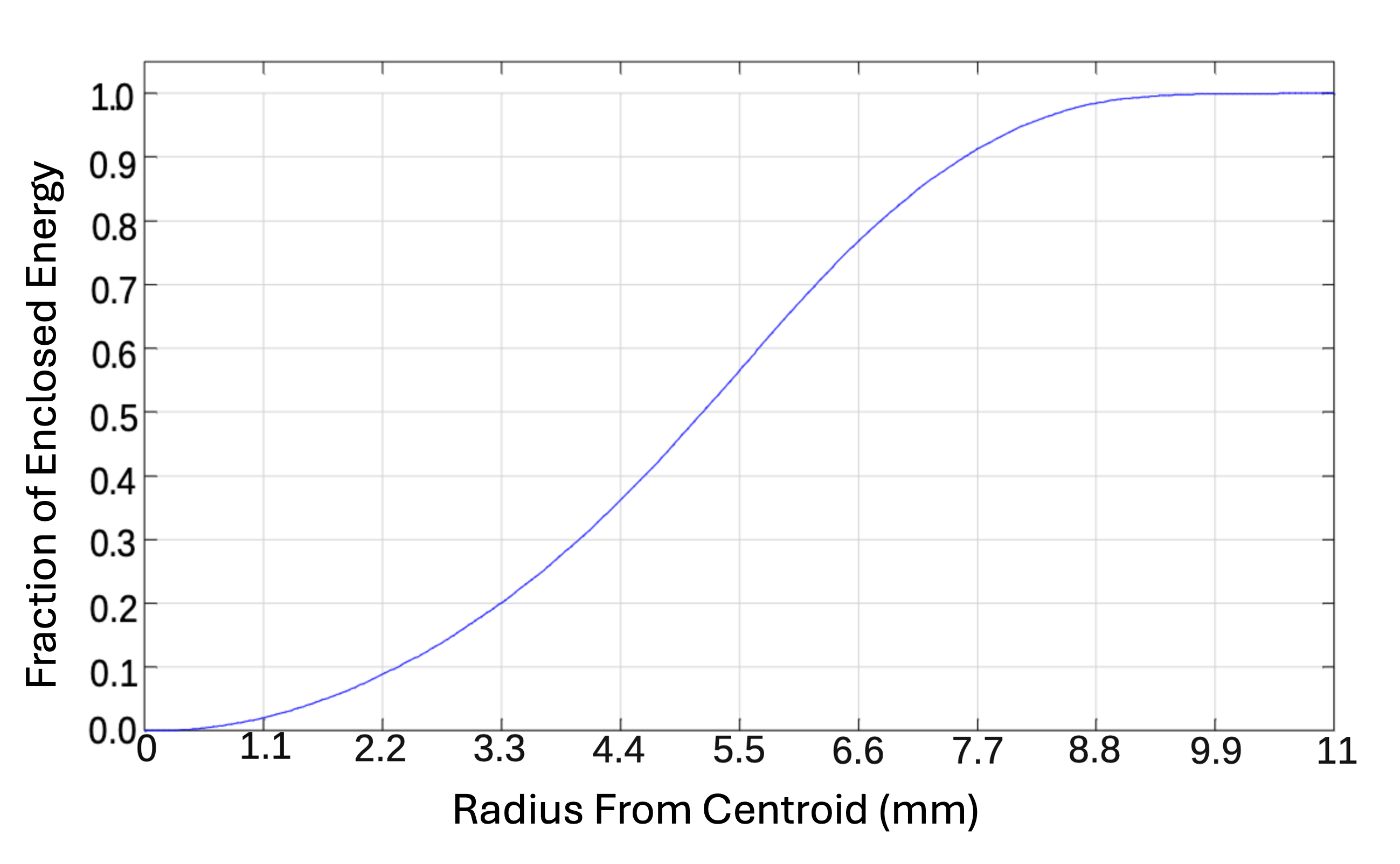} 
\caption{\label{fig:polychromator_features} Left: Footprint diagram
  of the PMT active area. Right: Encircled energy diagram of the PMT
  active area.  
  %\sa{The numbers on the right hand side plot should be larger.}\mg{I only have a png of this figure, unfortunately.}\md{Markus, I can re-generate this plot if needed by graph-importing the numbers, as I did for other plots in this work}\mg{Have succeeded to re-generate the right picture.}
  }
\end{figure}

In Figure~\ref{fig:polychromator_features} (left), the footprint of five sample points simulated by Zemax, one in the centre and four at the edges of the input optical fibre, is shown together with a circle corresponding to the 22~mm active area of the PMT detector. In Figure~\ref{fig:polychromator_features} (right), the enclosed energy fraction calculated for a uniform circular object of 8~mm in diameter, such as the input fibre, is shown. The total energy emitted by the fibre is
collected and focalized on the PMT area. 

\begin{figure}[H]
\begin{adjustwidth}{-\extralength}{0cm}
\centering
    \includegraphics[width=0.48\linewidth]{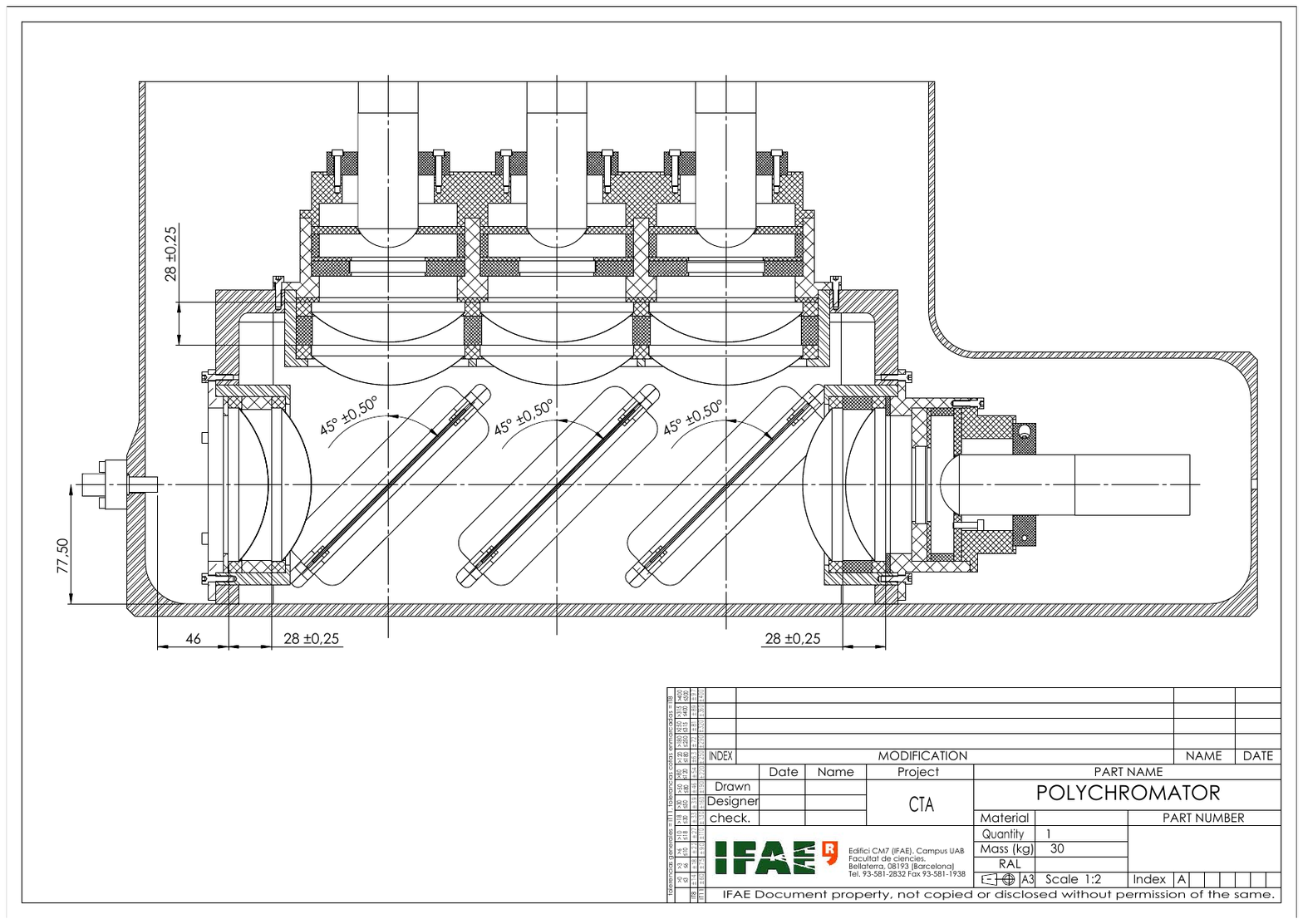} 
    \includegraphics[width=0.47\linewidth]{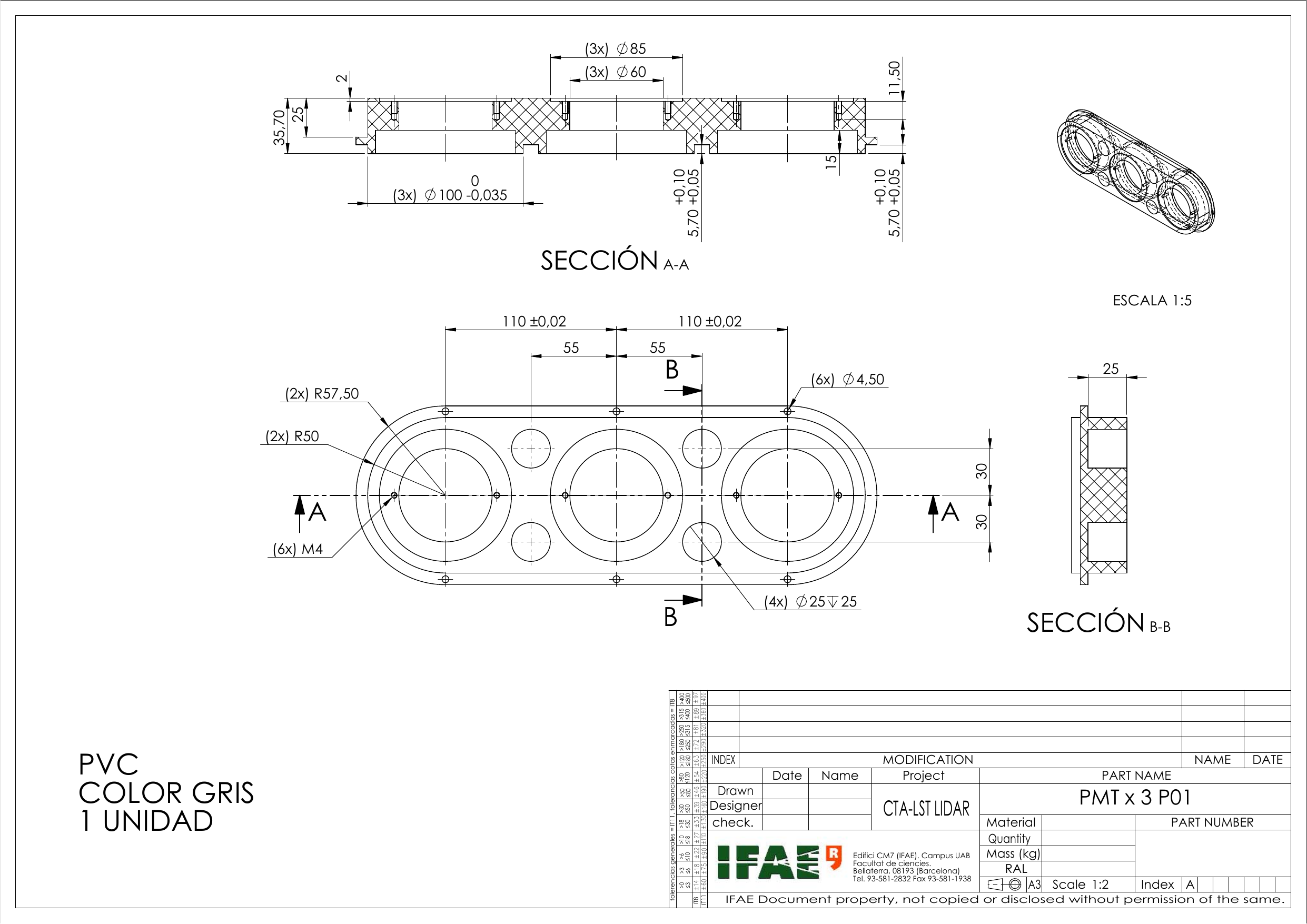}
\end{adjustwidth}
\caption{Design of the polychromator (\mdpi{left}) and the PMT holder (\mdpi{right}).\label{fig:polychromatortechdesign}}
%\md{I made it in wide page so that it stays inline with the text, for easier reading. In case you don't like it, switch back to normal. This is allowed by the journal btw} 
\end{figure}

Finally, Fig.~\ref{fig:polychromatortechdesign} shows the mechanical design of the polychromator developed in the IFAE engineering division. 

\begin{figure}[h!]
\centering
\begin{overpic}[width=0.8\textwidth,clip]{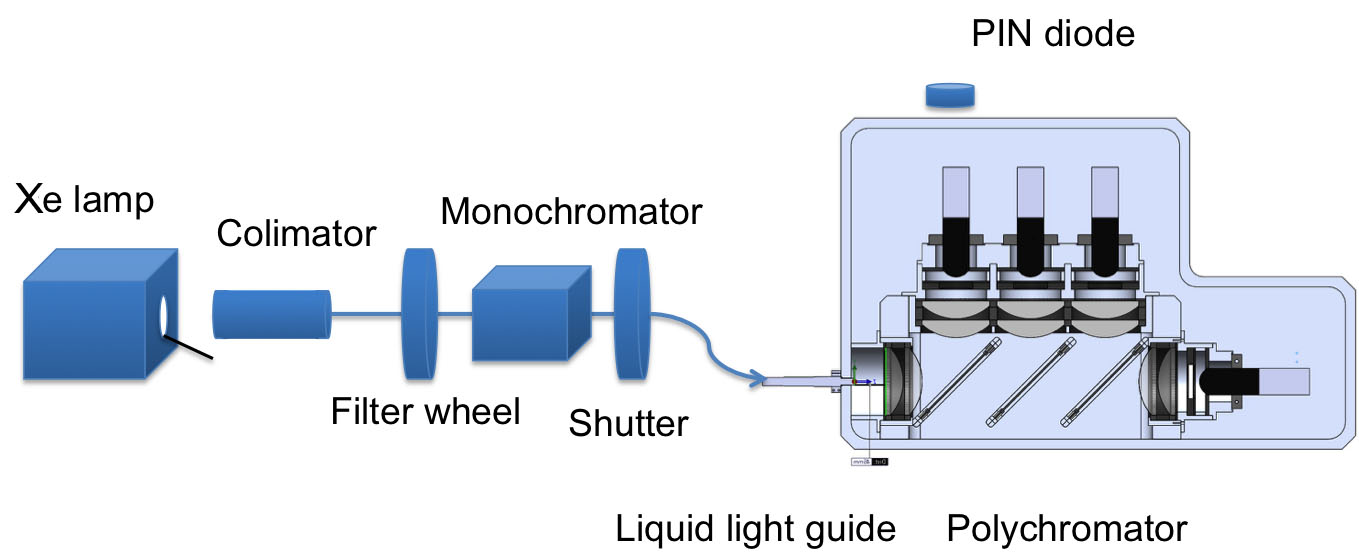}
  \put (70, 37) {\colorbox{white}{\parbox{0.15\linewidth}{\footnotesize PMT}}}
  \end{overpic}
  \caption{Sketch of the setup for polychromator tests: the light of a calibrated Xenon lamp passes through a collimator and a filter wheel with either an empty hole or a broadband filter. Then, it passes through a grating monochromator and a shutter and gets coupled to the LLG, which is connected at the other end to the polychromator unit. The current of the four PMTs is read out by a picoammeter (not shown in the figure). 
  %\sa{I would recommend a more descriptive caption.} 
  \label{fig:PolyIllustration} 
  }
  \end{figure}

\bigskip
\textbf{Light leakage tests} 
Light leakage tests are 
%of vital importance for the reliability of the system. 
used to control whether the Raman signals, which are more than two orders of magnitude dimmer than the elastic signals, may be affected by external or internal leakage of light.
%into the PMTs of the two Raman channels. 
%The light leakage test setup of the polychromator consisted of a Xenon lamp, collimators, a monochromator with shutter, the LLG and the polychromator box~\citep{Ballester:2019}. PIN diodes instead of PMTs have used as light detectors for these tests. 
%The light com- ing out of the Xe lamp (see Figure 4.13 for the spectrum) passed through a black box with collimator and filter sets, and then the wavelengths were selected with a monochromator. Just outside this black box, there was a shutter for background noise measurements.
For the tests, the collimated light from a stabilized Xenon lamp was wavelength-selected with the help of a grating monochromator. To discriminate unwanted harmonics, the light was first passed through a broadband filter. A shutter allows to remotely switch on and off the monochromatic light. The light was then fed into the LLG and coupled to the polychromator unit (see Fig.~\ref{fig:PolyIllustration}). Finally, PMT currents were recorded using a picoammeter. Series of measurements with an open shutter were taken, immediately followed by background estimates using the closed shutter, until sufficient statistics had been accumulated.

\begin{figure}[h!]
\centering
\includegraphics[width=0.485\linewidth]{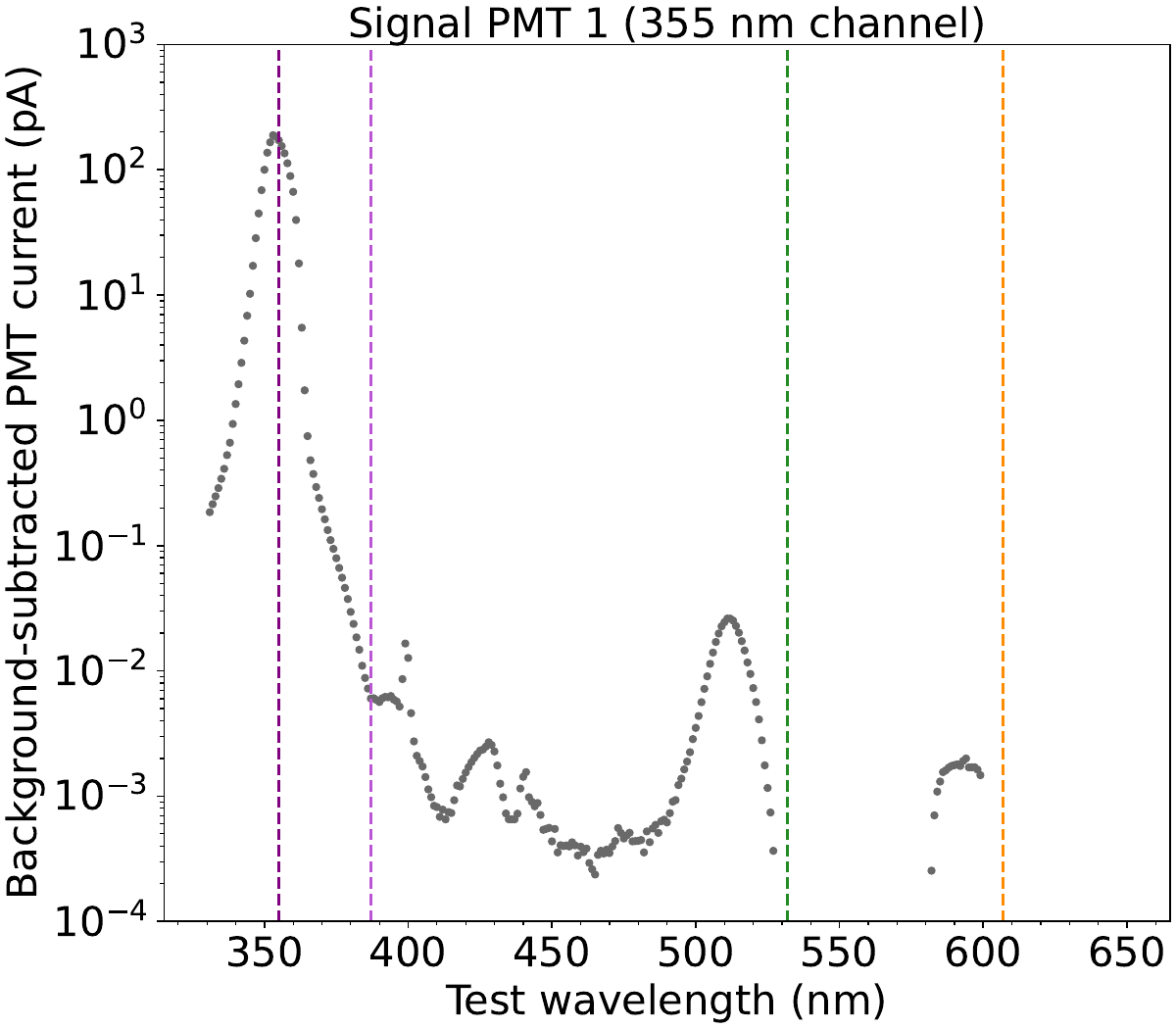} 
\includegraphics[width=0.485\linewidth]{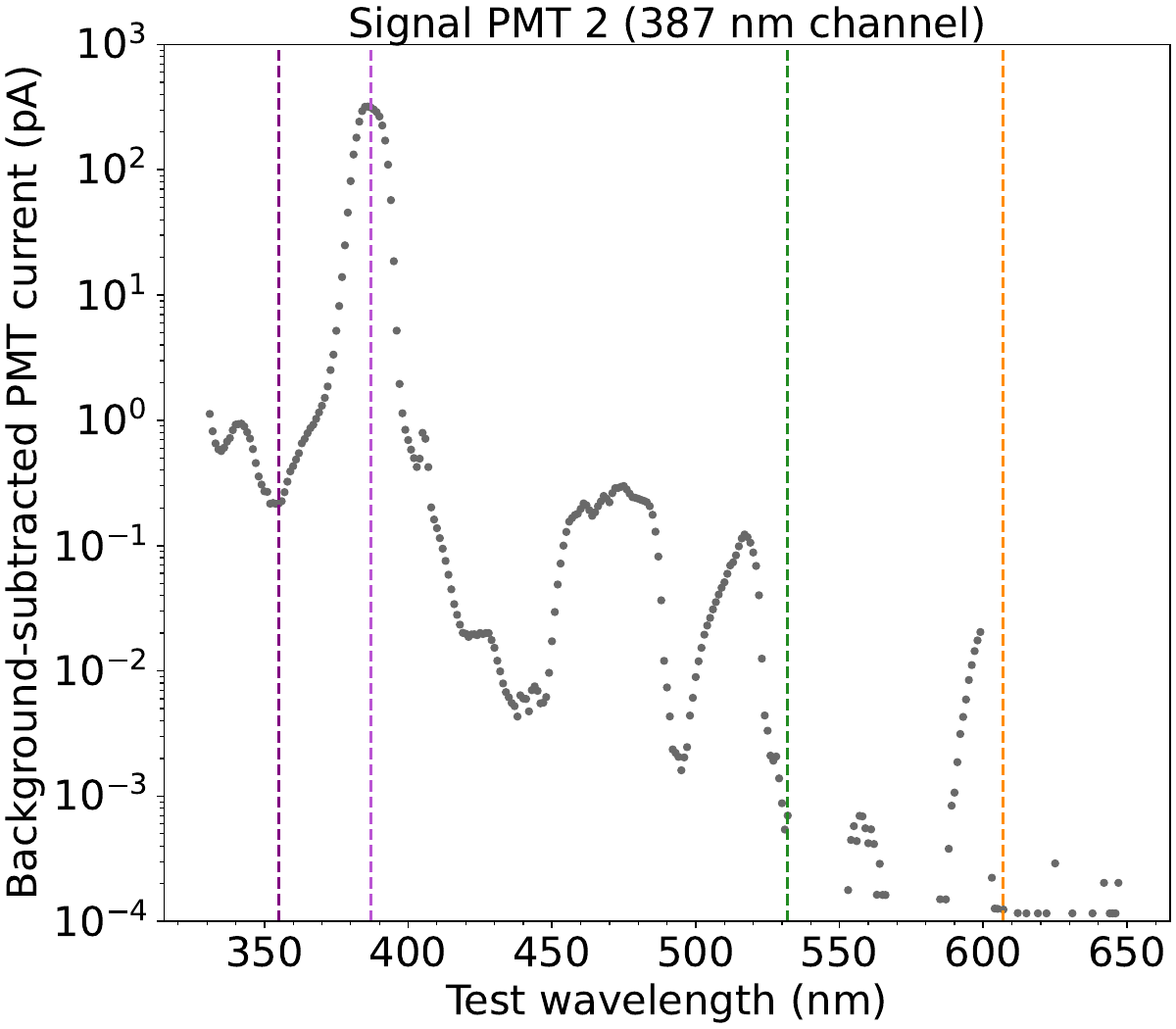} 
\includegraphics[width=0.485\linewidth]{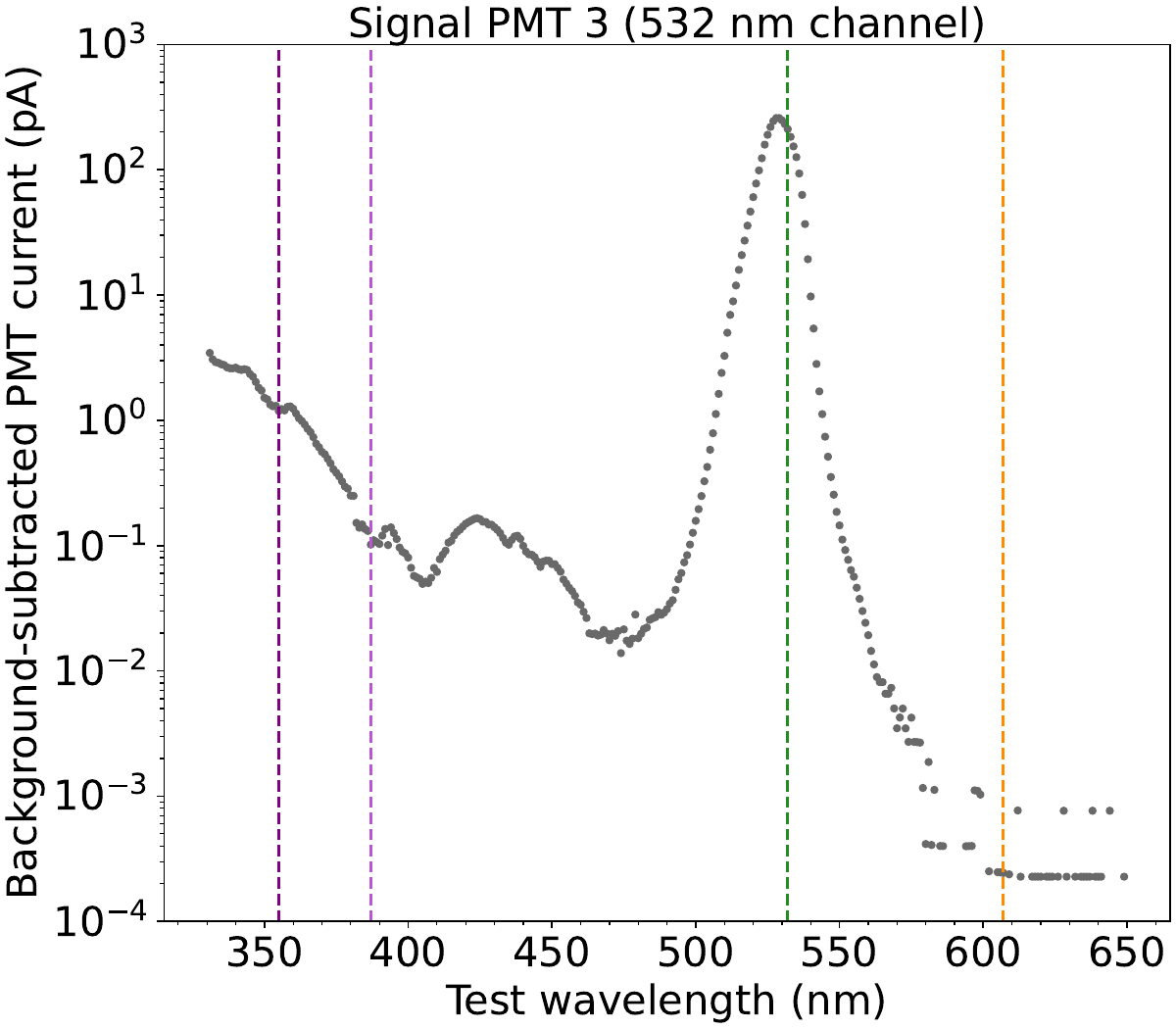} 
\includegraphics[width=0.485\linewidth]{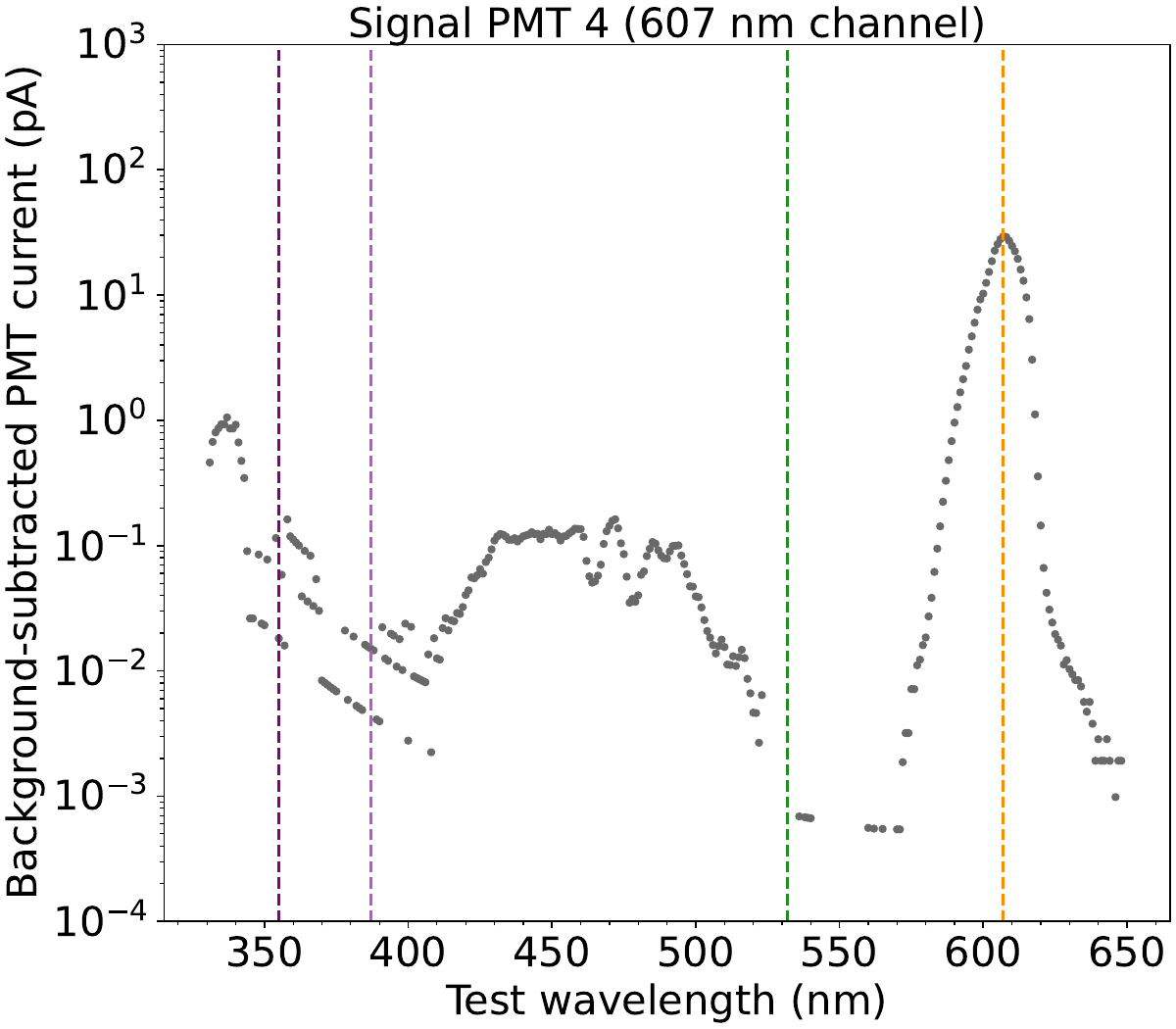} 
\caption{Results of a wavelength scan for the four channels. The dashed coloured lines show the four LIDAR wavelengths. 
%\mg{This figure corresponds to Merve's thesis Fig. 4.14, but explicitly shown in log-scale to highlight the possible leakages. Additionally, I have corrected for the Xenon lamp spectrum (her Fig. 4.13) and the PIN diode quantum efficiency used to register the spectrum of Merve's Fig. 4.13. These corrections do not alter the overall picture: there seems to be leakage of the order of 10$^{-5}$. The files used are called "\#1PolyCharFinal\_300\_650\_1\_open\_pmt.lvm" and "\#1PolyCharFinal\_300\_650\_1\_shut\_pmt.lvm", supposedly taken by Scott and Merve with the Keithley. I suppose that first the full scan with a closed shutter was taken and then another scan with an open shutter. Unfortunately, the "open" file shows jumps exactly from 599 nm to 600 nm, that is when the filter is switched. The "shut" file reproduces the background for the $>$600 nm region of the "open" file, so I guess that the "shut" file was taken after the "open" one, with the filter kept set to the $>$600 nm configuration. For this reason, I have guessed the background from the "open" file from reasonable regions, mostly just before the 599 nm "jump". Finally, I believe that Merve may have erroneously used $\mu$A in her Fig. 4.14, instead of nA.} 
\label{fig:PolyChar} }
\end{figure}

\begin{figure}[h!]
\centering
\includegraphics[width=0.485\linewidth]{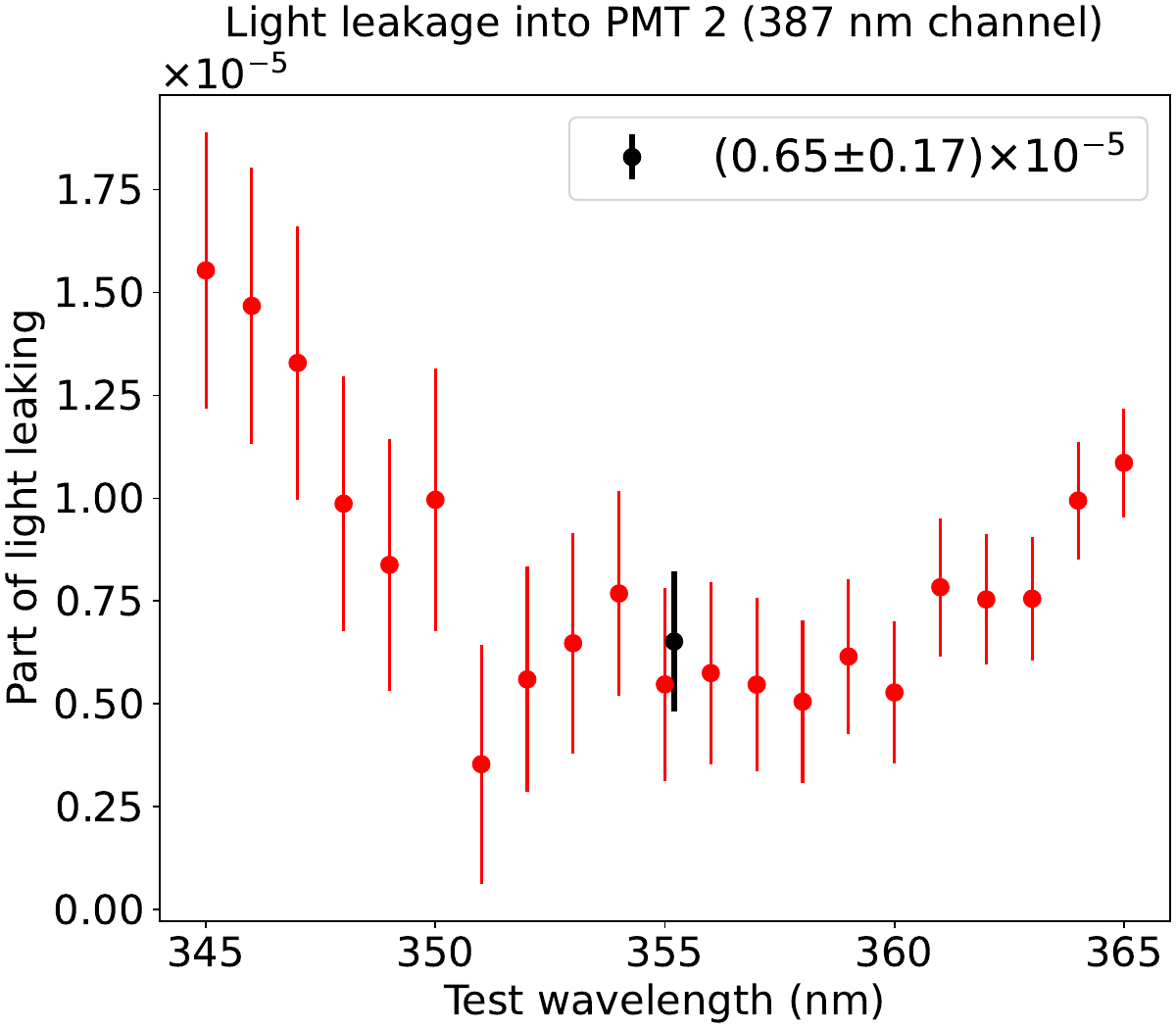} 
\includegraphics[width=0.485\linewidth]{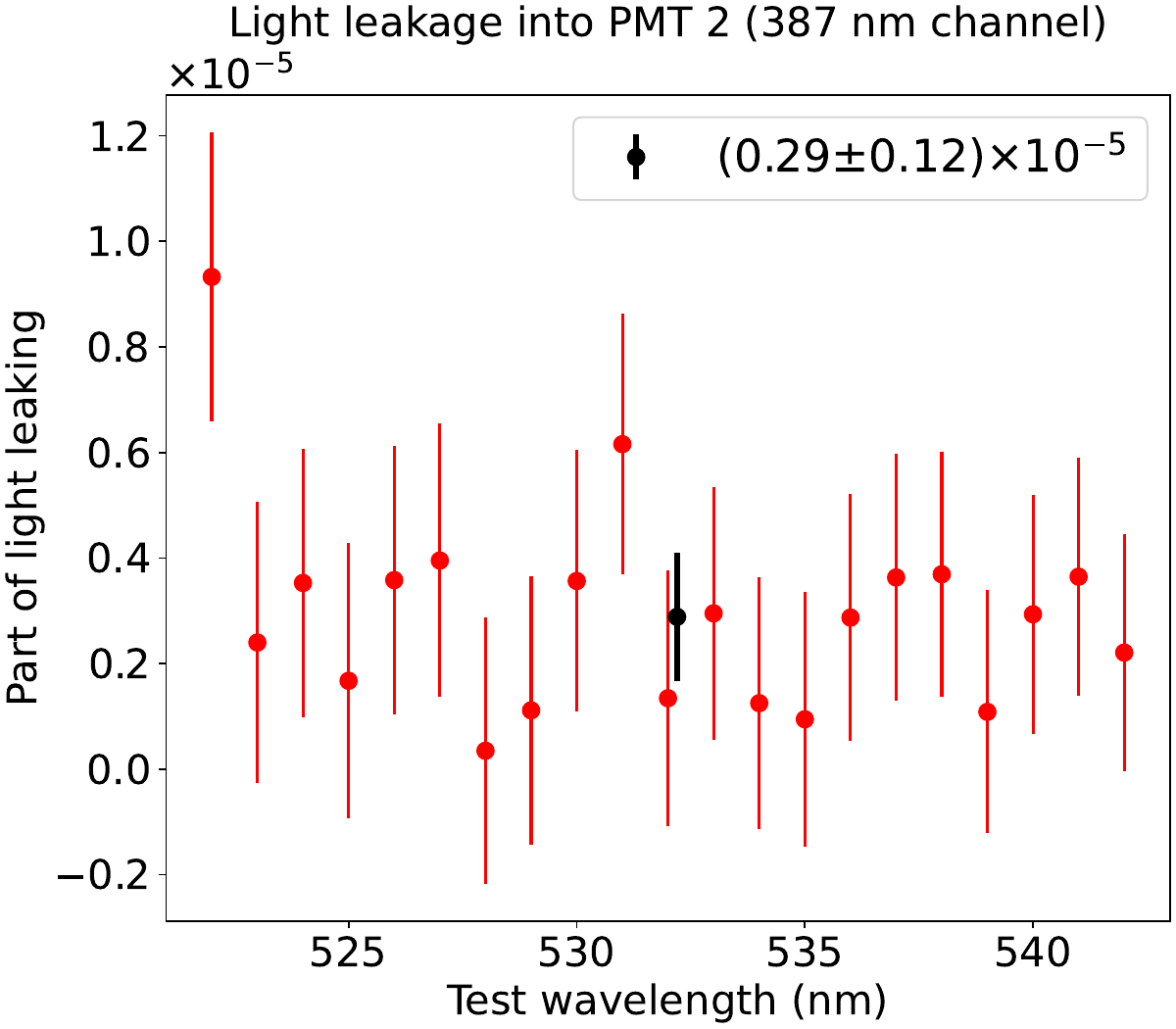} 
\includegraphics[width=0.485\linewidth]{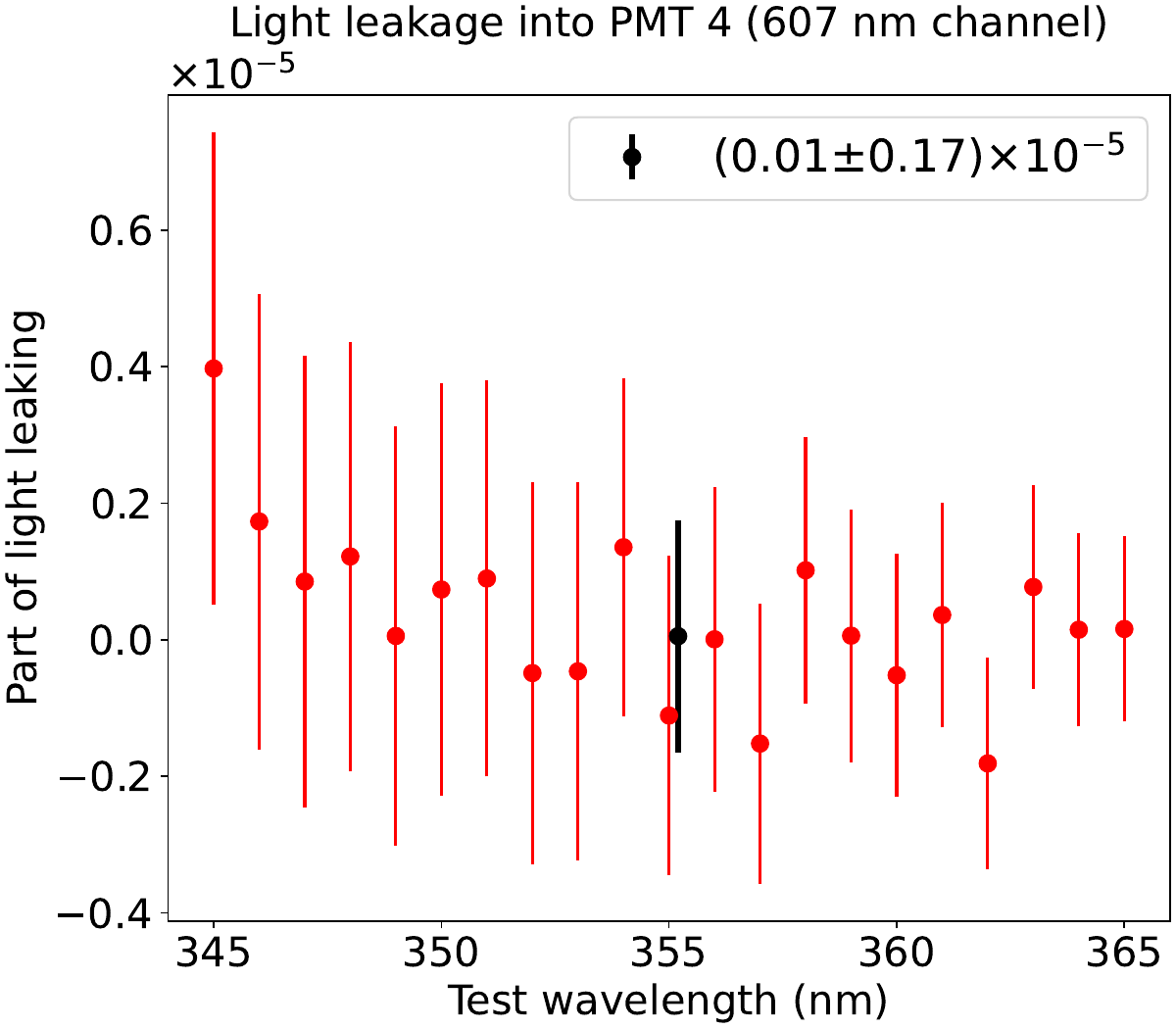} 
\includegraphics[width=0.485\linewidth]{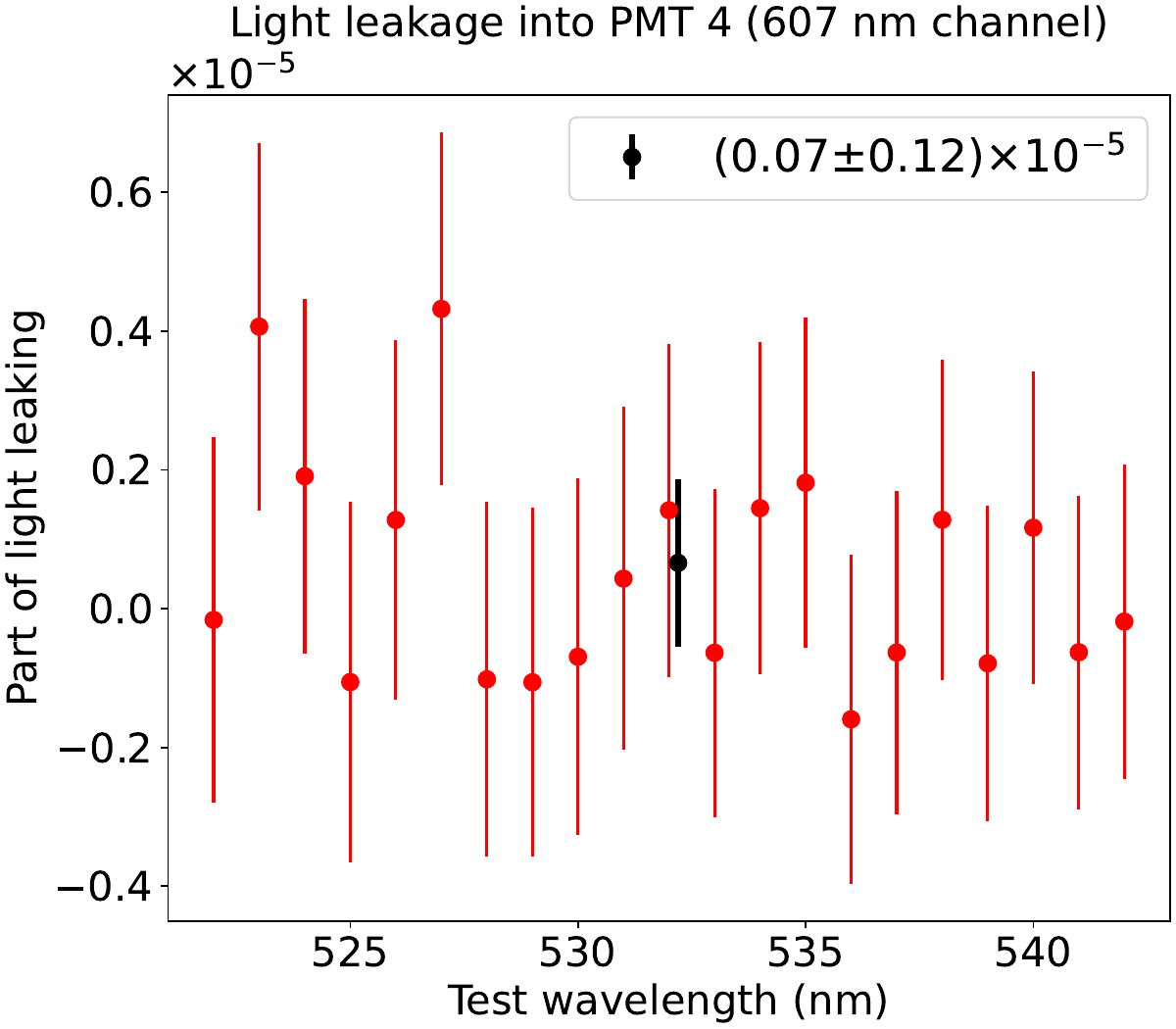} 
\caption{Results of a wavelength scan around the elastic lines (355~nm, left, and 532~nm, right) for the two Raman channels (387~nm, top, 607~nm, below). The black entries show the weighted average of ten entries around the central elastic wavelength. 
%\mg{This figure corresponds to Merve's thesis Fig. 4.15. Additionally, I have corrected for the Xenon lamp spectrum (her Fig. 4.13) and the PIN diode quantum efficiency used to register the spectrum of Merve's Fig. 4.13. The files used are called "\#2PMT\_around\_355\_test\_open.lvm" and "\#2PMT\_around\_355\_test\_shut.lvm", supposedly taken by Scott and Merve with the Keithley one week later than the previous data. Here, much better statistics are used (20 measurements for each wavelength), but still one full shut scan after a full open scan. In any case, the observed leakage of $\sim 6\times$10$^{-6}$  is a bit lower than that observed in the previous figure, but incompatible with Merve's claimed "signal-to-noise ratio" of 2.2$\cdot$10$^6$ (i.e., leakage of 4.5$\cdot$10$^{-7}$ claimed by Merve or the $<2\cdot$10$^{-7}$ claimed in \citet{Gaug:RamanLIDAR:2019} from the Pin diode measurements.}\aco{We agreed on not mentioning the proceeding but comment that this value is not detrimental because it will disturb only in the case when CTAO is not observing.}
\label{fig:PolyCharLeakage} }
\end{figure}

Figure~\ref{fig:PolyChar} shows the measurements made for a full wavelength scan from 300~nm to 650~nm. The individual measurement points have been corrected for the Xenon lamp spectrum to match the polychromator's response to a flat photon spectrum. Nevertheless, the tiny background-subtracted currents far from the main response peaks may be inaccurate in this figure, because of the relatively long duration of the full-wavelength scan, followed by the background measurements with the closed shutter. Furthermore, slight changes in background PMT currents were observed with different filter wheel settings (changed at 600~nm), which add to the inaccuracy of the small residual currents. 
%Blue 355nm, orange 387nm, green 532nm, red show the signal measurement results on 607nm channels, with peak points corresponding to 238 $\mu$A, 437 $\mu$A, 341 $\mu$A, 25 $\mu$A, respectively.

To address these issues, improved measurements were designed specifically for the light leakage from the strong elastic lines into the Raman channels. 
%For this reason, 
Figure~\ref{fig:PolyCharLeakage} shows measurements made with the PMTs of the 387~nm and 607~nm Raman channels to test the presence of additional PMT current when the polychromator is illuminated with wavelengths around 355~nm and 532~nm. 
%These tests were carried out with  monochromator wavelength scans from 522~nm to 542~nm and 345~nm to 365~nm, respectively, with shutter sequentially open and closed. This scan was repeated 1000 times \mg{Still TBC. For the moment, I have analyzed a file with 1000 measurements of each wavelength (in total 10000 for 10 wavelengths) followed afterwards by a closed shutter measurement}. 
%As a result, 1000 on, 1000 off measurements were taken for each wavelength.
Figure~\ref{fig:PolyCharLeakage} (top) shows the observed relative part of the signal from the elastic channels leaking into the 387~nm Raman channel. Likewise, Figure~\ref{fig:PolyCharLeakage} (below) shows the results of the 607~nm Raman channel. We observe a residual light leakage of (6.5$\pm$0.2)$\times$10$^{-6}$ from the 355~nm elastic line into the 387~nm Raman channel and (3$\pm$1)$\times$10$^{-6}$ from the 532~nm elastic line into the 387~nm Raman channel, respectively. 
Leakage from the elastic lines into the 607~nm channel was not observed and can be excluded above
% 0.01 + 1.64*0.17 = 0.3
% 0.07+1.64*0.12
3$\times$10$^{-6}$ (95\%~CL) from both elastic lines.

\begin{figure}[h!]
\centering
\includegraphics[width=0.19\linewidth]{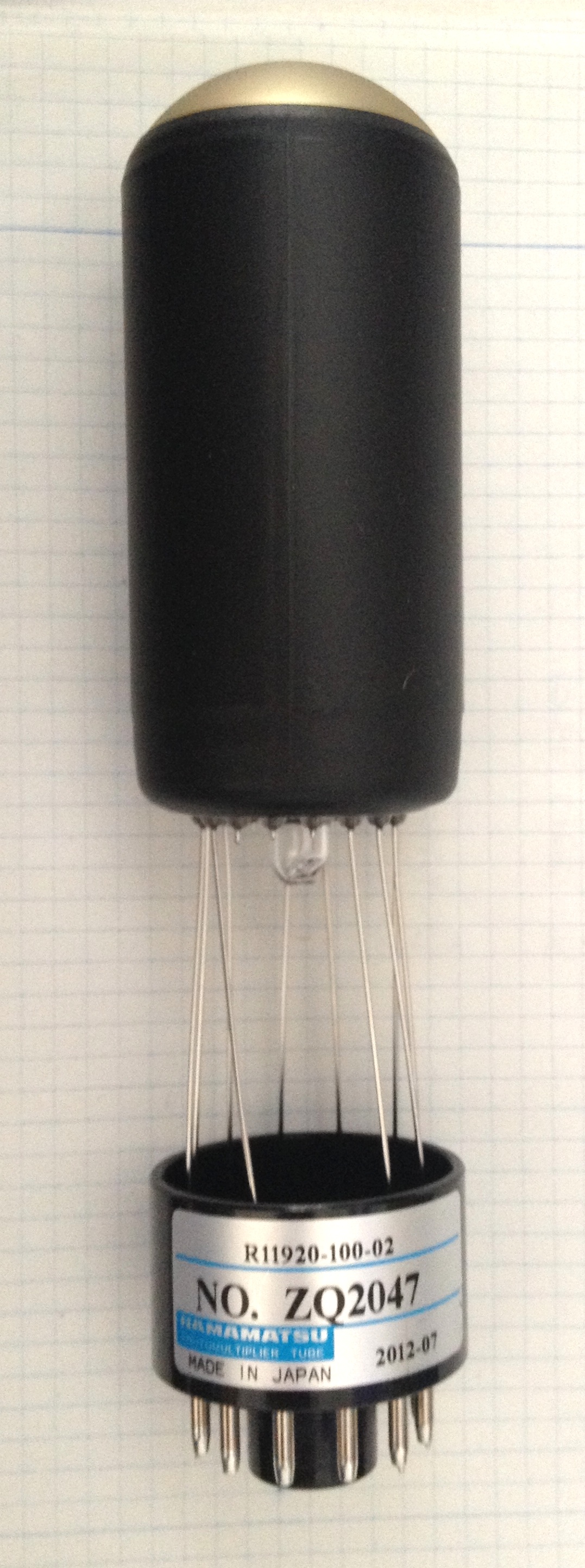}
\includegraphics[width=0.78\linewidth]{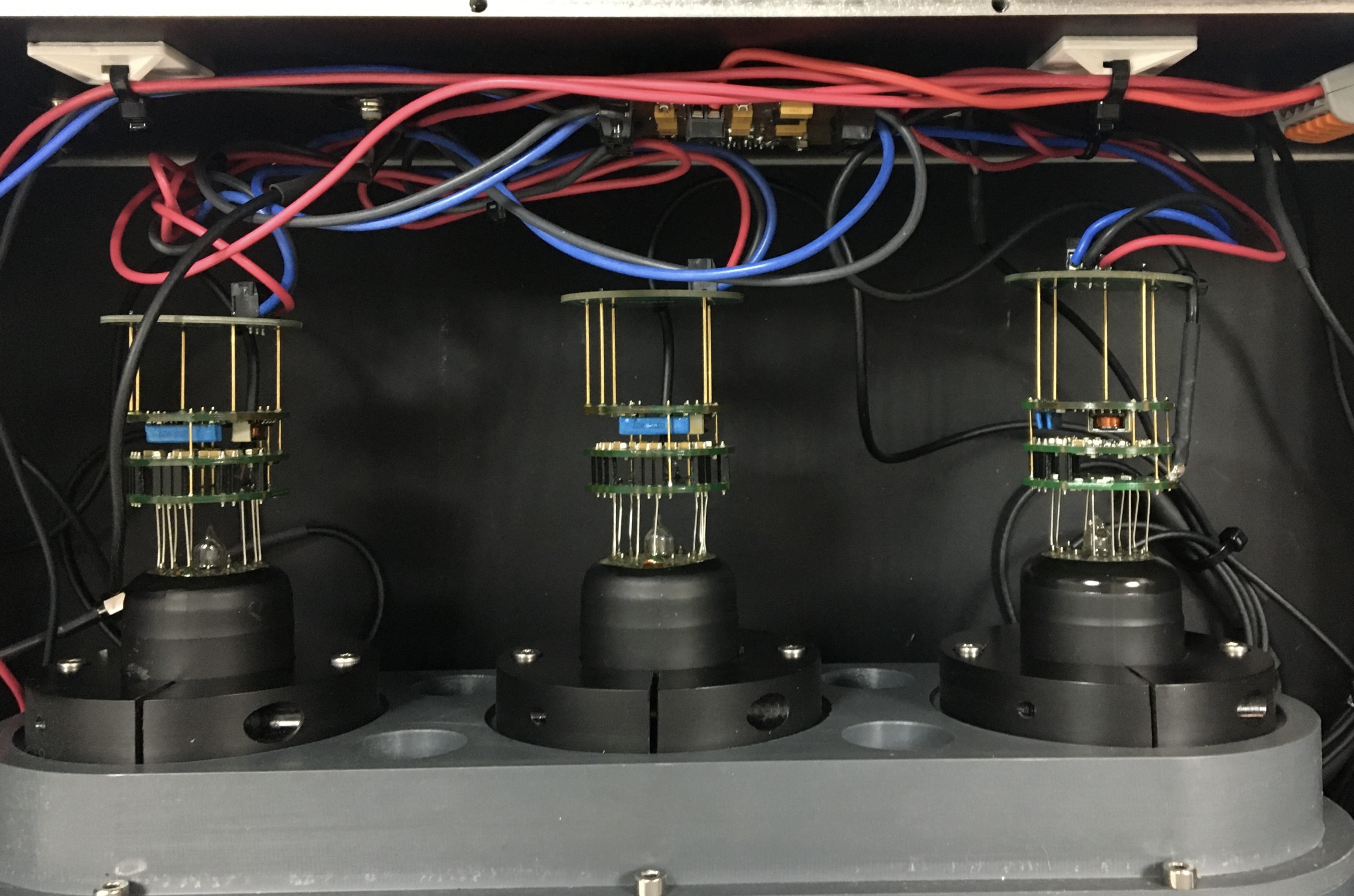}
\includegraphics[width=0.98\linewidth]{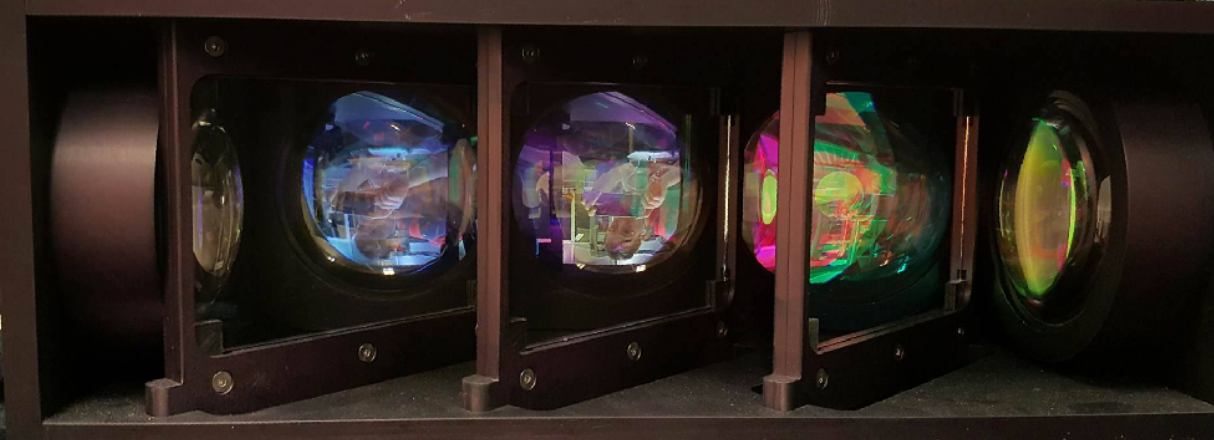}
\caption{Top left: A picture of the \sapocol{Hamamatsu} R11920-100. Top right: A photo of the %\sapocol{\st{current}} 
PMTs inside the polychromator. Below: The light concentrated onto the PMTs photocathodes.\label{fig:currentPMTsphoto}}
\end{figure}

\bigskip
\mdpi{We collect several transmittance factors contributing to the light collection efficiency of the polychromator in Table~\ref{tab:calibpower2}. We report the mirror reflectivity, the light transmittance of the LLG, the dichroic mirrors, and the interference filters, as well as the photon detection efficiency (\gls{pde}) of the PMTs. We compute the total transmittance and its uncertainty from these values.}

\begin{table}
\centering
\begin{tabular}{ccp{8cm}}
\toprule
Parameter & Value  & Comments \\
\bottomrule
% 0.9^2 * pi - 
\multicolumn{3}{l}{355 nm}\\
\midrule
$R$        & 0.95  &  after re-aluminization, otherwise $<$0.3 %\mg{check this number}
\\
$\epsilon$ &  $0.34\pm0.04$   & $T_\mathrm{LLG} = 0.5\pm0.05$,  $R_\mathrm{DM1,polych.} = 0.85\pm0.03$, $T_\mathrm{2LCs} = 0.81\pm0.04$ %\citep{technicalpaper} 
\\
\textit{PDE} & $0.42\pm0.03$  & \citep{toyama} \\
$T(355~\mathrm{nm})$ &  0.13$\pm$0.02   &   \\
\toprule
\multicolumn{3}{l}{387 nm}\\
\midrule
$R$        & 0.96  &  after re-aluminization,  otherwise $<$0.3  %\mg{check this number}
\\
$\epsilon$ &  $0.31\pm0.04$   & $T_\mathrm{LLG} = 0.57\pm0.05$, $T_\mathrm{DM1,polych.} = 0.70\pm0.03$, $R_\mathrm{DM2,polych.} = 0.93\pm0.03$, $T_\mathrm{2LCs} = 0.83\pm0.04$ %~\citep{technicalpaper} %\mg{check these numbers} 
\\
\textit{PDE} & $0.43\pm0.03$  & \citep{toyama} \\
$T(387~\mathrm{nm})$ &  0.12$\pm$0.02   &   \\
\toprule
\multicolumn{3}{l}{532 nm}\\
\midrule
$R$        & 0.97  &  after re-aluminization,  otherwise $<$0.3  %\mg{check this number}
\\
$\epsilon$ &  $0.31\pm0.03$   & $T_\mathrm{LLG} = 0.55\pm0.05$,  $T_\mathrm{DM1,polych.} = 0.85\pm0.03$, $T_\mathrm{DM2,polych.} = 0.85\pm0.03$,
$R_\mathrm{DM3,polych.} = 0.93\pm0.02$, $T_\mathrm{2LCs} = 0.85\pm0.03$ %~\citep{technicalpaper}  %\mg{check these numbers}
\\
\textit{PDE} & $0.13\pm0.03$  & \citep{toyama} \\
$T(532~\mathrm{nm})$ &  0.035$\pm$0.009  &   \\
\toprule
\multicolumn{3}{l}{607 nm}\\
\midrule
$R$        & 0.97  &  after re-aluminization,  otherwise $<$0.3 %\mg{check this number}
\\
$\epsilon$ &  $0.14\pm0.02$   & $T_\mathrm{LLG} = 0.42\pm0.04$,  $T_\mathrm{DM1,polych.} = 0.85\pm0.03$, $T_\mathrm{DM2,polych.} = 0.80\pm0.03$,
$T_\mathrm{DM3,polych.} = 0.55\pm0.05$, $T_\mathrm{2LCs} = 0.87\pm0.03$ %~\citep{technicalpaper}  %\mg{check these numbers}
\\
\textit{PDE} & $0.04\pm0.01$  & \citep{toyama} \\
%$\beta_\mathrm{mol}(r=0)$ & $1.55\cdot 10^{-6} \cdot \dfrac{P}{P_s}\cdot\dfrac{T_s}{T}$~m$^{-1}$ &   \\
$T(607~\mathrm{nm})$ &  0.05$\pm$0.01  &   \\
\bottomrule
\end{tabular}
\caption{Transmittance parameters for the wavelengths of interests. \label{tab:calibpower2} }
\end{table}

\bigskip
\textbf{Photomultiplier tubes photosensors} %\md{Must be completed/discussed (Markus)}
Lidars require photon counting and signal amplification. The custom solution is based on the use of photomultiplier tubes (\glspl{pmt}). However, the supremacy of PMTs is currently being challenged by photon sensors rapidly spreading in popularity, the silicon photomultipliers (\glspl{sipm}), which are becoming a valid alternative because of their high photon detection efficiency, low operating voltage, and installation flexibility. SiPMs are photosensors composed of microscopic diode cells assembled in matrices of thousands to reach sizes of a few millimetres in diameter. 
%They are valid alternatives to PMTs where compactness is required and because they operate at low voltage.   
%These devices are advancing fast in the field of fast photodetectors. 
In comparison to PMTs, in addition to their low operating voltage, they provide additional advantages such as high efficiency, insensitivity to magnetic fields, and robustness against bright ambient light. 
%The compact geometrical sizes of SiPMs allow to easily shape them in different types of cameras.   
Their main drawbacks are their small size, higher optical cross-talk, longer signal duration, temperature dependence, and sensitivity to red light compared to PMTs. 

Considering that the photon beam opens up widely on its way from the focal plane through the polychromator, a sensitive area of the order of a few cm$^2$ is required. %in order not to lose stray light. 
Currently, single SiPM matrices can hardly reach 1~cm$^2$. The combination of several SiPMs, albeit possible, would require a complication in terms of electronics, which led us to discard this option in favour of the traditional PMT. 
%Considering the aperture of the photon beam at the entrance of the photosensor, solid-state photosensors such as silicon photomultipliers (SiPMs) were soon discarded, because they were available only with a limited size of the order of 1~cm$^2$, and the combination of several SiPMs would have required additional electronics. The standard solution for photosensors for atmospheric remote sensing in 2008 was photomultiplier vacuum tubes (PMTs). 
For the choice of PMTs, we chose those in use in actual IACTs, specifically those chosen for the first Large Size Telescope (\gls{lst}1) of CTAO, under construction at the ORM at the time of purchase. Their sensitivity overlaps with our wavelength range of interest. This guaranteed reduced acquisition and characterization costs.
%The used PMTs are Hamamatsu R1924A, 1~inch diameter, with spectral response in the range from 300 to 650~nm. \md{Add photon detection efficiency? add more data?} For their use, the operation ambient temperature should be in the range from \emph{-30$^\circ$\,C -- +50$^\circ$\,C}. \md{Put a picture of the PMT housing within the bench?}
The pBRL used three identical Hamamatsu 8-dynode PMTs of type R11920-100-20~\cite{Orito:2011} for the UV and green lines (see Fig.~\ref{fig:currentPMTsphoto} (left)). 
%and the 
%\md{I would remove the next sentence as it's a repetition:}These are of the same type as those installed in the LST-1 camera, the choice was motivated by convenience.
A second PMT type, the Hamamatsu H10425-01\footnote{\url{https://www.hamamatsu.com/eu/en/product/optical-sensors/pmt/pmt-module/current-output-type/H10425-01.html}} has been purchased for the red line at 607~nm, but not yet %never 
integrated into the system. %\md{Before we say there was no option, here was say slightly differently, maybe we can homogeneize}
Since the polychromator system had been designed for 1.5$^{\prime\prime}$ PMTs, the latter was also selected to be of the same size but with a higher quantum efficiency at 607~nm. Table~\ref{tab:r11920} summarizes several technical parameters of the R11920-100-20 PMT.

\begin{table}[h!]
    \centering
    \begin{tabular}{llp{5cm}}
    \toprule 
     Parameter & Value & Notes \\
     \midrule
     Photocathode diameter & 38.1~mm & 1.5$^{\prime\prime}$\\
     Length of PMT        & 49.7~mm & PMT only 43~mm, connectors: 6.7~mm \\
     Length of HV base    & 64.0~mm & \\
     Dynodes    &   8    &   \\
     Power      &  +5V/GND & HV produced internally in PMT base by a Cockroft-Walton circuit \\ 
     Control voltage & +0.9~--~+1.5V & Correspondance of 1/1000 \\
     QE@355~nm  &  41\%  & %R11920-100
     \protect\cite{toyama,Orito:2011}\\
     QE@387~nm  &  42\%  & %R11920-100
     \\
     QE@532~nm  &  13\% & %R11920-100
     \\
     QE@607~nm  &  4\%  & %R11920-100
     \\
 %    QE@607~nm  &  4\%  & H10425-01\\
     Gain       & (2--3)$\cdot$10$^5$ &  %R11920-100 
     at HV$\approx$ 1200 V~\protect\cite{toyama} \\
% gain       & 10$^6$ &  H10425-01 at control voltage $\approx$1.1 V (datasheet) \\
     Pulse width &  2.0~ns  & %R11920-100 
     at HV$\approx$1500~V \protect\cite{toyama,victortfg}\\
% pulse width &  3.0~ns  & H10425-01 at control voltage $\approx$1 V (datasheet) \\
     Afterpulse rate & $<$10$^{-3}$ & %R11920-100 
     for $\geq$1 p.e. \protect\cite{toyama} \\
     & $<$2$\cdot$10$^{-4}$ & %R11920-100 
     for $\geq$2 p.e. \protect\cite{toyama} \\
         \bottomrule
    \end{tabular}
    \caption{Selection of technical properties of the PMT Hamamatsu R11920-100-20. }
    \label{tab:r11920}
\end{table}

\bigskip
\textbf{Polychromator assembly} 
The polychromator was built and assembled at the IFAE engineering division. It is housed in a large box with dimensions of 760~mm$\times$550~mm$\times$170~mm, which is additionally enclosed in an aluminium outer box that prevents the possible leakage of stray light into the unit and protects it from electronic noise.
%In addition to the central part, it is housed in an aluminium outer box, which prevents the possible leakage of stray light into the unit and protects it from electronic noise.
Together with the optical system and the photomultipliers, the entire unit weighs about 30~kg.

The polychromator unit has five connectors: a separate coaxial connector for the signal line of each of the four PMTs and a DS9 connector, which carries the 5V and GND feeding lines, plus the four control voltage lines for the PMTs.  

The signal lines are connected to the readout electronics through 10~m coaxial cables. In a previous version of the pBRL, the signals were preamplified by a factor of 10 and deamplified again before entering the readout, a solution later discarded given the relatively large pulse width of the single photoelectron produced by the Hamamatsu R11920. 

The control voltages of the PMTs are produced by a separate small electronic unit carrying corresponding digital-to-analogue converters (\glspl{dac}s), which allow user-controlled setting of the HV of each PMT individually. 
 
\subsection{Readout Electronics}
\label{sec:licel}
%\md{We should mention that the latest two modules are different and twice as fast, as well as they provide a measurement of the variance of the signal. Markus to update} 

%\mg{A photo of the Licel rack would be nice, maybe also of the individual two module types.} \aco{Where are the Licel modules? Not at the LIDAR container.}
%\aco{Photo of the LICEL rack to be added.}

\begin{comment}
\begin{figure}
\centering
\includegraphics[width=0.8\linewidth]{Figures/sec3/LicelModules.jpg}
 \caption{A picture of the Licel rack, with two modules installed on the left side, and the control and trigger module on the right side. In the meantime, two further modules have been added to the system.  \label{fig:LicelModules}}
\end{figure}
\end{comment}

Although the readout electronics can be custom-made, the Licel company\footnote{\url{www.licel.com}} offers a highly successful commercial solution. The Licel Optical Transient Recorder (\gls{lotr}) is a powerful data acquisition system, specially designed for remote sensing applications. Table~\ref{tab:licel_specs} summarizes the LOTR specifications provided by the manufacturer. %Figure~\ref{fig:LicelModules} shows a picture of the rack holding two LOTR modules and the control and trigger module.
%To meet the demanding requirements of optical signal detection, a new concept was developed to reaching the best dynamic range together with high temporal resolution at fast signal repetition rates. 
%Analog detection of the PMT current and PMT single photon counting is combined in one acquisition system. 
%The combination of a 12-bit A/D converter (at 40 MHz) with a 250 MHz fast photon counting system increases the dynamic range of the acquired signal
%substantially compared to conventional systems. 
%Signal averaging is performed by specially designed ASIC's.
% which outperform any CISC- or RISC-processor based solution. 
%A high speed data interface to the host computer allows readout of the acquired signal even between two laser shots. 
%The implementation of this concept makes the Licel transient recorder the state of the art solution for all applications where fast and accurate detection of photomultiplier, photodiode or other electrical signals is required at high repetition rates.

\begin{table}[htp]
\centering
\begin{tabular}{lp{8cm}}
\toprule
%\multicolumn{2}{l}{In/Outputs} \\
%\midrule
%Signal input: & BNC, 50~$\Omega$, front panel \\
%Trigger:  &	BNC, 50~$\Omega$, front panel \\
%Host    I/O: & 50-pol bus, back panel \\
%Indicators:  &  Analog input range \\
%             & Signal over-range \\
%             & Trigger and Host I/O \\
%\midrule
\multicolumn{2}{l}{Environmental:} \\
\midrule
%Power:	               & 230V/50 Hz or 110V /60Hz, 12 W per transient recorder \\
Operating temperature: & +10$^\circ$C to + 40$^\circ$C \\
Storage Temperature:   & --30$^\circ$C to + 70$^\circ$C \\
Humidity:	       & 0 to 95\%, non-condensing. \\
\midrule
\multicolumn{2}{l}{Dimensions:} \\
\midrule
Rack:	       & 448.6$\times$311.5$\times$361~mm housing \\                & for 19'' rack mounting \\ 
LOTR Modules   & RF-shielded cassette 6U (3U mounted in 6U) \\
\multicolumn{2}{l}{Acquisition:} \\
\midrule
Signal input range:    & 0... -500 mV \\
A/D Resolution:	       & 12 Bit / 16 Bit \\
Sampling rate:	       & 20 MSamples/s (40 MSamples/s) \\
Spatial resolution: & 7.5 m (3.75) m \\
Bandwidth:                & DC-10 (20) MHz \\
A/D differential nonlinearity: &   typ. 0.65 (0.5) LSB \ \  max. 1.25 (3) LSB  at 25$^\circ$ C \\
A/D integral nonlinearity:     &   typ. 1 (3) LSB at 25$^\circ$ C \\
Spurious free dynamic range:   & 74 (88) dB \\
S/N single shot:               & 66 (74) dB at 100~mV input range \\
Memory depth:	               & 16384 (32768) bins \\
Summation memory:              & 2 channels, up to 4094 acquisitions \\
Max. photon counting rate   &  250 (800) MHz \\
\bottomrule
\end{tabular}
\caption{Licel Optical Transient Recorder specifications for the 20~MHz recorders. In brackets are the values corresponding to the newer 40~MHz modules. }
\label{tab:licel_specs}
\end{table}

The LOTR is made up of a fast transient digitiser with onboard signal averaging, a discriminator for single-photon detection, and a multi-channel scaler combined with preamplifiers for both systems.  
For analog detection the signal is amplified 
%according to the input range selected 
and digitized by an A/D converter. 
A hardware adder is used to write the summed signal into a 24-bit wide RAM. 
Three LOTR versions have been purchased: a 12-bit-20~MHz recorder for the 355~nm channel, a 16-bit-20~MHz for the 387~nm Raman line, and 16-bit-40~MHz versions for the other lines.
The latter two also register the standard deviation of the analogue signal. 
%Depending on whether trigger A or B is used \md{not clear}, the signal is added to RAM A or B, which allows acquisitions of two repetitive channels if these signals can be measured sequentially.
%At the same time, 
The signal part in the high-frequency domain is amplified and a 250(800)~MHz fast discriminator detects single photon events above a selectable discriminator threshold voltage ranging from 0 to -25~mV in 64 steps for the 12-bit-20~MHz (16-bit-40MHz) version, respectively. 
%64 different discriminator levels ranging from 0 to -25~mV
%and two different preamplifier settings 
%can be selected using the acquisition software. 
The photon counting signal is written to a 16-bit wide summation RAM which allows averaging of up to 4094 acquisition cycles.
%The photon-counting acquisition system includes a fast three-stage preamplifier and a discriminator with 64 threshold levels, controlled by the host computer. 
%With a maximum count rate of 250 MHz, single photon counting is pushed to new limits when the selected photomultipliers are used. 
A time resolution of 50(25)~ns without any dead time or overlap between two memory bins is reached by using a continuous counter together with a multichannel scaler %burnt into the silicon 
of an ASIC custom-designed by the provider.

The LOTR is completely software controlled and interfaced by a custom-written software module of the \textit{LIDAR Client} (\gls{licli}) programmed in \textit{Java}. 
Input ranges for analogue and photon counting acquisition, discriminator levels, and the number of active bins can be selected. 
The acquired analogue and photon counting signals for both summation memories can be read out separately. 
Data are transferred via a 2$\times$16 bit interface to a National Instruments DIO-32-HS family (PC) interface card.
%Data transfer rates are 800 kB (PC/486 DX2-66) using DOS and 500 kB using LabVIEW. 
%Up to 16 transient recorders can be controlled by one interface card. 
A custom \textit{ Python} interface was created to read, convert into \textit{FITS} format and visualize the data. 
%It can be run on PC/Windows platforms. 
%Software drivers and acquisition programs for PC/Linux are supplied.

\subsection{The Laser}
\label{sec:laser}

The pBRL uses a Brilliant Nd:YAG laser from the company Quantel\footnote{\url{https://www.quantel-laser.com}}. 
%see Fig.~\ref{fig:Laser} (left). 
It is a pulsed 10~Hz flashlight-pumped and \sapocol{actively} Q-switched class~IV laser with a fundamental wavelength of 1064~nm and energy per pulse of up to 400~mJ. The pulse intensity can be changed by the user through manipulation of the \sapocol{Pockels cell} Q-switch delay. 
A second and third harmonic generator at 532~nm (200~mJ per pulse) and 355~nm (100~mJ per pulse) are added to the main body of the laser. They are assembled in compact modules, including the non-linear crystals and a removable set of dichroic mirrors. 
Phase matching for the second and third harmonics is obtained by a simple mechanical adjustment (adjustment screw accessible from the top of the module). %A schematic view is shown in Fig.~\ref{fig:Laser} (right). 
For the pBRL application, we are only interested in the second and third harmonics, which exit the laser coaxially. Since there is no automatic way to mask the 1064~nm pulse, a configuration has been chosen where all three wavelengths exit the same output hole. In a later stage, the undesired 1064~nm line is therefore removed by the two dichroic guiding mirrors; see Sect.~\ref{sec:dichroicmirrors}. The general characteristics of the laser are collected in Tables~\ref{tab:laser} and~\ref{tab:laser2}.
%\md{Mention the dichroic mirror of the guiding arms} 
Since its purchase in 2010, the laser was maintained several times: the flashlamp needed to be replaced once, and the deionized water and filters were exchanged three times. The laser broke and was repaired in 2017 by the company \textit{Proton Laser}\footnote{\sapocol{Proton Laser Applications S. L. has ceased to exist.}} (Spain), who also measured that the output power had dropped to 250~mJ per pulse at 1064~nm. 

\begin{comment}
\begin{figure}[h!]
\centering
%\subfloat[The Quantel Nd:YAG 1064 Laser]{\label{fig:Laser}
    \includegraphics[width=0.48\linewidth]{Figures/sec3/Laser.pdf}%}
%\subfloat[Different harmonic configuration for the laser output. The pBRL operates in upper configuration without dichroics.]{\label{fig:Configuration}
    \includegraphics[width=0.48\linewidth]{Figures/sec3/Configuration.pdf}%}
    \caption{Left: The Quantel Nd:YAG 1064 Laser modules. Right: Different harmonic configurations for the laser output. The pBRL operates in the upper configuration without dichroic.}
    \label{fig:Laser}
    %\label{fig:brilliant}
\end{figure}
\end{comment}

\begin{table}[h!t]
\begin{center}
\begin{tabular}{l l c c}
\toprule
\textbf{General characteristics}\\
\midrule
Pulse repetition rate & 10 Hz \\\addlinespace[0.1cm]
%\midrule
Power drift &  3$\%$\\\addlinespace[0.1cm]
%\midrule
 Pointing Stability &  $<$ 75 $\mu$rad \\\addlinespace[0.1cm]
%\midrule
Pulse jitter (1064 nm)    & $\pm$ 0.5 ns  \\\addlinespace[0.1cm]
%\midrule
 Beam divergence full-angle (1064 nm) & 0.5 mrad \\\addlinespace[0.1cm]
%\midrule
 Beam waist diameter (1064nm) & 6 mm\\
\bottomrule
\end{tabular}
\end{center}
\caption{Main characteristics of the pBRL Brilliant Nd:YAG laser from the Quantel company.\label{tab:laser}}
\end{table}

\begin{table}[h!t]
\begin{center}
\begin{tabular}{ccccc}
\toprule
\multicolumn{5}{c}{\textbf{Wavelength-dependent parameters}}\\
\midrule
Wavelength & Nominal energy     & Maximum       & Pulse    & Energy stability\\
           &  per pulse & average power & duration & shot-to-shot\\
(nm) & (mJ) & (W) & (ns) & (\%)\\
\midrule
1064 & 360 & 3.6 & $\sim$ 5 & $\pm$ 2 (0.6) \\
(after repair)    &  250 & 2.5  &         &         \\
\midrule
532 & 180 & 1.8 & $\sim$ 4 & $\pm$ 4 (1.3) \\
(after repair) & 128  & 1.3  &         &         \\
\midrule
355 & 100 & 1 & $\sim$ 4 & $\pm$ 6 (2)\\
(after repair) & 80  &  0.8  &         &         \\
\bottomrule
\end{tabular}
\end{center}
\caption{Characteristics for main pulse and harmonics.\label{tab:laser2}}
\end{table}
%\mg{new measurements - Camilla}

\begin{figure}[h!]
\centering
%\subfloat[The Quantel Nd:YAG 1064 Laser]{\label{fig:Laser}
    \includegraphics[width=0.45\linewidth]{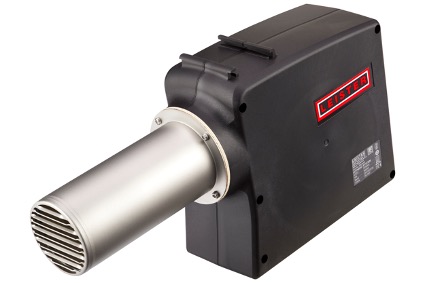}%}
    \includegraphics[width=0.55\linewidth]{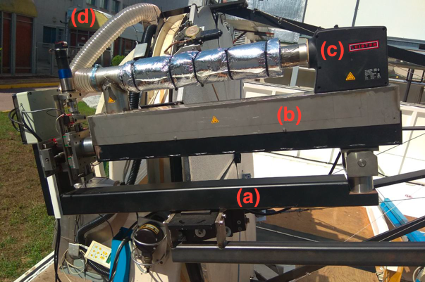}%}
    
    \caption{Left: \sapocol{The industrial hot air blower system.} %Hot wind system. 
    Right: Picture of the laser arm: (a) laser arm structure, (b) laser head housing, (c) hot air blower, and (d) laser cooling unit heating hose.}
    \label{fig:Hot-wind}
    %\label{fig:brilliant}
\end{figure}

\bigskip\textbf{Temperature Control} The laser head is a monolithic,  temperature-stabilized block which ensures the alignment of the resonator mirrors. The temperature is controlled by a water loop that goes through a water/air heat exchanger. 
The cooling group is an independent unit that cools the systems using a closed loop of deionized water. This temperature-regulated water also provides thermal stabilization $\pm $1$^\circ$C of the oscillator structure. 
For proper operation of the laser, the ambient temperature should lie between 18 $^\circ$C and 28$^\circ$C. Quantel does not guarantee that beam quality stays within specifications outside this temperature range. 
%Beam quality could remain even down to +15 $^\circ$C, but the company can not certify it. 
%The temperature stabilization is $\pm $1$^\circ$C and, while ambient temperature remains in the 18 $^\circ$C - 28$^\circ$C range, there will be no consequences for laser operation. 
Because the pBRL is regularly operated at night at ambient temperatures well below 18$^\circ$C, falling even below 0$^\circ$C, we equipped the system with additional heating, which can be optionally operated when the temperatures are so low that the cooling group unit could not heat the water to its operating temperature. The heater consists of a 3700~W Hotwind system from the company \textit{Leister Process Heat} with a stepless adjustable heat output and air volume through a potentiometer. The \sapocol{industrial hot air blower} %hot wind 
is fed into a hose that delivers warm air both to the cooling unit and to the laser head (see Fig.~\ref{fig:Hot-wind} (right)).
%\mg{Here, one or two pictures of the system would be nice} \aco{Only able to find this one...}
During operation, the system is only turned on to speed up the cooling unit, achieving the operating temperature during cold nights. Once that temperature is achieved, the \sapocol{industrial hot air blower} %hot wind 
is automatically switched off, allowing the cooling group unit maintain the water temperature without further assistance, even during freezing nights. 

\bigskip\textbf{Near-field spot size}

The waist diameter of the beam, according to the manufacturer, is 6~mm at 1064 nm. As we are mainly interested in the second (532 nm) and third (355 nm) harmonics, we measured the spot size in the near-field at these two wavelengths. %The beam divergence is 0.5 mrad, we would asssume that all wavelengths present the same value. 
Measurements were carried out by pointing the laser at a lead target at about 2.5~m distance and capturing the images with a Canon EOS 1000D camera~\citep{lopezphd}, which were later digitized, each pixel was assigned a distance, and numerical integrals were performed to obtain a measurement of the relative amount of encircled energy~\citep{lopezphd}.  

\begin{figure}[h!]
\centering
\includegraphics[width=0.75\textwidth]{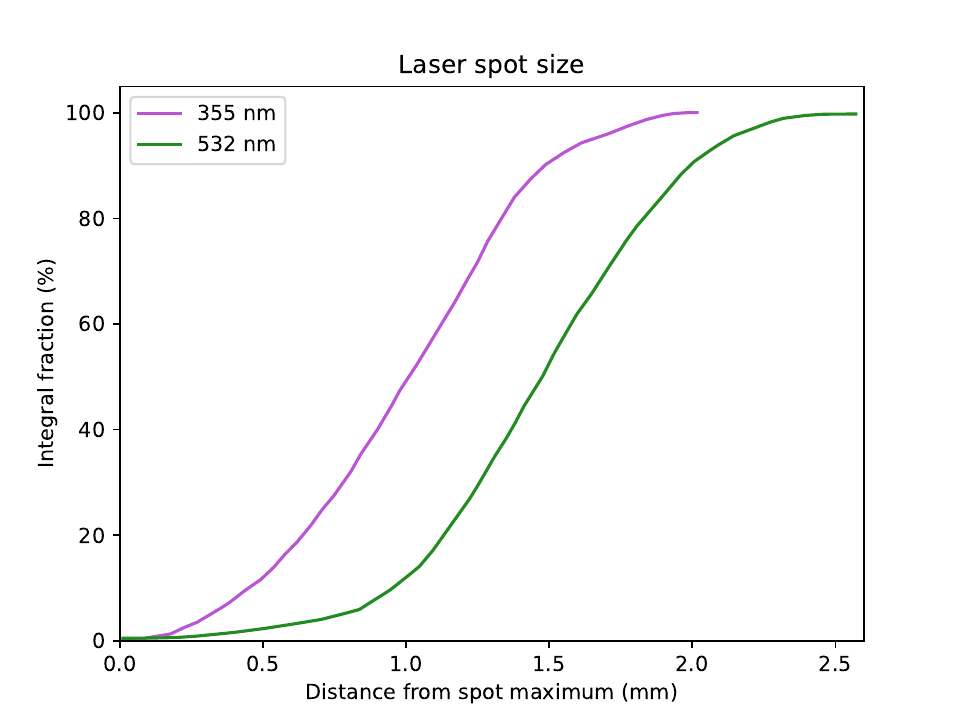}
\caption{\label{fig:laser_spotsize} Spot size for the 355 nm and 532 nm lines of the Brilliant laser.}
\end{figure}

For the 355~nm (532~nm) wavelength, about 80\% of the light is enclosed in a circle of 2.6~mm (3.6~mm) of diameter, and 90\% falls in a circular shape of about 3~mm diameter (4~mm). Most of the light (99.9\%) is contained in a spot not much larger than 4~mm (5~mm) in diameter. 
%On the other hand, for the 532~nm harmonic, about 80\% of the light is enclosed in a circle of 3.6 mm of diameter, 90\% in an area of about 4~mm of diameter. 99.9\% of the light is contained in a circular spot of 5~mm diameter. 
The spot profiles are shown in Fig.~\ref{fig:laser_spotsize}.

\begin{comment}
 \begin{figure}[h!]
\centering
\includegraphics[width=0.8\textwidth]{Figures/sec3/laser_spotsize.png}
\caption{\label{fig:laser_spotsize} Spot size for the 355 nm and 532 nm lines of the Brilliant laser. \md{532 nm line missing}\mg{Don't have the original data from Alicia, we could digitize her figures 2.9 and 2.10 and plot them nicely again} \aco{I can do it.}}
\end{figure}   
\end{comment}

\bigskip\textbf{Far-field beam divergence}

\begin{comment}
\begin{figure}
\centering
\includegraphics[width=0.7\linewidth]{./Figures/sec3/Picture_LaserSpotTests.jpg}
\caption{\label{fig:Laserspottests}Location and setup of the long-range laser beam divergence tests. \md{IMHO this figure does not say anything about the setup and could be removed} \aco{Agree.}}
\end{figure}
\end{comment}

The beam divergence test was carried out at the bus terminal of the Universitat Aut\`onoma de Barcelona Campus where it was possible to reach distances larger than 80~m between the laser and a target sheet of graph paper mounted on a white panel. Pictures of the laser signal reflected on the paper were taken in RAW mode with a Canon EOS~1000D and a Nikon~D5000. The beam spot was observed from very small angles concerning the perpendicular incidence. A Neutral Density (\gls{nd}) filter was added in front of the camera to avoid saturation of the images. Only the Bayer-Green points were extracted from non-saturated pictures. All image pixels were converted to physical units by comparing the distance in image pixels of the outer edges of the white panel with its measured real size. The analysis algorithm\footnote{script available at \url{https://github.com/mgaug/LIDAR-tools/blob/main/diameters_from_image.py}} is described in the following~\citep{phdthesis-maggio}.

A perfectly Gaussian beam produces irradiance that decreases monotonically with radius from the beam axis. 
In the case of real laser beams, the irradiance may not be uniform around the beam axis, introducing some arbitrariness in the definition of the beam profile. We used here a definition of the beam diameter that is based on the concept of encircled energy:
the major and minor axes of an ellipse around a central point at which the encircled energy (expressed as summed and background-subtracted) image content has fallen to (1-1/$e^2$) of the total. 
\begin{comment}
This definition turns out to be equivalent to the one based on irradiance in the case of Gaussian beams, since: 
\begin{equation}
e^{-r^2} = \dfrac{\int_0^r e^{-r'2}\mathrm{d}r'}{\int_0^\infty e^{-r'2}\mathrm{d}r'} \label{eq:em2} 
\end{equation}
\end{comment}

The analysis algorithm was designed to fit such ellipses to the images. First,
the images were fitted with a two-dimensional asymmetric Gaussian of height $I_0$, with variable centre coordinates $x_0$ and $y_0$, the widths of the major and minor axes, $\sigma_x$, $\sigma_y$ and a rotation angle $\alpha$, plus an offset $I_\mathrm{off}$: %\md{I removed the numbers of second equations. For me those three equations could go into one line}

\begin{equation}
I(x,y)= I_\mathrm{off} + I_0 \cdot \exp\left( -\dfrac{a \cdot (x-x_0)^2 +c \cdot(y-y_0)^2 +2b\cdot(x-x_0)·(y-y_0)}{2} \right) \label{eq:ellipse}
\end{equation}
with:
\begin{equation}
a = \dfrac{\cos(\alpha)^2}{\sigma_x^2} + \dfrac{\sin(\alpha)^2}{\sigma_y^2}; \quad
%\nonumber \\
b = -\dfrac{\sin(2\alpha)}{2\sigma_x^2} + \dfrac{\sin(2\alpha)}{2\sigma_y^2}; \quad
%\nonumber \\
c =  \dfrac{\sin(\alpha)^2}{\sigma_x^2} + \dfrac{\cos(\alpha)^2}{\sigma_y^2} 
%\nonumber
\end{equation}

\begin{figure}[htp]
\centering
\includegraphics[width=0.8\linewidth]{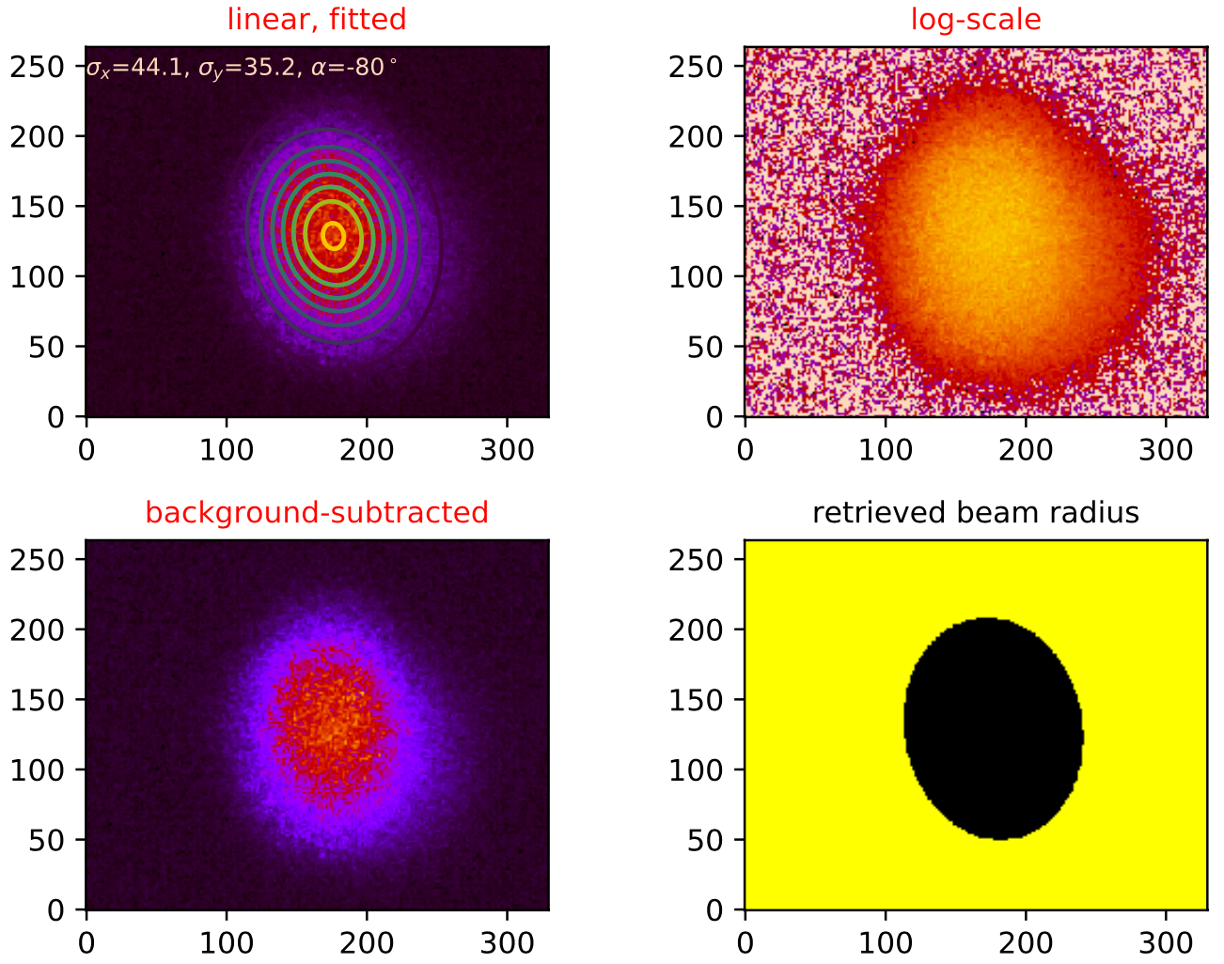}
\caption{\label{fig:laseranalyzedpicture}The upper left part shows a laser spot image in linear scale, fitted with the 2D-elliptic Gaussian. The upper right picture presents the same plot on a logarithmic scale and without the retrieved concentric ellipses. The lower left panel shows the background-subtracted image and the lower right displays the ellipse that contains $(1- 1/e^2)$ of the total normalized distribution.}
\end{figure}

The background was evaluated considering the outside regions of the fitted ellipse, at 5\,$\sigma_{x,y}$ from the centre, up to a suitably chosen cut-off point of the image. The cutout was chosen by eye on the basis of a constant number of entries, coinciding with a constant image colour. After subtraction of the background, the image was normalized and integrated into ellipses of the same axis ratio and rotation angle from ($x_0, y_0$) to the point where the integral reaches $(1-e^{-2})$. At that point, the major and minor axes were evaluated. Subsequently, the full-angle beam divergences $\Theta_{x,y}$ were calculated.  
Finally, the same procedure was used for ellipses of any size (always with the same axis ratio, rotation angle, and centre) and the image intensity was computed for ellipses of increasing half-axes (see Fig.~\ref{fig:laseranalyzedpicture}).

Table~\ref{tab:spotsizes} shows the results of the analysis of the full sample of images taken from the laser spot. 

\begin{figure}[htp]
\centering
\includegraphics[width=0.47\linewidth]{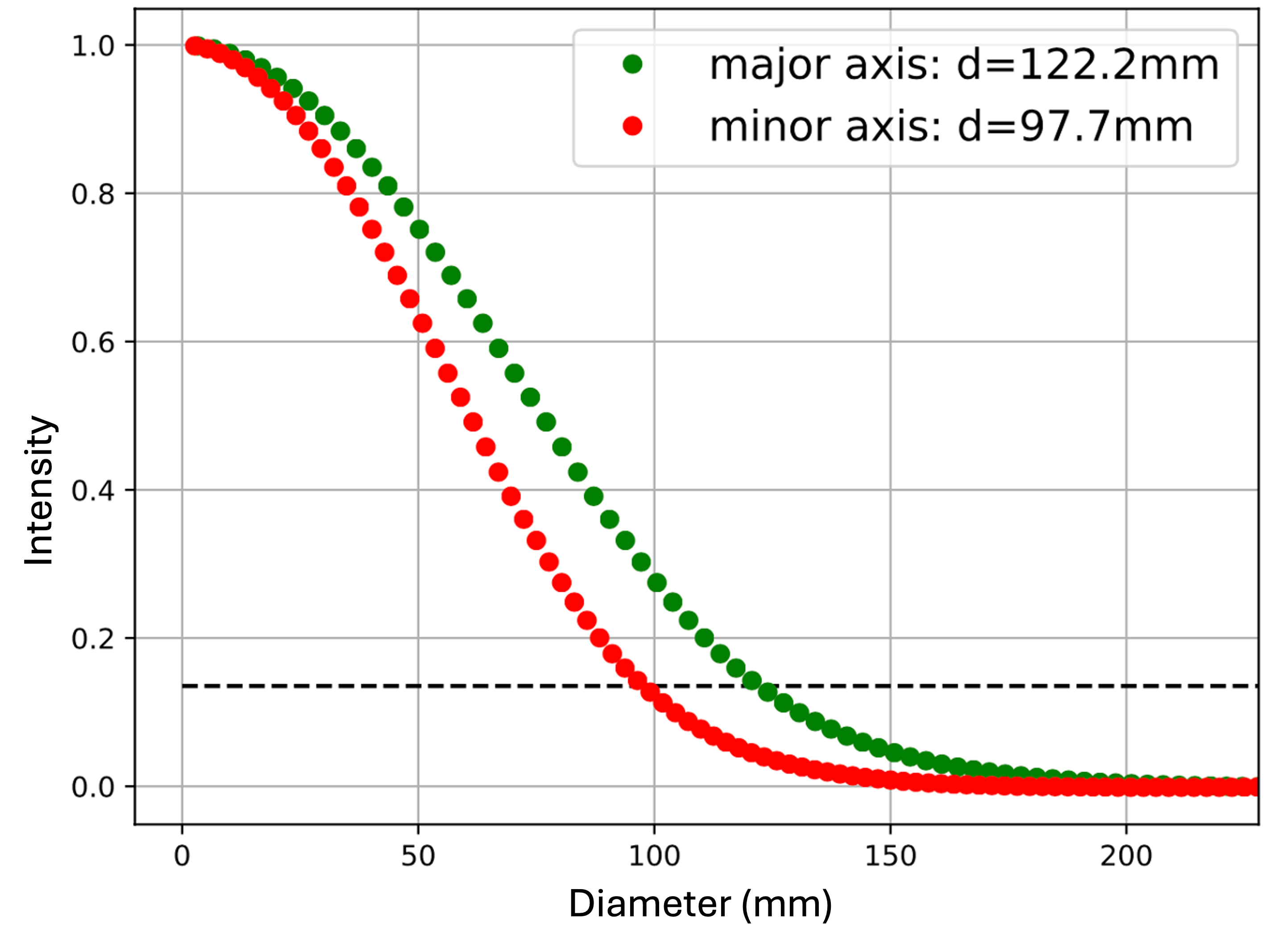}
\includegraphics[width=0.505\linewidth,trim={0cm 0cm 1cm 0cm},clip]{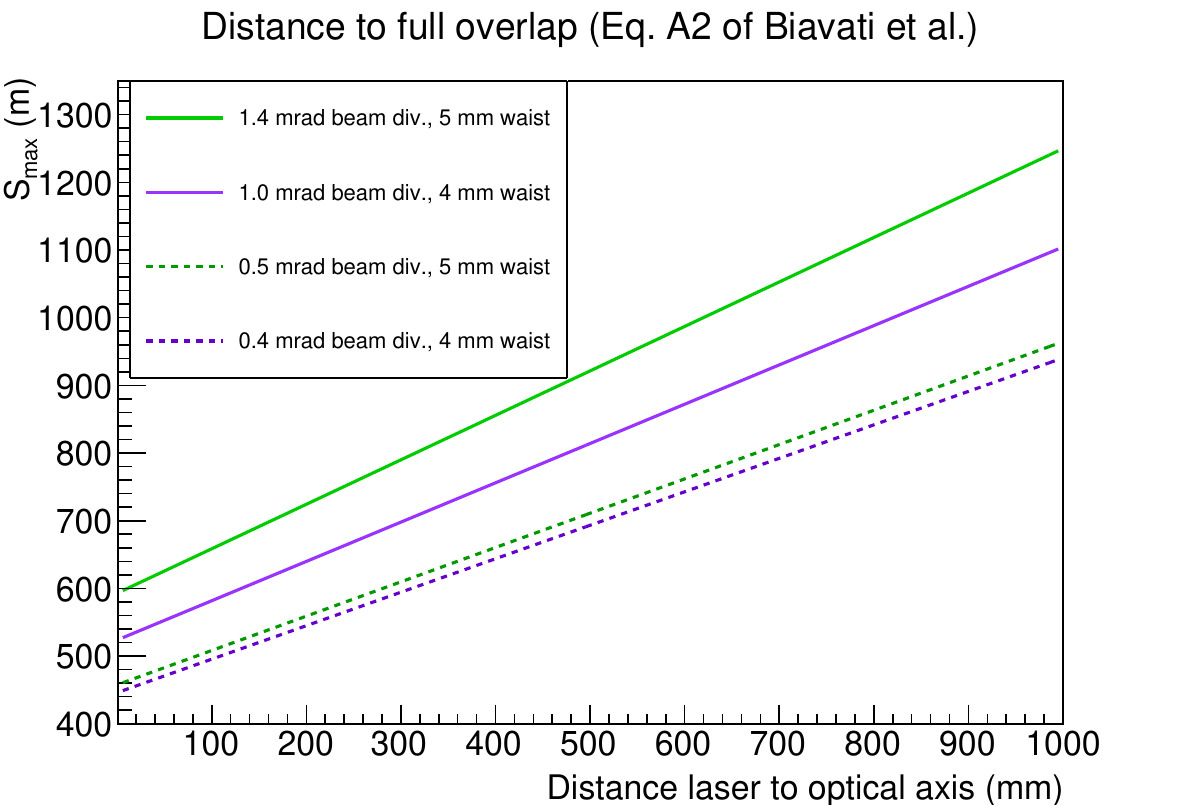}
\caption{\label{fig:intensitycoverage}Left: Intensity coverage of the background subtracted and normalized medium-intensity non-saturated image, as a function of the fitted ellipse’s major and minor axes. 
\sapocol{The dashed black line shows $e^{-2}$, that is the nominal reference for the opening angle of a Gaussian laser beam. }
Right: Distance to full overlap~\citep{Biavati:2011} for the green and UV elastic lines, for the nominal laser beam divergences (dashed lines) and the measured ones (full lines).}
\end{figure}

\begin{table}[htp]
\centering
\begin{tabular}{cc|cc|c}
\toprule
$d_\mathrm{maj}$ & $\Theta_\mathrm{maj}$ & $d_\mathrm{min}$ & $\Theta_\mathrm{min}$ &  $\alpha$ \\
(mm) & (mrad) & (mm) & (mrad) & ($^\circ$) \\
\midrule
$120\pm3$ & $1.39\pm0.04$ &  $90\pm8$ & $1.06\pm0.09$ & $102\pm5$ \\
\bottomrule
\end{tabular}
\caption{\label{tab:spotsizes}
Results of the eight non-saturated laser spot images analyzed in the far field: major axis of the spot ellipse ($d_\mathrm{maj}$), beam divergence along the major axis ($\Theta_\mathrm{maj}$), minor axis of the spot ellipse ($d_\mathrm{min}$), beam divergence along the minor axis ($\Theta_\mathrm{min}$), rotation angle of the ellipse ($\alpha$). See \citet{phdthesis-maggio} for further details.}
\end{table}

The beam divergences were measured at different laser intensities. We observed that the beam divergences obtained are significantly larger (by a factor of 2.6 along the major axis, and a factor of 1.7 wider along the minor axis) than the ones claimed by the manufacturer (unless the absolutely lowest laser intensity was used).
%\md{Before no mention of intensity}). 
The ratio between the major and minor axes is greater than 1.25 and increases as the laser intensity becomes lower.
%\md{Before no mention of intensity}). 
Moreover, the beam appears rotated by about 10$^\circ$ with respect to the vertical axis. 
%, although the form of the laser spot shows often further substructures, additionally to its elliptic shape.

The degradation in beam quality might have been due to the repair carried out in 2017.  
%\md{which} 
%or manipulations of the small dichroic mirrors in the second and third harmonic generators \md{what do you mean/}.
Since we did not test the beam quality in the far field \textit{before} the laser repair, we cannot compare with its original quality. The measured beam shape divergence increases the distance to full overlap of the pBRL by up to 150~m, as shown in Fig.~\ref{fig:intensitycoverage} (right). 

%\aco{I am a bit lost with this subsection, it's a bit messy and I think I am missing some explanations. }

%The images have been analyzed using a ROOT macro, after conversion to FITS format.  They are treated as 2D histograms and then retrieved into an array of numbers. After extracting a background image (a picture without the artificial laser star), the program searches for the spot maximum by scanning the image and calculates the center of gravity of the spot. Finally, different containment radii (enclosing 50\%, 80\%, 90\%, 95\%, 99\% and 99.9\% of the light in a circular shape) are calculated and displayed.  For the conversion from pixel size to millimeters, we used the marks on the metric paper and calculated their distance in pixels: one millimeter corresponds to approximately 48 pixels.\\

\bigskip\textbf{Laser arm}

\begin{figure}[h!t]
 \centering
 \includegraphics[width=0.67\textwidth]{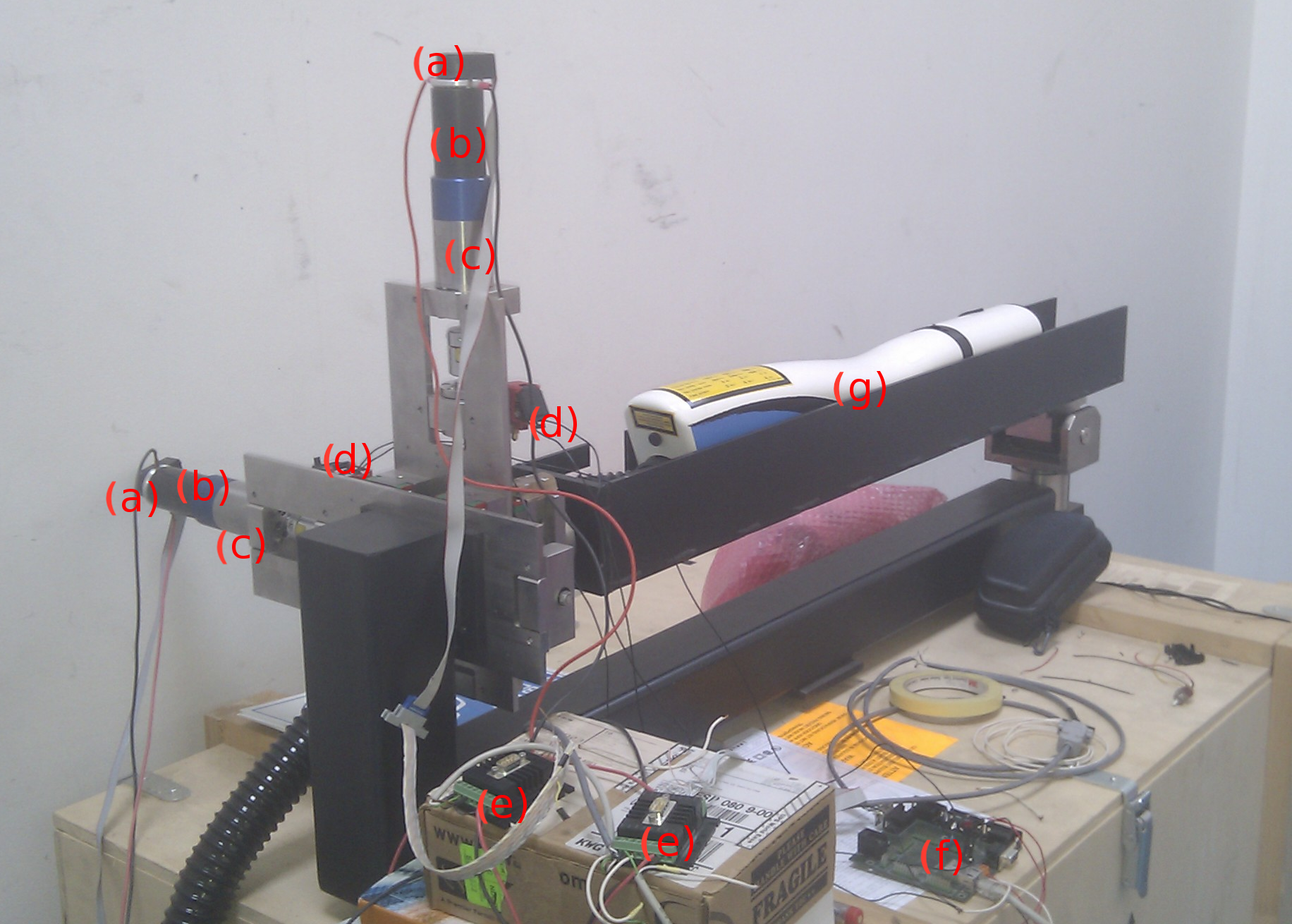}
 \includegraphics[width=0.32\textwidth]{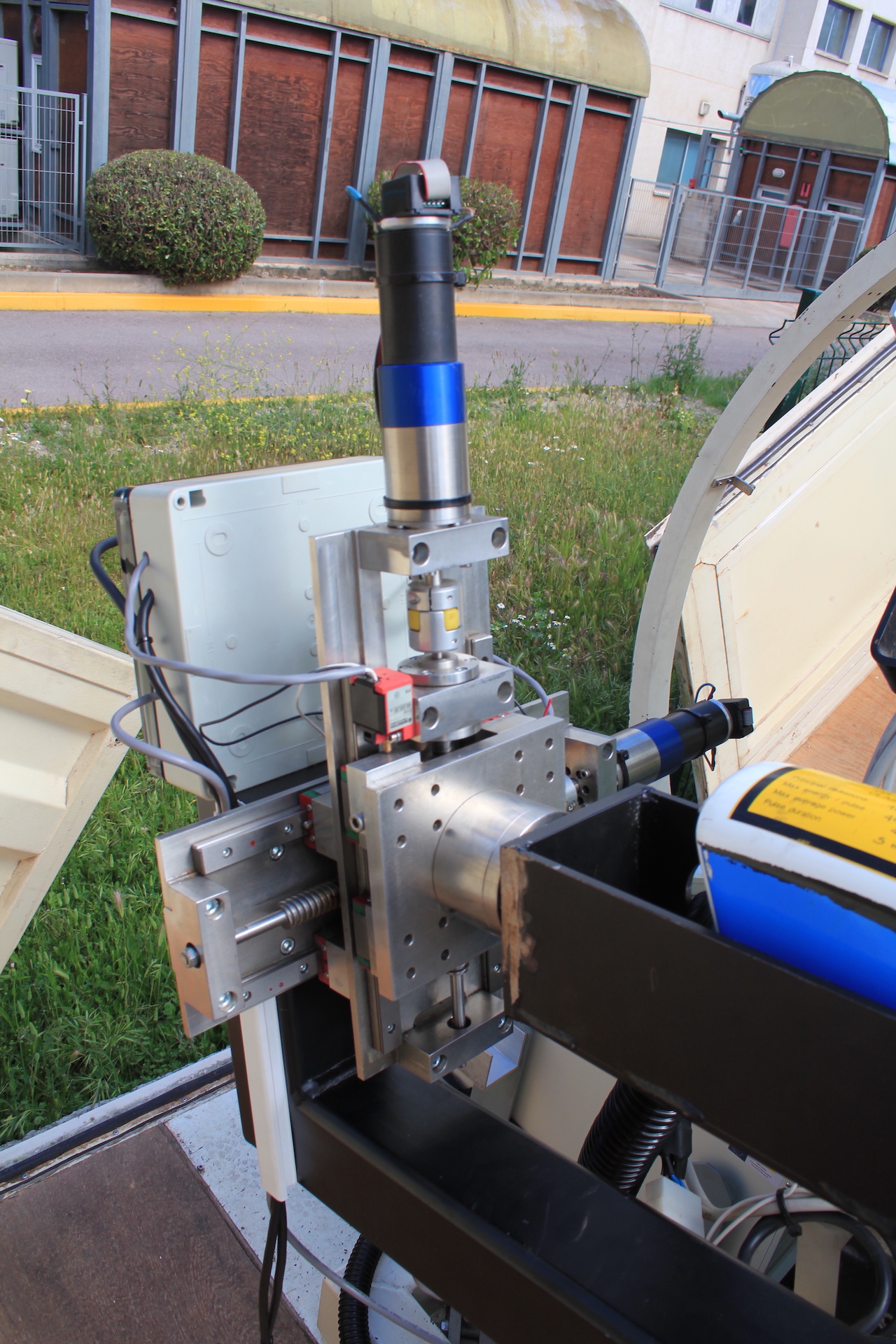}
  \caption{Left: The laser arm in the laboratory: the encoder (a), the motors (b), the 111:1 reduction (c), the final switches (d), the driver (e), the control board (f) and the laser arm control (g). Right: The laser arm mounted on the telescope. %\aco{I could not find the original.}%\md{the left figure is a bit shady}\aco{Yes, do we have a better one?}
  \label{fig:laserarm}
}
\end{figure}

Fig.~\ref{fig:laserarm} shows the laser arm designed to control the correct alignment between the laser and the telescope. It is based on a $XY$ table, designed to point with a precision better than 1~mrad. To move two different axes, two DC~motors of type Faulhaber~3863H024C are attached to an endless screw of 1~cm/rev, one for each degree of freedom, controlled by encoder of type Faulhaber~HEDL5540. %\aco{Unclear.}
%The full system is shown in Fig.~\ref{fig:laserarm}.
A 111:1 reduction is used to improve the resolution.  To fix the initial position of the table, two final switches are used. To control the entire laser arm, a control board with Ethernet interface has been developed.
%The laser control board has two serial ports and one Ethernet port. 
%The last is used to communicate with a central computer. One of the serial ports are used to control the unit laser control. 
%As already stated earlier (see section~\ref{sec:laser}), the laser has the possibility to be controlled via serial port. 
%The other port is used to control the motors. Both motors are connected to the drivers. 
%Further, the drivers are connected in cascade by serial communication. 
%We only need a port to connect both drivers. 
%The drivers are intelligent enough to make a soft start and soft stop of the motors. All the system is controlled by sockets.

\subsection{The Laser Dichroic Guiding Mirrors}
\label{sec:dichroicmirrors}

\begin{figure}[h!]
\centering
\includegraphics[width=0.389\linewidth,trim={3cm 19cm 3cm 15cm},clip]{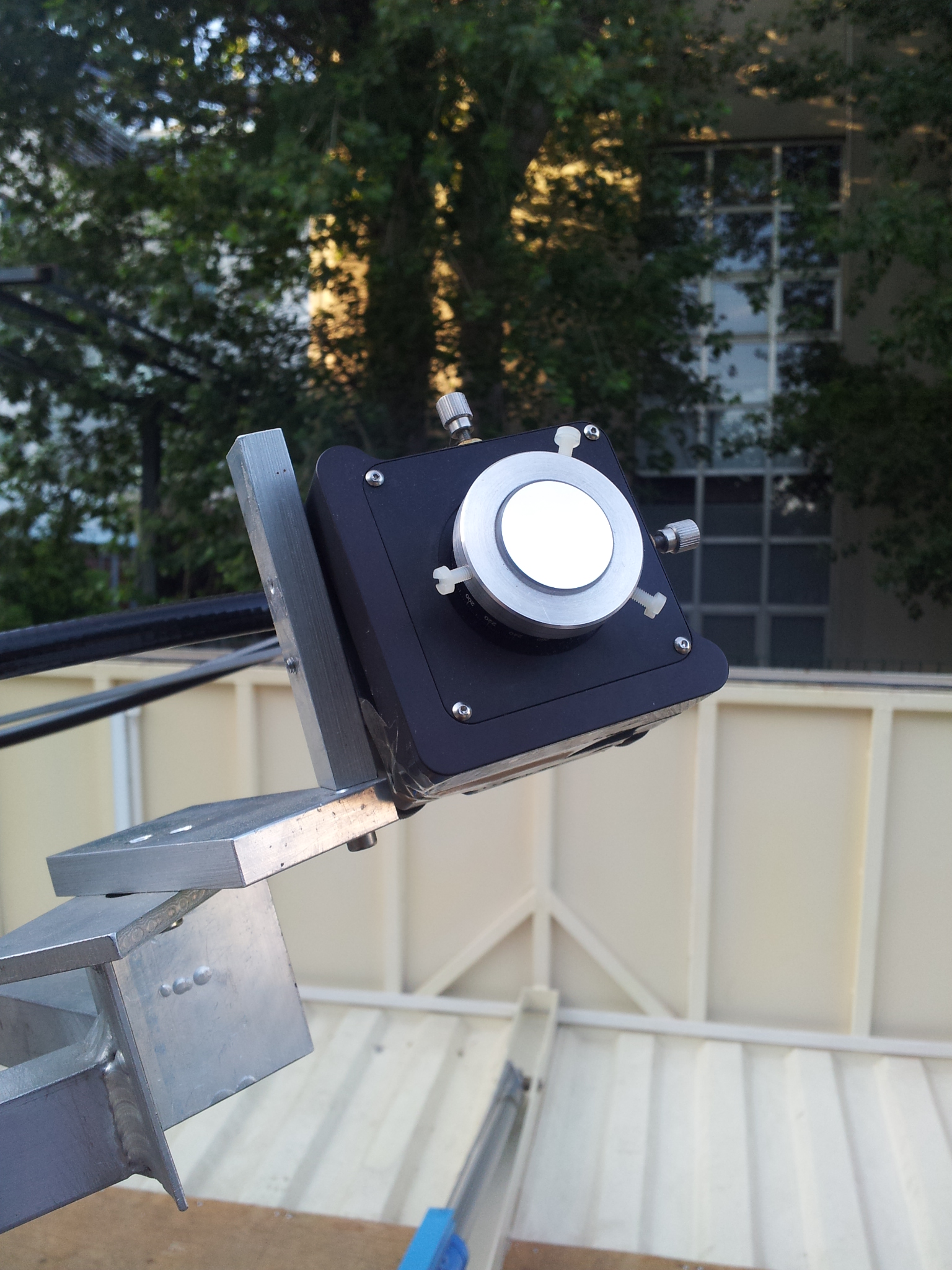}
\includegraphics[width=0.418\linewidth]{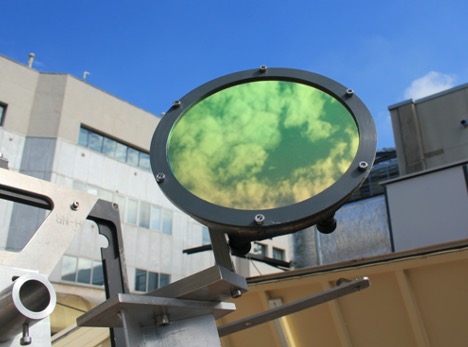}
\includegraphics[width=0.15\linewidth,trim={3cm 0cm 9cm 6cm},clip]{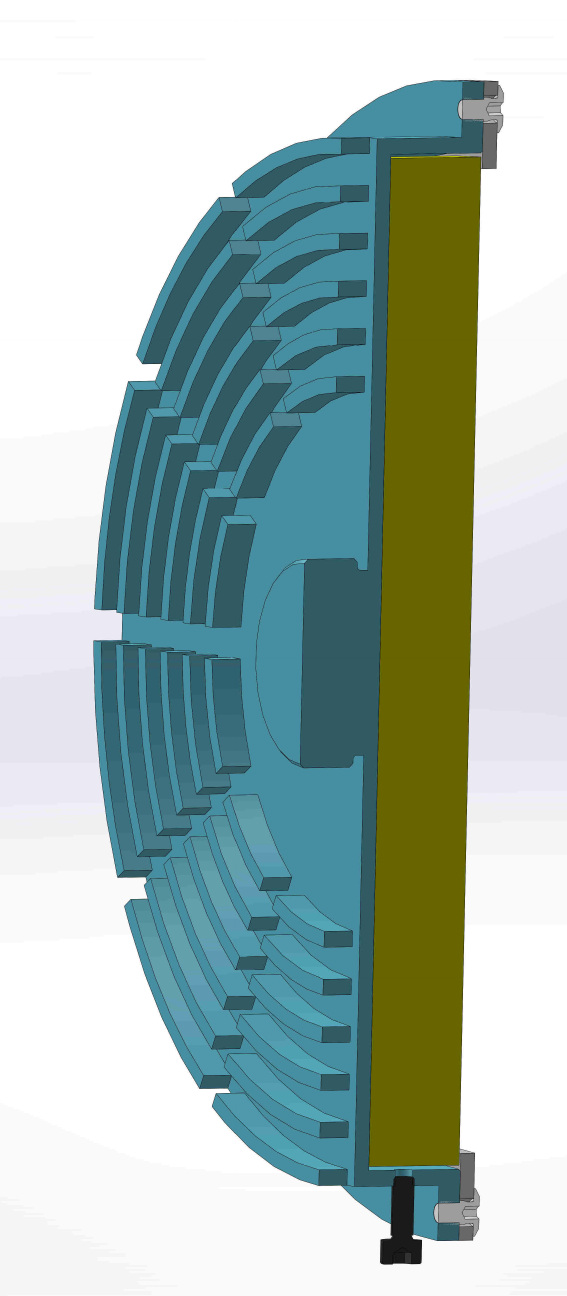}
\caption{\label{fig:guidingmirrors} Pictures of the first guiding mirror 1$^{\prime\prime}$ fused silica mirrors of type MI1050-SBB (left) and the final custom-made ones (center). Right: Design of the guiding mirror holder, with heat dissipator.}
\end{figure}

%\begin{figure}[h!]
%\centering
%\caption{\label{fig:guidingmirrorsholder}}
%\end{figure}

The laser is mounted on one side of the telescope structure (see Fig.~\ref{fig:lidar_scheme2}) at about 1.1~m distance from the optical axis of the telescope. In order to make the laser light co-axial, 
%(see Sect.\ref{sec:biaxiality}) 
we engineered a guiding arm system completed with mirrors to guide the laser light. This system also allows for a) precise adjustment of the laser beam orientation and b) filtering of the laser 1064~nm line, which may damage the light guide (see Sect.~\ref{sec:llg}).

The arm holds the guiding mirrors and connects with the focal plane. For simplicity, the exact inclination of these mirrors needs to be adjusted by hand and fixed with screws.  Space restrictions inside the container did not allow us to have a perpendicular design, and the arms are mounted at specific angles of $61.1\pm 0.3^\circ$ 
(see Fig.~\ref{fig:lidar_scheme2}).

We purchased standard 1$^{\prime\prime}$ fused silica mirrors of type MI1050-SBB from the company Precision Photonics\footnote{\url{http://www.precisionphotonics.com/}} (surface flatness $l$/10, damage threshold of $>$1~J~cm$^{-2}$ at 355 nm and reflectivity $>$99\% (see Fig.~\ref{fig:guidingmirrors} (left) and~\citep{Barcelo:2011stu}). We realized, however, that mirrors of only 1$^{\prime\prime}$ diameter were insufficient to allow carrying out a pre-alignment of the laser beam (see also Sect.~\ref{sec:calib:pointing}), particularly because the second dichroic allowed only a margin of $\sim$5~mrad for the fine-adjustment by the remote-controlled laser arm. The beam needed to be pre-adjusted to that precision beforehand by manual movement of the mirrors, a task impossible to achieve in one night, even by an experienced person. After a thorough study~\citep{Eudald:thesis}, we found that a mirror of at least 10~cm diameter was necessary to achieve the required $\sin(\alpha) \cdot D/214~\mathrm{cm} \approx 25$~mrad margin of operation for the pre-alignment. 

%At the start of this project, 1$^{\prime\prime}$ fused silica mirrors of type MI1050-SBB were purchased from the company Precision Photonics\footnote{\url{http://www.precisionphotonics.com/}} (surface flatness $l$/10, damage threshold of $>$1~J~cm$^{-2}$ at 355 nm and reflectivity $>$99\% (see Fig.~\ref{fig:guidingmirrors} (left) and~\citep{Barcelo:2011stu}). During the first laser alignment attempts, we realized, however, that mirrors of only 1$^{\prime\prime}$ diameter were insufficient to allow carrying out a pre-alignment of the laser beam (see also Sect.~\ref{sec:calib:pointing}), particularly because the second dichroic allowed only a margin of $\sim$5~mrad for the fine-adjustment by the remote-controlled laser arm. The beam needed to be pre-adjusted to that precision beforehand by manual movement of the mirrors, a task impossible to achieve in one night, even by an experienced person. After a thorough study~\citep{Eudald:thesis}, we found that a mirror of at least 10~cm diameter was necessary to achieve the required $\sin(\alpha) \cdot D/214~\mathrm{cm} \approx 25$~mrad margin of operation for the pre-alignment. 

\begin{figure}[h!]
\centering
\includegraphics[width=0.75\linewidth, angle=1.2]{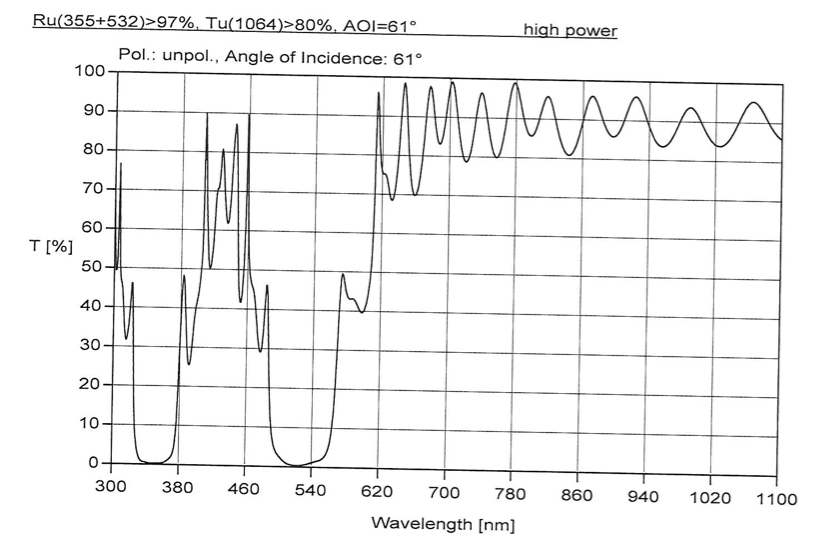}
\caption{\label{fig:guidingmirror_transmission}Transmission of the dichroic guide mirrors, as a function of wavelength.}
\end{figure}

The final 5$^{\prime\prime}$ diameter dichroic guiding mirrors have been designed such that they resist high laser power, have reflectivity $>$97\% for the 355~nm and 532~nm wavelengths at an angle of incidence of 61$^\circ$ (see Fig.~\ref{fig:guidingmirror_transmission})  and a transmission $>$80\% at 1064~nm and high surface flatness $l/8$ over 2$^{\prime\prime}$ and surface quality 20-10. The dichroic mirrors were designed by the company Optoprim\footnote{\url{https://www.optoprim.it/}} (Milano, Italy) and manufactured by the company Laser Components\footnote{\url{https://www.lasercomponents.com/en/}}(Olching, Germany). Figure~\ref{fig:guidingmirror_transmission} shows the final transmission spectrum achieved from guide mirrors, which is much higher than the requirements. The mirrors were mounted on a metallic custom-designed structure by the IFAE engineering division and allowed the adjustment of the mirrors on two axes by hand. %\md{Mirror mount to be described }

%The alignment procedure requires \md{complete} and it's documented in~\citet{Eudald:thesis}. 

\subsection{Stability of the Guide Mirror Structure and Mount}
\label{sec:stabilitydichroics}

The stability of the structure holding the dichroic guide mirrors was tested for telescope movements in \sapocol{elevation}
%zenith 
and azimuth, as well as for different temperatures~\citep{Eudald:thesis}. 

For these tests, we placed a millimeter paper in front of the second guide mirror and fixed it so that it did not move during the tests. Then the center of the laser spot was marked on the millimeter paper when the telescope was in the parking position. Throughout all tests, the same position of the XY-table was maintained, as will be the case during normal operation, once the system has been aligned. 

The tests consisted of taking photographs from the laser spot at varying telescope \sapocol{elevation} %zenith 
positions and fixed azimuth and at varying telescope azimuth positions at fixed \sapocol{elevation}
%zenith 
angles. 
%This is how we check the movement of the laser arm depending on the two angles of freedom of the chassis.
%For the azimuth fixed in the parking position
%, the zenith was moved from parking up to 600 motor paces in a downward zenith angle direction, approximately 20◦, with intervals of 200 motor paces. Figure 7.1 shows the four spot photos for each zenith position.
We observed that varying the \sapocol{elevation angle by 70$^\circ$}
%zenith angle by 70$^\circ$ 
led to a movement of the laser spot by up to 4.5~mm, corresponding to 4.5~mm/(0.84~m + 1.3~m)$\approx$2~mrad, i.e. slightly less than half the field of view of the LLG pinhole. These movements were also tested for possible hysteresis or temperature dependencies, but none could be detected within the sensitivity limit of 1~mm, i.e. 0.5~mrad. The exact displacement could be reproduced after several days. 
No movement of the laser spot could be detected during an approximately 250$^\circ$ rotation in azimuth executed, leading to a limit of $<$1~mm, corresponding to $<$0.5~mrad.
%, when the telescope moves in the azimuthal direction. This displacement is within the uncertainty predicted by the engineers when designing the original CLUE telescope.  
In addition, in this case, no hysteresis could be detected. 

All in all, beam alignment may be necessary %\md{but not with the brl as said above?} 
to be carried out at various \sapocol{elevation} %zenith 
angles, and the results stored in lookup tables to correct for the observed sagging of the arm. Movements in azimuth should not cause any alteration of the alignment.

\subsection{Short-range System}
\label{sec:shortrange}

\begin{figure}[h!]
\centering
\includegraphics[width=0.55\linewidth, angle=90]{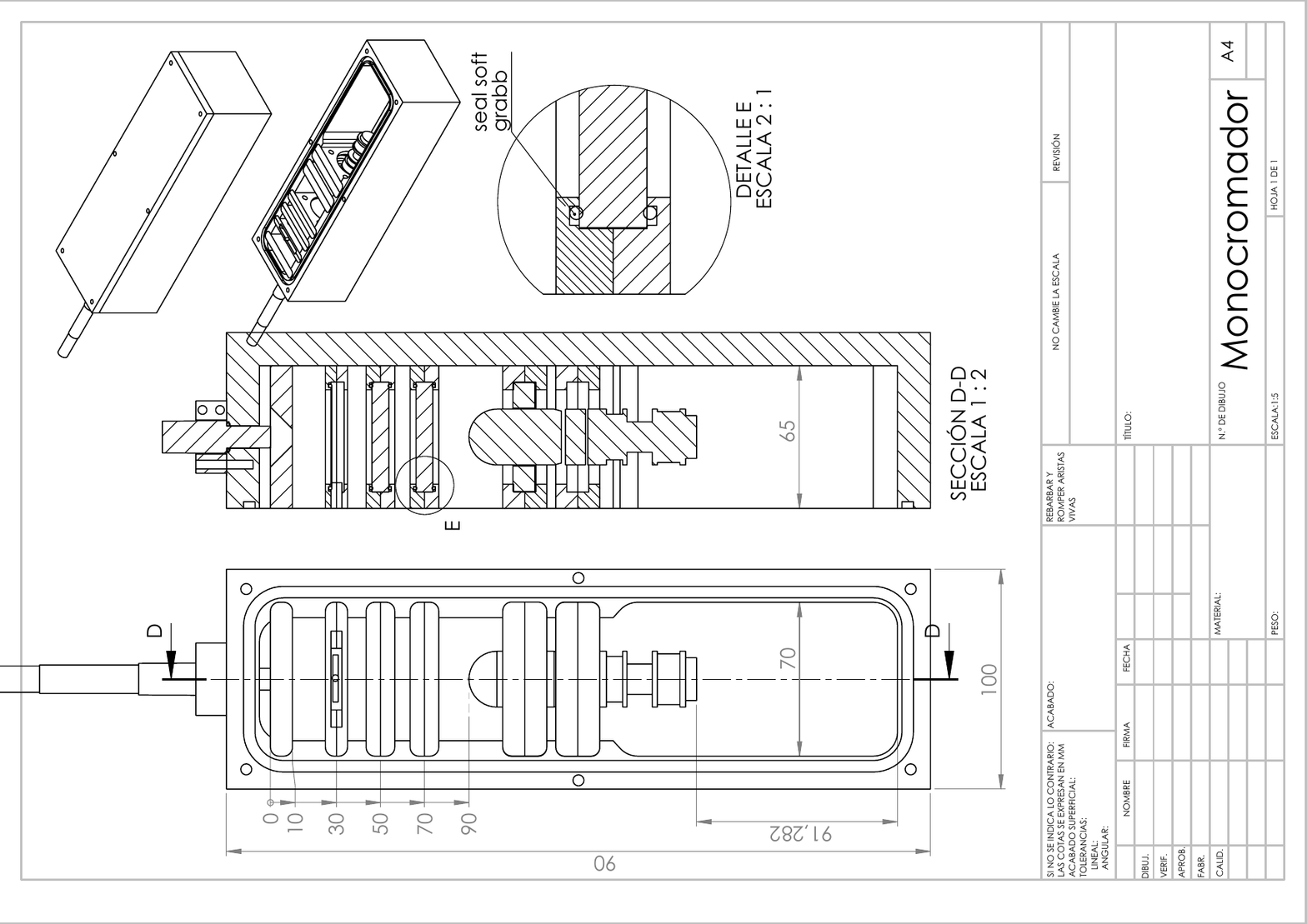} 
  \caption{Design of the protective box for the near-range  optics and PMT.\label{fig:design_monochromator}}
\end{figure}

\begin{figure}[htp]
    \centering
    \includegraphics[width=0.2\linewidth]{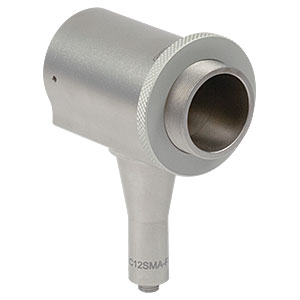}
      \includegraphics[width=0.2\linewidth]{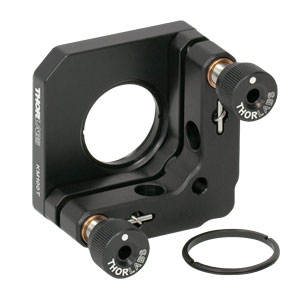}
        \includegraphics[width=0.58\linewidth,trim={0cm 9cm 8cm 8cm},clip]{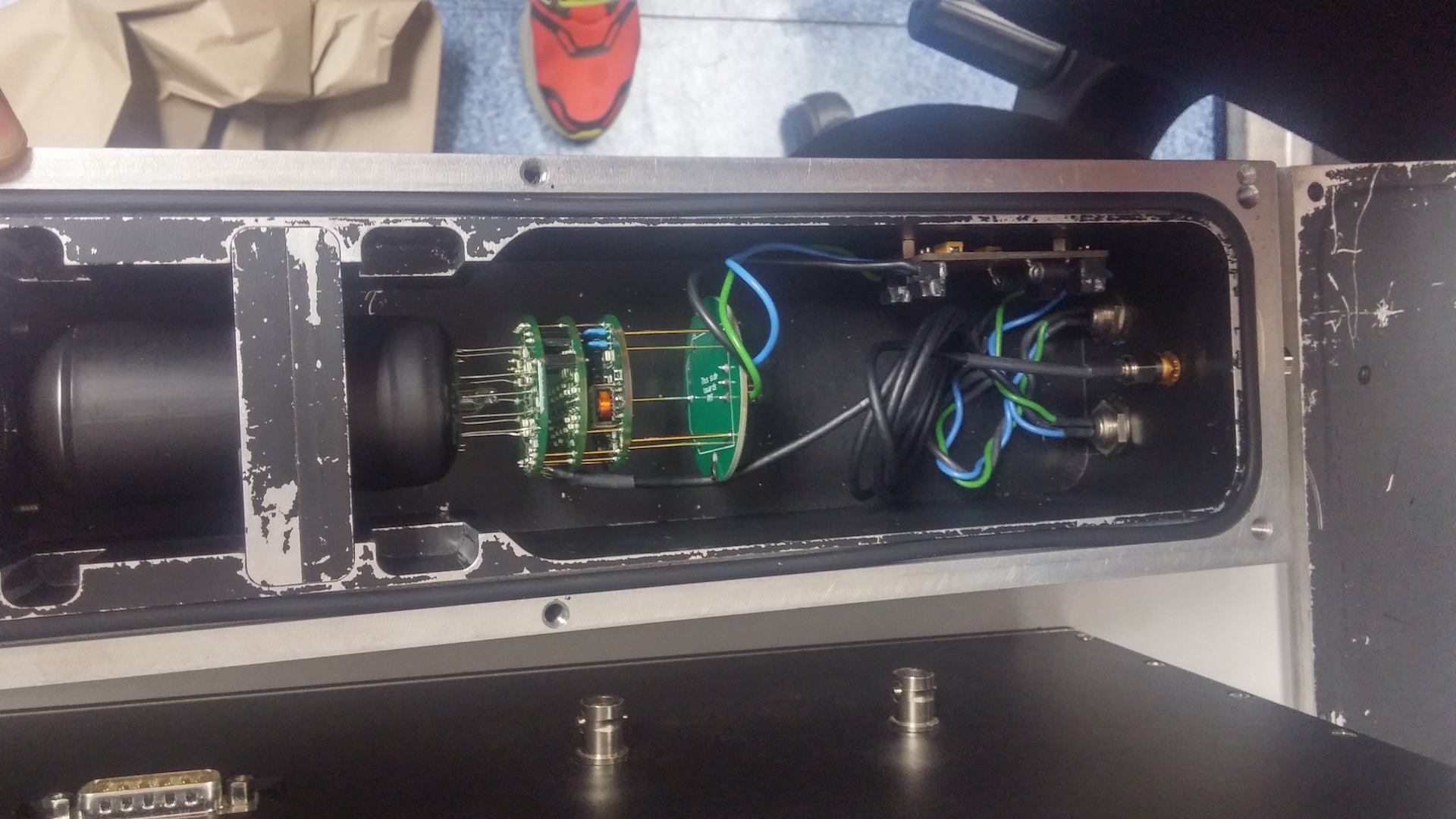}
    \caption{Components of the near-range system. The parabolic mirror (left), the support to mount the mirror on the telescope structure (center) and the near-range PMT housed in a protective box (right). \label{fig:nr_components}}
\end{figure}

In order to increase sensitivity at distances below 500~m from the ground where most of the aerosols of the nocturnal boundary layer are found with an approximately exponential decrease of density with height at the ORM~\citep{Fruck:2022igg}, in a range where the signals are already strongly affected by saturation, the pBRL has been equipped with additional near-range optics.  
%that allow to determine the full ground layer transmission reliably.
The near-range optics is composed of a Thorlabs RC12SMA-F01 mini-telescope\footnote{\url{https://www.thorlabs.com/thorproduct.cfm?partnumber=RC12SMA-F01}} held by a standard mount KM100T\footnote{\url{https://www.thorlabs.de/thorproduct.cfm?partnumber=KM100T}} \sapocol{(see Fig.~\ref{fig:design_monochromator} and~\ref{fig:nr_components})}. The near-range mini-telescope collimator is composed of a parabolic mirror used to focus light onto a 1.5~mm optical fibre of type FT1500~UMT\footnote{\url{https://www.thorlabs.com/thorproduct.cfm?partnumber=FT1500UMT}} (NA: 0.39, transmission from 300~nm to 1200~nm) screwed to the collimator with an SMA connector. The fibre transmits the signal to a PMT housed within a protective box with interference filter~\citep{calpe:thesis}. The box allows placing up to two interference filters in front of the PMT and hence makes it suitable to observe one of the two elastic lines.

With a field-of-view of $\sim$30~mrad, the near-range system can achieve full overlap basically from a few meters of distance~\citep{Biavati:2011}. 
%With only the use of the elastic channel, the full overlap is already reached at 30~m, consistent with the theoretical expectation~\citep{Biavati:2011}, see also Eq.~\ref{eq:short_range}.

%\begin{figure}[htp]
%    \centering
%    \includegraphics[width=0.5\linewidth]{Figures/sec3/Transmission.png}
%    \caption{Near-range fibre transmission as a function of wavelength. \label{fig:nr_transmission}}
%\end{figure}

%\clearpage
\section{Commissioning, Calibration and Monitoring\label{sec:calibration}}
%\md{Maybe the title should be commissioning? Or only calibration? Certainly not monitoring I wol}
In \S\ref{sec:tech_design} we have detailed the technological aspects of the pBRL. In this section, we discuss activities needed to characterize or calibrate the pBRL as a whole before putting it into operation.

\subsection{Calibration of the Pointing and Beam Alignment}
\label{sec:calib:pointing}

\begin{figure}[h!t]
\centering
\includegraphics[width=0.67\linewidth]{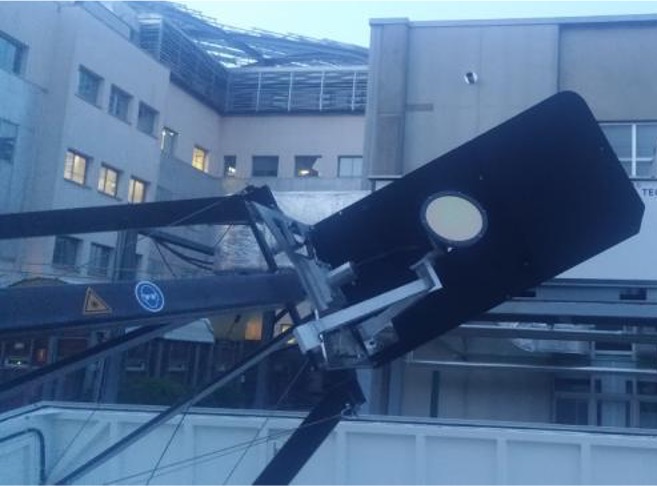}
\includegraphics[width=0.28\linewidth]{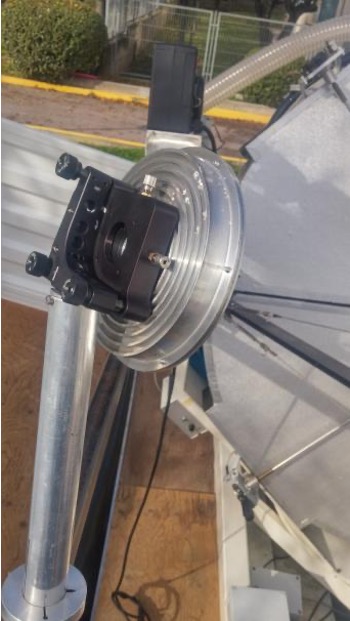}\caption{\label{fig:blackprotection}Left: Picture of the black protection shield deployed to alignment position. Right: Picture of the first dichroic mirror with its fixation pole (lower left part of the image). The screw to untighten the movement of the pole and mirror is found on the aluminum ring surrounding the pole, visible on the very lower left part of the image.}
\end{figure}

%A way to perform a pre-alignment of the laser beam is demonstrated in Fig.~\ref{fig:prealignment}.
The alignment procedure must be performed during the night and takes about two hours. It requires a sheet of white paper, a set of Allen keys, and an oscilloscope.
%You need the following tools:
%• a sheet of white paper
%• a set of Allen keys
%• the oscilloscope
%• about two hours of time
A dedicated black light protection shield has been developed for the purpose of beam alignment, which protects the surroundings from spurious laser light (see Fig.~\ref{fig:blackprotection}, left). 
%Screw it to the hole prepared for that position.
%2.Move the arm to X = 1.800.000, Y = 1.800.000

First, the telescope has to be moved to the lowest elevation (highest zenith) allowed by the end-switch to operate the laser, and the laser has to be operated at the lowest light intensity. That intensity can be considered eye-safe.  
%3. Move the telescope to the minimum zenith position, with which the laser can be operated (should be 215 steps now).
%4.Power on the laser with 0\% intensity
With the white paper and the low-power laser beam, the first laser guide dichroic mirror is aligned by hand using its fixation screws (see Fig.~\ref{fig:blackprotection}, right) to point the laser beam towards the centre of the second guide mirror. 
%5. Hold a white paper at the exit of the laser and check, whether you can see the green spot, otherwise gradually increase intensity until you can see the blinking spot on the paper. DO NOT OPERATE THE LASER WITH INTENSITY GREATER THAN 10\% AT THIS STAGE
%6.Now move the white paper towards the first dichroic mirror maintaining the green spot in the middle and check at which part of the mirror the spot incides.
%7. If the spot does not hit the mirror, or if it hits the mirror close to its edge, move the laser arm to center the spot. The spot should not hit exactly the center of the mirror.
%8.Now follow the same steps 4-5 with the second dichroic mirror.
%9.If the spot does not hit the second dichroic mirror, open the lower screw of the first dichroic mirror’s fixation pole (see Figure 3 )and turn the first mirror around such that the laser spot is centered on the second dichroic. Tightly fix the screw again.

\begin{figure}[h!t]
\centering
\includegraphics[width=0.67\linewidth,trim={4.5cm 0cm 2.5cm 0cm},clip]{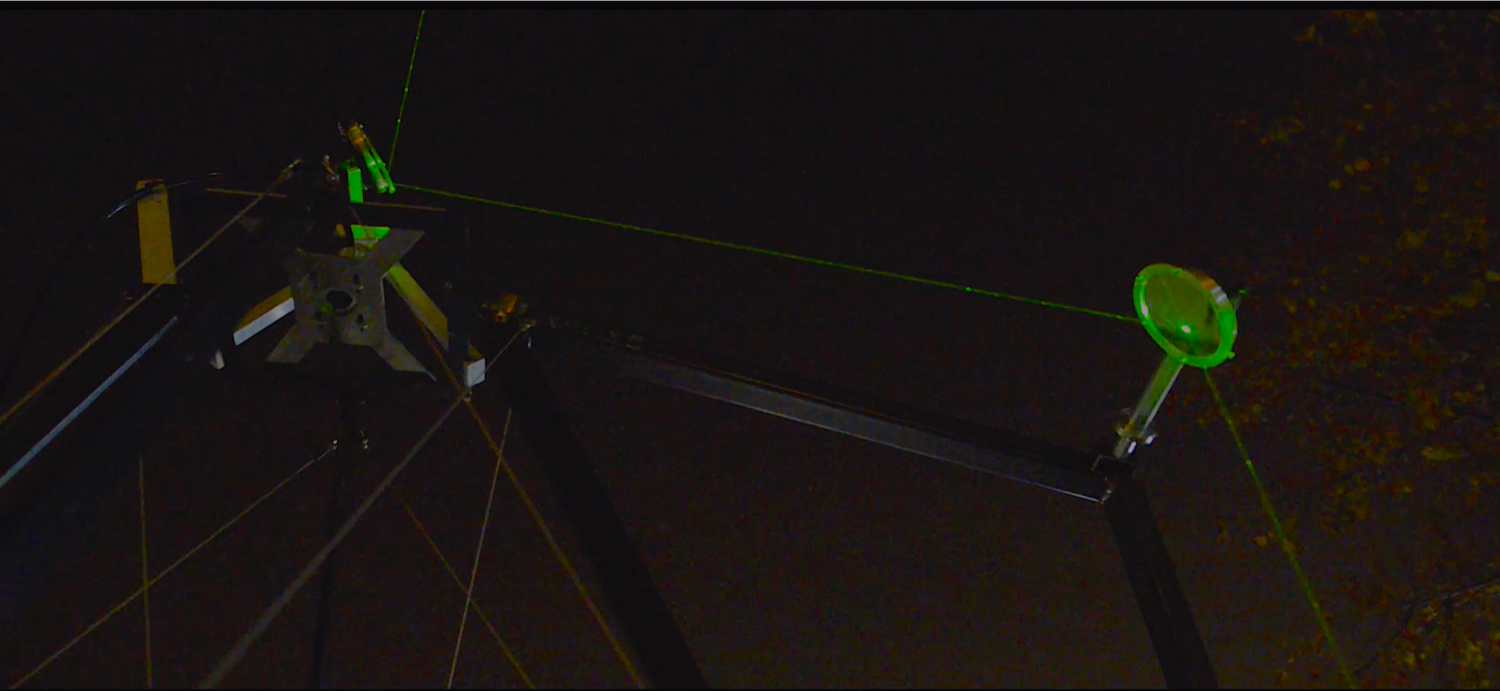}
\includegraphics[width=0.32\linewidth]{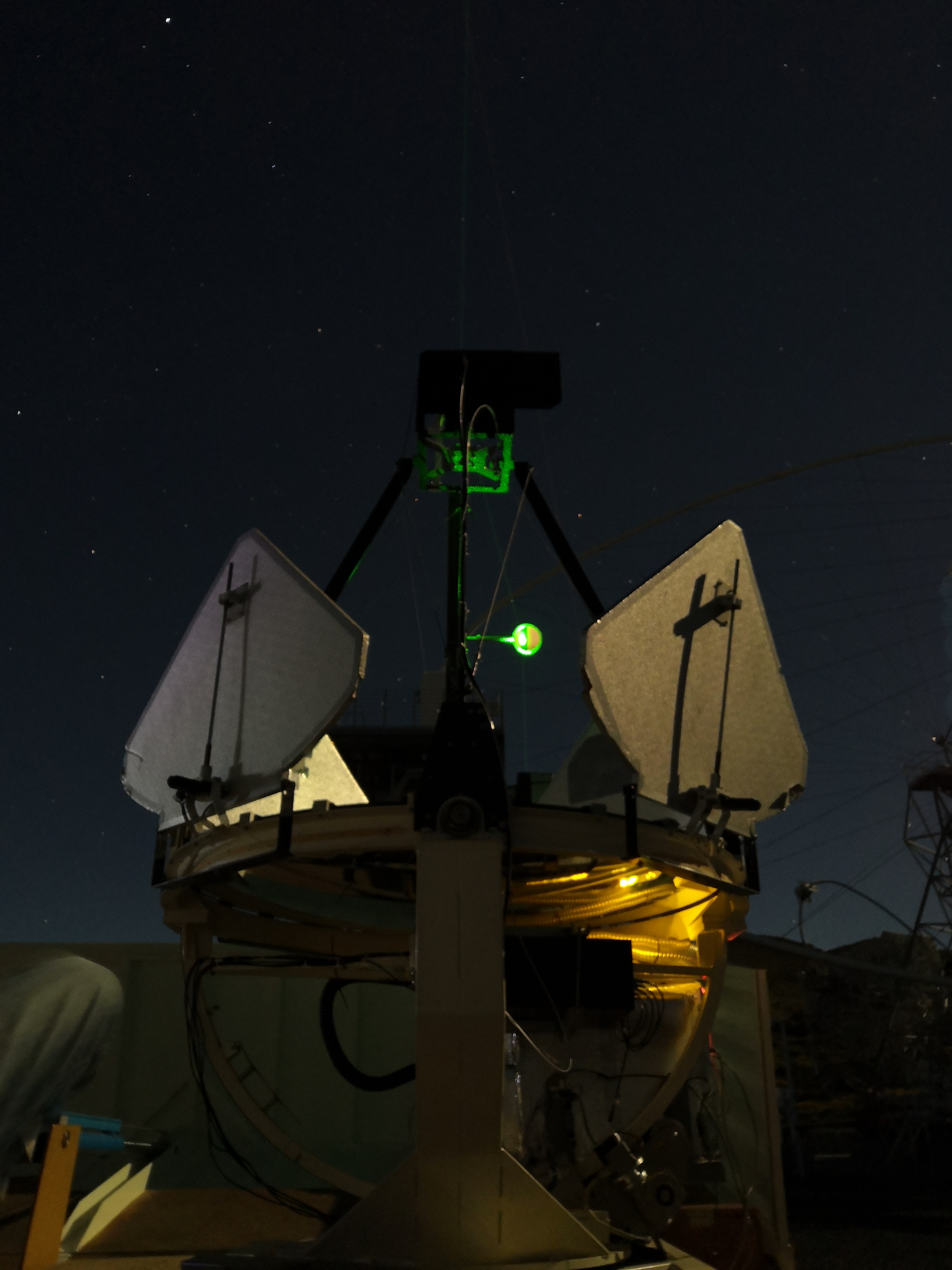}
\caption{\label{fig:laserlight}Images of the laser light path visible during beam alignment.}
\end{figure}

Subsequently, the telescope is moved to \sapocol{90$^\circ$ elevation} %the zenith %\md{it was already there}\mg{No, it was at the lowest elevation allowed by the end-switch, that is about 60 deg zenith} 
and the laser intensity is turned to 100\%. With this, a vertical green laser beam is visible to the eye in the atmosphere above the telescope (see Fig.~\ref{fig:laserlight}). 
%10. Pause the laser and move the telescope to zenith
%11. Switch the laser to 100\% intensity. You should now see the laser light reflected by both dichroics and sent up to the atmosphere.
%\md{Vague from here to end of paragraph} 
If the telescope is exactly pointing to the zenith, 
a possible inclination of the laser beam can be discerned by the human eye, and in such a case, the second dichroic guide mirror must be slightly moved into the opposite direction after the laser is switched off and the telescope has been moved into parking position again. 
%12. Check a possible inclination of the laser beam from all four sides.
%13. If the beam is seen inclined, switch off the laser, move the telescope to parking position and cautiously move the second dichroic mirror to counter-act the inclination. You may have to unfasten and tighten a screw to do so (see Figure 3). Move the telescope back to zenith and switch the laser on again.
This step may be repeated several times until the beam is found to exit the telescope exactly vertically. 
%14. You may have to repeat steps several times, until you cannot see any inclination of the beam any more.

\begin{figure}[h!t]
\centering
\includegraphics[width=0.51\linewidth,trim={2.5cm 0.5cm 3cm 1.5cm},clip]{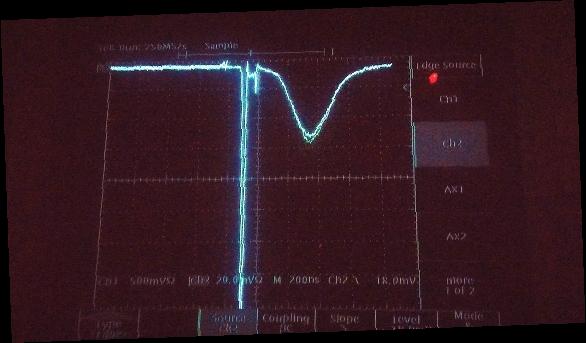}
\includegraphics[width=0.42\linewidth,trim={2cm 2.3cm 4cm 0.8cm},clip]
{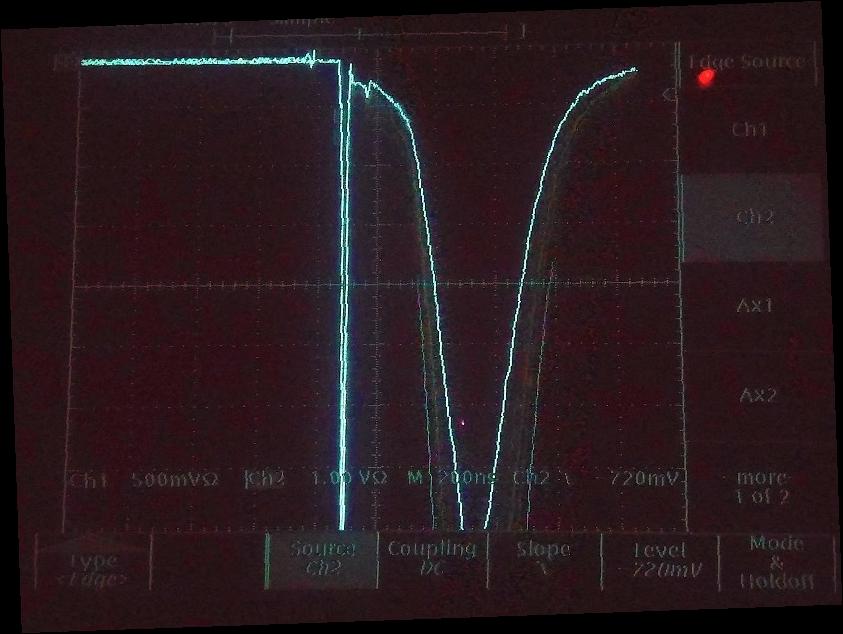}
\caption{\label{fig:signaloscilloscope}Pictures of the oscilloscope signal, before the LIDAR is fully aligned (left) and once it is aligned (right). The first short pulse is due to multiple reflections of the laser beam on the different telescope parts, of which a part finally ends up in the focal point.}
\end{figure}

At this point of provisional alignment, the 355~nm elastic line cable exiting the polychromator unit shall be monitored with an oscilloscope (see Fig.~\ref{fig:signaloscilloscope}). 
%\md{dont get this:}\aco{Same.} 
A strong direct backscatter pulse from the light scattering 3-4 times off the optical and mechanical elements of the LIDAR and ending in the LLG pinhole can always be seen if the PMT is operative.   
%15. Now, get the oscilloscope, unplug the BNC connector with the cable labelled “355 nm” from the Licel and connect it to channel 1 of the oscilloscope. Check that the termination of that line is set to 50$\Omega$.
%16. Ramp up the HVs of the PMTs to 1500. 17.Open the petals of the telescope.
%18. You should now trigger on a negative signal about 10 nm wide, and a small part on the right side of the main peak (see Figure 5). If not, please check that the petal are open and the PMT voltages set correctly.

From that moment on, the laser arm can be moved remotely and the backscatter signal increased. Since the LIDAR is coaxial, there is always a small backscatter signal visible, which increases with better alignment, even in the absence of clouds. 
%19. Now start to move the arm in X-axis, in steps of 100.000. A second person should check that the spot does not move outside the second dichroic mirror.
%20.If you move the arm in the right direction, the right part of the signal, after the main peak, becomes larger and particularly wider.
At a given moment, the signal "explodes", that is, when the LIDAR starts to become aligned. 
%21. At a given moment, that part of the signal almost “explodes”, that is when the lidar becomes aligned.
From then on, the right (far-distance) part of the signal can still be increased until it reaches a maximum. 
%22.If you see that right part of the signal continuing to increase and widen, but the spot on the second dichroic moves outside the mirror, you have to repeat steps 12-
Unfortunately, sometimes that part of the alignment procedure leads to the laser spot moving towards the edge of the second dichroic guide mirror.  In such a case, if the beam is found to exit the lower part of the second dichroic guide mirror, that mirror needs to be moved slightly to the lower side, i.e. counterclockwise. In contrast, if the beam exits the second dichroic on the upper edge, the mirror needs to be rotated clockwise.
%13. If the beam is found to move out of the lower edge of the second dichroic, you have to very slightly move the second dichroic to the lower side, i.e. counter- clock-wise. If it exits the second dichroic to the upper edge, move the mirror clock-wise.
%23.Repeat steps 19-22, until you cannot increase the signal anymore by moving the spot close to the edge of the mirror. You should have observed the “explosion” at this point.
That procedure is followed with both axes of the XY-table moving the laser arm until the backscatter signal is maximum, and both axes can be moved by about the same range until the signal starts to decrease. This range is about 300.000 steps wide. 
%24.Repeat steps 19-22 for the Y-axis of the arm.
%25. After having the “exploded” signal, check the range of parameter within which the signal remains “exploded”. That range should be about 300.000 arm steps wide. Choose the position in the middle of that range, for both axes separately and note down their position.
%26.The lidar is now aligned, switch off the laser. 27.Close the petals
%28. Move the telescope to parking position.
%29. Put the black protection shield back to its normal position.

A similar procedure involving only remotely controlled movements of the XY-table controlling the laser arm can be followed for different telescope \sapocol{elevation} %zenith 
angles, and the lookup table for the 
\sapocol{elevation-dependent} %zenith-dependent 
alignment generated. 

We observed that such an alignment remained stable during several months of operation under conditions of varying temperature and humidity; hence the procedure is tedious but once achieved, stable. 

\subsection{Calibration of the Timing}

The pBRL readout is triggered by the Q-switch of the laser, which our Brilliant model allows us to use for external triggering of the Licel readout. The trigger signal is the same for all readout modules. Therefore, one could expect that all channels provide a signal with exactly the same time offset. Unfortunately, this is not the case: we observed up to half a microsecond difference in the internal time delay between the 12-bit and the 16-bit models in the analogue line, and up to 100~ns in the photon counting lines. Within a same Licel Optical Transient Recorder (LOTR), %\aco{acronym of ??}\mg{was defined on page 26, shall we re-define it here?}, 
the internal time delay between the readout of the analogue and photon counting readout may amount to up to half a microsecond for the old 20-MSample models, but less than 50~ns for the new 40-MSample modules, which shows a considerable improvement. 

Given that the light path from the laser to the dichroic guide mirrors is unprotected, a strong signal pulse is recorded at time $T\approx 0$ due to multiple scattering of the light from the guide mirrors towards the primary mirror structure. Even if only a tiny part of the laser light ends up in the LLG, a strong peak is nevertheless observed (see Fig.~\ref{fig:signaloscilloscope}) marking the signal start. That peak is visible in the amplitude recording, but not in the photon counting channel, where the signal readout starts a few tens of nanoseconds after that peak. For this reason, the direct laser backscatter peak serves for absolute calibration of the all-time delays, but not for the relative ones, for which clouds visible in both analogue and photon channels are needed. 

\subsection{Calibration of the Telescope Pointing}

\begin{figure}
\centering
\includegraphics[width=0.75\textwidth]{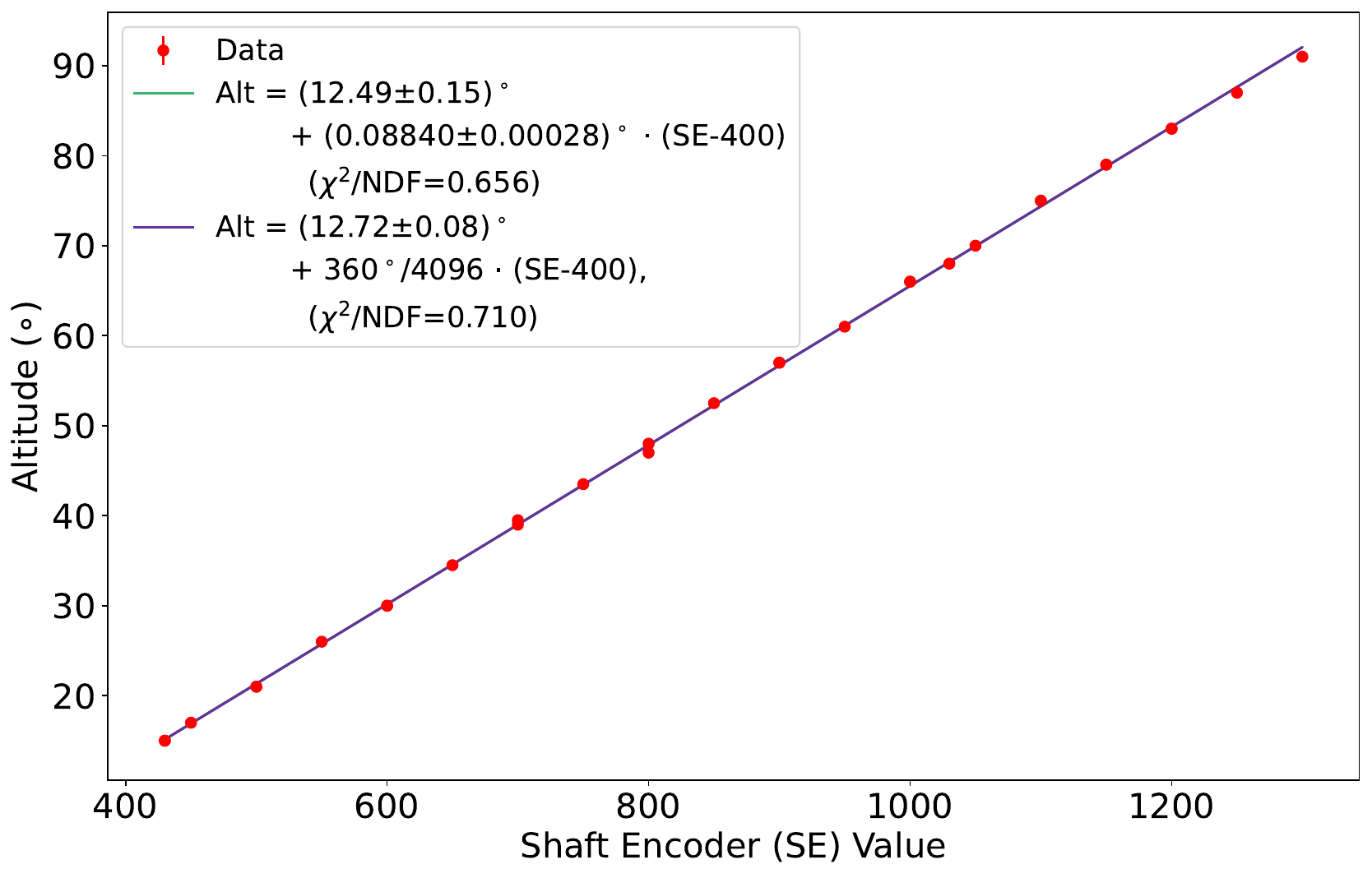}
\caption{Calibration of the telescope altitude vs. the altitude motor's shaft encoder values. %right: the tripod fixed to provide the $2f$ location of the domestic laser.
%\mg{Azimuth calibration missing}
\aco{Azimuth calibration to be included after measurements.}
\label{fig:pointingcalibration}} 
\end{figure}

The telescopes position is controlled by two 12-bit shaft encoders (\gls{se}), one for each axis. This leads to an angular step size of $360^\circ/4096 \approx 0.0879^\circ$ per SE step. Nevertheless, an absolute calibration is necessary. 
The calibration of the shaft encoder values on the altitude axis was performed with an inclinometer with an accuracy of about 0.5$^\circ$ located on the main arm, which holds the plate in the focal plane. A scan through different altitudes yields the calibration recovered, shown by the red points in Fig.~\ref{fig:pointingcalibration}. A fit with fixed slope yielded a $\chi^2$/NDF of 0.71, compatible with the data. 
To test for possible sagging of the telescope structure,  a fit with variable slope was also tried, which yielded $(0.0884\pm0.0003)^\circ$ per SE  step, 1.8~standard deviations from the expected value of 0.0879 and a marginally improved $\chi^2$/NDF of 0.66, hinting at negligible sagging of the telescope structure.
That calibration has a precision of better than 0.2$^\circ$ and an accuracy of 0.5$^\circ$. Its accuracy can be improved by pointing the telescope at \textit{Polaris} in such a way that its image is correctly centred at the focal point. 

\subsection{Calibration of the Discriminator Thresholds}

\begin{figure}
\centering
\includegraphics[width=0.46\textwidth,trim={45cm 35cm 60cm 20cm},clip]{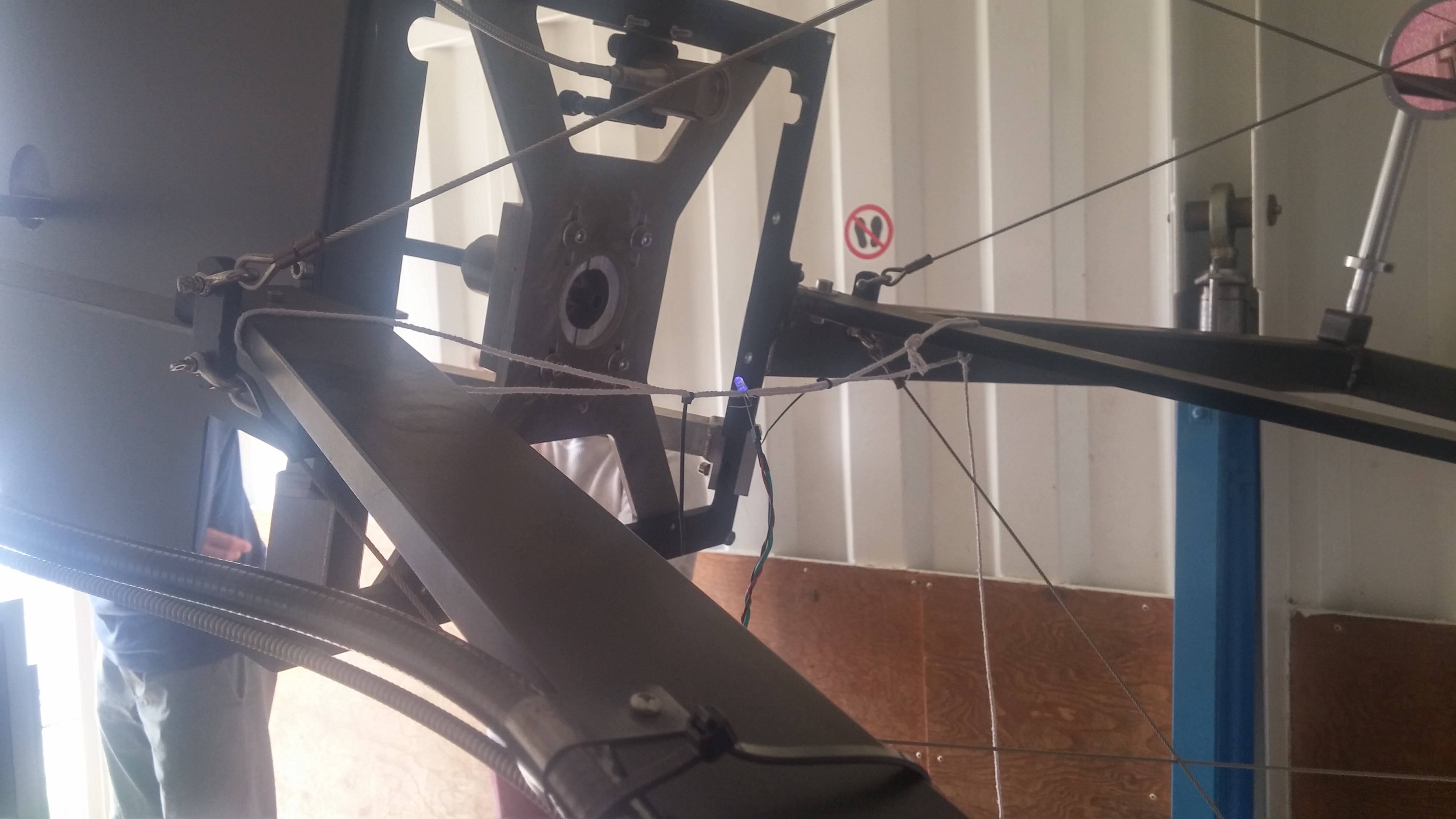}
\includegraphics[width=0.5\textwidth,trim={0cm 0.4cm 0.1cm 0cm},clip]{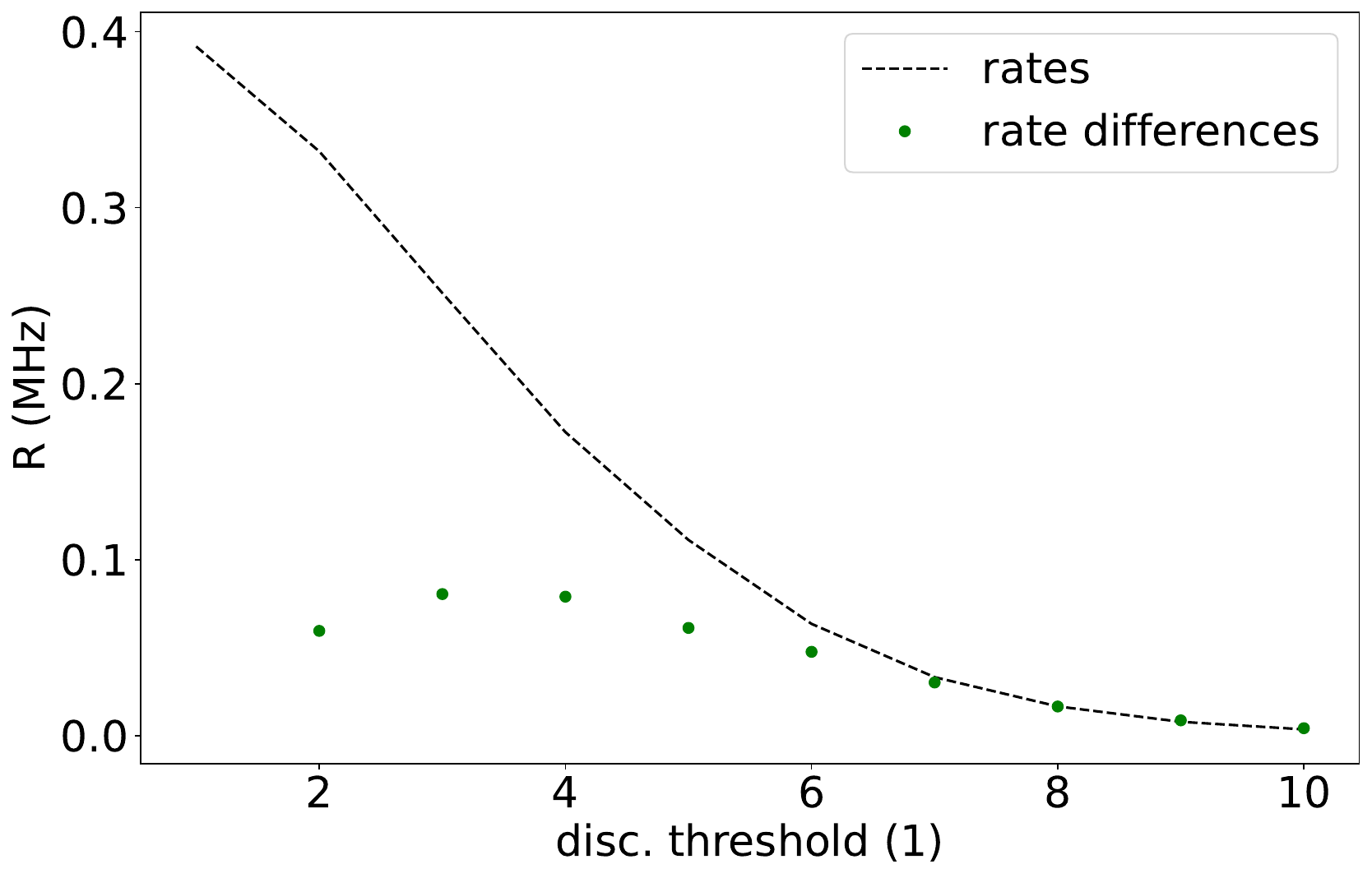}
\caption{Left: Picture of the setup used to calibrate discriminator setting. Right: A rate scan of one of the channels; the mean single photo-electron pulse height is found at around a discriminator threshold of 3--4. \label{fig:DiscCalibration}} 
\end{figure}

To calibrate the discriminator thresholds, we used the \textit{pulse height distribution} method as suggested by \textit{Licel}\footnote{see, e.g., \url{https://licel.com/pulseheight.html}.}. For this purpose, a stable UV LED was located in front of the pBRL's focal point (see Fig.~\ref{fig:DiscCalibration} left), and average photon counting rates were recorded, as a function of the discriminator threshold. \textit{Licel} recommends using, as a starting point, a discriminator level of 8 (corresponding to $\sim$3~mV), suitable for PMTs with a typical gain of $\sim$10$^6$ for LIDAR applications. However, our PMTs were designed for a different purpose~\citep{Mirzoyan:2017} and have a factor of approximately 4--5 lower gain. This has led to an optimum discriminator setting between 1 and 2 (see Fig.~\ref{fig:DiscCalibration} right), at the edge of the possibilities of the LOTR. The photon-counting channel is still operational at such low discriminator levels, due to the even lower PMT and electronic noise levels.     

\subsection{Monitoring of the Telescope Reflectivity}

Even with a proper coating, the overall reflectivity of the mirror is subject to degradation. 
If a coating of quartz is applied, this degradation should not exceed a few
percent per year; however, this number depends on the quality of the coating. 
In general, it is advisable to calibrate the mirror reflectivity over time, 
of the order of once every few months. 
In order to do so, one can easily measure the surface reflectivity (as opposed to the on-focus reflectivity). 
The reason is that for surfaces with small roughness, as this mirror is (and in
general solid glass mirrors are), the surface reflectivity is a good
estimator of the on-focus reflectivity. Possible dispersion of light
out of the focal spot is, on the contrary, more complex to measure
because of the calibration of the setup for this measurement. 

\section{pBRL Operation}
\label{sec:operation}

\aco{To include the reliability of the system.}
%\mg{Apart from operation, we should also mention the reliability of the system, at least a list of the main found that have caused down time at La Palma.}

After being awarded the status of a CTAO Pathfinder, the pBRL was operated from March 2021 to April 2022 at the Observatorio Roque de Los Muchachos (\gls{orm}) on the Canary Island of La Palma (Spain). It was located within the restricted area of the LST1 telescope~\citep{Mazin:2021}. %protected by a fence. 
This location was chosen because of existing infrastructure, such as flat terrain in a fenced area, easy access to power and the network. The pBRL was set so that no interference was possible with the operation of the LST1 telescope, which included additional limit switches for the movement of the telescope to prevent any laser light from hitting the telescope or the camera access tower. Moreover, operation of the pBRL was only allowed during moonlit nights when the LST1 did not make observations. 

\begin{figure}[h!t]
    \centering
    \includegraphics[width=0.75\linewidth]{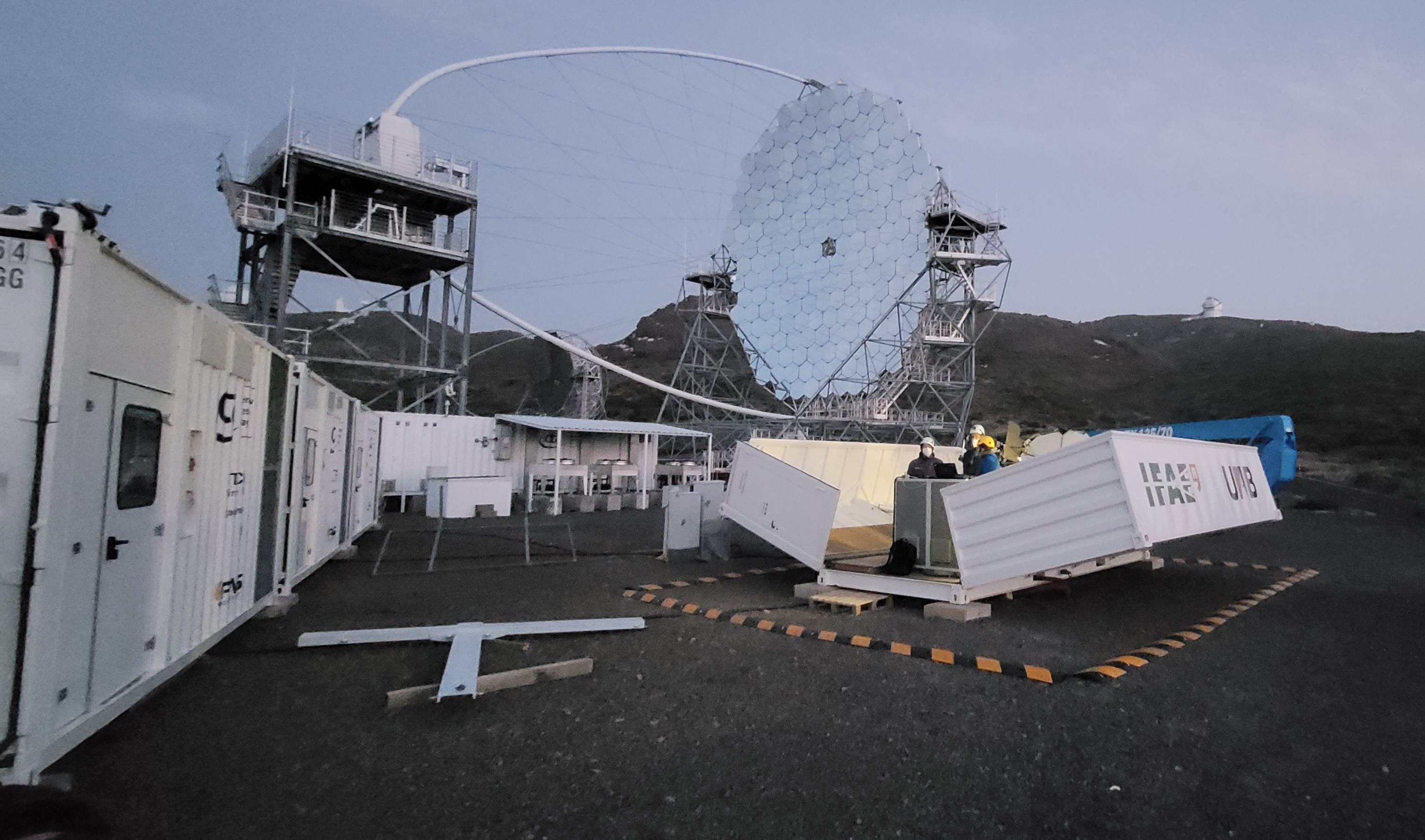} \hfill
    \caption{A photo of the pBRL deployed into the LST1 area with the container open.}
    \label{fig:pBRL_atLST1}
\end{figure}

Additionally, in order to avoid interference with other telescopes at the ORM, any operation of the pBRL that involved laser shooting was granted only during the evening, no later than 10 minutes before astronomical twilight, or in the morning, starting 10 minutes after astronomical twilight. Although the pBRL was fully operable from within the container, for the comfort of the operators, a semi-remote connection was established to the Counting House of the nearby MAGIC telescopes~\citep{magicperformance1}. The LST consortium allowed us also to install monitoring webcams for improved safety of the semi-remote operations. A photo of the pBRL deployed in the LST1 area is shown in Fig.~\ref{fig:pBRL_atLST1}.

After the end of the pBRL test phase at the ORM, the system was prepared for transport back to the campus of the Universitat Aut\`{o}noma de Barcelona.
In the following subsections, basic information about the pBRL operation is provided.

\subsection{Command-line Interface Semi-remote Operation and GRAFANA Monitoring}

Operational commands for the pBRL were given through a custom-made Command Line Interface (\gls{cli}) called \texttt{licli}. The \texttt{licli} commands for the Linux-based server that controls the pBRL operation can be issued from a laptop inside the BRL container, synchronized over the ethernet or through a remote server. In addition to single action raw commands, complex operational commands can be prepared for operator use, while experts use raw commands. 
%\md{Do we need to say more}

%\subsection{GRAFANA monitoring}
\begin{figure}[h!t]
    \centering
    \includegraphics[width=0.9\linewidth]{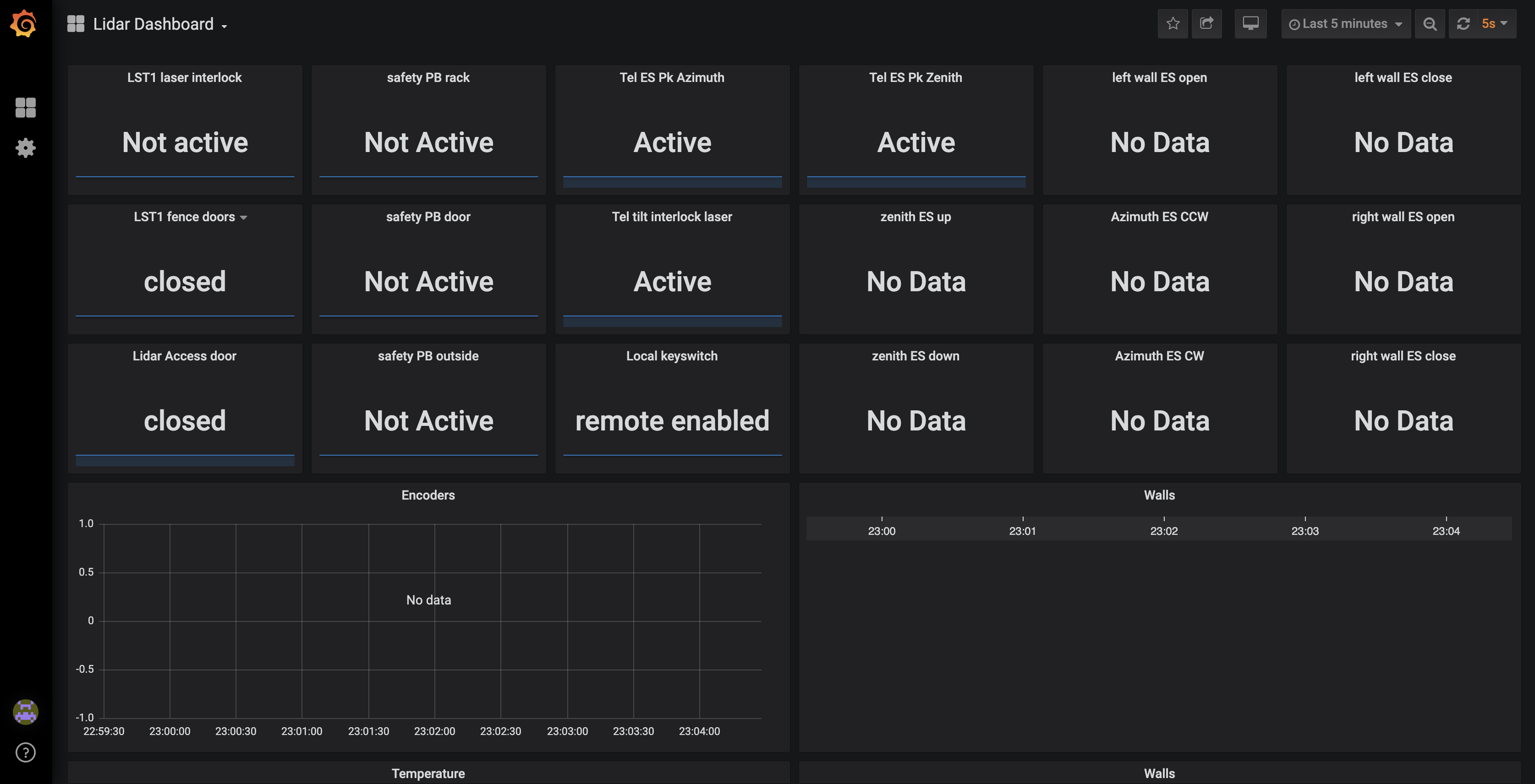}
    \caption{The GRAFANA BRL monitor. \aco{To be updated with a picture of when the telescope was operating. }%\md{Asked P. Calisse for a picture with colors on}
    }
  \label{fig:grafana}
\end{figure}

The status of several pBRL subsystems was monitored by GRAFANA (Fig. \ref{fig:grafana}) monitors\footnote{https://grafana.com/}. %\sa{I would add a link to GRAFANA.}
%The monitors can be accessed from \url{https://grafana.control.ifae.es/} \mg{Suggest to remove this website. A reader of the article should never connect to it, no?}. 
From GRAFANA, one can monitor the \texttt{laser interlock} and \texttt{LST1 fence doors} because the pBRLs could operate only if the interlocks of the hosting area of LST1 were not active. Furthermore, it could be checked if the pBRL was in zenith and azimuth park position (\texttt{Tel ES Pk Azimuth} and \texttt{Tel ES Pk Zenith} indicators) and if the container walls were open or closed (\texttt{left wall ES open} and \texttt{left wall ES closed} indicators). 
An additional indicator tells whether the pBRL has moved out of the laser interlocks and is safely allowed to operate the laser (\texttt{Tel tilt interlock laser}).
%\texttt{zenith ES up} means...
%\texttt{Azimuth ES CCV} means...
%\texttt{right wall ES open} means...
% \texttt{Lidar Access door} refers to the interlock due to the BRL container door. If it is open, operations are not permitted. CLOSED allows operation.
%\texttt{safety PB outside}
%\texttt{Local keyswitch}
% \texttt{zenith ES down}
% \texttt{Azimuth ES CV}
% \texttt{right wall ES close}
%\texttt{safety PB rack} and \texttt{safety PB door} \md{don't know what this is}
Other controls are of minor interest for this paper.
%\mg{I remember that there was an agreement with the MAGIC collaboration that shifters may overlook the area of the pBRL and intervene in case of any emergency, right? In the end, these "emergencies" consisted mainly of closed interlock from LST1 fences, etc.} \aco{MD and OBl do you remember?}

\subsection{Preparation}
\aco{To be checked by IFAE engineers.} 

Before starting, the weather had to be checked to ensure it complied with the conditions established within the operational limits of the pBRL: relative humidity $<85$\%, wind speed $<36$~km/h, no rain. 
Preparation started by % opening a website with the pBRL webcams
%\footnote{\url{https://lst1.iac.es/webcams/lidar/}}\mg{That website is protected by a password, and the LIDAR is anyhow not there anymore. I would remove this footnote.} \aco{agree} and
checking the area around the pBRL for obstacles or personnel with the help of the pBRL webcams and an onsite inspection. 
%Not only can the mechanics of the container cause damage, but also the laser shooting. 
There were four webcams available for pBRL operation: three located on the outer perimeter %\mg{correct?}\aco{IFAE engineers can you confirm?} 
and one inside the container. They were continuously taking images and have special sensitivity for night mode. %\mg{specify the model? These were all four IR webcams?}\aco{Same, engineers to confirm and provide the model used}.

Communication with the system was verified from the semi-remote or local CLI. 
%\mg{Are we sure that remote and local operation did not mutually exclude one another? This is actual a very strong requirement from CTAO.}

A GRAFANA terminal was opened on a web browser%\footnote{\url{https://grafana.control.ifae.es/}}\mg{Suggest to remove this website. A reader of the article should never connect to it, no?} 
and all interlock indicators are checked to be set correctly. 
The system then received a start-up command that included:
\begin{enumerate}
            \item Heat-up of the laser. The \sapocol{industrial hot air blower}
            %Hot Wind 
            was turned on if the outside temperature required it. The laser operating temperature of 32$^{\circ}$C 
            was reached in about 5--10 minutes depending on the outside temperature. 
            \item Open the container walls. This operation took about a minute.
            \item Initialization of the laser arm and movement to the aligned position. 
            %\md{I forgot here}
            \item Telescope preparation. The telescope system was moved out of the park position to a \sapocol{elevation angle of about 32$^\circ$.} %zenith angle of about 58$^\circ$.  
            In order to test full movement control, it was also moved back and forth in azimuth of a few degrees. %(concretely, to 7~deg and back to 0~deg). 
            Once the test was successful, the protective petals of the mirrors were folded open.
            \item The LICEL electronics %and the read-out system 
            were turned on. 
            \item Ramp up the High Voltage of the PMTs to pre-defined settings. 
            \item Power off the \sapocol{industrial hot air blower}.
            %Hot Wind. 
            When the laser reached the final temperature, the external heating system was switched off.
            \item Power on the laser. The laser was initialized without firing and set to 5\% laser power. 
        \end{enumerate}
        The start-up procedure could take about 10-15 minutes. 
        
        %After this, the system was ready for data-taking except for the LLG shutter protection which needed to be opened, also controlled via CLI. %To improve safety, we check for airplanes passages above the pBRL location\footnote{ \url{https://www.flightradar24.com/28.71,-17.94/12}}.  \mg{Really?? This is a no-fly zone...}

\subsection{Data Taking}
Data taking of the pBRL was remotely controlled and commands were issued with the CLI. 
%These are the main steps:
\begin{enumerate}
    \item Open the LLG shutter protection.
    %\item Ramp up the PMT High Voltage to pre-defined settings. The lifetime of a PMT also depends on its usage so HV operations are limited to when the PMT is under use. \mg{OK, but a few minutes are needed for stabilization of the PMT. Was that respected?}\mg{The HV was ramped up only once, not before every data set taken..?}
\item Set the laser power to the default %desired 
power of 80\%, define the number of shots (default: 1000) and fire the laser.
\item After taking a data set, the laser was set to \textit{pause} mode.
\item Repeat steps 2 to 3 as often as possible within the time limits or if weather conditions allow. 
   \end{enumerate}
   During such operations, GRAFANA and the webcams are monitored. A \textit{status} CLI also allowed monitoring of subsystems by specific status requests. In the end, the correct storage of the data taken on the remote server is checked.

\subsection{System Shut-down}
At the end of the operations, the LLG shutter protection was closed,  the \sapocol{industrial hot air blower}
%Hot Wind 
(if still on) \mg{To confirm if the Hot Wind needed to be restarted during operation after the laser cooling had reached the 32$^\circ$C temperature.} 
and the laser turned off, the petals were closed, the telescope was parked, and the container walls were closed. The whole system was shut down with a unique command. Raw data were automatically prepared for transfer and an automatic analysis was performed on the last images taken.

\section{Summary and Conclusions}
\label{sec:conclusions}

The Cherenkov Telescope Array Observatory (\gls{ctao}) requires accurate measurements of atmospheric conditions on its two sites and includes Raman LIDARs (RLs), among other common array elements~\citep{Gaug:2017Atmo,Prester:2024} for continuous atmospheric characterization.  These RLs need to be able to point at, or close to, the observed field of view of the CTAO telescopes and operate at wavelengths covered by the observed spectrum of Cherenkov light in the range from 300~nm to 700~nm. 

Such a RL needs to characterize the optical properties, mainly the vertical aerosol optical depths (\mdpi{\glspl{vaod}}) of the ground layer, clouds, and possible stratospheric debris. For that reason, an accurate determination of aerosol optical depths on short ranges is needed, as well as the ability to reach altitudes of at least 20~km. 

We have presented the prototype of a RL solution for that task, based on a pointable 1.8~m telescope within a standard 20~ft maritime container cut into two halves, whose walls can be opened sideways. That system had been refurbished from the discontinued CLUE experiment~\citep{Peruzzo:1990ia,Alexandreas:1995}, of which ten copies had been built.  

We have shown that it is possible to design and construct a prototype RL (the pBRL) suitable for the needs of CTAO from such a CLUE telescope. Whenever possible, original parts were maintained and new hardware items purchased that could be easily obtained. Not all have been optimized yet for their final use, like the PMTs.
Container walls and motors have been proven to work stably for a time period of at least 20~years. Nevertheless, the two container and two telescope motors are not available anymore on the market and shall be exchanged in the future by more modern versions, which incorporate also torque monitoring. 

We have chosen a Newtonian mirror configuration, which we found superior to a Cassegrain solution, due to the large shadows introduced by a necessarily large secondary mirror, needed to achieve the relatively small magnification required for a low region of full overlap. The latter is also achieved through a coaxial configuration in which the laser light exits the system along the line of sight of the telescope. 
The telescope structure has proven reliable to such an extent that laser alignment was not lost, even after several months of movements and operation, and passing through various temperature cycles. However, the point spread of the image in the focal point had been found degraded, after the pBRL came back from a one-year test period on La~Palma. That degradation was probably due to high stress on the mirror after inappropriate handling of the container during transport. Given that a spare mirror exists, we do not foresee any showstopper for this reason, however, all mirrors must be unmounted during transport and installed and aligned again, once the material has arrived on site.

A Nd:YAG laser of $\sim$100~mJ pulse energy, a repetition rate of 10~Hz (from the company Quantel) and frequency doubling and tripling was purchased and configured so that all three wavelengths exit coaxially the same window. That laser was repaired after years of use and showed slightly degraded output power and beam quality after repair. The laser light is guided into its coaxial configuration by two dichroic mirrors, optimized for an incidence angle of 61$^\circ$ and the removal of the undesired fundamental 1064~nm line. 
An alignment system based on moving the laser arm, instead of the second dichroic guide mirror, has proven possible, at the expense of the need for relatively large 10~cm guide mirrors. Nevertheless, the alignment procedure is tedious and shall be improved with the use of a steerable second guide mirror based on piezoelectric motors.

Given the relatively large optical aberrations of the 1.8~m telescope of $\sim$3~mm radius in the focal point and the large mirror size, a large numerical aperture and a large diameter light transport system were needed. We have shown that it is possible to use an 8~mm diameter liquid light guide (LLG) from the company Lumatec for such a purpose and construct a 10~cm optical system (the polychromator) based on 10~cm lens couples made of N-BK7 and three custom-designed dichroic mirrors, together with 10~nm wide interference filters for the clean separation and detection of four wavelengths: 355~nm, 387~nm, 532~nm and 607~nm. Light leakage (possibly from LLG internal fluorescence) from the 355~nm line into the 387~nm channel $<10^{-5}$ (95\% CL) and from the 532~nm line of $<5\times 10^{-6}$ (95\% CL) can be guaranteed. This level is acceptable for the reconstruction accuracy of ground-layer VAODs typically found at the CTAO sites, similarly to that for optically thin clouds. Light leakage into the second 607~nm Raman line is always $<3 \times 10^{-6}$ (95\% CL). Transmission of the LLG has been shown to degrade slightly by about 10\% over ten years, a value that is found acceptable over a projected lifetime of 15~years. Further stability was gained through the use of a shutter in front of the LLG, which protects the system from light when not used and, moreover, acts as a field stop when opened. 
\sapocol{Unfortunately, the LLG has shown serious degradation in transmission over the past 10~years and will certainly need to get replaced once in the 15-years projected lifetime of the LIDAR. For this purpose, a spare LLG will need to be provided.}
%\aco{What about the near-range solution?}

The light is finally detected by identical 1.5$^{\prime\prime}$ PMTs, available from the Large Sized Telescope~\citep{Mazin:2021} and readout by commercial Licel Optical Transient Recorder (LOTR) modules. These PMTs have a gain of $<3\times 10^5$ and were found to operate at the limit of the LOTR's photon-counting threshold range. They need to be replaced by PMTs of higher gain and adapted for each wavelength channel. 

A laser in the range of 100~mJ pulse energy in combination with a 1.8~m mirror produces large saturation of the elastic lines in the amplitude readout of the LOTR up to $\sim$300~m distance, even with a PMT gain considerably smaller than 10$^6$. That distance is still lower than the distance to the full overlap of $\sim$450~m with the current laser employed, but it can be easily improved with a laser of smaller beam divergence. In such a scenario, PMTs with gating capabilities may be considered for the two elastic lines. 
A final version shall therefore incorporate a laser of slightly smaller power, but a higher repetition rate instead.
To further lower the range of full overlap, a near-range system has been incorporated into the pBRL, which uses a commercial paraboloid mini-telescope fed into a standard POF fibre with a SMA connector. That system has been operated with only one wavelength read out so far. 

The pBRL has been brought to the Observatorio del Roque de los Muchachos (\gls{orm}) for a one-year test campaign, during which all safety measures,  semi-remote operation and stability of the system could be tested.  
%\aco{Maybe mention that we were able to operate it from semi-remote?}

A complete description of the data analysis carried out on all test data taken with this system is outside the scope of this manuscript and will be presented in a follow-up article~\citep{performancepaper}.

\printglossary[title=List of abbreviations,nonumberlist] %, toctitle=Acronyms, ]%nonumberlist prevents the page number from appearing along with the term

%%%%%%%%%%%%%%%%%%%%%%%%%%%%%%%%%%%%%%%%%%
\authorcontributions{
%OB
Conceptualization, 
ACO, MD, MG; 
%methodology, 
%D.D.P., J.Eb., M.G.; 
software, 
OBa, JB, ACO, SMC, VDD, EFP, RGr, MG, ALO, CM, MM, OM, VRM, DR;  
%J.El., S.K., M.G., A.B.; 
formal analysis, 
OBa, JB, ACO, SMC, VDD, MD, EFP, MG, ALO, CM, MM, OM, VR; 
data acquisition, 
OBl, JB, PGC, ACO, SMC, MD, EFP, MG, RGr, ALO, CM, MM, OM, RVM, DR, SS, SUR, MZa, MZi;
hardware contribution, %\md{to be added?};
OBa, JB, RGa, OM, SS, MZa; 
%J.Eb., M.G., M.K.; 
%investigation, 
%D.D.P, M.K.; 
%resources, 
%; 
%data curation,
%M.G., J.Eb.,P.J., S.K., J.El., M.K. ; 
writing---original draft preparation, 
ACO, MD, MG;
writing---review and editing, all authors; 
%visualization, 
%M.G., J.Eb., A.B.; 
%supervision, 
%D.D.P.; 
project administration, 
OBl, MD, LF, MG, MM;
%D.D.P., M.M.; 
funding acquisition, 
OBl, MD, LF, MG, MM, SS;
%D.D.P, M.M., R.M., M.G., J.Eb.
All authors have read and agreed to the published version of the manuscript.}

\funding{%This research was funded by the Spanish grants PID2022-139117NB-C41 and PID2022-139117NB-C43, funded by MCIN/AEI/10.13039/501100011033/FEDER, UE, and the Departament de Recerca i Universitats de la Generalitat de Catalunya (grant SGR2021 00607). MD acknowledges funds from the 2012 "Bando Giovani Studiosi" of the University of Padova.
This project has received funding from the European Union's Horizon Europe Research and innovation programme under Grant Agreement No 101131928; by the Spanish grants PID2022-139117NB-C41 and PID2022-139117NB-C43, funded by MCIN/AEI/10.13039/501100011033/FEDER, UE, the Departament de Recerca i Universitats de la Generalitat de Catalunya (grant SGR2021 00607), and by "ERDF A way of making Europe",  the CERCA program of the Generalitat de Catalunya and by the European Union NextGenerationEU/PRTR. In Slovenia, it was funded by the Slovenian Research And Innovation Agency, grants P1-0031, J1-3011 and I0-E018. M.D. acknowledges funds from the 2012 "Bando Giovani Studiosi" of the University of Padova.   R.Gr.~acknowledges funding from the FSE under the program Ayudas predoctorales of the Ministerio de Ciencia e Innovación  PRE2020-093561.
}

\dataavailability{The data supporting the conclusions of this article will be made available by the authors on request.}

\acknowledgments{This work would have been impossible without the support of our colleagues from the MAGIC and LST collaboration and the CTAO Consortium, which we gratefully acknowledge.
We thank the Instituto de Astrof\'isica de Canarias for the excellent working conditions at the Observatorio del Roque de los Muchachos on La Palma. 
We also thank the funding agencies and institutions mentioned in the above section (Funding) for the financial support.}

\conflictsofinterest{The authors declare no conflicts of interest. The funders had no role in the design of the study; in the collection, analysis or interpretation of the data; in the writing of the manuscript; or in the decision to publish the results.}

\newpage
\begin{adjustwidth}{-\extralength}{0cm}
%\printendnotes[custom] % Un-comment to print a list of endnotes

\reftitle{References}

% Please provide either the correct journal abbreviation (e.g. according to the “List of Title Word Abbreviations” http://www.issn.org/services/online-services/access-to-the-ltwa/) or the full name of the journal.
% Citations and References in Supplementary files are permitted provided that they also appear in the reference list here. 

%=====================================
% References, variant A: external bibliography
%=====================================
\bibliography{biblio}

\begin{thebibliography}{999}

\bibitem[{Acharya} \em{et~al.}(2013){Acharya}, {Actis}, {Aghajani}, {Agnetta},
  {Aguilar}, et~al.]{ctaconcept}
{Acharya}, B.S.; {Actis}, M.; {Aghajani}, T.; {Agnetta}, G.; {Aguilar}, J.;
  et~al.
\newblock {Introducing the CTA concept}.
\newblock {\em Astroparticle Physics} {\bf 2013}, {\em 43},~3--18.
\newblock {\url{https://doi.org/10.1016/j.astropartphys.2013.01.007}}.

\bibitem[{The Cherenkov Telescope Array Consortium} \em{et~al.}(2019){The
  Cherenkov Telescope Array Consortium}, {Acharya}, {Agudo}, {Samarai},
  {Alfaro}, {Alfaro}, {Alispach}, {Alves Batista}, {Amans}, and
  et~al.]{ScienceCTA:2019}
{The Cherenkov Telescope Array Consortium}.; {Acharya}, B.S.; {Agudo}, I.;
  {Samarai}, I.A.; {Alfaro}, R.; {Alfaro}, J.; {Alispach}, C.; {Alves Batista},
  R.; {Amans}, J.P.; et~al..
\newblock {\em {Science with the Cherenkov Telescope Array}}; {World
  Scientific},  2019; p. 364,
  \href{http://xxx.lanl.gov/abs/1709.07997}{{\normalfont [1709.07997]}}.
\newblock Edited by the CTA Consortium, {\url{https://doi.org/10.1142/10986}}.

\bibitem[{Schmuckermaier} \em{et~al.}(2023){Schmuckermaier}, {Gaug}, {Fruck},
  {Moralejo}, {Hahn}, et~al.]{Schmuckermaier:2023huo}
{Schmuckermaier}, F.; {Gaug}, M.; {Fruck}, C.; {Moralejo}, A.; {Hahn}, A.;
  et~al.
\newblock {Correcting Imaging Atmospheric Cherenkov Telescope data with
  atmospheric profiles obtained with an elastic light detecting and ranging
  system}.
\newblock {\em A\&A} {\bf 2023}, {\em 673},
  \href{http://xxx.lanl.gov/abs/2302.12072}{{\normalfont [2302.12072]}}.
\newblock {\url{https://doi.org/10.1051/0004-6361/202245787}}.

\bibitem[{Fruck}(2015)]{fruck:phd}
{Fruck}, C.
\newblock {The Galactic Center resolved with MAGIC and a new technique for
  Atmospheric Calibration}.
\newblock PhD thesis, {Technische Unversit\"at M\"unchen, Germany}, Arcisstr.
  21, 80333 M\"unchen, Germany,  2015.
\newblock {Available at
  \url{https://mediatum.ub.tum.de/doc/1237928/document.pdf}}.

\bibitem[{de Naurois}(2000)]{Naurois:2000a}
{de Naurois}, M.
\newblock {L'exp{\'e}rience CELESTE: Reconversion d'une centrale solaire pour
  l'astronomie gamma. Premi{\`e}re observation de la N{\'e}buleuse du Crabe et
  du Blazar Markarian~421 entre 30 et 300 GeV}.
\newblock PhD thesis, Universit{\'e} Pierre et Marie Curie - Paris VI,  2000.
\newblock {Available at \url{http://tel.archives-ouvertes.fr/tel-00004261}}.

\bibitem[{{Navas}, S. et al.} and {{(Particle Data Group)}}(2024)]{NavasPDG}
{{Navas}, S. et al.}.; {{(Particle Data Group)}}.
\newblock 34 Passage of Particles Through Matter.
\newblock {\em Phys. Rev. D} {\bf 2024}, {\em 110},~030001.
\newblock See also
  \url{https://pdg.lbl.gov/2024/AtomicNuclearProperties/HTML/air_dry_1_atm.html}.

\bibitem[{Hillas} and {Patterson}(1990)]{HILLAS:JPGPP1990a}
{Hillas}, A.M.; {Patterson}, J.R.
\newblock {Characteristics and brightness of Cerenkov shower images for
  gamma-ray astronomy near 1 TeV}.
\newblock {\em J. Phys. G-Nucl. Part. Phys.} {\bf 1990}, {\em 16},~1271--1281.
\newblock {\url{https://doi.org/10.1088/0954-3899/16/8/022}}.

\bibitem[{Bernl{\"o}hr}(2000)]{Bernloehr:2000}
{Bernl{\"o}hr}, K.
\newblock {Impact of atmospheric parameters on the atmospheric Cherenkov
  technique}.
\newblock {\em Astroparticle Physics} {\bf 2000}, {\em 12},~255--268,
  \href{http://xxx.lanl.gov/abs/astro-ph/9908093}{{\normalfont
  [astro-ph/9908093]}}.
\newblock {\url{https://doi.org/10.1016/S0927-6505(99)00093-6}}.

\bibitem[{Munar-Adrover} and {Gaug}(2019)]{Munar:2019}
{Munar-Adrover}, P.; {Gaug}, M.
\newblock {Studying molecular profiles above the Cherenkov Telescope Array
  sites}.
\newblock In Proceedings of the European Physical Journal Web of Conferences,
  2019, Vol. 197, {\em European Physical Journal Web of Conferences}, p. 01002.
\newblock {\url{https://doi.org/10.1051/epjconf/201919701002}}.

\bibitem[{Ebr} \em{et~al.}(2019){Ebr}, {Mandat}, {Pech}, {Chytka}, {Jurysek},
  {Prouza}, {Jane{\v{c}}ek}, {Tr{\'a}vn{\'\i}{\v{c}}ek}, {Bla{\v{z}}ek},
  {Bulik}, {Cieslar}, {Suchenk}, {Rizi}, {Pietropaolo}, {Iarlori}, {Aramo},
  {Valore}, {Di Pierro}, {Vallania}, {Depaoli}, {Will}, {Gaug}, {Font},
  {Ma{\v{s}}ek}, {Eli{\'a}{\v{s}}ek}, {Jelinek}, and {Karpov}]{Ebr:2019ICRC}
{Ebr}, J.; {Mandat}, D.; {Pech}, M.; {Chytka}, L.; {Jurysek}, J.; {Prouza}, M.;
  {Jane{\v{c}}ek}, P.; {Tr{\'a}vn{\'\i}{\v{c}}ek}, P.; {Bla{\v{z}}ek}, J.;
  {Bulik}, T.;  et~al.
\newblock {Characterization of atmospheric properties at the future sites of
  the Cherenkov Telescope Array}.
\newblock In Proceedings of the 36th International Cosmic Ray Conference
  (ICRC2019),  2019, Vol.~36, {\em International Cosmic Ray Conference}, p.
  667,  \href{http://xxx.lanl.gov/abs/1909.08088}{{\normalfont [1909.08088]}}.

\bibitem[{Fruck} \em{et~al.}(2022){Fruck}, {Gaug}, {Hahn}, {Acciari},
  {Besenrieder}, {Dominis Prester}, et~al.]{Fruck:2022igg}
{Fruck}, C.; {Gaug}, M.; {Hahn}, A.; {Acciari}, V.; {Besenrieder}, J.; {Dominis
  Prester}, D.;  et~al.
\newblock {Characterizing the aerosol atmosphere above the Observatorio del
  Roque de los Muchachos by analysing seven years of data taken with an GaAsP
  HPD-readout, absolutely calibrated elastic LIDAR}.
\newblock {\em MNRAS} {\bf 2022}, {\em 515},~4520--4550,
  \href{http://xxx.lanl.gov/abs/2202.09561}{{\normalfont [2202.09561]}}.
\newblock {\url{https://doi.org/10.1093/mnras/stac1563}}.

\bibitem[{Garc{\'\i}a-Gil} \em{et~al.}(2010){Garc{\'\i}a-Gil},
  {Mu{\~n}oz-Tu{\~n}{\'o}n}, and {Varela}]{GarciaGil:2010}
{Garc{\'\i}a-Gil}, A.; {Mu{\~n}oz-Tu{\~n}{\'o}n}, C.; {Varela}, A.M.
\newblock {Atmosphere Extinction at the ORM on La Palma: A 20~yr Statistical
  Database Gathered at the Carlsberg Meridian Telescope}.
\newblock {\em PASP} {\bf 2010}, {\em 122},~1109,
  \href{http://xxx.lanl.gov/abs/1009.4056}{{\normalfont
  [arXiv:astro-ph.IM/1009.4056]}}.
\newblock {\url{https://doi.org/10.1086/656329}}.

\bibitem[{Kremser} \em{et~al.}(2016){Kremser}, {Thomason}, {von Hobe},
  {Hermann}, {Deshler}, {Timmreck}, {Toohey}, Stenke, Schwarz, Weigel,
  Fueglistaler, Prata, Vernier, Schlager, Barnes, Antuña-Marrero, Fairlie,
  Palm, Mahieu, Notholt, Rex, Bingen, Vanhellemont, Bourassa, Plane, Klocke,
  Carn, Clarisse, Trickl, Neely, James, Rieger, Wilson, and
  Meland]{Kremser:2016}
{Kremser}, S.; {Thomason}, L.W.; {von Hobe}, M.; {Hermann}, M.; {Deshler}, T.;
  {Timmreck}, C.; {Toohey}, M.; Stenke, A.; Schwarz, J.P.; Weigel, R.;  et~al.
\newblock Stratospheric aerosol—Observations, processes, and impact on
  climate.
\newblock {\em Reviews of Geophysics} {\bf 2016}, {\em 54},~278--335,
  \href{http://xxx.lanl.gov/abs/https://agupubs.onlinelibrary.wiley.com/doi/pdf/10.1002/2015RG000511}{{\normalfont
  [https://agupubs.onlinelibrary.wiley.com/doi/pdf/10.1002/2015RG000511]}}.
\newblock {\url{https://doi.org/https://doi.org/10.1002/2015RG000511}}.

\bibitem[{{Aleksi{\'c}}, J. et al.}(2012)]{Magic:performance}
{{Aleksi{\'c}}, J. et al.}.
\newblock {Performance of the MAGIC stereo system obtained with Crab Nebula
  data}.
\newblock {\em Astroparticle Physics} {\bf 2012}, {\em 35},~435--448,
  \href{http://xxx.lanl.gov/abs/1108.1477}{{\normalfont
  [arXiv:astro-ph.IM/1108.1477]}}.
\newblock {\url{https://doi.org/10.1016/j.astropartphys.2011.11.007}}.

\bibitem[{{Aharonian}, F. et al.}(2006)]{Hess:performance}
{{Aharonian}, F. et al.}.
\newblock {Observations of the Crab nebula with HESS}.
\newblock {\em Astrononmy \& Astrophsics} {\bf 2006}, {\em 457},~899--915,
  \href{http://xxx.lanl.gov/abs/arXiv:astro-ph/0607333}{{\normalfont
  [arXiv:astro-ph/0607333]}}.
\newblock {\url{https://doi.org/10.1051/0004-6361:20065351}}.

\bibitem[{Gaug}(2017)]{Gaug:2017Atmo}
{Gaug}, M.
\newblock {CTA Atmospheric Calibration}.
\newblock In Proceedings of the European Physical Journal Web of Conferences,
  2017, Vol. 144, {\em European Physical Journal Web of Conferences}, p. 01003.
\newblock {\url{https://doi.org/10.1051/epjconf/201714401003}}.

\bibitem[{Ballester} \em{et~al.}(2019){Ballester}, {Blanch}, {Boix}, {Bregeon},
  {Brun}, {{\c C}olak}, {Doro}, {Da Deppo}, {Font}, {Gabella}, {Garc{\'\i}a},
  {Gaug}, {Maggio}, {Mart{\'\i}nez}, {Mart{\'\i}nez}, {Munar-Adrover}, {Ramos},
  {Rivoire}, {Stanic}, {Villar}, {Vasileiadis}, {Wang}, and
  {Zavrtanik}]{Ballester:2019}
{Ballester}, O.; {Blanch}, O.; {Boix}, J.; {Bregeon}, J.; {Brun}, P.; {{\c
  C}olak}, S.M.; {Doro}, M.; {Da Deppo}, V.; {Font}, L.; {Gabella}, O.;  et~al.
\newblock {Raman LIDARs for the atmospheric calibration along the line-of-sight
  of CTA}.
\newblock In Proceedings of the Proceedings 36$^\mathrm{th}$ International
  Cosmic Ray Conference -ICRC2019- Madison, USA,  2019, Vol. PoS(ICRC2019)814.

\bibitem[{Ebr} \em{et~al.}(2021){Ebr}, {Karpov}, {Eli{\'a}{\v{s}}ek},
  {Bla{\v{z}}ek}, {Cunniffe}, {Ebrov{\'a}}, {Jane{\v{c}}ek}, {Jel{\'\i}nek},
  {Jury{\v{s}}ek}, {Mand{\'a}t}, {Ma{\v{s}}ek}, {Pech}, {Prouza}, and
  {Tr{\'a}vn{\'\i}{\v{c}}ek}]{Ebr:2021}
{Ebr}, J.; {Karpov}, S.; {Eli{\'a}{\v{s}}ek}, J.; {Bla{\v{z}}ek}, J.;
  {Cunniffe}, R.; {Ebrov{\'a}}, I.; {Jane{\v{c}}ek}, P.; {Jel{\'\i}nek}, M.;
  {Jury{\v{s}}ek}, J.; {Mand{\'a}t}, D.;  et~al.
\newblock {A New Method for Aerosol Measurement Using Wide-field Photometry}.
\newblock {\em Astronomical Journal} {\bf 2021}, {\em 162},~6,
  \href{http://xxx.lanl.gov/abs/2101.03074}{{\normalfont [2101.03074]}}.
\newblock {\url{https://doi.org/10.3847/1538-3881/abf7b1}}.

\bibitem[{Doro} \em{et~al.}(2013){Doro}, {Gaug}, {Blanch}, {Font}, {Garrido},
  {Lopez-Oramas}, and {Martinez}]{doro:2013}
{Doro}, M.; {Gaug}, M.; {Blanch}, O.; {Font}, L.; {Garrido}, D.;
  {Lopez-Oramas}, A.; {Martinez}, M.
\newblock {Towards a full Atmospheric Calibration system for the Cherenkov
  Telescope Array}.
\newblock In Proceedings of the {Proceedings of the 33$^\mathrm{rd}$ ICRC, Rio
  de Janeiro},  2013, p. 0151,
  \href{http://xxx.lanl.gov/abs/1307.3406}{{\normalfont [1307.3406]}}.

\bibitem[{Otarola} \em{et~al.}(2019){Otarola}, {Dumas}, {Gaug}, {Benn},
  {Mu{\~n}oz-Tu{\~n}{\'o}n}, {Castro-Almaz{\'a}n}, {Cabrera-Lavers}, and
  {Hinton}]{otarola:2019}
{Otarola}, A.; {Dumas}, C.; {Gaug}, M.; {Benn}, C.; {Mu{\~n}oz-Tu{\~n}{\'o}n},
  C.; {Castro-Almaz{\'a}n}, J.; {Cabrera-Lavers}, A.; {Hinton}, J.
\newblock {Enhanced Laser Traffic Control System Operation Mode}.
\newblock In Proceedings of the Proceedings AO4ELT6 - June 9-14, 2019,
  Qu{\'e}bec City, Canada,  2019.
\newblock Available at \url{http://ao4elt6.copl.ulaval.ca/proceedings.html}.

\bibitem[{Gaug} and {Doro}(2018)]{Gaug:2018zfj}
{Gaug}, M.; {Doro}, M.
\newblock {Impact of Laser Guide Star facilities on neighbouring telescopes:
  The case of GTC, TMT, VLT and ELT lasers and the Cherenkov Telescope Array}.
\newblock {\em Mon. Not. Roy. Astron. Soc.} {\bf 2018}, {\em 481},~727--748,
  \href{http://xxx.lanl.gov/abs/1808.03820}{{\normalfont
  [arXiv:astro-ph.IM/1808.03820]}}.
\newblock {\url{https://doi.org/10.1093/mnras/sty2188}}.

\bibitem[{Bregeon} \em{et~al.}(2016){Bregeon}, {Compin}, {Rivoire},
  {Sanguillon}, and {Vasileiadis}]{Bregeon:2016}
{Bregeon}, J.; {Compin}, M.; {Rivoire}, S.; {Sanguillon}, M.; {Vasileiadis}, G.
\newblock An elastic lidar system for the H.E.S.S. Experiment.
\newblock {\em Nuclear Instruments and Methods in Physics Research Section A:
  Accelerators, Spectrometers, Detectors and Associated Equipment} {\bf 2016},
  {\em 819},~60--66.
\newblock {\url{https://doi.org/https://doi.org/10.1016/j.nima.2016.02.087}}.

\bibitem[{Abreu} \em{et~al.}(2012){Abreu} et~al.]{Auger:2012}
{Abreu}, P.;  et~al.
\newblock {The Rapid Atmospheric Monitoring System of the Pierre Auger
  Observatory} {\bf 2012}.
\newblock  \href{http://xxx.lanl.gov/abs/1208.1675}{{\normalfont
  [arXiv:astro-ph.HE/1208.1675]}}.

\bibitem[{Rizi} \em{et~al.}(2019){Rizi}, {Hernandez}, {Iarlori}, and
  {Pietropaolo}]{Rizi:2019}
{Rizi}, V.; {Hernandez}, C.M.; {Iarlori}, M.; {Pietropaolo}, E.
\newblock {The Auger Raman Lidar: several years of continuous observations}.
\newblock In Proceedings of the European Physical Journal Web of Conferences,
  2019, Vol. 197, {\em European Physical Journal Web of Conferences}, p. 02003.
\newblock {\url{https://doi.org/10.1051/epjconf/201919702003}}.

\bibitem[{Tomida} \em{et~al.}(2011){Tomida}, {Tsuyuguchi}, {Arai}, {Benno},
  {Chikawa}, et~al.]{Tomida:2011cb}
{Tomida}, T.; {Tsuyuguchi}, Y.; {Arai}, T.; {Benno}, T.; {Chikawa}, M.;  et~al.
\newblock {The atmospheric transparency measured with a LIDAR system at the
  Telescope Array experiment}.
\newblock {\em Nucl. Instrum. Meth.} {\bf 2011}, {\em A654},~653--660,
  \href{http://xxx.lanl.gov/abs/1109.1196}{{\normalfont
  [arXiv:astro-ph.IM/1109.1196]}}.
\newblock {\url{https://doi.org/10.1016/j.nima.2011.07.012}}.

\bibitem[{Gaug} \em{et~al.}(2022){Gaug}, {Hahn}, {Acciari}, {Besenrieder},
  {Dominis Prester}, {Dorner}, {Fink}, {Font}, {Fruck}, {Mi{\'c}anovi{\'c}},
  {Mirzoyan}, {Pavleti{\'c}}, {Schmuckermaier}, and {Will}]{Gaug:2022}
{Gaug}, M.; {Hahn}, A.; {Acciari}, V.; {Besenrieder}, J.; {Dominis Prester},
  D.; {Dorner}, D.; {Fink}, D.; {Font}, L.; {Fruck}, C.; {Mi{\'c}anovi{\'c}},
  S.;  et~al.
\newblock {Seven years of quasi-continuous LIDAR data}.
\newblock {\em Journal of Physics: Conference Series} {\bf 2022}, {\em
  2398},~012010.
\newblock {\url{https://doi.org/10.1088/1742-6596/2398/1/012010}}.

\bibitem[{B{\"o}senberg} and {Hoff}(2017)]{Bosenberg:2017}
{B{\"o}senberg}, J.; {Hoff}, R.
\newblock Plan for the implementation of the GAW Aerosol Lidar Observation
  Network GALION.
\newblock Tech. Rep. WMO/GAW 178, World Meteorological Organization,  2017.

\bibitem[{Berj\'on} \em{et~al.}(2019){Berj\'on}, {Barreto}, {Hern\'andez},
  {Yela}, {Toledano}, and {Cuevas}]{Berjon:2019}
{Berj\'on}, A.; {Barreto}, A.; {Hern\'andez}, Y.; {Yela}, M.; {Toledano}, C.;
  {Cuevas}, E.
\newblock {A 10-year characterization of the Saharan Air Layer lidar ratio in
  the subtropical North Atlantic}.
\newblock {\em Atmospheric Chemistry and Physics} {\bf 2019}, {\em
  19},~6331--6349.
\newblock {\url{https://doi.org/10.5194/acp-19-6331-2019}}.

\bibitem[{Hellemeier} \em{et~al.}(2019){Hellemeier}, {Yang}, {Sarazin}, and
  {Hickson}]{Hellemeier:2019}
{Hellemeier}, J.A.; {Yang}, R.; {Sarazin}, M.; {Hickson}, P.
\newblock {Weather at selected astronomical sites - an overview of five
  atmospheric parameters}.
\newblock {\em Monthly Notices of the Royal Astronomical Society} {\bf 2019},
  {\em 482},~4941--4950.
\newblock {\url{https://doi.org/10.1093/mnras/sty2982}}.

\bibitem[Otarola and Hickson(2017)]{Otarola:2017}
Otarola, A.; Hickson, P.
\newblock Study of cirrus clouds and implications for the variability of laser
  guide star intensity and fratricide effects.
\newblock {\em Proceedings of the Fifth AO4ELT Conference, Puerto de La Cruz,
  Tenerife, Canary Islands, Spain (June 25-30, 2017)} {\bf 2017}.
\newblock {\url{https://doi.org/10.26698/AO4ELT5.0028}}.

\bibitem[{Laken} \em{et~al.}(2016){Laken}, {Parviainen}, {Garc{\'\i}a-Gil},
  {Mu{\~n}oz-Tu{\~n}{\'o}n}, {Varela}, {Fernandez-Acosta}, and
  {Pall{\'e}}]{Laken:2016}
{Laken}, B.A.; {Parviainen}, H.; {Garc{\'\i}a-Gil}, A.;
  {Mu{\~n}oz-Tu{\~n}{\'o}n}, C.; {Varela}, A.M.; {Fernandez-Acosta}, S.;
  {Pall{\'e}}, P.
\newblock {Thirty Years of Atmospheric Extinction from Telescopes of the North
  Atlantic Canary Archipelago}.
\newblock {\em Journal of Climate} {\bf 2016}, {\em 29},~227--240.
\newblock {\url{https://doi.org/10.1175/JCLI-D-14-00600.1}}.

\bibitem[{Lombardi} \em{et~al.}(2010){Lombardi}, {Zitelli}, {Ortolani},
  {Ghedina}, {Garcia}, {Molinari}, and {Gatica}]{Lombardi:2010}
{Lombardi}, G.; {Zitelli}, V.; {Ortolani}, S.; {Ghedina}, A.; {Garcia}, A.;
  {Molinari}, E.; {Gatica}, C.
\newblock New dust measurements at ORM, and comparison with Paranal
  Observatory.
\newblock {\em Ground-based and Airborne Telescopes III} {\bf 2010}, {\em
  7733},~77334G.
\newblock {\url{https://doi.org/10.1117/12.856981}}.

\bibitem[{Sicard} \em{et~al.}(2010){Sicard} et~al.]{Sicard:2010}
{Sicard}, M.;  et~al.
\newblock Results of site testing using an aerosol, backscatter lidar at the
  Roque de los Muchachos Observatory.
\newblock {\em Monthly Notices of the Royal Astronomical Society} {\bf 2010},
  {\em 405},~129--142.
\newblock {\url{https://doi.org/DOI: 10.1111/j.1365-2966.2010.16468.x}}.

\bibitem[{Garrido} \em{et~al.}(2013){Garrido}, {Gaug}, {Doro}, {Font},
  {L\'opez-Oramas}, and {Moralejo}]{Garrido:2013}
{Garrido}, D.; {Gaug}, M.; {Doro}, M.; {Font}, L.; {L\'opez-Oramas}, A.;
  {Moralejo}, A.
\newblock {Influence of atmospheric aerosols on the performance of the MAGIC
  telescopes}.
\newblock In Proceedings of the {Proceedings of the 33$^\mathrm{rd}$ ICRC, Rio
  de Janeiro},  2013, p. 0465.
\newblock {arXiv:1308.0473}, {\url{https://doi.org/10.48550/arXiv.1308.0473}}.

\bibitem[{Sobczy{\'{n}}ska} and {Bednarek}(2014)]{Sobczynska:2014}
{Sobczy{\'{n}}ska}, D.; {Bednarek}, W.
\newblock {Influence of clouds on the parameters of images measured by {IACT}
  at very high energies}.
\newblock {\em Journal of Physics G: Nuclear and Particle Physics} {\bf 2014},
  {\em 41},~125201.
\newblock {\url{https://doi.org/10.1088/0954-3899/41/12/125201}}.

\bibitem[{Dominis Prester} \em{et~al.}(2024){Dominis Prester}, {Ebr}, {Gaug},
  {Hahn}, {Babi{\'c}}, {Eli{\'a}{\v s}ek}, {Jane{\v c}ek}, {Karpov}, {Kolarek},
  {Manganaro}, and {Mirzoyan}]{Prester:2024}
{Dominis Prester}, D.; {Ebr}, J.; {Gaug}, M.; {Hahn}, A.; {Babi{\'c}}, A.;
  {Eli{\'a}{\v s}ek}, J.; {Jane{\v c}ek}, P.; {Karpov}, S.; {Kolarek}, M.;
  {Manganaro}, M.;  et~al.
\newblock {Characterisation of the Atmosphere in Very High Energy
  Gamma-Astronomy for Imaging Atmospheric Cherenkov Telescopes}.
\newblock {\em Universe} {\bf 2024}, {\em 10}.
\newblock {\url{https://doi.org/10.3390/universe10090349}}.

\bibitem[{Keckhut} \em{et~al.}(1990){Keckhut}, {Chanin}, and
  {Hauchecorne}]{Keckhut:1990}
{Keckhut}, P.; {Chanin}, M.L.; {Hauchecorne}, A.
\newblock {Stratosphere temperature measurement using Raman lidar}.
\newblock {\em Appl. Opt.} {\bf 1990}, {\em 29},~5182--5186.
\newblock {\url{https://doi.org/10.1364/AO.29.005182}}.

\bibitem[{Rees} \em{et~al.}(2000){Rees}, {von Zahn}, {von Cossart}, {Fricke},
  {Eriksen}, and {McKay}]{Rees:2000}
{Rees}, D.; {von Zahn}, U.; {von Cossart}, G.; {Fricke}, K.; {Eriksen}, W.;
  {McKay}, J.
\newblock {Daytime lidar measurements of the stratosphere and mesosphere at the
  alomar observatory}.
\newblock {\em Advances in Space Research} {\bf 2000}, {\em 26},~893--902.
\newblock Advances in Remote Sensing of the Atmosphere from Space and from the
  Ground,
  {\url{https://doi.org/https://doi.org/10.1016/S0273-1177(00)00027-2}}.

\bibitem[Wandinger(2005)]{Wandinger:2005}
Wandinger, U.
\newblock {Raman Lidar}. In {\em {Lidar Range-Resolved Optical Remote Sensing
  of the Atmosphere}}; Weitkamp, C., Ed.; {Springer Science+Business Media,
  Inc.}: {233 Spring Street, New York, NY 10013, USA},  2005; chapter~9, pp.
  241--271.

\bibitem[{Avdikos}(2015)]{Avdikos:2015}
{Avdikos}, G.
\newblock {Powerful Raman Lidar systems for atmospheric analysis and
  high-energy physics experiments}.
\newblock In Proceedings of the European Physical Journal Web of Conferences,
  2015, Vol.~89, {\em European Physical Journal Web of Conferences}, p. 04003.
\newblock {\url{https://doi.org/10.1051/epjconf/20158904003}}.

\bibitem[{Gerding} \em{et~al.}(2016){Gerding}, {Kopp}, {H\"offner},
  {Baumgarten}, and {L\"ubken}]{Gerding:2016}
{Gerding}, M.; {Kopp}, M.; {H\"offner}, J.; {Baumgarten}, K.; {L\"ubken}, F.J.
\newblock {Mesospheric temperature soundings with the new, daylight-capable IAP
  RMR lidar}.
\newblock {\em Atmospheric Measurement Techniques} {\bf 2016}, {\em
  9},~3707--3715.
\newblock {\url{https://doi.org/10.5194/amt-9-3707-2016}}.

\bibitem[{Klanner} \em{et~al.}(2021){Klanner}, {H\"oveler}, {Khordakova},
  {Perfahl}, {Rolf}, {Trickl}, and {Vogelmann}]{Klanner:2021}
{Klanner}, L.; {H\"oveler}, K.; {Khordakova}, D.; {Perfahl}, M.; {Rolf}, C.;
  {Trickl}, T.; {Vogelmann}, H.
\newblock {A powerful lidar system capable of 1\,h measurements of water vapour
  in the troposphere and the lower stratosphere as well as the temperature in
  the upper stratosphere and mesosphere}.
\newblock {\em Atmospheric Measurement Techniques} {\bf 2021}, {\em
  14},~531--555.
\newblock {\url{https://doi.org/10.5194/amt-14-531-2021}}.

\bibitem[{Winker} \em{et~al.}(2003){Winker}, {Pelon}, and
  {McCormick}]{WinklerCALIPSO:2003}
{Winker}, D.M.; {Pelon}, J.R.; {McCormick}, M.P.
\newblock {CALIPSO mission: spaceborne lidar for observation of aerosols and
  clouds}.
\newblock In Proceedings of the Lidar Remote Sensing for Industry and
  Environment Monitoring III; Singh, U.N.; Itabe, T.; Liu, Z., Eds.
  International Society for Optics and Photonics, SPIE,  2003, Vol. 4893, pp. 1
  -- 11.
\newblock {\url{https://doi.org/10.1117/12.466539}}.

\bibitem[{M{\"u}ller} \em{et~al.}(2010){M{\"u}ller}, {Weinzierl}, {Petzold},
  {Kandler}, {Ansmann}, {M{\"u}ller}, {Tesche}, {Freudenthaler}, {Esselborn},
  {Heese}, {Althausen}, {Schladitz}, {Otto}, and {Knippertz}]{Mueller:2010}
{M{\"u}ller}, D.; {Weinzierl}, B.; {Petzold}, A.; {Kandler}, K.; {Ansmann}, A.;
  {M{\"u}ller}, T.; {Tesche}, M.; {Freudenthaler}, V.; {Esselborn}, M.;
  {Heese}, B.;  et~al.
\newblock {Mineral dust observed with AERONET Sun photometer, Raman lidar, and
  in situ instruments during SAMUM 2006: Shape-independent particle
  properties}.
\newblock {\em Journal of Geophysical Research: Atmospheres} {\bf 2010}, {\em
  115},
  \href{http://xxx.lanl.gov/abs/https://agupubs.onlinelibrary.wiley.com/doi/pdf/10.1029/2009JD012520}{{\normalfont
  [https://agupubs.onlinelibrary.wiley.com/doi/pdf/10.1029/2009JD012520]}}.
\newblock {\url{https://doi.org/https://doi.org/10.1029/2009JD012520}}.

\bibitem[{{The CTA Consortium}}(2010)]{cta}
{{The CTA Consortium}}.
\newblock Design Concepts for the Cherenkov Telescope Array,  2010.

\bibitem[{Pappalardo} \em{et~al.}(2004){Pappalardo}, {Amodeo}, {Pandolfi},
  {Wandinger}, {Ansmann}, {B\"{o}senberg}, {Matthias}, {Amiridis}, {De Tomasi},
  {Frioud}, {Iarlori}, {Komguem}, {Papayannis}, {Rocadenbosch}, and
  {Wang}]{Pappalardo:2004}
{Pappalardo}, G.; {Amodeo}, A.; {Pandolfi}, M.; {Wandinger}, U.; {Ansmann}, A.;
  {B\"{o}senberg}, J.; {Matthias}, V.; {Amiridis}, V.; {De Tomasi}, F.;
  {Frioud}, M.;  et~al.
\newblock {Aerosol lidar intercomparison in the framework of the EARLINET
  project. 3. Raman lidar algorithm for aerosol extinction, backscatter, and
  lidar ratio}.
\newblock {\em Appl. Opt.} {\bf 2004}, {\em 43},~5370--5385.
\newblock {\url{https://doi.org/10.1364/AO.43.005370}}.

\bibitem[{Zenteno-Hern{\'a}ndez} \em{et~al.}(2021){Zenteno-Hern{\'a}ndez},
  {Comer{\'o}n}, {Rodr{\'\i}guez-G{\'o}mez}, {Mu{\~n}oz-Porcar}, {D'Amico}, and
  {Sicard}]{Zenteno-Hernandez:2021}
{Zenteno-Hern{\'a}ndez}, J.A.; {Comer{\'o}n}, A.; {Rodr{\'\i}guez-G{\'o}mez},
  A.; {Mu{\~n}oz-Porcar}, C.; {D'Amico}, G.; {Sicard}, M.
\newblock {A Comparative Analysis of Aerosol Optical Coefficients and Their
  Associated Errors Retrieved from Pure-Rotational and Vibro-Rotational Raman
  Lidar Signals}.
\newblock {\em Sensors} {\bf 2021}, {\em 21}.
\newblock {\url{https://doi.org/10.3390/s21041277}}.

\bibitem[{She}(2001)]{She:2001}
{She}, C.Y.
\newblock {Spectral structure of laser light scattering revisited: bandwidths
  of nonresonant scattering lidars}.
\newblock {\em Appl. Opt.} {\bf 2001}, {\em 40},~4875--4884.
\newblock {\url{https://doi.org/10.1364/AO.40.004875}}.

\bibitem[{Gao}(2012)]{Gau:phd}
{Gao}, F.
\newblock {Study of Processes in Atmospheric Boundary Layer over Land-Sea
  Transition Interface Using Scanning Lidar}.
\newblock PhD thesis, University of Nova Gorica,  2012.
\newblock Available at
  \url{https://repozitorij.ung.si/IzpisGradiva.php?id=1289}.

\bibitem[Alexandreas \em{et~al.}(1995)Alexandreas et~al.]{Alexandreas:1995}
Alexandreas, D.;  et~al.
\newblock Status report of CLUE.
\newblock {\em Nuclear Instruments and Methods in Physics Research A} {\bf
  1995}, {\em 360},~385--389.

\bibitem[Bartoli \em{et~al.}(2001)Bartoli et~al.]{Bartoli:2001gt}
Bartoli, B.;  et~al.
\newblock {Observation of the moon shadow using a new reconstruction technique
  in the CLUE experiment}.
\newblock {\em Nuovo Cim. C} {\bf 2001}, {\em 24},~669--674.

\bibitem[{Vasileiadis} \em{et~al.}(2020){Vasileiadis}, {Brun}, {Gabella},
  {Rivoire}, {Avdikos}, {Louridas}, {Georgoussis}, and
  {Papayannis}]{Vasileiadis:2020}
{Vasileiadis}, G.; {Brun}, P.; {Gabella}, O.; {Rivoire}, S.; {Avdikos}, G.;
  {Louridas}, A.; {Georgoussis}, G.; {Papayannis}, A.
\newblock {Design and Development of a Raman Lidar for Cherenkov Gamma Array
  Experiments}.
\newblock In Proceedings of the European Physical Journal Web of Conferences,
  2020, Vol. 237, {\em European Physical Journal Web of Conferences}, p. 07006.
\newblock {\url{https://doi.org/10.1051/epjconf/202023707006}}.

\bibitem[{Doro} \em{et~al.}(){Doro} et~al.]{Doro:2012}
{Doro}, M.;  et~al.
\newblock A Raman LIDAR for atmospheric calibration of the Cherenkov Telescope
  Array.
\newblock In preparation.

\bibitem[{Barreto} \em{et~al.}(2022){Barreto}, {Cuevas}, {Garc\'{\i}a},
  {Carrillo}, {Prospero}, {Ili\'c}, {Basart}, {Berj\'on}, {Marrero},
  {Hern\'andez}, {Bustos}, {Ni\v{c}kovi\'c}, and {Yela}]{Barreto:2022}
{Barreto}, A.; {Cuevas}, E.; {Garc\'{\i}a}, R.D.; {Carrillo}, J.; {Prospero},
  J.M.; {Ili\'c}, L.; {Basart}, S.; {Berj\'on}, A.J.; {Marrero}, C.L.;
  {Hern\'andez}, Y.;  et~al.
\newblock {Long-term characterisation of the vertical structure of the Saharan
  Air Layer over the Canary Islands using lidar and radiosonde profiles:
  implications for radiative and cloud processes over the subtropical Atlantic
  Ocean}.
\newblock {\em Atmospheric Chemistry and Physics} {\bf 2022}, {\em
  22},~739--763.
\newblock {\url{https://doi.org/10.5194/acp-22-739-2022}}.

\bibitem[{Spinhirne} \em{et~al.}(1995){Spinhirne}, {Rall}, and
  {Scott}]{spinhirne1995}
{Spinhirne}, J.D.; {Rall}, J.A.R.; {Scott}, V.S.
\newblock {Compact Eye Safe Lidar Systems}.
\newblock {\em Laser Review} {\bf 1995}, {\em 23},~112--118.

\bibitem[{Eisele} and {Trickl}(1997)]{Eisele:1997}
{Eisele}, H.; {Trickl}, T.
\newblock {Second Generation of the IFU Stationary Tropospheric Ozone Lidar}.
\newblock In Proceedings of the Advances in Atmospheric Remote Sensing with
  Lidar; Ansmann, A.; Neuber, R.; Rairoux, P.; Wandinger, U., Eds.; Springer
  Berlin Heidelberg: Berlin, Heidelberg,  1997; pp. 379--382.
\newblock {\url{https://doi.org/10.1007/978-3-642-60612-0_91}}.

\bibitem[{M\"uller} \em{et~al.}(2014){M\"uller}, {Hostetler}, {Ferrare},
  {Burton}, {Chemyakin}, {Kolgotin}, {Hair}, {Cook}, {Harper}, {Rogers},
  {Hare}, {Cleckner}, {Obland}, {Tomlinson}, {Berg}, and
  {Schmid}]{Mueller:2014}
{M\"uller}, D.; {Hostetler}, C.A.; {Ferrare}, R.A.; {Burton}, S.P.;
  {Chemyakin}, E.; {Kolgotin}, A.; {Hair}, J.W.; {Cook}, A.L.; {Harper}, D.B.;
  {Rogers}, R.R.;  et~al.
\newblock {Airborne Multiwavelength High Spectral Resolution Lidar (HSRL-2)
  observations during TCAP 2012: vertical profiles of optical and microphysical
  properties of a smoke/urban haze plume over the northeastern coast of the
  US}.
\newblock {\em Atmospheric Measurement Techniques} {\bf 2014}, {\em
  7},~3487--3496.
\newblock {\url{https://doi.org/10.5194/amt-7-3487-2014}}.

\bibitem[Peruzzo \em{et~al.}(1991)Peruzzo et~al.]{Peruzzo:1990ia}
Peruzzo, L.;  et~al.
\newblock {CLUE: Cherenkov light ultraviolet experiment: Preliminary results
  and future plans}.
\newblock {\em Italian Phys. Soc. Proc.} {\bf 1991}, {\em 28},~423--430.

\bibitem[Baillon \em{et~al.}(1989{\natexlab{a}})Baillon et~al.]{Baillon:1989a}
Baillon, P.;  et~al.
\newblock An improved method for manufacturing accurate and cheap glass
  parabolic mirrors.
\newblock {\em Nuclear Instruments and Methods in Physics Research A} {\bf
  1989}, {\em 276},~{492--495}.

\bibitem[Baillon \em{et~al.}(1989{\natexlab{b}})Baillon et~al.]{Baillon:1989b}
Baillon, P.;  et~al.
\newblock Production of 300 paraboloidal mirrors with high reflectivity for use
  in the Barrel RICH counter in DELPHI at LEP.
\newblock {\em Nuclear Instruments and Methods in Physics Research A} {\bf
  1989}, {\em 277}.

\bibitem[{S{\'a}nchez~Alonso}(2011)]{sanchez:tesina}
{S{\'a}nchez~Alonso}, C.
\newblock {Caracteritzaci\'o del LIDAR de CTA}.
\newblock Bachelor's thesis, Universitat Autonoma de Barcelona,  2011.
\newblock Available at
  \url{https://ctan-lidar-pathfinder.ung.si/public/MasterTheses/Sanchez_MSc_2011.pdf}.

\bibitem[{L\'opez-Oramas}(2014)]{lopezphd}
{L\'opez-Oramas}, A.
\newblock {Multi-year Campaign of the Gamma-Ray Binary LS I +61$^\circ$ 303 and
  Search for VHE Emission from Gamma-Ray Binary Candidates with the MAGIC
  Telescopes}.
\newblock PhD thesis, Universitat Aut{\`o}noma de Barcelona,  2014.
\newblock Available at \url{ddd.uab.cat/record/127519/}.

\bibitem[{Pesci}(1995)]{Pesci:1995}
{Pesci}, A.
\newblock PhD thesis, {University of Padova},  1995.

\bibitem[{L{\'o}pez-Oramas}(2010)]{lopez:master}
{L{\'o}pez-Oramas}, A.
\newblock {Development and description of a Raman LIDAR}.
\newblock Master's thesis, {Universitat Aut{\`o}noma de Barcelona},  2010.
\newblock Available at
  \url{https://ctan-lidar-pathfinder.ung.si/public/MasterTheses/Alicia_Lopez-Oramas_MSc_2010.pdf}.

\bibitem[{Riu~Molinero}(2020)]{victortfg}
{Riu~Molinero}, V.
\newblock {Posada en marxa del UAB-IFAE Raman LIDAR}.
\newblock Bachelor's thesis, Universitat Aut{\`o}noma de Barcelona,  2020.
\newblock Available at
  \url{https://ctan-lidar-pathfinder.ung.si/public/BachelorTheses/Riu_Molinero_2020t.pdf}.

\bibitem[{Biavati} \em{et~al.}(2011){Biavati}, {Di~Donfrancesco}, {Cairo}, and
  {Feist}]{Biavati:2011}
{Biavati}, G.; {Di~Donfrancesco}, G.; {Cairo}, F.; {Feist}, D.G.
\newblock {Correction scheme for close-range lidar returns}.
\newblock {\em Applied Optics} {\bf 2011}, {\em 50},~5872--82.
\newblock {\url{https://doi.org/10.1364/AO.50.005872}}.

\bibitem[{Wandinger} and {Ansmann}(2002)]{wandinger:2002}
{Wandinger}, U.; {Ansmann}, A.
\newblock Experimental determination of the lidar overlap profile with Raman
  lidar.
\newblock {\em Applied Optics} {\bf 2002}, pp. 511--514.

\bibitem[{Wiencke} \em{et~al.}(2017){Wiencke}, {Rizi}, {Will}, {Allen},
  {Botts}, {Calhoun}, {Carande}, {Claus}, {Coco}, {Emmert}, {Esquibel},
  {Grillo}, {Hamilton}, {Heid}, {Iarlori}, {Klages}, {Kleifges}, {Knoll},
  {Koop}, {Mathes}, {Menshikov}, {Morgan}, {Patterson}, {Petrera}, {Robinson},
  {Runyan}, {Sherman}, {Starbuck}, {Wakin}, and {Wolf}]{Wiencke:2017}
{Wiencke}, L.; {Rizi}, V.; {Will}, M.; {Allen}, C.; {Botts}, A.; {Calhoun}, M.;
  {Carande}, B.; {Claus}, J.; {Coco}, M.; {Emmert}, L.;  et~al.
\newblock {Joint elastic side-scattering LIDAR and Raman LIDAR measurements of
  aerosol optical properties in south east Colorado}.
\newblock {\em Journal of Instrumentation} {\bf 2017}, {\em 12},~P03008.
\newblock {\url{https://doi.org/10.1088/1748-0221/12/03/P03008}}.

\bibitem[{Da~Deppo} \em{et~al.}(2012){Da~Deppo}, {Doro}, {Blanch}, {Font},
  {Lopez}, {Gaug}, and {Martinez}]{Deppo:2012}
{Da~Deppo}, V.; {Doro}, M.; {Blanch}, O.; {Font}, L.; {Lopez}, A.; {Gaug}, M.;
  {Martinez}, M.
\newblock {Preliminary optical design of a polychromator for a Raman LIDAR for
  atmospheric calibration of the Cherenkov Telescope Array}.
\newblock In Proceedings of the Proc. SPIE Optical Systems Design,  2012, Vol.
  8550, p. 85501V.
\newblock {\url{https://doi.org/10.1117/12.979789}}.

\bibitem[{Toyama} \em{et~al.}(2015){Toyama}, {Hanabata}, {Hose}, {Menzel},
  {Mirzoyan}, {Nakajima}, {Takahashi}, {Teshima}, and {Yamamoto}]{toyama}
{Toyama}, T.; {Hanabata}, Y.; {Hose}, J.; {Menzel}, U.; {Mirzoyan}, R.;
  {Nakajima}, D.; {Takahashi}, M.; {Teshima}, M.; {Yamamoto}, T.
\newblock {Evaluation of the basic properties of the novel 1.5in. size PMTs
  from Hamamatsu Photonics and Electron Tubes Enterprises}.
\newblock {\em Nuclear Instruments and Methods in Physics Research Section A:
  Accelerators, Spectrometers, Detectors and Associated Equipment} {\bf 2015},
  {\em 787},~280 -- 283.

\bibitem[{Orito} \em{et~al.}(2011){Orito}, {Ohoka}, {Aoki}, and
  {Awane}]{Orito:2011}
{Orito}, R.; {Ohoka}, H.; {Aoki}, M.; {Awane}, Y.
\newblock {Development of PMT clusters for CTA-LST camera}.
\newblock In Proceedings of the proceeding of ICRC2011, Beijing,  2011, Vol.~9,
  pp. 171--174.
\newblock {\url{https://doi.org/10.7529/ICRC2011/V09/1091}}.

\bibitem[{Maggio}(2021)]{phdthesis-maggio}
{Maggio}, C.
\newblock Indirect search for WIMPS Dark Matter with the MAGIC telescopes.
\newblock PhD thesis, Universitat Autonoma de Barcelona, Spain,  2021.
\newblock Available at \url{https://www.tdx.cat/handle/10803/671998}.

\bibitem[{Barcel\'o} \em{et~al.}(2011){Barcel\'o}, {Blanch}, {Boix},
  {Bourgeat}, {Compin}, {Doro}, {Eizmendi}, {Font}, {Garrido}, {Glass},
  {Gra{\~n}ena}, {Lopez-Oramas}, {Mart\'inez}, {Moralejo}, {Rivoire}, {Royer},
  {S\'anchez}, {Valvin}, and {Vasileiadis}]{Barcelo:2011stu}
{Barcel\'o}, M.; {Blanch}, O.; {Boix}, J.; {Bourgeat}, M.; {Compin}, M.;
  {Doro}, M.; {Eizmendi}, M.; {Font}, L.; {Garrido}, D.; {Glass}, D.;  et~al.
\newblock {Development of Raman Lidars made with former CLUE telescopes for
  CTA}.
\newblock In Proceedings of the {32nd International Cosmic Ray Conference},
  2011, Vol.~9, p.~22.
\newblock {\url{https://doi.org/10.7529/ICRC2011/V09/0408}}.

\bibitem[{Font}(2014)]{Eudald:thesis}
{Font}, E.
\newblock {Alineament de l’UAB-IFAE Raman LIDAR}.
\newblock Bachelor's thesis, Universitat Aut\`onoma de Barcelona,  2014.
\newblock Available at
  \url{https://ctan-lidar-pathfinder.ung.si/public/BachelorTheses/Font_2014n.pdf}.

\bibitem[{Calpe}(2017)]{calpe:thesis}
{Calpe}, O.
\newblock {CTA, lidar i near range}.
\newblock Bachelor's thesis, Universitat Autonoma de Barcelona,  2017.
\newblock Available at
  \url{https://ctan-lidar-pathfinder.ung.si/public/BachelorTheses/Calpe_Blanch_2017a.pdf}.

\bibitem[{Mirzoyan} \em{et~al.}(2017){Mirzoyan}, {M{\"u}ller}, {Hose},
  {Menzel}, {Nakajima}, {Takahashi}, {Teshima}, {Toyama}, and
  {Yamamoto}]{Mirzoyan:2017}
{Mirzoyan}, R.; {M{\"u}ller}, D.; {Hose}, J.; {Menzel}, U.; {Nakajima}, D.;
  {Takahashi}, M.; {Teshima}, M.; {Toyama}, T.; {Yamamoto}, T.
\newblock {Evaluation of novel PMTs of worldwide best parameters for the CTA
  project}.
\newblock {\em Nuclear Instruments and Methods in Physics Research A} {\bf
  2017}, {\em 845},~603--606.
\newblock {\url{https://doi.org/10.1016/j.nima.2016.06.080}}.

\bibitem[{Mazin} \em{et~al.}(2021){Mazin}, {Abe}, {Aguasca}, {Agudo},
  {Antonelli}, {Aramo}, {Armstrong}, Artero, Asano, Ashkar, Aubert, Baktash,
  Bamba, Baquero~Larriva, Baroncelli, Barres~de Almeida, Barrio, Batkovi{\'c},
  Becerra~Gonzalez, Bernardos, Berti, Biederbeck, Bigongiari, Blanch, Bonnoli,
  Bordas, Bose, Bulgarelli, Burelli, Buscemi, Cardillo, Caroff, Carosi, Cassol,
  Cerruti, Chai, Cheng, Chikawa, Chytka, Contreras, Cortina, Costantini,
  Dalchenko, De~Angelis, de~Bony~de Lavergne, Deleglise, Delgado,
  Delgado~Mengual, Della~Volpe, Depaoli, Di~Pierro, Di~Venere, D{\'\i}az,
  Dominik, Dominis~Prester, Donini, Dorner, Doro, Els{\"a}sser, Emery,
  Escudero, Fiasson, Foffano, Fonseca, Freixas~Coromina, Fukami, Fukazawa,
  Garcia, Garcia~L{\'o}pez, Giglietto, Giordano, Gliwny, Godinovic, Green,
  Grespan, Gunji, Hackfeld, Hadasch, Hahn, Hassan, Hayashi, Heckmann, Heller,
  Herrera~Llorente, Hirotani, Hoffmann, Horns, Houles, Hrabovsky, Hrupec, Hui,
  H{\"u}tten, Inada, Inome, Iori, Ishio, Iwamura, Jacquemont,
  Jim{\'e}nez~Mart{\'\i}nez, Jouvin, Jurysek, Kagaya, Karas, Katagiri, Kataoka,
  Kerszberg, Kobayashi, Kong, Kubo, Kushida, Lamanna, Lamastra, Le~Flour,
  Longo, Lopez-Coto, L{\'o}pez-Moya, L{\'o}pez-Oramas, Luque-Escamilla,
  Majumdar, MAKARIEV, Mandat, Manganaro, Mannheim, Mariotti, Marquez, Marsella,
  Mart{\'\i}, Martinez, Mart{\'\i}nez, Martinez, Marusevec, Mas, Maurin, Mazin,
  Mestre~Guillen, Mi{\'c}anovi{\'c}, Miceli, Miener, Miranda, Miranda,
  Mirzoyan, Mizuno, Molina, Montaruli, Monteiro, Moralejo, Morcuende, Moretti,
  Morselli, Mrakovcic, Murase, Nagai, Nakamori, Nickel, Nieto, Nievas,
  Nishijima, Noda, Nosek, N{\"o}the, Nozaki, Ohishi, Ohtani, Oka, Okazaki,
  Okumura, Orito, Otero-Santos, Palatiello, Paneque, Paoletti, Paredes,
  Pavleti{\'c}, Pech, Pecimotika, Poireau, Polo, Prandini, Prast, Priyadarshi,
  Prouza, Rando, Rhode, Rib{\'o}, Rizi, Rugliancich, Ruiz, Saito, Sakurai,
  Sanchez, {\v S}ari{\'c}, Saturni, Scherpenberg, Schleicher, Schubert,
  Sch{\"u}ssler, Schweizer, Seglar~Arroyo, Shellard, Sitarek, Sliusar, Spolon,
  Stri{\v s}kovi{\'c}, Strzys, Suda, Sunada, Tajima, Takahashi, Takahashi,
  Takata, Takeishi, Tam, Tanaka, Tateishi, Tejedor, Temnikov, Terada, Terzic,
  Teshima, Tluczykont, Tokanai, Torres, Travnicek, Truzzi, Vacula,
  VAZQUEZ~ACOSTA, VERGUILOV, Verna, Viale, Vigorito, Vitale, Vovk, Vuillaume,
  Walter, Will, Yamamoto, Yamazaki, Yoshida, Yoshikoshi, and
  Zari{\'c}]{Mazin:2021}
{Mazin}, D.; {Abe}, H.; {Aguasca}, A.; {Agudo}, I.; {Antonelli}, L.A.; {Aramo},
  C.; {Armstrong}, T.; Artero, M.; Asano, K.; Ashkar, H.;  et~al.
\newblock {Status and results of the prototype LST of CTA}.
\newblock {\em PoS} {\bf 2021}, {\em ICRC2021},~872.
\newblock {\url{https://doi.org/10.22323/1.395.0872}}.

\bibitem[{Aleksi{\'c}} \em{et~al.}(2016){Aleksi{\'c}}, {Ansoldi}, {Antonelli},
  {Antoranz}, {Babic}, {Bangale}, {Barcel{\'o}}, {Barrio}, {Becerra
  Gonz{\'a}lez}, {Bednarek}, {Bernardini}, {Biasuzzi}, {Biland}, {Bitossi},
  {Blanch}, {Bonnefoy}, {Bonnoli}, {Borracci}, {Bretz}, {Carmona}, {Carosi},
  {Cecchi}, {Colin}, {Colombo}, {Contreras}, {Corti}, {Cortina}, {Covino}, {Da
  Vela}, {Dazzi}, {DeAngelis}, {De Caneva}, {De Lotto}, {de O{\~n}a Wilhelmi},
  {Delgado Mendez}, {Dettlaff}, {Dominis Prester}, {Dorner}, {Doro}, {Einecke},
  {Eisenacher}, {Elsaesser}, {Fidalgo}, {Fink}, {Fonseca}, {Font}, {Frantzen},
  {Fruck}, {Galindo}, {Garc{\'{\i}}a L{\'o}pez}, {Garczarczyk}, {Garrido
  Terrats}, {Gaug}, {Giavitto}, {Godinovi{\'c}}, {Gonz{\'a}lez Mu{\~n}oz},
  {Gozzini}, {Haberer}, {Hadasch}, {Hanabata}, {Hayashida}, {Herrera},
  {Hildebrand}, {Hose}, {Hrupec}, {Idec}, {Illa}, {Kadenius}, {Kellermann},
  {Knoetig}, {Kodani}, {Konno}, {Krause}, {Kubo}, {Kushida}, {La Barbera},
  {Lelas}, {Lemus}, {Lewandowska}, {Lindfors}, {Lombardi}, {Longo},
  {L{\'o}pez}, {L{\'o}pez-Coto}, {L{\'o}pez-Oramas}, {Lorca}, {Lorenz},
  {Lozano}, {Makariev}, {Mallot}, {Maneva}, {Mankuzhiyil}, {Mannheim},
  {Maraschi}, {Marcote}, {Mariotti}, {Mart{\'{\i}}nez}, {Mazin}, {Menzel},
  {Miranda}, {Mirzoyan}, {Moralejo}, {Munar-Adrover}, {Nakajima}, {Negrello},
  {Neustroev}, {Niedzwiecki}, {Nilsson}, {Nishijima}, {Noda}, {Orito},
  {Overkemping}, {Paiano}, {Palatiello}, {Paneque}, {Paoletti}, {Paredes},
  {Paredes-Fortuny}, {Persic}, {Poutanen}, {Prada Moroni}, {Prandini},
  {Puljak}, {Reinthal}, {Rhode}, {Rib{\'o}}, {Rico}, {Rodriguez Garcia},
  {R{\"u}gamer}, {Saito}, {Saito}, {Satalecka}, {Scalzotto}, {Scapin},
  {Schultz}, {Schlammer}, {Schmidl}, {Schweizer}, {Sillanp{\"a}{\"a}},
  {Sitarek}, {Snidaric}, {Sobczynska}, {Spanier}, {Stamerra}, {Steinbring},
  {Storz}, {Strzys}, {Takalo}, {Takami}, {Tavecchio}, {Tejedor}, {Temnikov},
  {Terzi{\'c}}, {Tescaro}, {Teshima}, {Thaele}, {Tibolla}, {Torres}, {Toyama},
  {Treves}, {Vogler}, {Wetteskind}, {Will}, and {Zanin}]{magicperformance1}
{Aleksi{\'c}}, J.; {Ansoldi}, S.; {Antonelli}, L.A.; {Antoranz}, P.; {Babic},
  A.; {Bangale}, P.; {Barcel{\'o}}, M.; {Barrio}, J.A.; {Becerra Gonz{\'a}lez},
  J.; {Bednarek}, W.;  et~al.
\newblock {The major upgrade of the MAGIC telescopes, Part I: The hardware
  improvements and the commissioning of the system}.
\newblock {\em Astroparticle Physics} {\bf 2016}, {\em 72},~61--75,
  \href{http://xxx.lanl.gov/abs/1409.6073}{{\normalfont [1409.6073]}}.
\newblock {\url{https://doi.org/10.1016/j.astropartphys.2015.04.004}}.

\bibitem[{Grau} \em{et~al.}(){Grau}, {Zivec}, et~al.]{performancepaper}
{Grau}, R.; {Zivec}, M.;  et~al.
\newblock A 1.8~m class pathfinder Raman lidar for the Northern Site of the
  Cherenkov Telescope Array -- Performance.
\newblock In preparation.

\end{thebibliography}

% If authors have biography, please use the format below
%\section*{Short Biography of Authors}
%\bio
%{\raisebox{-0.35cm}{\includegraphics[width=3.5cm,height=5.3cm,clip,keepaspectratio]{Definitions/author1.pdf}}}
%{\textbf{Firstname Lastname} Biography of first author}
%
%\bio
%{\raisebox{-0.35cm}{\includegraphics[width=3.5cm,height=5.3cm,clip,keepaspectratio]{Definitions/author2.jpg}}}
%{\textbf{Firstname Lastname} Biography of second author}

% For the MDPI journals use author-date citation, please follow the formatting guidelines on http://www.mdpi.com/authors/references
% To cite two works by the same author: \citeauthor{ref-journal-1a} (\citeyear{ref-journal-1a}, \citeyear{ref-journal-1b}). This produces: Whittaker (1967, 1975)
% To cite two works by the same author with specific pages: \citeauthor{ref-journal-3a} (\citeyear{ref-journal-3a}, p. 328; \citeyear{ref-journal-3b}, p.475). This produces: Wong (1999, p. 328; 2000, p. 475)

%%%%%%%%%%%%%%%%%%%%%%%%%%%%%%%%%%%%%%%%%%
%% for journal Sci
%\reviewreports{\\
%Reviewer 1 comments and authors’ response\\
%Reviewer 2 comments and authors’ response\\
%Reviewer 3 comments and authors’ response
%}
%%%%%%%%%%%%%%%%%%%%%%%%%%%%%%%%%%%%%%%%%%
\end{adjustwidth}
\end{document}